\documentclass[12pt,a4paper,twoside]{book}
\usepackage{Preamble}
\usepackage{thcover}

\author{Stefano Pierini}

\newcommand{\frenchtitle}{\\\vspace{2em} \'Etude exp\'erimentale des
	nanocristaux de p\'erovskite comme sources de photons uniques pour la
	photonique quantique int\'egr\'ee}
\title{Experimental study of perovskite nanocrystals as single photon sources
	for integrated quantum photonics \frenchtitle}
\date{}
\begin{document}
%uncomment the following to create a front page
\frontcover

\maketitle
\frontmatter
\dominitoc
\tableofcontents
\chapter*{Acknowledgments}
Research is a collective process of discovery and transmitting knowledge. With no-doubts, my PhD project was not possible without the help of all the people I worked with.
I would like to thank them all, no matter their role, for their contribution in this project. All of them were important. I would like to mention some of those whose contributions have been most relevant.

First of all, my PhD directors Christophe Couteau and Alberto Bramati, who found the funding to make this project possible, believed in my abilities and guided me in this path. Alberto, moreover, has followed me scientifically and humanly, in every step I have taken since I decided to continue my studies outside Italy, and I want to thank him for that. I would also like to acknowledge the other permanents of the group, Quentin and Elisabeth: they didn't hesitate to give me help and suggestions any time I needed.

In understanding the experiment, the help of the previous Ph.D. student of the group, Maxime, was essential. He spent all the needed time in training me, from the most technical skills like fiber pulling to the most simple ones, like were to find stuff in the lab.

I also want to thank my colleagues, Giuseppe, Chendjie and Marianna, working with them was a privilege and a pleasure. We shared hours in the lab and we discussed on how to solve problems.

A good working environment is the key to stay productive and motivated: from this point of view I was lucky. For this reason I would like to thank Jean-Philippe, Aurthur, Aurelie, Giovanni, Anne, Quentin, Murad, Ferdinand, Tom, Rajiv, Giacomo.

My studies were allowed by a permitted by different collaboration, among them the most fruitful was the one with Emmanuel Lhuillier, at the INSP. I thank him for all the time he spent in the developing perovskites nanocrystlas object of my research and replying to all my questions on them.

Finally I want to thank Ermelinda, with whom I not only spent all my free time, but also shared every detail of this work from the beginning and who helped me to overcome all the difficulties encountered, scientific and human.

%\linespread{1.2} %change interline spacing
%\renewcommand{\theequation}{\thechapter.\arabic{equation}}

\mainmatter
\chapter{Introduction}
%The necessity for new quantum emitters
\minitoc
\section{Preface}
%The necessity for new quantum emitters
%\minitoc
\subsection{Motivation for this study}

Perovskites were originally studied for solar-cell
applications~\cite{park2015perovskite}, but
the
interest in these material is recently increased in the quantum optics
community. The versatility of perovskite nanocrystal, the fact that they have
been shown to emit single photon at
low~\cite{rainoSingleCesiumLead2016,huo2020Optical} and
room~\cite{park2015Room}
temperature, the possibility
to tune their emission playing on their composition, makes them interesting
emitters for quantum optics applications.

As other kind of colloidal quantum dots, these emitters can be fabricated
without heavy facilities, using low-cost wet-chemistry techniques, and this in
contrast to other emitters, such single defects in nanodiamonds and epitaxial
quantum dots, that require heavy fabrication facilities.

Despite several kind of colloidal quantum dots present similar features and are
nowadays well known, the appropriate choice of composition makes perovskite
nanocrystals able to emit in the near infrared range, fact that is difficult to
obtain with other kind of colloidal nanocrystals.

Perovskites nanocrystals currently produced have a main limitation: the optical
stability. Previous attempts to improve the photostabilty of these emitters involved
polymer encapsulation~\cite{chenInfluencePMMAAllInorganic2019,raino2019Underestimated},
alumina encapsulation using atomic layer deposition  and surface
passivation~\cite{panAirStableSurfacePassivatedPerovskite2015}.
In this work, in the second chapter, I describe a fabrication method to
produce perovskites nanocrystals with a better optical stability than
previously
achieved in literature, underling at the same time the role of the dilution on
the stability of the sample.

The improved stability allows to use these emitters for further studies. In
particular, guided structure as tapered nanofibers allow to collect the
emitted photons  in a efficient way using near field coupling and opening the path to
obtain a compact, integrated guided source of single photons. While the coupling with
other kind of emitters, such
atoms~\cite{le2005spontaneous,klimov2004spontaneous} and solid state
emitters~\cite{fujiwara2011highly,yalla2012Fluorescence,%
	joos2018polarization,vorobyovCouplingSingleNV2016},
 had already been
shown in literature, this is the first time this coupling has been achieved
using perovskite nanocrystals.

\subsection{Structure of the manuscript}
The manuscript is divided in four parts.

In the first chapter I will introduce basic concepts useful to understand the experimental work presented in the following. This involves
a short introduction on quantum mechanics and quantum optics, useful to define single photon emitters and their interest. %This is followed by a
%description of what single photon emitters are from a theoretical and
%experimental point of view: 
In particular, I explain why experimentally we
consider to be dealing with a single photon emitter any time that the photon
statistic shows that the second order autocorrelation function respects the
following relationship $\gdz<0.5$. I also briefly introduce various single photon emitters,
specifically solid state single photon emitters, detailing different analysis
that can be performed on these emitters in order to characterize them. I
conclude the chapter with an introduction to guided and integrated optics,
presenting its advantages respect to the free-space optics.

In the second chapter I describe the experimental details of my project. In
particular. I introduce perovskite nanocrystals as single photon emitters, and present
the setup I used to study their properties. I detail two different fabrication
methods for these nanocrystals, and present a systematic study of their emission
properties and stability.

In the third chapter I present the nanofiber platform, which will be coupled with a
single nanocrystal. After an introduction on fibers and nanofibers theory, I
describe the technique and the setup used to fabricate the nanofibers. Finally I
explain the technique used to depose a single perovskite nanocrystal on a
tapered nanofiber and show a measurement proving that the nanocrystal emits
single photons directly inside the nanofiber.

In the fourth chapter I show some outlooks of these system. Firstly I
discuss the possible improvements of the emitters that can be done by better
controlling the perovskite nanocrystals fabrication. Also the use of different
emitters, such single defects inside nanodiamonds, is envisioned.

Finally the manuscript is concluded with a french resume and
an appendix describing the main code used for the data analysis.

In this chapter I  introduce the theory that underlies our research:
I will review the fundamental concepts on which our experiments are based on and their interpretation.
\section{Quantum mechanics}
The theory of Quantum Mechanics was developed at the beginning of the XXth century in
order to explain different phenomena impossible to explain with a classical
approach. This brief introduction in which some basic concepts of quantum
mechanics are recalled was inspired by  two classical textbooks: the book
written by
\textcite{griffiths2018introduction} and the one written by
\textcite{sakurai2011Modern}.

In order to understand the differences between the classical and the quantum
theories, it can be interesting to consider them from an axiomatic point of view.
This is not the usual way to proceed, as most authors prefers to present the
physics from a historical point of view, but it gives a rapid insight on the
differences of the two theories.
It is possible to define the classical theory as a theory in which the following axioms are
valid:
\begin{enumerate}
	\item at any time, the state of a particle can be specified by two
	variables $x(t)$ and $p(t)$ that forms a point in the phase-space
	\item Any variable can be expressed as a function of $x$ and $p$
	\item The ideal measurement of a variable $\omega$  will result in
	$\omega\of{x,p}$ if the particle is in the state ($x$,$p$), and the
	measurement does not perturb the state.
	\item State variables follow the Hamilton equations:
	\begin{equation}\dot{x}=\frac{\partial H}{\partial p},\qquad
	\dot{p}=-\frac{\partial H}{\partial x}\end{equation}

\end{enumerate}

In quantum mechanics partially different assumptions are made:
\begin{enumerate}
	\item at any time, the state of a particle can be specified as a vector
	$\ket{\psi\of{t}}$
	\item The classical variables of quantum mechanics are replaced by the
	hermitian operators $X$ and $P$, with the following matrix elements:
	 \begin{subequations}
	 	\begin{gather}
	 		\braket{x}{X}{x'} = x \delta \of{x-x'}\\
	 		\braket{x}{P}{x'} = -i \hbar \dfrac{\dif}{\dif x} \delta
	 		\of{x-x'}
	 	\end{gather}
	 \end{subequations}
 	To obtain the operator $\Omega$ corresponding to a generic variable
 	$\omega$, we use the classical definition with the substitution
 	$$ \Omega\of{X,P}= \omega\of{x\rightarrow X, p\rightarrow P} $$
	\item The ideal measurement of a variable corresponding to $\omega$  will
	results in 	one of the eigenvalue $\omega$ of $\Omega\of{x,p}$ with a
	probability $P\of{\omega} \propto \abs{\left<\omega\middle|\psi\right>}^2$.
	After the measurement, the system is projected on the state $\ket{\omega}$.
	\item The state variables follow the Schr\"{o}dinger equation:
	\begin{equation}i \hbar \frac{\partial}{\partial
	t}\ket{\psi(t)}=H\ket{\psi(t)}\end{equation} where H is the quantum
	hamiltonian, obtained from the classical one with the substitution
	$H\of{X,P}= H\of{x\rightarrow X, p\rightarrow P}$

\end{enumerate}

From these axioms, two specific points can be underlined.
The first point is that the state is not anymore defined by two
variables ($x$ and $p$), but rather by a state-vector $\ket{\psi}$. Usually
$\ket{\psi}$ lives in a Hilbert space with infinite dimensions.

The state vector contains a huge amount of information that is the starting point for the evolution of the system. The downside is that the measure process has the
effect to destroy the biggest part of the information.

The other point to underline is that the measurement process is completely different
from the classical one. If we measure $\omega$ in classical mechanics we obtain, for
a particle in the state ($x_0$,~$p_0$), $\omega\of{x_0,p_0}$: and this is quite
straightforward. In the quantum mechanics case, we need first of all to obtain the
operator $\Omega$ corresponding to $\omega$ with the substitution given by postulate 2: $\Omega=\omega\of{x\rightarrow X, p\rightarrow P}$. After that we need
to determine the eigenvalues $\omega_i$ and eigenvectors $\ket{\omega_i}$ of
$\Omega$ and to expand the vector $\ket{\psi}$ on the basis of eigenvectors of
$\Omega$.
The measure will return as a result $\omega_i$ for a certain $i$; the probability to
obtain it is given by:
\begin{equation}
	P\of{\omega_i}= \abs{\left< \omega_i \middle| \psi \right>}^2
\end{equation}.
What is counter-intuitive in the quantum mechanical approach is that in the general case,
the result of a single measurement is not completely defined by the initial state of the system and by
the measurement itself but, with the same initial conditions it will return different results with different probabilities. For no-specific reason we
obtain from the measurement $\omega_i$ and not $\omega_j$.
In addition the measurement perturbs the system: if we repeat the measurement of
$\Omega$ we will obtain again the same $\omega_i$ obtained before. The system now is
in the state $\ket{\psi'}=\ket{\omega_i}\left< \omega_i \middle| \psi \right>$.

The fact that the measure perturbs the system can appear as an evidence:
when  a measure is performed in the laboratory, there is always something that perturbs the system
in some ways. However, in the specific case of quantum mechanics, we are speaking about an ideal
measure: no matter how careful the measure will be, there is no way to measure
$\Omega$ without projecting the system in an eigenstate of $\Omega$. This behavior, known as the the collapse of the wave function  is probably the most surprising behavior in quantum mechanics.

It is also useful to note that, due to the collapse of the wave function, position and momentum of a particle cannot be measured simultaneously. This consequence of the postulate is known as Heisenberg's uncertainty principle, and can be formalized with the following inequality
\begin{equation}
\label{eq:uncertainty}
\sigma_{x} \sigma_{p} \geq \frac{\hbar}{2}
\end{equation}
In equation~\eqref{eq:uncertainty} $\sigma_{x}$ and $\sigma_{p}$ are the incertainties on the position and on the momentum, while $\hbar$ is the  reduced Planck constant.
In consequence of the Heisenberg's uncertainty principle is impossible to follow the trajectory of a given particle at any time, even in principle. If multiple identical particles are present, as their trajectory cannot be followed, it is not possible to distinguish a particle from another one. Particles that, even in principle, cannot be distinguished are called indistinguishable. \label{sec:indistiguibili}

This revolution of physics, born at the beginning of the XXth century, is at the
basis of the explanation for several  particular phenomena; the most known is
maybe the photoelectric effect.

Nowadays, many applications are based on knowledge coming from the quantum theory, but do not use the quantum mechanical properties in elaborating the information. For example, modern electronics devices uses transistors, whose behaviour can be explained only with the use of quantum mechanics. However the information is stored and processed in a classical way. Recently, many  applications that
directly create and manipulate quantum states to process information have been developed.
Between them we can name the following ones.
\begin{description}
	\item[\textbf{Quantum Key Distribution}:] It is based on the \textit{no-cloning
	theorem} \cite{wootters1982Single} witch states that it is impossible to
	make a copy of an unknown quantum state leaving the first one unaltered.
	From this, it is possible to establish a protocol to exchange secure keys and makes it possible to know whether a key has been intercepted by someone
	else. As opposed to classical cryptography systems that are based on the
	fact that to decrypt a certain information is  too expensive, in
	terms of money and time, quantum cryptography based on quantum key
	distribution relies on a fundamental physical law and is for this reason
	way more secure. Moreover, we know now that possessing a faster computer could allow to decrypt classical
	cryptography, while this is not valid for quantum cryptography.
	\item[\textbf{Quantum simulation}:] The idea here is to use a quantum system to
	simulate another one. This is particularly useful, due to the complexity of
	quantum mechanics, to obtain information on systems that are too complex
	to be simulated with a classical computer.
	\item[\textbf{Quantum computation}:] A quantum computer uses quantum
	superposition and quantum interference to solve computational problems that
	are hard to solve with a classical computer. This does not mean that
	problems are impossible to solve with classical computers, but that the
	required time for solving a given problem  will be too long: the ability to solve them in a
	reasonable amount of time with a quantum computer is called quantum
	supremacy or quantum advantage. In 2019 \textcite{arute2019Quantum} working at Google claimed to
	have been able to solve with a quantum computer a problem that was
	impossible to solve with a classical one. However, there are still practical
	problems for the physical
	implementations of quantum computers, mainly connected to quantum
	decoherence and error correction, that still need to be solved.
\end{description}
The potential of the applications cited above has raised a growing interest
from various public and private founders and all these fields are very active.

However, in order to implement these quantum technologies, there is a need to produce so-called
quantum objects, systems with quantum properties that can store a quantum state and can be manipulated in
different ways. This implies to have a system (for example a single photon,
or a single atom) with an observable $A$ that has two different
eigenvectors, called $\ket{0}$ and $\ket{1}$ corresponding to two different
eigenvalues $a_0$ and $a_1$. If we measure $A$ on a generic $\ket{\psi}$
state we obtain either $a_0$ or $a_1$. In analogy this can be regarded as the
equivalent of a bit of information in classical computers, where the
information is normally encoded in two different values of tension, for example
\SI{5}{\volt} and \SI{0}{\volt}: one is indicated conventionally with $1$ and
the other with $0$. The huge difference in quantum information is that while
the result of a measurement is either $a_0$ or $a_1$, the state
$\ket{\psi}$ is generically given by
\begin{equation}
\label{eq:ch1-psi}
	\ket{\psi}=\dfrac{1}{\sqrt{\,\abs{\alpha}^2+\abs{\beta}^2}} \tonda{\alpha
	\ket{0}+
	\beta
	\ket{1}},
\end{equation}
with $\alpha, \beta \in \mathbb{C}$, where $\mathbb{C}$ is the set of
complex numbers. We note here that we have ignored any observable that is
orthogonal to $A$, that will not be affected in any way by the measurement
process.
\begin{figure}[tbh]
	\centering
	\includegraphics[width=0.4\linewidth]{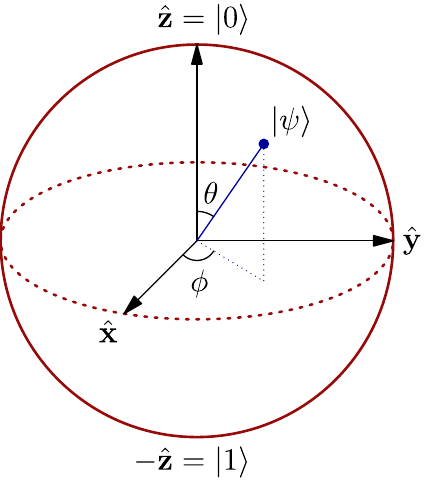}
	\caption{Block representation of a qubit $\ket{\psi}$. All the points on
		the sphere correspond to a possible state vector
		$\ket{\psi}$
		of the qubit. $\theta=0$ corresponds to the state $\ket{0}$ while
		$\theta=\pi$ corresponds to the state $\ket{1}$.}
	\label{fig:ch1_blochsphere}
\end{figure}
As the global phase is irrelevant, it is possible to rewrite
equation~\eqref{eq:ch1-psi} as
\begin{equation}
	\ket{\psi} = \cos \theta \ket{0} + e^{i\phi} \sin \theta \ket{1}
\end{equation}
with $\theta, \phi \in \mathbb{R}$ and $0 \leq \theta \leq \pi/2$ and $0 \leq
\phi \leq 2\pi$.
The state $\ket{\psi}$ is usually called \textit{qubit}, from \textit{quantum bit} in
analogy with classical computation and it is usually visualized as a vector in
the Bloch sphere, represented in figure~\ref{fig:ch1_blochsphere}.
All the points of the sphere corresponds to a possible state vector
$\ket{\psi}$, in particular the top and the bottom point, respectively when
$\theta =0$ and $\theta = \pi$ are the corresponding states $\ket{0}$
and $\ket{1}$.

The described formalism as the advantage to be independent of the physical
realization: this makes it possible to advance in theoretical and mathematical
research without paying too much attention of the specific platform used in practice.

Nowadays, many platforms are being investigated such as trapped ions, trapped neutral atoms, superconducting qubits or electron spins in semiconductors. However, single photons and quantum photonics present a particular interest as they are ideal
carriers for quantum information and as such holds a special place amongst the various platforms studied across the globe for quantum technologies.
%sphere

\section{Quantum Optics}
Quantum optics is the study of light and light-matter interaction from the point of view of quantum mechanics.

It normally uses the formalism of second quantization, in which not only the
physical quantities but also the fields are considered as operators. In the
first quantization, each particle is associated with a wavefunction $\ket{\psi}$ with
the result of having often more than one way to write the same global
wavefunction. For example if we have two indistinguishable particles, the
following notations are equivalent:
\begin{equation}
	 \psi_1
	\otimes \psi_2 \hspace{2cm} \psi_2 	\otimes \psi_1
\end{equation}
To remove this ambiguity, they need to be symmetrized or anti-symmetrized,
depending if the particles have a fermionic or a bosonic behaviour.
In the second quantization approach we do not consider which particle is in
which state but how many particles are in each state: this provides a more
efficient way to deal with the problem. We can denote the state as:
\begin{equation}
	\ket{n_1,n_2, \cdots, n_\alpha, \cdots}.
\end{equation}
This notation holds also for the case in which no particle is present, i.e. the vacuum,
indicated as $\ket{0}=\ket{0,0,0,\cdots,0,\cdots}$. In addition, we can
consider the case in which only one state contains particles
($\ket{0,0,0,\cdots,n_\alpha,\cdots}$) and this is called a single-mode Fock state.

Without entering into the details, it is now possible to define a photon as an excitation of the electromagnetic field from the quantification of Maxwell's equations. When we count single photons, in practice, we
consider a Fock state.

%In the notation of the second quantization, it is also useful to define the
%creation and annihilation operator

\section{Single photon emitters}
\subsection{Theory and definition}
\label{sec:sing_ph_em_theory}

By \textit{single photon emitters} we indicate any object that emits not more
than one photon at each given time. Theoretically the simplest single photon
emitter we can think to is a two level system: when it is excited from the
ground state to the excited state it needs a certain amount of time to relax again in
the ground state while emitting a photon. Until this photon is emitted, it is not
possible to excite it anymore. In practice this object has a more complex
structure, with more energy bands and levels and ideally one energy gap between them. We can then optically excite the emitter with a more energetic light that can be absorbed by the emitter. The recombination of the excited carriers is divided into two parts: a radiative recombination, when the photon is emitted, and a non-radiative recombination where some energy is lost
somehow.
This leads to the fact that the emitted photon is usually at a different wavelength than the excitation beam, allowing an easy way to separate them by filtering the light using spectral selective optical elements.

\begin{figure}[tbp]
  \centering
  \includegraphics[width=0.5\linewidth]{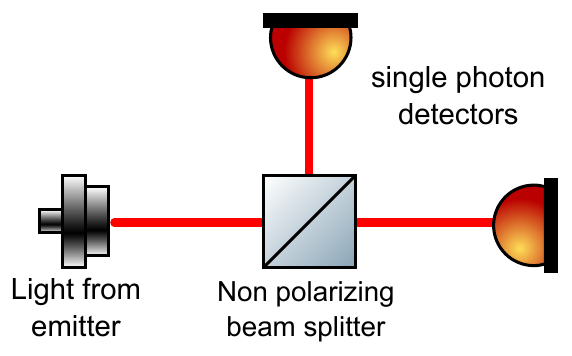}
  \caption{Schematic setup to measure the \gdt{}. The non-polarizing beam
    splitter divides the beam in two equal parts. The light is then detected by
     avalanche photodiode (APD) detectors.}
  \label{fig:g2_scheme}
\end{figure}

To characterize the single photon emission of an object, we perform a
measurement using the setup represented in figure~\ref{fig:g2_scheme}.
Before defining rigorously what is happening, let us try to explain it in easy
words. Let us imagine photons coming from the emitter acting like a particle for this example and passing through a 50/50 non-polarizing beam splitter: if we have only one
photon, it can not be divided and it will go in the first detector or in the
second one but never in both. On the other hand, if we have many photons coming at the same
time, there will be approximately half of them going to the first side and half
going to the other one. Then, if we look to the function that describes the number of coincidences, i.e. the  number of photons detected by the two detectors as a function of the delay
$\tau$ from one detected photon on one detector and one on the other one, it has to
go to zero at zero delay when we are in presence of a true single photon emitter.
This simple argument provides a qualitative insight in the single photon emission but is not enough for a full
understanding of this behavior.
% Generally we can distinguish evidence three cases, represented in
% figure FIG add:
% \begin{itemize}
% \item coherent emission, as a for a laser, in which each photon is emitted and is
%   not correlated to the arrival time of other photons
% \item bunched light, in which the photons have a bigger probability to arrive
%   in groups
% \item anti-bunched light, as in single photon emitters.
% \end{itemize}
% It is interesting to notice that both the first two cases have a semi-classical
% explanation, while the third has not.
To explain it better we need to introduce a mathematical description,
starting from a classical point of view.

\subsection{Semiclassical approach}

The semi-classical theory uses the quantum theory to describe the interaction
of light with the detector and with the atoms and the classical theory
for the treatment of the field~\cite{teich_i_1988}. It gives us a better
understanding of the single photon emission.

Let $\vec{x}$ be a point in space-time, $\vec{x}=\tonda{\vec{r},t}$. In the framework of the semiclassical theory, we can represent the light field as a complex
signal $V(\vec{x})$ as well as the light intensity as $I\of{\vec{x}}=\abs{V(x)}^2$.

% At a certain point $\vec{x}$ it is useful to
% define:
% \begin{itemize}
% \item $P\of{I}$, the probability density of the intensity $I(\vec{x})$
% \item $\mean{I}$, the ensemble average of the intensity
% \item $\Var{I}$, the variance of the intensity
% \end{itemize}.

To characterize the fluctuations in two points in space-time, we can use the
amplitude correlation function and the intensity correlation function:
\begin{subequations}
  \label{eq:corr_func_nn}
  \begin{equation}
    G^{\tonda{1}}\of{\vec{x_1},\vec{x_2}}=\mean{
      V^\ast\tonda{\vec{x_1}}
          V\tonda{\vec{x_2}}
    }
  \end{equation}
  \begin{equation}
    G^{\tonda{2}}\of{\vec{x_1},\vec{x_2}}=\mean{
      I\tonda{\vec{x_1}}
          I\tonda{\vec{x_2}}
    }
  \end{equation}
\end{subequations}
and their normalized versions
\begin{subequations}
  \label{eq:corr_func}
  \begin{equation}
    \label{eq:am_corr_func}
    g^{\tonda{1}}\of{\vec{x_1},\vec{x_2}}
    =
    \dfrac{
      G^{\tonda{1}}\of{\vec{x_1},\vec{x_2}}
    }{
      \sqrt{\mean{I\of{\vec{x_1}}}
      \mean{I\of{\vec{x_2}}}}
    }
  \end{equation}
  \begin{equation}
    \label{eq:int_corr_function}
    g^{\tonda{2}}\of{\vec{x_1},\vec{x_2}}
    =
    \dfrac{
      G^{\tonda{2}}\of{\vec{x_1},\vec{x_2}}
    }{
      {\mean{I\of{\vec{x_1}}}
      \mean{I\of{\vec{x_2}}}}
    }.
  \end{equation}
\end{subequations}

At this point it is useful to recall the Cauchy-Schwarz inequality:
\newtheorem*{Cauchy}{Cauchy-Schwarz inequality}
\begin{Cauchy}
  Given two vectors $\vec{u}$ and $\vec{v}$ of a inner product space, the following inequality is true
  \begin{equation}
    \abs{\vec{u}\cdot\vec{v}}\leq\abs{\vec{u}}\abs{\vec{v}}
    \label{eq:cauchy-schwarz}
\end{equation}
\end{Cauchy}
Making use of this relation, and considering that the expression contains
functions of real numbers and the inner product between these
functions, it is possible to show that
\begin{equation}
  \label{eq:g2_only_bunching}
  g^{\tonda{2}}\of{\vec{x_1},\vec{x_2}} \leq
  \sqrt{g^{\tonda{2}}\of{\vec{x_1},\vec{x_1}}g^{\tonda{2}}\of{\vec{x_2},\vec{x_2}}}
\end{equation}

Experimentally, we measure the function $g^{(2)}\of{\tau}$ defined as
\begin{equation}
  \label{eq:g2_tau}
  g^{(2)}\of{\tau}=g^{(2)}\of{\vec{r},t,\vec{r},t+\tau}.
\end{equation}
From equation~\ref{eq:g2_only_bunching} and  equation~\ref{eq:g2_tau} we
can conclude that for classical light
\begin{equation}
  \label{eq:g2_only_bunching_tau}
  g^{(2)}\of{0}\geq g^{(2)}\of{\tau}
\end{equation}
In case of $\gdz{}$ is bigger than $\gdt{}$ for any $\tau>0$ we say that the light is bunched, if $\gdt{}$ is constant the light is called unbunched, while if $\gdz{}$ is smaller than $\gdt{}$ for $\tau>0$ the light is called antibunched~\cite{teich_i_1988}. Due to the normalization, $\lim_{\tau \to \infty} \gdt{}=1$, then the light is bunched if $\gdz{}>1$.
In other words, in the semiclassical approach the $g^{(2)}$ function can only be
bunched or unbunched. An ideal coherent light is unbunched, and $\gdt{}=1$ for any $\tau$, while the thermal light from a light bulb is bunched.

\subsection{Quantum theory of coherence}
In the quantum theory of coherence~\cite{glauber_quantum_1963} correlation
functions are expressed in terms of the electric field operators
$\hat{E}^+\of{\vec{x}}$ and $\hat{E}^-\of{\vec{x}}$, that represent the positive
and negative frequency parts, respectively. Indicating by $\hat{\rho}$ the
density matrix operator, we can redefine  functions~\ref{eq:corr_func_nn} in
the following way:
\begin{subequations}
  \begin{equation}
    G^{\tonda{1}}\of{\vec{x_1},\vec{x_2}}=\Tr\graffa{\hat{\rho}
      \hat{E}^-\of{\vec{x_1}}
      \hat{E}^+\of{\vec{x_2}}
    }
  \end{equation}
  \begin{equation}
     G^{\tonda{2}}\of{\vec{x_1},\vec{x_2}}=\Tr\graffa{\hat{\rho}
       \hat{E}^-\of{\vec{x_1}}
       \hat{E}^-\of{\vec{x_2}}
       \hat{E}^+\of{\vec{x_1}}
       \hat{E}^+\of{\vec{x_2}}
    }
  \end{equation}
\end{subequations}
and their normalized versions $g^{\tonda{1}}\of{\vec{x_1},\vec{x_2}}$ and
$g^{\tonda{2}}\of{\vec{x_1},\vec{x_2}}$ can be defined by analogy with the
semi-classical case. The main difference in this case is that we cannot interpret
anymore $g^{\tonda{2}}\of{\vec{x_1},\vec{x_2}}$ as a normalized statistical
correlation function of the optical intensity; for this reason we cannot apply
anymore the Cauchy-Schwarz inequality~\eqref{eq:cauchy-schwarz}. As a conclusion,
in the quantum case the $g^{\tonda{2}}\of{\tau}$ function defined by analogy by
the function~\eqref{eq:g2_tau} is not anymore bounded to be at least one when
$\tau=0$. Finding $g^{\tonda{2}}\of{0}<1$ is indeed a behavior that can be
explained only in quantum theory of coherence and for that reason we usually
address to it as  ``quantum light''.

\subsection{Single photon emission}
As we defined in the previous section, any antibunched light has to be regarded
as ``quantum light'', as there is no classical explanation for it. In the
presence of quantum emission we can look at the $g^{\tonda{2}}\of{\tau}$ function in more
depth, to understand how many photons are emitted.
In particular it is possible to show~\cite{loudon2000quantum} that:
\begin{equation}
g^{(2)}\of{\tau}\geq 1-\frac{1}{\mean{n}}
\end{equation}
where $\mean{n}$ is the mean photon-number. This is valid for any $\mean{n}\geq
1$.
From the previous equation we deduce that only $g^{(2)}\of{0}=0$ can ensure that $\mean{n}\leq1$ and describe a ``pure'' single photon emission. Experimentally, in practice, we measure the value of $g^{(2)}\of{0}$ to conclude on the quality of a single photons emitter. The smaller
$g^{(2)}\of{0}$ is, the less probable is to have more than one photon emitted at each time.
Anyway if
\begin{equation}
g^{(2)}\of{0}<0.5\label{eq:single_photon_inequality}
\end{equation}
the mean number of photons
emitted is smaller than~$2$ and we can consider to deal with a single photon emitter. In the following, any emitter that satisfies the
inequality~\ref{eq:single_photon_inequality} will be considered as a single photon
emitter.

\subsection{Experimental limitations}
In the theoretical analysis, we never had the need to introduce the setup
represented in figure~\ref{fig:g2_scheme}. For the theory, indeed, it is only
needed to measure the photon statistics emitted, and a unique single photon
detector with no beam-splitter should be enough. In order to perform a correct
measurement of $g^{\tonda{2}}\of{0}$ this detector should be able to detect two
photons arriving at the same time: this could be possible with photon-resolved
single photon detectors, but it is not possible with common
avalanche-photodiode single photon detectors we use (Excelitas SPCM-AQRH-14-FC).
In our case, after a photon arrives onto one detector, there is a reset time in which the detector is off until it is capable of receiving and detecting the next photon: this dead-time is about \SI{25}{\nano\second}.

To avoid such problem the most common solution is to split the signal into two
parts using two different APDs, one for the so-called start signal and another one for the so-called
stop signal. This method has the disadvantage to reduce the rate of collection
of the data but allows to measure correctly the $g^{\tonda{2}}\of{0}$. It is
important, for the future understanding, to keep in mind that in any
experimental measurement of the function~\eqref{eq:g2_tau}, $\tau$ will be considered as the delay between a start event detected by one APD and a stop event
detected by the other APD.

\subsection{Examples of single photon emitters}

An ideal single photon source should be able to produce deterministically
on-demand indistinguishable single photons (as defined in section~\ref{sec:indistiguibili}).
In practice, multiple approaches have been studied to realize such a source,
each of them presenting advantages and disadvantages.

We can subdivide them in two groups~\cite{eisaman2011Inviteda}:
\begin{description}
	\item[\textbf{Stochastic single photon generation}]: in this approach, one of the most interesting techniques is to use a laser and a non linear crystal in order to obtain
		pairs of photons using the so-called spontaneous parametric down-conversion effect (SPDC).The laser intensity has to be weak in order to generate in average not more than one pair of photons.  The advantage of this technique is the possibility
		to use one photon of the pair to know when they are generated, and for this reason
		they are called heralded single photons. Upon this technique rely various
		devices used for quantum key distribution applications.
		Unfortunately, such device produces single photons only when we
		require from them a low photon rate. In addition, they are not
		deterministic as the time at which the photon pairs will be emitted is not
		known. Different solutions have been proposed to tackle this problems, such as
		using multiplexed photons to increase the single-pair emission rate
		without increasing the probability of multiple pair
		emission~\cite{migdall2002Tailoring,
			fitch2003Photonnumber,jeffrey2004Periodic,shapiro2007Ondemand}
		or storing the emitted photons to use them on-demand at later times.
	\item[\textbf{Deterministic single photon sources}:] in this approach, already mentioned, the
		aim is to find a single photon source that can work as a deterministic
		single photon emitter. Multiple alternatives have been explored such as single neutral atoms, single trapped ions, single molecules,
		different kind of quantum dots, single color centers in diamonds and
		nanodiamonds.
\end{description}

During my work I studied different emitters from the second group, and I will concentrate on such systems in the following. Indeed, differently from an attenuated laser that has a poissonian statistics, these sources can provide true single photons.
%
% The most important part of my work was
% the study of perovskites colloidal quantum dots, described
% in more details in chapter~\ref{chap:perovskites}. Nevertheless, in the course of my PhD, I also studied and used other kinds of colloidal quantum dots, such as dot-in-rod \ch{CdS}/\ch{CdSe}
% nanocrystals as well as nanodiamonds with single defects.
% \textbf{(I feel there is a 'mise en forme' issue here between the 2 groups and the fact that you spend time describing these emitters i nthe following)}
I proceed to describe some different kinds of \textit{deterministic single photon sources}.
\subsubsection{Single neutral atoms}
They have the advantage of repeatability as each atom is identical to the
others (they are slightly 'different' in practice due to different environments), but they are challenging to operate with. The most used atoms for this
application are alkali atoms~\cite{kuhn2002Deterministic,mckeever2004Deterministic,
	hennrich2004Photon,wilk2007PolarizationControlled,
	hijlkema2007Singlephoton,dayan2008Photon,aoki2009Efficient}

They are usually used in a regime of strong coupling with a cavity that allows to efficiently collect emitted photons. The atoms are first trapped in a magneto-optical trap; then the trap is switched off and the atoms falls inside a
cavity with the help of gravity. Once they fell into the cavity, another optical
trap is turned on to keep them in place.  This operation is, of course,
experimentally challenging and space consuming, which poses limitation to future
scalability of this source for quantum technologies.

Such atoms usually are three level systems, with two ground states $\ket{g_1}$ and
$\ket{g_2}$ and an excited state $\ket{e}$. The optical cavity is resonant to
the transition $\ket{g_1} \rightarrow \ket{e}$, so that the excited state,
thanks to the Purcell effect\cite{purcell1995spontaneous}, tends to decay in the
$\ket{g_1}$ state. The pumping laser is set to the transition $\ket{g_2}
\rightarrow \ket{e}$, while the population is moved from the state $\ket{g_1}$
to the state $\ket{g_2}$ via stimulated Raman adiabatic
passage~\cite{vitanov2017Stimulated}.

\subsubsection{Single ions}
The single photon emission mechanism is similar to the one for single atoms,
with the difference that single ions can be trapped using a radio-frequency ion
trap. The trapping is simpler in respect to the optical trapping needed for single
atoms, and the position of the ion can be set with a precision of few
nanometers.

As in the case for atoms, ions are identical and the reproducibility of emitted
photons is ensured.
Multiple ions can also be controlled in a singular
trap\cite{kielpinski2002Architecture,riebe2008Deterministic,home2009Complete},
but it is difficult to efficiently collect the emitted photons.

\subsubsection{Quantum dots}
Quantum dots semiconductors can be split into two different groups:
\begin{itemize}
	\item quantum dots grown by epitaxial methods on a bulk material;
	\item colloidal quantum dots, obtained in solution using a chemical
	synthesis.
\end{itemize}
The mechanism is different in respect to single atoms/ions. In this case, the
small size results in a discrete level structure. When excited by a pumping laser, one or more
excitons are produced. The excitation can be performed optically, usually
non-resonantly with the emission, or electrically.
Usually some internal mechanisms allow for the non-radiative recombination
of most of the excitons produced, except one, resulting thus in a single photon
emission.
In the next sections, I will describe in more details the strong and weak points of
colloidal quantum dots.

\subsubsection{Color centers in diamonds and nanodiamonds}
Newer than the others in the single photon emitters use, they are based on a defect
in the crystal structure of diamond~\cite{jelezko2006Single}.
The emission mechanism is similar to the quantum dots, the defect creates a
multiple level energy system that can be excited out (or in-) of resonance in order to
produce single photons. The main advantage, compared to colloidal quantum dots, is that
they are optically stable. In addition, differently from epitaxial quantum dots, they can emit at room temperature.
They are naturally present in diamond, but they can also be artificially
created at a certain place using laser writing
techniques\cite{hadden2018Integrated}.
Diamond has a high refraction index and photon extraction can be
difficult. In order to overcome this issue and use this kind of sources with glass-based platforms, single defects in nanodiamonds are being
developed. One of the main problems with this approach is that they are not easy to fabricate.
Controlling the composition and the nanometric size  of nanodiamonds is challenging.
The easier defect to create in nanodiamonds is the Nitrogen-Vacancy (NV)-center and multiple
studies have shown their capabilities as single photon emitters. At room temperature, the
emission spectrum of a \ch{NV} center is very broad. In
addition, crystal strain can affect the emission wavelength and that pose some issues when we want to couple them together.

Other kinds of defects, such as the Silicon-Vacancy (\ch{SiV}) center, have also shown single
photon emission \cite{wang2006Single}, but
are even more difficult to produce in nanodiamond shape \cite{neu2011Single} and, as
in the NV case, they can
suffer from crystal strain present in the lattice~\cite{lindnerStrongly}.
\subsubsection*{} %\textbf{(why a big gap here?)}
\vspace{-1cm}
As I have shown, different kinds of single photon emitters exist.
Existing works at the LKB and L2n focus on the study
of \ch{CdSe/CdS} rod-shaped quantum dots.
During my thesis, I worked with \ch{CdSe/CdS} rod-shaped quantum dots even though I concentrated my
studies on perovskites quantum dots, a new kind of quantum dots from
perovskite nanocrystals. I describe them in more details in
chapter~\ref{chap:perovskites}. I also had the opportunity to perform some preliminary studies using \ch{NV-} and \ch{SiV-} color centers in nanodiamonds. These emitters and their characteristics are described in more details in chapter~\ref{chap:outlook}.

%\section{Colloidal quantum dots}
\section{Common characteristics of solid state single photon emitters}

A solid state single photon emitter is an object of nanometric size made by a semiconductor material or that presents a semiconductor-like level structure.
When it is excited with the correct wavelength, an electron can jump from the
valence band to the conduction band.  In the relaxation process, there is emission of light and the color emission depends on the energy gap.
When one electron is in the conduction band, it leaves a corresponding hole in
the valence band. We can now define the notion of exciton: an exciton is a neutral
quasi-particle, formed by the bounding of an electron and a hole. It can be
seen as an elementary excitation of condensed matter.

Different mechanisms can be present in a solid state single photon emitter that allows only a single
photon to be emitted at a given time: the most important of them for colloidal quantum dots is probably
the Auger Effect.
% Quantum dots can be produced in different manner and one of the most
% interesting way to produce them is a chemical production that ends with a
% colloid of quantum dots in a given solvent, and the quantum dots produced in
% this way are called colloidal quantum dots. Different kind of quantum dots can
% be produced with chemical synthesis, but the advantage of this production
% technique is that it is feasible in a standard chemistry lab and not require
% the facilities needed for other production systems, like epitaxial growth in
% example.

%Different shape and composition of quantum dots have been studied in literature;

In this section I will explain the mechanisms common to most of the solid states single photon emitters,
making reference, when useful, to some studies performed at the LKB in the past.
% with \ch{CdSe/CdS} rods
% that are a promising kind of colloidal quantum dots capable, among other
% things,
% to emit polarized single photons.

\subsection{Non radiative recombination}
As opposed to radiative recombination processes, a non radiative
recombination allows the annihilation of an exciton without the emission of
photons. These processes are present in quantum dots, as well as in
semiconductors in general, and are important to understand the dynamics of the system.
The most important ones are:
\begin{itemize}
	\item Auger recombination,
	\item exciton-phonon interaction,
	\item charge trapping.
\end{itemize}

I briefly describe these effects below, starting form the most important in our
case of colloidal quantum dots, the Auger recombination.

\subsubsection{Auger recombination}
\begin{figure}
	\centering
	\subfloat[]{\label{fig:Augerion1}\includegraphics[width=0.4\linewidth]
		{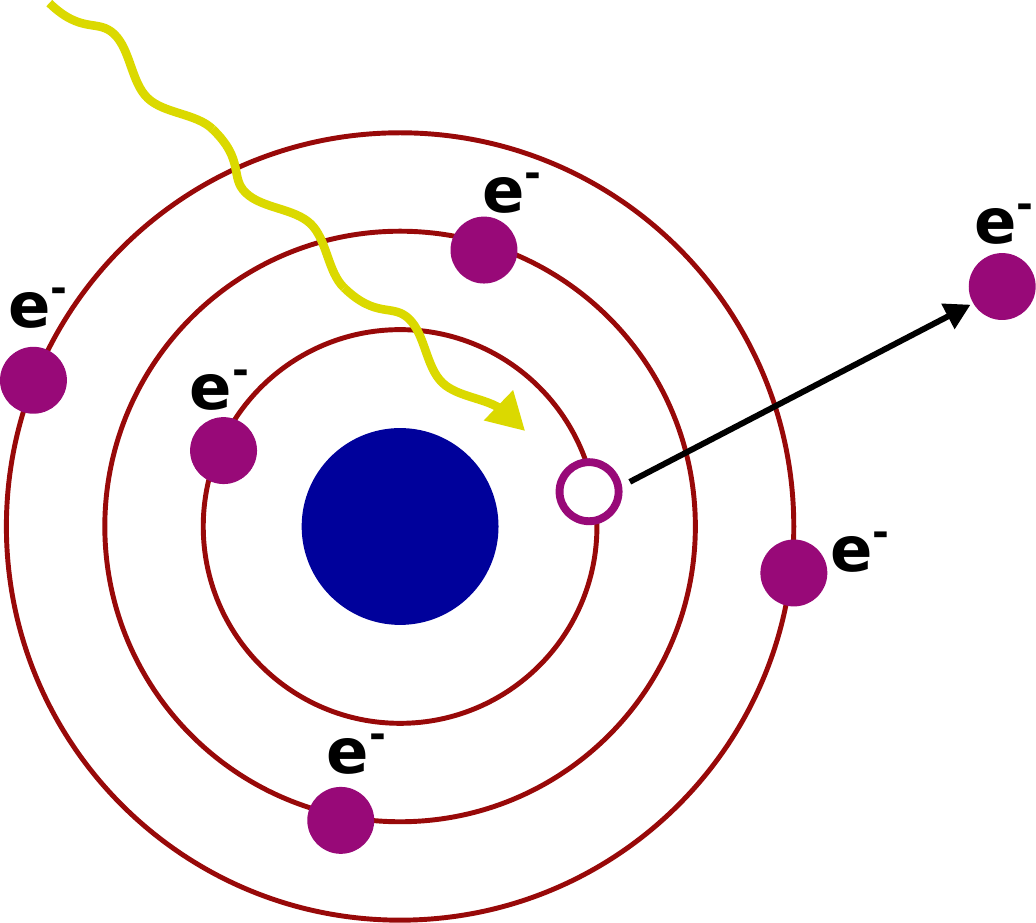}}\qquad\qquad
	\subfloat[]{\label{fig:Augerion2}\includegraphics[width=0.4\linewidth]
 		{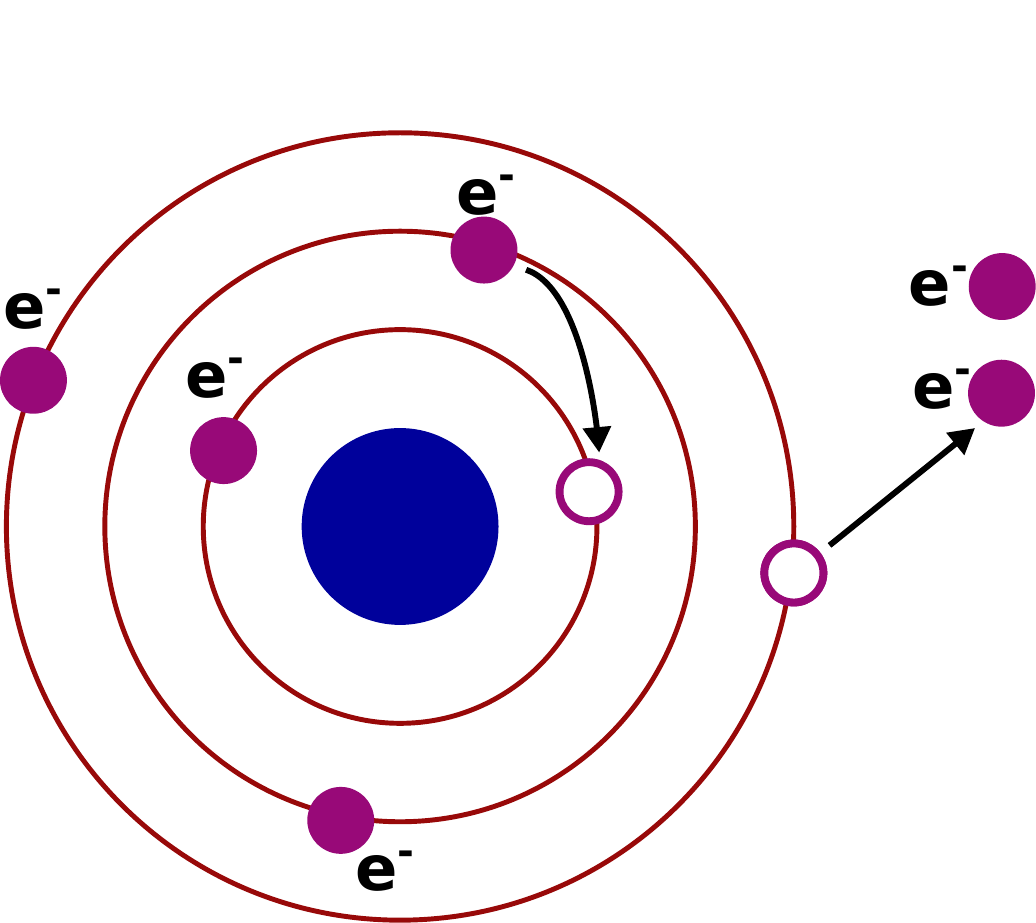}}
	\caption{Mechanism of Auger ionization: \protect\subref{fig:Augerion1}~The
	atom is ionized by a radiation (i.e. a $\gamma$ ray) and looses
	one of its electrons of the inner shell;
		\protect\subref{fig:Augerion2}~One outer electron relax to the hole
		left by the missing electron giving its energy to an outer electron
		and the atom is ionized for the second time.}
	\label{fig:Augerion}
\end{figure}

Auger recombination is a mechanism present in semiconductors analogous to the Auger
effect~\cite{meitner1922Ueber} in atoms.
The Auger effect\cite{meitner1922Ueber} is the second ionization of an atom
due to the energy
transferred by an electron of the atom itself: the process is shown in
figure~\ref{fig:Augerion}.
It happens in two steps:
\begin{itemize}
	\item The atom is first ionized, i.e. by an ionizing radiation and as a
	consequence of this ionization, a hole is present in one of its inner shells
	\item One electron relaxes from an outer level to the hole while another
	``collects'' its energy and is emitted
\end{itemize}

In Auger recombination there is a similar effect: an exciton in a semiconductor
recombines giving to a carrier in the valence band its energy. As opposed to
the Auger ionization, this time the carrier that collects the charge stays in the
conduction band at a higher level state and no-ionization happens. The carrier
can then relax non radiatively to the ground state of the conduction band.

\begin{figure}
	\centering
	\subfloat[]{\label{fig:AugerR1}\includegraphics[width=0.3\linewidth]
		{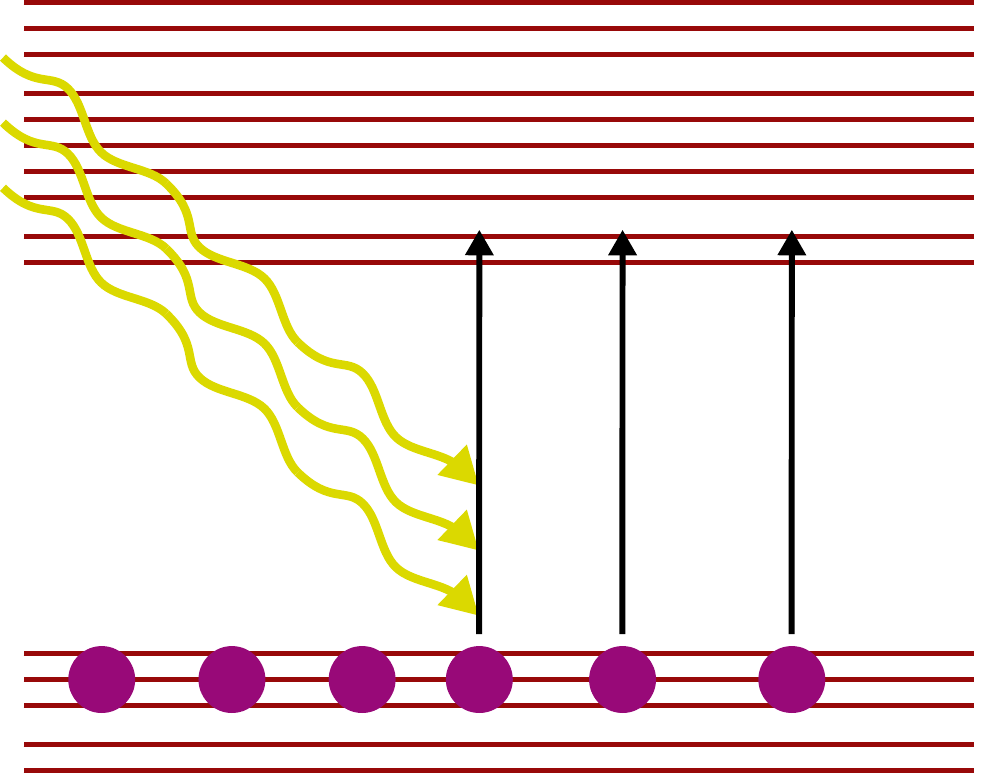}}\qquad
	\subfloat[]{\label{fig:AugerR2}\includegraphics[width=0.3\linewidth]
		{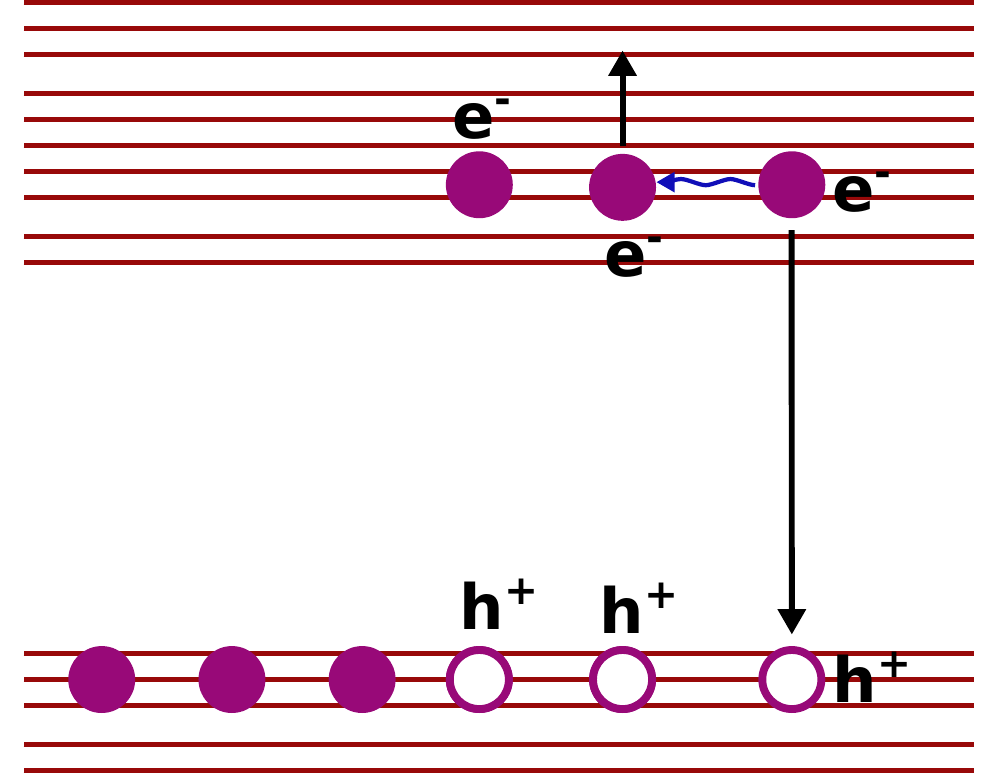}}\\
	\subfloat[]{\label{fig:AugerR3}\includegraphics[width=0.3\linewidth]
		{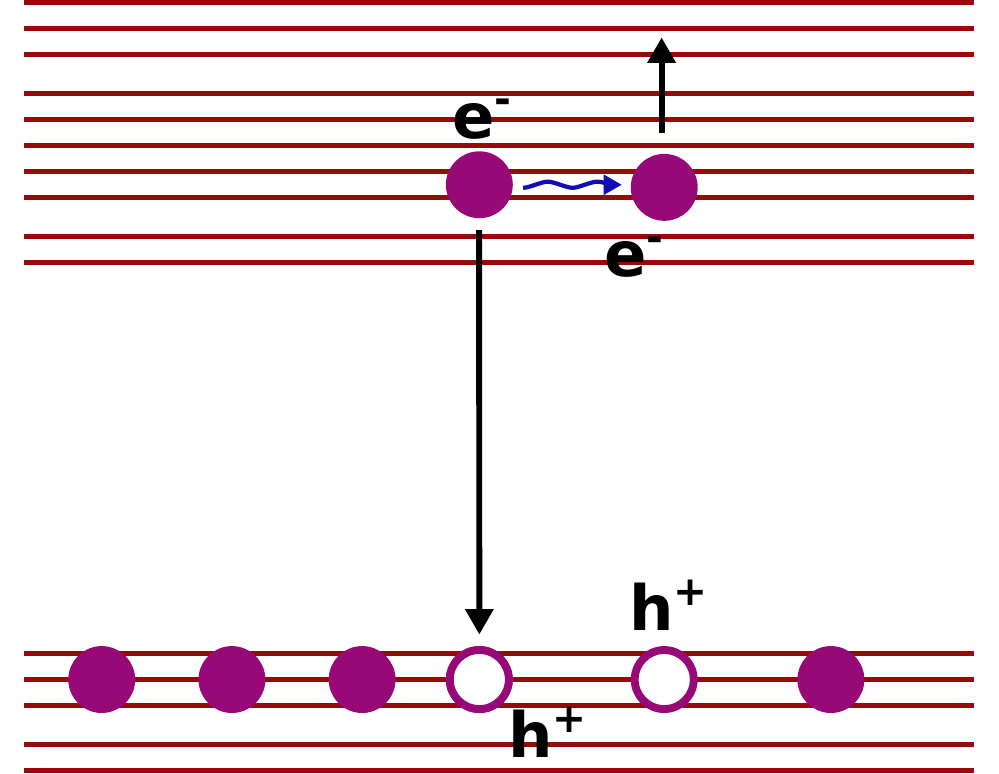}}\qquad
	\subfloat[]{\label{fig:AugeR4}\includegraphics[width=0.3\linewidth]
		{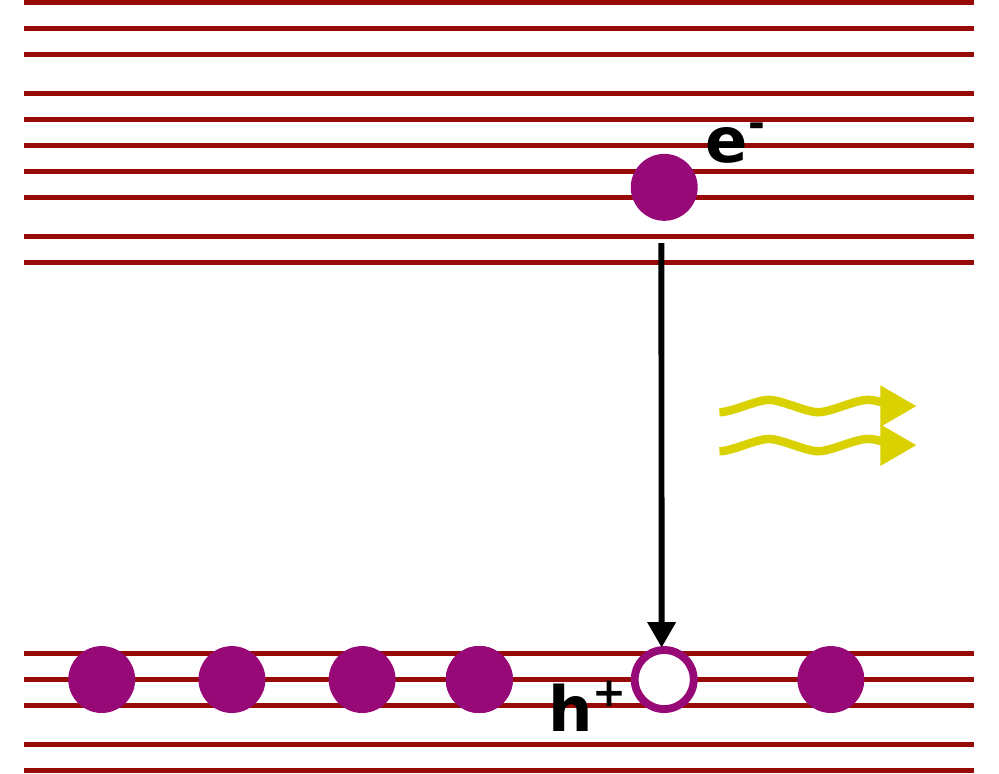}}
	\caption{Mechanism for Auger recombination: \protect\subref{fig:AugerR1}~The
		semiconductor is excited by light and multiple excitons are created;
		\protect\subref{fig:AugerR2} and \protect\subref{fig:AugerR3}~until
		mutiple excitons are present, one exciton can relax non-radiatively
		passing the energy to another one; finally
		\protect\subref{fig:AugerR3}~the last exciton relaxes radiatively by
		emitting a single photon with the energy of the gap.
	}
	\label{fig:AugerR}
\end{figure}

This mechanism is important in quantum dots as it often guarantees
single photon emission. Indeed, if more than one exciton is created, the
excitons in addition to the first one, can recombine non radiatively giving
their energy to the first one. The Auger recombination is usually a very efficient mechanism: until the Auger recombination is possible, it is predominant on the radiative recombination. When only one exciton is present, there are no
more carriers in the conduction band to which the energy of the exciton can be
passed on, and the Auger effect is not possible anymore; this time, the exciton
relaxes radiatively with the emission of one photon. The scheme of this mechanism
is shown in figure~\ref{fig:AugerR}.

It has been shown for different kinds of quantum dots that this mechanism
scales linearly with the volume of the
quantum dot~\cite{klimov2000Quantization,robel2009Universal}. This property, known
as `V-scaling', is the main reason of why Auger recombination is so effective in
colloidal
quantum dots.

\subsubsection{Interaction of excitons with phonons}
A phonon is a quasi-particle that represents the elementary vibrational
excitation of a lattice.
We can distinguish two different cases: one is the case of intraband relaxation
and the other one is the case of interband relaxation.

\begin{description}
	\item[\textbf{Intraband relaxation}:] the fact that electrons and holes can
	relax giving energy in a semiconductor seems normal, although with a deeper thinking, it is quite surprising. Indeed, as well as for
	electrons and holes, phonon energy levels are also discrete due
	to the quantum confinement. As a consequence, except in the unlikely case of
	energy overlap, the transfer of energy from an exciton to a phonon should be
	prohibited.
	On the contrary, it was observed that the thermal relaxation is even
	accelerated by the quantum confinement, with a relaxation time that can be ten times faster in a nanocrystal than in the bulk
	material~\cite{vezzoli2013Experimental} in \ch{CdS/CdSe}. This is due to
	the fact that this process is mediated by the Auger energy transfer from
	electrons to holes. The process is usually very fast, on the order of
	hundreds of femtoseconds (compared to some nanoseconds for the radiative
	decay) and this is the reason why all the emission happens at the lowest
	energy and we see only one emission peak.
	\item[\textbf{Interband relaxation}:] phonons have an effect also in
	interband relaxation, in two distinct ways. First, they affect the
	spectrum by broadening the emission peak and secondly they create a
	non-radiative relaxation channel that is in competition with the radiative
	one.
\end{description}

\subsubsection{Charge trapping}
\label{sec:Chargetrap}
A carrier can be trapped onto a defect of the lattice or at the surface of the
nanocrystal. When this happens, it creates additional energy levels within the
gap of the semiconductor providing a non radiative decay path and, as
a consequence, no photon is emitted by the nanocrystal.

A trapped state can last from tens of picoseconds up to several seconds and its
desexcitation is usually non radiative.

This is however a reversible process, as opposed to chemical
reactions that provoke a permanent bleaching of the emission, i.e. the emitters do not emit light any more and they become completely dark.

\subsection{Saturation}
\label{sec:sat}
A typical effect present in deterministic single photon emitters is the saturation of the emitted intensity. In case of nanocrystals this is an effect of Auger recombination.
%\textbf{(isn't it always for a 2 level system?)}.
If we imagine to start at zero
excitation intensity, no excitons are created in the nanocrystal.
When the excitation power increases, one or more excitons are created in the
nanocrystal. At first, the probability to create one or more excitons
with a single pulse increases but at a certain point it is almost certain to
create one exciton and the excitation light pulse can create two or more of them. Increasing
the excitation power will create more excitons at the same time. As already mentioned, ideally only one of them can emit light while the others recombine
non radiatively. For this reason, the emitted intensity saturates and, up to a
certain point, exciting with a higher intensity does not imply a higher number
of photons emitted.
If we report on a graph the emitted intensity as a function of the excitation
intensity, we obtain a curve described by the equation:
\begin{equation}
	\label{eq:1-sat}
	P_{{PL}} \of{I}=P_{sat} \cdot \tonda{1-e^{-\frac{I}{I_{sat} }}}
\end{equation}

This equation is valid if the Auger recombination is efficient enough.
The saturation intensity $I_{sat}$ corresponds to the intensity at which, on average, an exciton is created for each excitation light pulse.
In case of a not-so-efficient Auger effect, the
equation~\eqref{eq:1-sat} needs to be modified adding a linear term and becomes
\begin{equation}
\label{eq:1-satl}
P_{{PL}} \of{I}=P_{sat} \cdot \tonda{1-e^{-\frac{I}{I_{sat} }}}+ B \cdot I
\end{equation}

\begin{figure}
	\centering
	\includegraphics[width=0.6\linewidth]{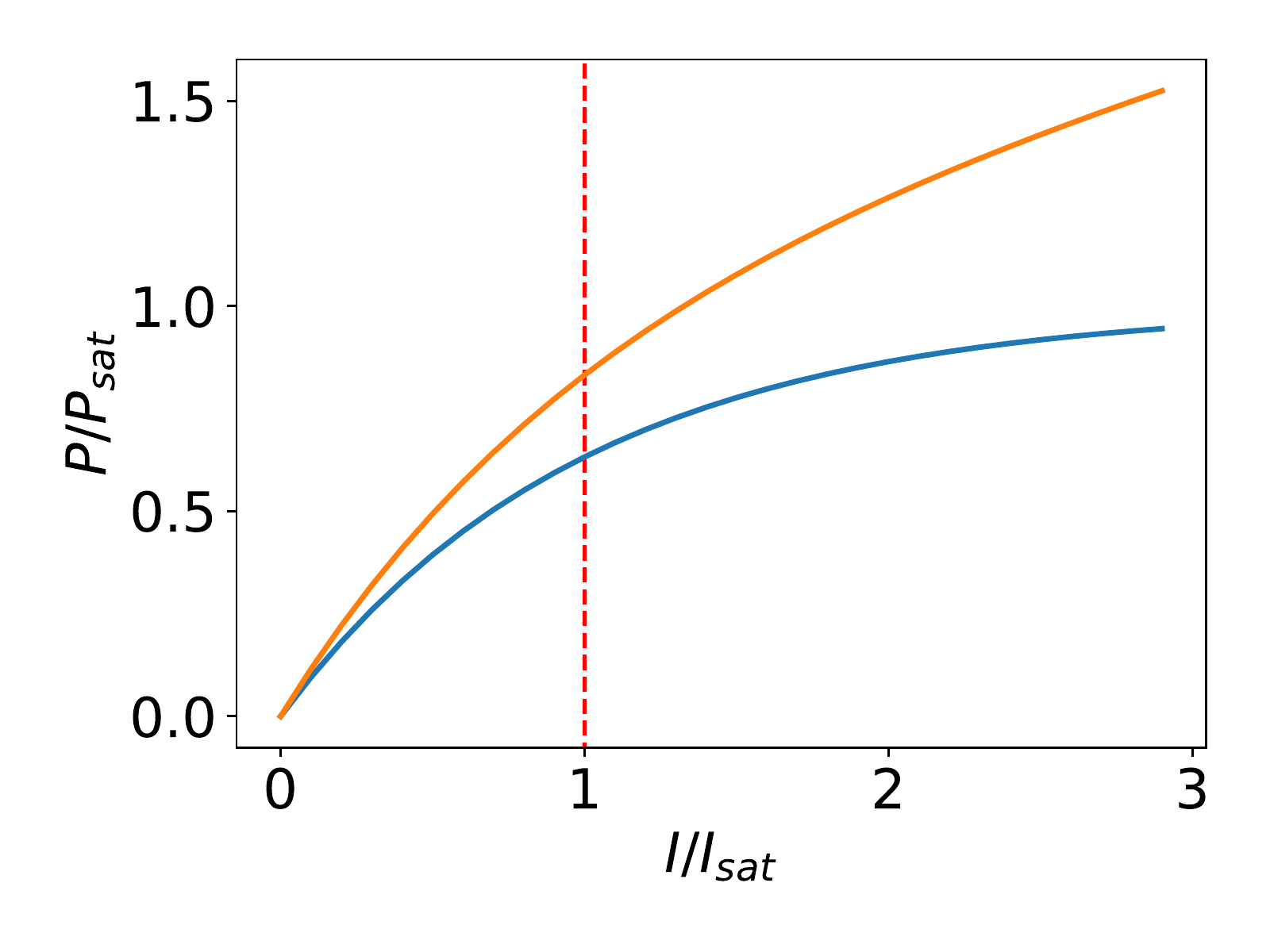}
	\caption{In blue, plot of the saturation curve described by equation
	\protect\eqref{eq:1-sat}. In orange, plot of the saturation curve described
	by equation \protect\eqref{eq:1-satl}, where $B=0.2 P_{sat}/I$.}
	\label{fig:ch1sat}
\end{figure}

The shape of the two curves is shown for comparison in figure~\ref{fig:ch1sat}.

\subsection{Emission Lifetime}
\label{sec:lifetime-ch1}

We have seen, talking about Auger Effect, that the decay rate $\gamma$ plays an
important role for single photon emission. It is thus possible to measure the
decay rate associated to the emission via a lifetime measurement corresponding to the spontaneous emission of the emitter. If we
call $\gamma_r$ the radiative decay rate, the lifetime of the emission $\tau_r$
is defined as:
\begin{equation}
	\tau_r=\frac{1}{\gamma_r}
\end{equation}

\begin{figure}
	\centering
	\includegraphics[width=0.6\linewidth]{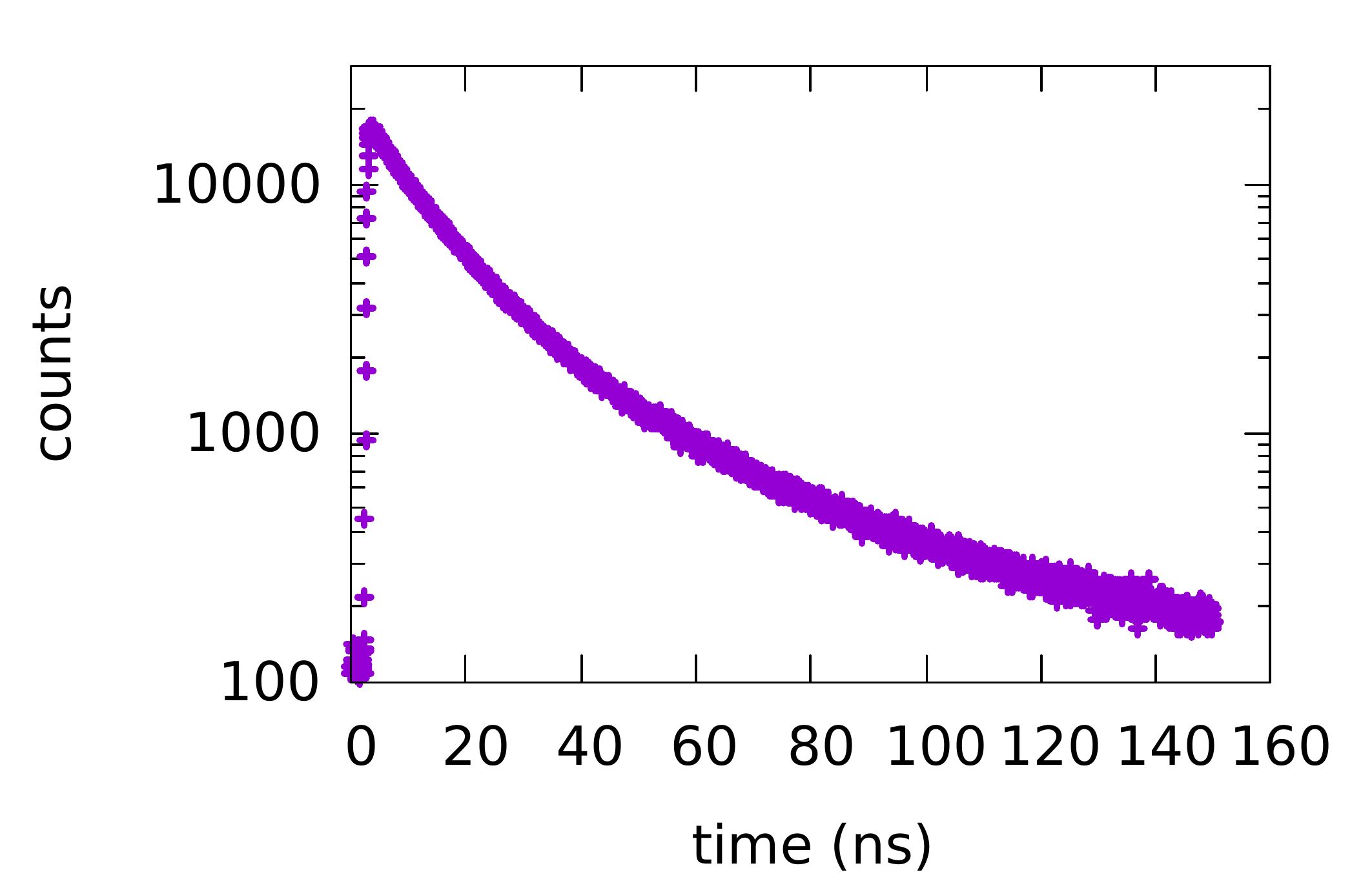}
	\caption{Lifetime histogram obtained from a single \ch{CdS}/\ch{CdSe} dot-in-rod nanocrystal.
	Each point represents the number of photons that is impinging onto the detector
	at a certain delay. The histogram can be fitted by an exponential decay
	function (equation~\protect\eqref{eq:exp_dec_func_s}).}
	\label{fig:lifeem5}
\end{figure}

This can be measured using a time correlated single photon counting (TCSPC)
measurement. In practice, exciting the emitter with a pulsed laser, we detect
at which time later from the excitation pulse the photon is emitted and we use it to create an
histogram in time (an example is represented in figure~\ref{fig:lifeem5}).
We can then fit it using an exponential decay function:
\begin{equation}
\label{eq:exp_dec_func}
I=A \cdot e^{-t/\tau_r}
\end{equation}
finding the lifetime of the emission $\tau_r$. If multiple emission states are
present, as often the case, multiple lifetimes are measured and
equation~\eqref{eq:exp_dec_func} becomes:
\begin{equation}
\label{eq:exp_dec_func_s}
I=\sum_{i} A_i \cdot e^{-t/\tau_i}
\end{equation}
where $i$ runs over all the radiative states. When the $\tau_i$ in~\eqref{eq:exp_dec_func_s} are of the same
order of magnitude, experimentally distinguishing between the different times can be
challenging.

\subsection{Blinking}
\label{sec:blinking-ch1}
Due to the charge trapping described in section~\ref{sec:Chargetrap}, the
emitter can become dark (or gray) for a certain time and recover its emission at a later stage. This phenomenon is known as blinking and has been largely studied.

There are different analysis that can be performed on blinking nanocrystals. I will detail some of them, knowing that they are not necessarily all applicable for our particular case.
\subsubsection{Binning analysis}
%todo add images here?
\begin{figure}
	\centering
	\includegraphics[width=0.8\linewidth]{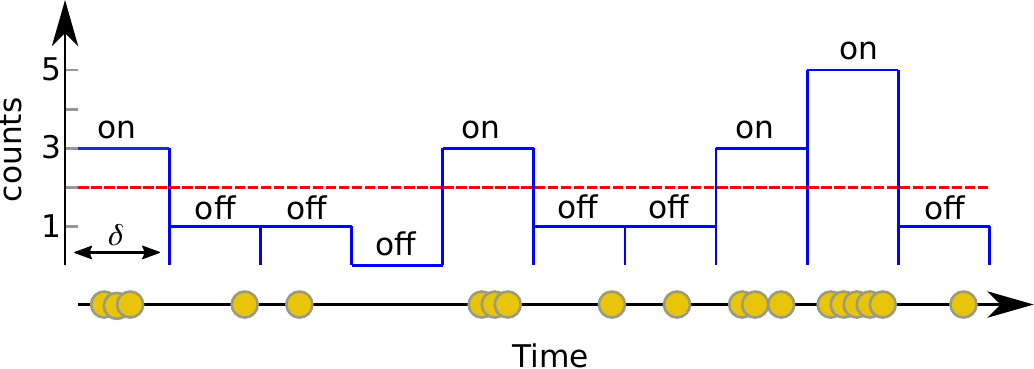}
	\caption{Exemplification of the creation of the blinking trace. Each bin of the histogram reports the number of photons (in yellow) arrived at the
	time delay $\delta$. A threshold is then chosen, I consider that the bin over
	the threshold are considered ``on'' while the bin under it are considered
	``off''.}
	\label{fig:blinkingtrace}
\end{figure}
One of the most traditional way to study the blinking is to perform a time
correlated single photon measurement or in other words to measure, with the
help of a single photon detector, if a photon arrives and when it arrives.
It is then possible to select a size for a temporal bin and to find out how many
photons arrived within a specific bin.
The result at the end is an histogram reporting for each time interval how many
photons arrived.
There are two different kinds of measurements that
can be performed depending on the instrument: one directly provides the histogram, while the other one provides
a list of photon arrival times with which it is possible to create the histogram.

Once we have the histogram,  we can inspect the distribution of the intensities in each time bin. We can then analyse if the distribution of the on-states and the distribution of the ``off''-states can easily be separated thus we can set a threshold in between. The probability for a state to stay ``off'' or ``on'' for a certain time is
measured and noted $\mathcal{P}_\mathrm{off}\of{t_\mathrm{off} > \tau }$ and
$\mathcal{P}_\mathrm{on}\of{t_\mathrm{on} > \tau }$.

In the case of the ``off''-states, the probability $\mathcal{P}_\mathrm{off}$
usually follows an inverse power-law distribution:
\begin{equation}
	\mathcal{P}_\mathrm{off} \propto \tau^{-\mu}
\end{equation}
The ``on'' duration is also a power-law distribution but in this case an exponential truncation is usually observed:
\begin{equation}
\mathcal{P}_\mathrm{off} \propto \tau^{-\mu} e^{-t/\tau_c}
\end{equation}
The estimation of the parameter $\mu$  gives important information on the blinking. It is possible to distinguish three different cases depending on the value of $\mu$:
\begin{itemize}
	\item[]$\mu <1$\\ in this case the average value and the variance of the
	distribution $1/\tau_\mu$ are not defined, as all the momenta of the
	distribution diverge;
	\item[]$1<\mu <2$\\ in this case, the mean value converges but not the
	variance;
	\item[]$\mu >2$\\ in this case, the mean value and the variance both converge.
\end{itemize}

Clearly the last case is the best one, while in the first case we cannot define a typical blinking time.

This kind of analysis is useful but it has some limitations. First of all, the bin size cannot be too small due to the need of having a good signal to noise
ratio and/or a finite temporal resolution of the detector. This fact imposes limits in the minimum time measured in the distribution and possible switching between the ``on'' and ``off'' states %\textbf{(what do you mean?)}
occurring in this time is averaged out. Another problem is that the result is
influenced by the choice of the bin size and by the threshold. It makes sense that the values of
$\mu$ and $\tau_c$ are affected by these choice. This dependence is more evident
when the ``on'' and ``off'' states are not well separated, while it is minimal
if they are separated.

When two different states are present in the blinking trace, setting a
threshold is also useful to separately  study the lifetime of each of the states.
This can be done when the acquisition card is able to record the arrival time of
each photon. In this case it is possible to make a post-selection on the lifetime
histogram events. We proceed as follows:
\begin{itemize}
	\item first of all, we set a threshold dividing ``on'' and
	``off'' states;
	\item we select only the events coming from one of the two states (i.e. the
	``on'' state);
	\item we create the histogram only taking in consideration the photons
	arrived when the emitter was in the selected state.
\end{itemize}
With the described procedure it is possible to separately obtain the decay rates of
the two states.

\subsubsection{Intensity auto-correlation function}
\begin{figure}
	\centering
	\includegraphics[width=0.7\linewidth]{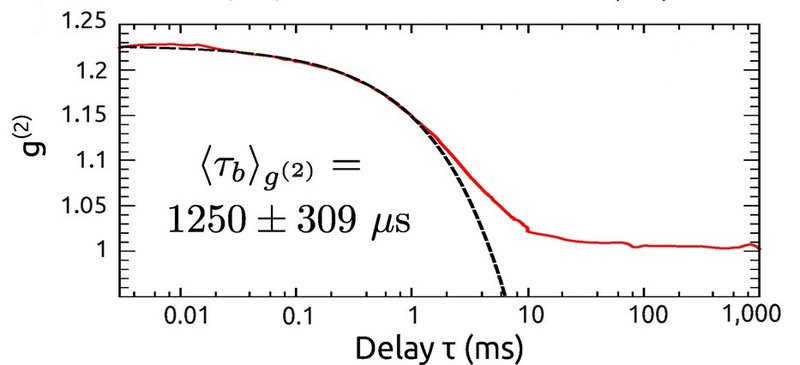}
	\caption{Example of \gd{} behavior as a function of the delay due to the
	blinking. The black dashed line is a fitting model using function~\protect\eqref{eq:g2blink}, while the continuous line is the
	experimental curve. \ccredit{manceau2018CdSe}}
	\label{fig:manceau2018g2}
\end{figure}
Another way to characterize the blinking properties is to use the intensity
autocorrelation function \gdt{} already described in
section~\ref{sec:sing_ph_em_theory}.
Due to the fluctuation of the intensity induced by the blinking, the emission is typically super-poissonian and shows photon bunching, except for low values of \gdz.
For timescales smaller than the exponential cutoff, it is possible to express the
\gdt{} function as~\cite{verberk2002Simple,verberk2003Photon,manceau2018CdSe}:
\begin{equation}
\label{eq:g2blink}
g^{(2)}(\tau)=B\left(1-A \tau^{1-\mu}\right).
\end{equation}
where $\mu$ is the highest exponent amongst $\mu_\mathrm{off}$ and
$\mu_\mathrm{on}$.

In equation~\eqref{eq:g2blink}, $B$ is the bunching value, i.e. the maximal value
reached by the \gdt{} (value that it takes at short timescales with respect to the characteristic blinking time). This model
does not take account the presence of photon antibunching (single photon
emission) that happens on much shorter times, i.e. on a nanosecond scale.
The $A$ coefficient is given by~\cite{verberk2003Photon}:
\begin{equation}
A=\frac{1}{\left\langle\tau_{b}\right\rangle} \frac{\tau_{\min
}^{\mu}}{\Gamma(2-\mu)}
\end{equation}
where $\mean{\tau_b}$ is the average duration of the bright ``on''-states,
${\tau_\mathrm{min}}$ is the minimal duration of an event and $\Gamma$ is the Euler
gamma function~\cite{euler1738progressionibus}.
By fitting the correlation function, we can obtain information on the parameters $B$, $A$ and
$\mu$. Unfortunately, the choice of the cutoff influences the result
of the fit for the last parameter. There are possibilities to overcome this problem,
one of them, as proposed by~\textcite{manceau2018CdSe} is to use the
information from the residual of the fit.

\subsubsection{Fluorescence lifetime  intensity distribution}
\begin{figure}
	\centering
	\subfloat[]{\label{fig:GallandA}\includegraphics[width=0.4\linewidth]
		{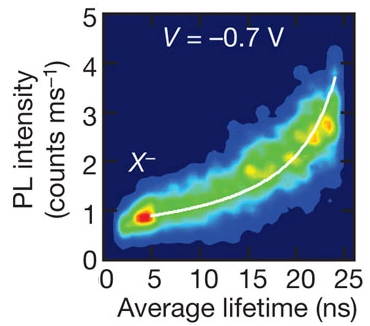}}\qquad\qquad
	\subfloat[]{\label{fig:GallandB}\includegraphics[width=0.4\linewidth]
		{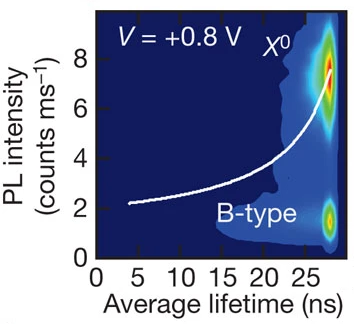}}
	\caption{FLID: \protect\subref{fig:GallandA}~in the presence of a type A
	blinking, the average lifetime depends on the emitted intensity;
		\protect\subref{fig:GallandB}~in the presence of type B
		blinking  the average lifetime does not depend
		on the emitted intensity.
		\ccredit{galland2011Two}}
	\label{fig:Galland}
\end{figure}
Often the signal is too low and the blinking too fast to allow the selection of
a single state: this was the case for the perovskite nanocrystals that I
studied and describe in the next chapter.

When it is impossible to select a single state in the blinking trace, it is more appropriate to talk about flickering \cite{park2015Room,galland2011Two}.
Usually, the study of this phenomenon is performed using the Fluorescence Lifetime Intensity
Distribution (FLID) introduced for the first time in~\citeyear{galland2011Two}
by \textcite{galland2011Two}. \label{sec:FLID}
This kind of study allows to distinguish between two different blinking types called A and B. As shown in figure~\ref{fig:Galland}, it is experimentally possible to distinguish between them as the A~type blinking presents a clear dependence with the emission intensity
and the fluorescence lifetime, while in B~type blinking the fluorescence
lifetime is independent of the emission intensity.

\textcite{galland2011Two} used this technique to study the blinking dependence
of the charged state, controlled with the use of a three electrode
electrochemical cell. This technique is useful also in neutral
conditions in order to identify the mechanisms involved in the blinking.

\begin{figure}
	\centering
	\includegraphics[width=0.7\linewidth]{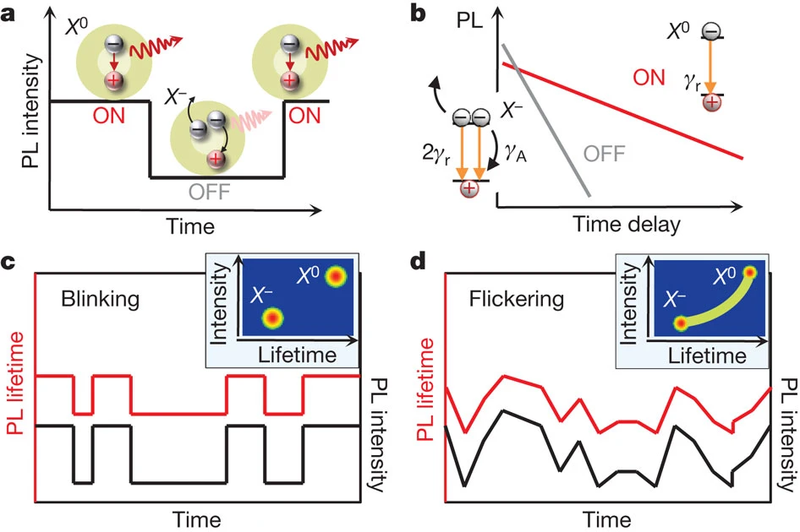}
	\caption{Mechanism of type A blinking: a)~the ``on'' periods corresponds to the
	neutral state, while the ``off'' period corresponds to the negatively charged excitation
	states. b)~Different mechanisms correspond to different measured lifetimes
	c)~in case of blinking between two well-separated states this behavior corresponds to two
	different spots in the FLID image, while (d)~in presence of flickering a
	curve is visible on the FLID
	\ccredit{galland2011Two}
	}
	\label{fig:mechanism-typea}
\end{figure}

\begin{figure}
	\centering
	\includegraphics[width=0.25\linewidth]{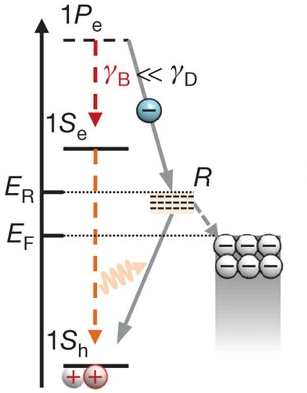}
	\caption{Mechanism of type~B blinking: The ``off'' periods are due to the
	presence of a relaxation center R that captures the electrons with a
	rate~$\gamma_D$ higher than the intraband radiative rate~$\gamma_r$.
	\ccredit{galland2011Two}}
	\label{fig:mechanism-typeb}
\end{figure}

Physically, different kinds of blinking correspond to different relaxation
mechanisms.
In the case of type~A blinking, the ``on''-to-``off'' switch happens when a carrier is
moved to a trap state: this trapping usually happens via thermal ionization or
Auger-assisted photo-ionization. The opposite process, ``off''-to-``on'' switch, happens when
the carrier is released, typically via a relaxation process. This is shown in
figure~\ref{fig:mechanism-typea}.
It goes differently for the case of type~B blinking as the intensity fluctuation is due to a fast trapping of high energy electrons followed by a non-radiative
recombination of the electron and the hole. The luminescence is decreased due
to this process, but it does not affect the lifetime of the emission as the
trapping process is not in direct competition with the radiative recombination
from the band edge. This mechanism is reported in
figure~\ref{fig:mechanism-typeb}.

\subsection{Polarization}
The polarization of a light beam is defined as the direction of oscillation of
the electrical field and lays always on the plane orthogonal to the propagation
direction. When the polarization changes are random we have non-polarized
light. In this case, even if in each point of the space and in each moment of
time, the polarization state is clearly defined, there is no relation with the
polarization state that the light will have in the same point in a subsequent moment.
This is the case for sunlight or in the case of incandescent lamps.

For various applications, it is more useful to work with polarized light, that is
light in which the polarization state follows a well defined temporal evolution, i.e. it is always on the
same direction as in case of  linearly polarized light. In order to describe such polarized light,
we can use three different parameters: the two transverse components of the electromagnetic field and
their relative phase. In the case where the light is not completely polarized, it is
necessary to add the degree of polarization. This way we obtain four
parameters, called the Stokes parameters. \label{sec:stokes}

\begin{figure}
	\centering
	\subfloat[]{\label{fig:1polell}\includegraphics[height=0.3\linewidth]
		{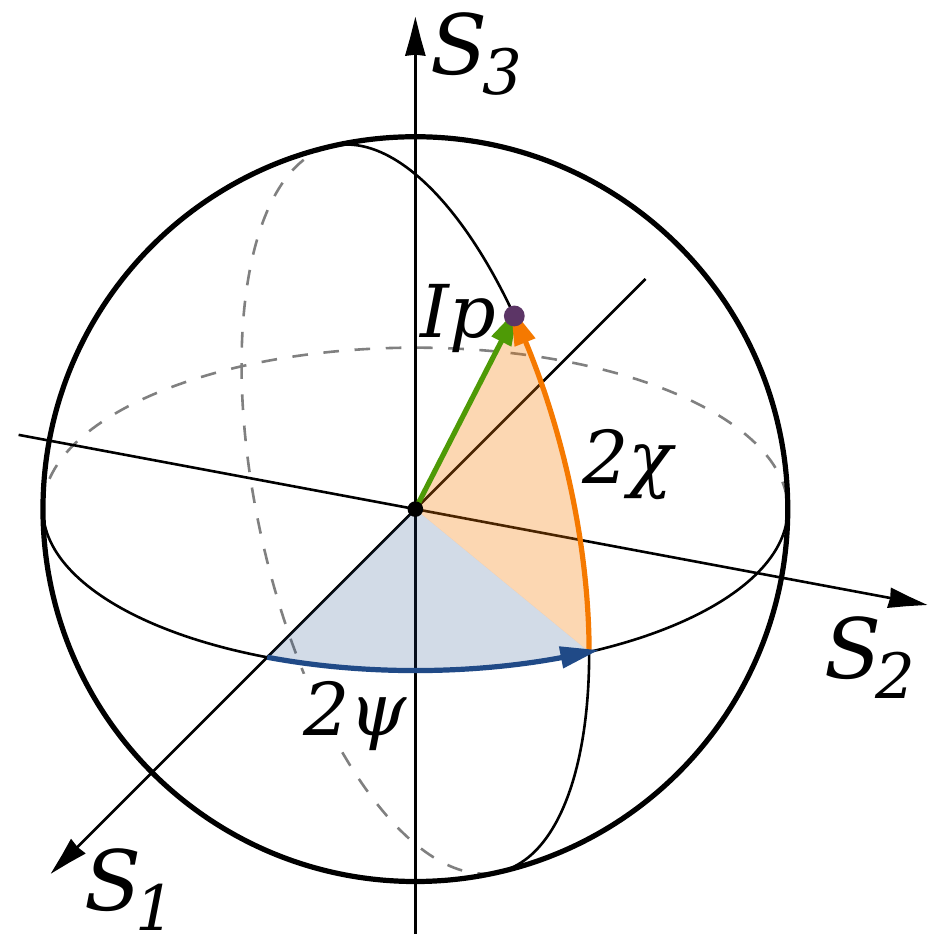}}\qquad\qquad
	\subfloat[]{\label{fig:1polsphere}\includegraphics[height=0.3\linewidth]
		{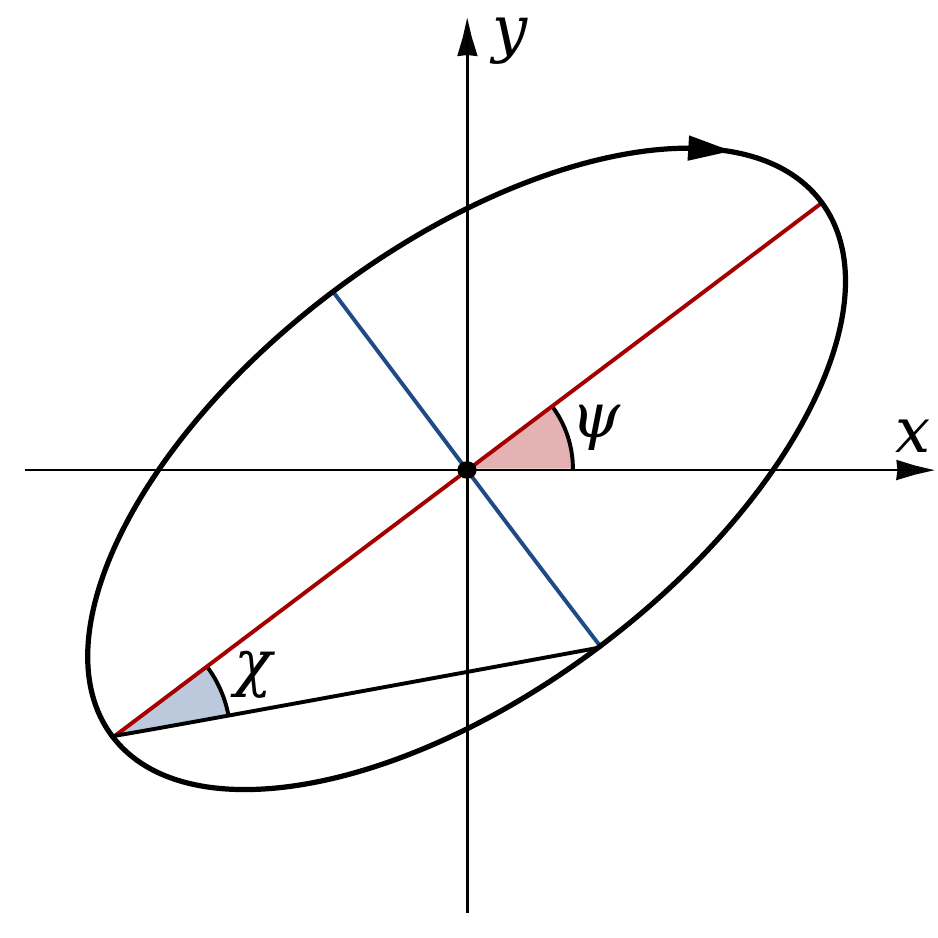}}
	\caption{Different graphical representations of the polarization state of light.
	\protect\subref{fig:1polell}~Poincare sphere: the polarization is
	represented as a point on a sphere, the coordinates of the point are in
	relation with the Bloch vector following the
	equations~\protect\eqref{eq:sphere2stokes}.
		\protect\subref{fig:1polsphere}~Polarization ellipse: two dimensional
		visualization of the polarization, easier to report of a flat support
		than the Poincare sphere it contains the same amount of information.
		\credit{Wikimedia Common}}
	\label{fig:1graphpol}
\end{figure}

There are several equivalent ways to represent the polarization of a
light beam. The most commonly used are the following three: the polarization ellipse, the Poincare
sphere and the Stokes parameters.
\begin{description}
	\item[\textbf{Stokes parameters}:]
	they allow to give a quantitative way of describing  the state of
	polarization. Usually they are indicated by $S_0$, $S_1$, $S_2$, $S_3$. They
	represent respectively:
	\begin{itemize}
		\item[$S_0$] is the intensity,
		\item[$S_0$] is the degree of horizontal/vertical polarization,
		\item[$S_2$] is the degree of polarization at $\pm \SI{45}{\degree}$,
		\item[$S_3$] is the degree of circular polarization.
	\end{itemize}
	 It is convenient to consider a monochromatic light propagating along the z axis with the two components of the electric field given by:
	\begin{subequations}
	    \begin{align}
	        E_{x}(z, t)&=E_{0 x} \cos \left(\omega t-\kappa z+\delta_{x}\right), \\
            E_{y}(z, t)&=E_{0 y} \cos \left(\omega t-\kappa z+\delta_{y}\right).
        \end{align}
	\end{subequations}
	In this case, the Stokes parameters, in terms of components of the electric field, are given by:
	\begin{subequations}
	  \begin{align}{l}
        S_{0}&=E_{0 x}^{2}+E_{0 y}^{2}, \\
        S_{1}&=E_{0 x}^{2}-E_{0 y}^{2}, \\
        S_{2}&=2 E_{0 x} E_{0 y} \cos( \delta), \\
        S_{3}&=2 E_{0 x} E_{0 y} \sin( \delta).
    \end{align}
	\end{subequations}
For the case of fully polarized light the following relation is valid:
	\begin{equation}
		S_0^2=S_1^2+S_2^2+S_3^2
	\end{equation}
	\item[\textbf{Poincare Sphere}:] the polarization state of the light can be expressed in spherical coordinates and is given in this case by a point on the so-called
	Poincare sphere. To obtain the coordinates of a point given the Stokes
	parameters we can use the following relations:
	\begin{equation}
	\label{eq:sphere2stokes}
		\begin{aligned}
			I &=S_{0} \\
			p &=\frac{\sqrt{S_{1}^{2}+S_{2}^{2}+S_{3}^{2}}}{S_{0}} \\
			2 \psi &=\arctan \frac{S_{2}}{S_{1}} \\
			2 \chi &=\arctan \frac{S_{3}}{\sqrt{S_{1}^{2}+S_{2}^{2}}}
		\end{aligned}
	\end{equation}
	\item[\textbf{Polarization ellipse}:] this representation is useful as it gives an easy way
	to represent graphically the polarization. As opposed to the Poincare
	Sphere, it is a two-dimensional representation, easier to report on a screen
	or on paper. The polarization is represented as an ellipse. When the  two
	semi-axis have the same length (then the ellipse coincides with a circle) the polarization is circular, when one semi-axis is equal to zero the polarization is linear (thus the ellipse is simply a segment).
\end{description}
The relation between the different graphical representations is shown in
figure~\ref{fig:1graphpol}.

Working with single photons, the polarization is an important property, especially when considering whether two photons are indistinguishable which means that they need to have the same polarization state. For this reason, for some
quantum applications, it is important to obtain  polarized single
photons. Working with quantum dots as single photons emitters, their
polarization properties are influenced by the  asymmetries in their structure
or in their shape. For example, \ch{CdS}/\ch{CdSe} dot in rods have a linearly polarized
emission due to the elliptical shape of the shell and, subsequently, of the
core~\cite{pisanello2010Dots,pisanello2010Room}. Nitrogen-Vacancy defects in nanodiamonds on the other hand have an
asymmetrical molecular structure due to the defect itself, that results in partially linear polarized emission\cite{kaiser2009Polarization}.

\section{Guided and integrated optics}
The advantages of guiding electrical fields and the enormous development
and amount of innovation that the ability to guide and control the electrical field has brought is quite clear. Firstly, the use of conductors in controlling it has given the possibility to use its energy for many different applications. This affected our life at the point that nowadays it is almost impossible to imagine a world without this technology. What followed was the ability to transport and manipulate information using an electrical field and it has given rise to  electronics and yet again changed the world with a revolution that is not yet finished. In the last two centuries, the life of people in the world has changed as never before.

It would seem that photonic applications are following the same path. They also contribute to changing our life as the electric and electronic revolutions did. The ability to transport light was initially used in medical applications (as explained in chapter~\ref{chap:fiber}), but it has recently shown its potential to control and manipulate information alongside electronics.
The world has been connected by optical fibers and it has turned out to be an efficient way to transport information: this has increased the necessity to create integrated optical circuits and connections.

\begin{figure}[tb]
	\centering
	\subfloat[]{\label{fig:bancoOttico}\includegraphics[height=0.4\linewidth]
		{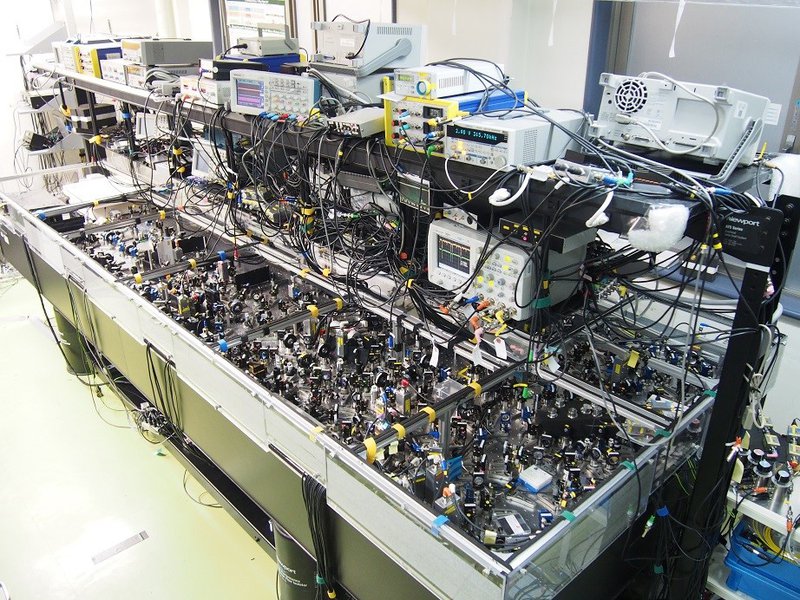}}\qquad
	\subfloat[]{\label{fig:chip}\includegraphics[height=0.4\linewidth]
		{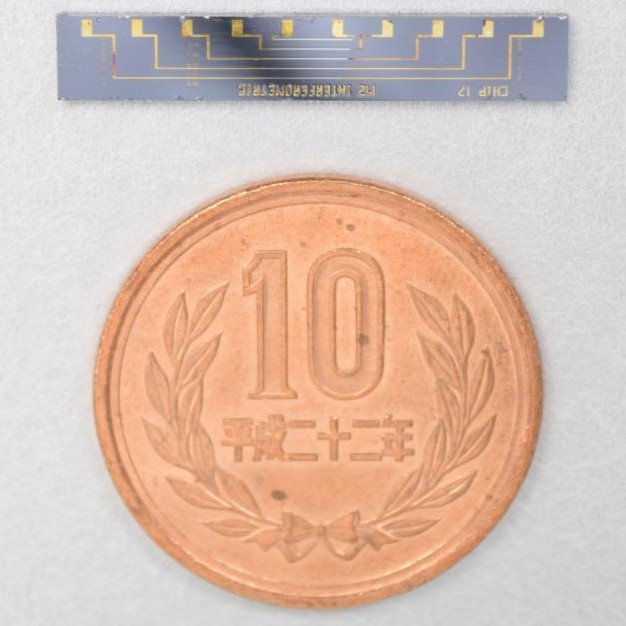}}
	\caption{\label{fig:bancoOtticovsChip}
		Difference between freespace~\protect\subref{fig:bancoOttico} and
		integrated~\protect\subref{fig:chip} optics. The free-space optics is flexible and historically well known, but requires fine alignment and wide space. On the contrary, integrated optics cannot be modified one fabricated, but it is small, stable and does not risk misalignment.
	}
\end{figure}

It is evident that for industry and practical applications, the integrated approach has some clear advantages over free-space optics:
\begin{itemize}
    \item it is stable and not affected by misalignment. It does not need to be realigned after temperature changes
    \item it is compact
    \item it can be industrially replicated, once the project made, in multitude of identical parts
\end{itemize}
These advantages are evident as shown by figure~\ref{fig:bancoOtticovsChip}.

In the domain of quantum information technologies, photons are one the most promising carriers of  information. As an example, for quantum key distribution, the information is usually encoded in the quantum state of a photon.
There is a clear interest in coupling single photons to guided and integrated optics.
Multiple approaches are available to reach this goal, the most classical one is to use a microscope objective to couple photons directly into an optical fiber.

During my Ph.D. I had the occasion to work on a different approach,  potentially more scalable: the near field coupling.

Multiple guided structures have been proposed in the last few years to be used for quantum technologies. We can classify them in three different groups~\cite{thylen2014Integrated}:
\begin{description}
\item[\textbf{Dielectric waveguides}:] in this category, we count all the structures that guide the light using total internal reflection. The light is guided in a medium with a higher refractive index with respect to the surrounding medium. The most important and known members of this group is the optical fiber, but other members, such as \IEW{}s (described in the last chapter) are part of this group.

\item[\textbf{Photonic crystal waveguides}:] in this kind of guides the light is confined thanks to the periodicity of the structure. Interestingly, this enables a sharp bending of the guide~\cite{engelen2006Effect, dutta2016Coupling} resulting in a slow group velocity for certain frequencies that can give origin to strong light-matter interactions. In addition, they have a narrow bandgap that allows the propagation of some selected frequencies only.

\item[\textbf{Surface plasmon polariton waveguides}:] these guides use the coupling of the electromagnetic field with the electron plasma of the conductor. The field is thus confined at a dielectric-conductor interface. They are interesting as they enable high confinement of the electromagnetic field~\cite{fang2015Nanoplasmonic}.
\end{description}
\begin{figure}
	\centering
	\subfloat[]{\label{fig:photonicwave}\includegraphics[width=0.45\linewidth]
		{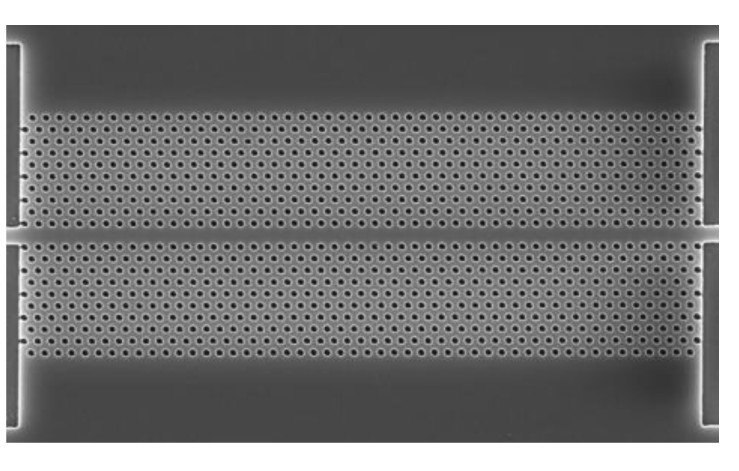}} \qquad
	\subfloat[]{\label{fig:plasmonwave}\includegraphics[width=0.45\linewidth]
		{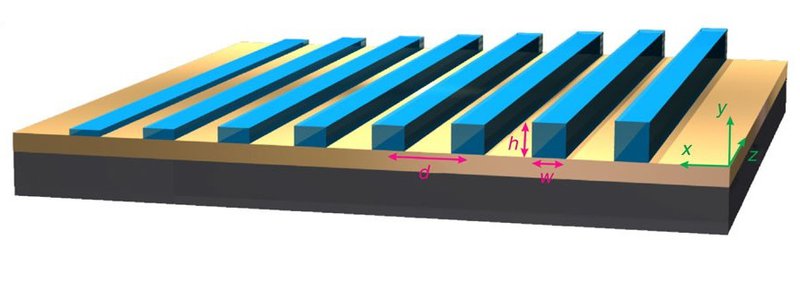}}
	\caption{\label{fig:wavegude_coparison}
		 Sanning electron microscopy image of a photonic crystal waveguide array~\protect\ref{fig:photonicwave} \ccreditnp{garcia-ruperez2010Labelfree} and scheme of an array of surface plasmon polariton waveguides (PMMA on gold surface)~\protect\ref{fig:plasmonwave} \ccreditnp{block2014Bloch}.
	}
\end{figure}
%todo mettere un immagine per ciascuna delle tre, eventualmente

We concentrated this PhD work on the study on two different kinds of dielectric waveguides. The main part of my work was dedicated to the optical tapered nanofibers, described in detail in chapter~\ref{chap:fiber}. I also studied the \IEW{} that could overcome some nanofiber limitations: they are described in chapter~\ref{chap:outlook}.

\clearpage
\begin{chrecap}
	In this chapter I presented the main concepts useful for understanding my PhD work:
	\begin{itemize}
		\item I introduced the basic concepts of quantum mechanics and explained why it is interesting
		to study the possibility of encode information in quantum states. % \textbf{(this sentence makes no sense)}.
		\item I introduced photons and the interest to use them in such
		applications.
		\item I defined a single photon emitter, explaining how it is possible to
		recognize its emission and to measure it.
		\item I defined the polarization of light and explained the notation to identify
		it.
		\item Single photon emitters can be of several kinds, each with specific
		properties.
		%I have shown them and in particular I focused on the
		%characteristics of colloidal quantum dots as single photon emitters.
		\item In the characterization of nanoscale single photon emitters
		properties, it is important to take into account multiple properties:
		\begin{itemize}
			\item their saturation intensity can give information on the
			mechanisms that generate the single photon emission;
			\item their fluorescence lifetime gives information on their
			emission rate, on the transition involved in the emission and on the state of the emitter;
			\item their blinking statistics gives information on the process
			that generates the emission.
		\end{itemize}
		\item The previous mechanisms are also important as they greatly influence
		the possible applications that a single photon emitter can have in the
		future
		\item I have briefly introduced the interest of an integrated optics approach and some of the structures that have been proposed for this approach.

	\end{itemize}
\end{chrecap}
\chapter{Perovskite nanocrystals}
\minitoc
\label{chap:perovskites}
\section{Introduction}
\label{sec:perov_intro}
\begin{figure}[tbh]
	\centering
	\subfloat[]{\label{fig:Naturalperovskite}\includegraphics[width=0.3\linewidth]
		{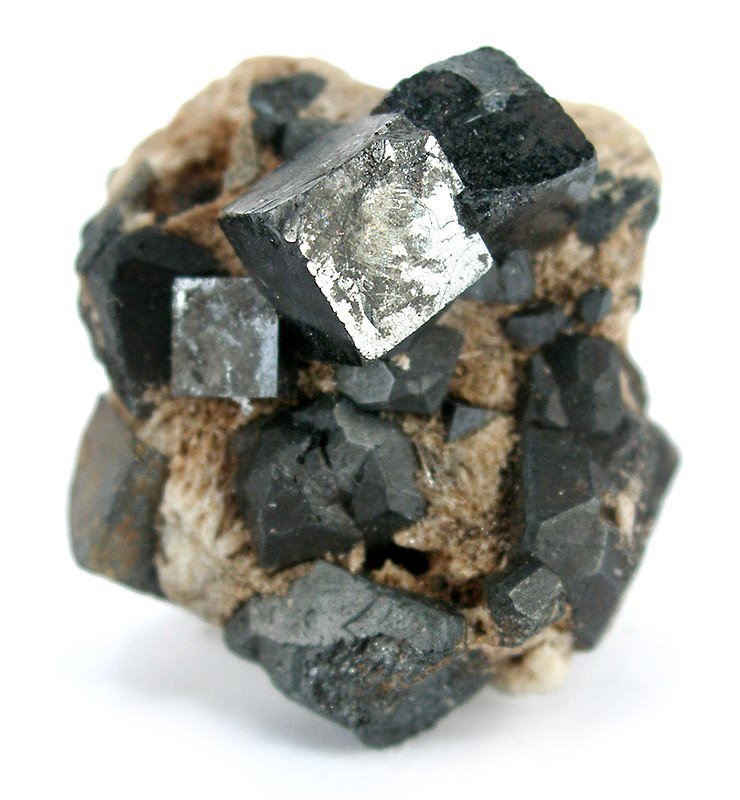}}\qquad\qquad
	\subfloat[]{\label{fig:PerovStruct}\includegraphics[width=0.3\linewidth]
		{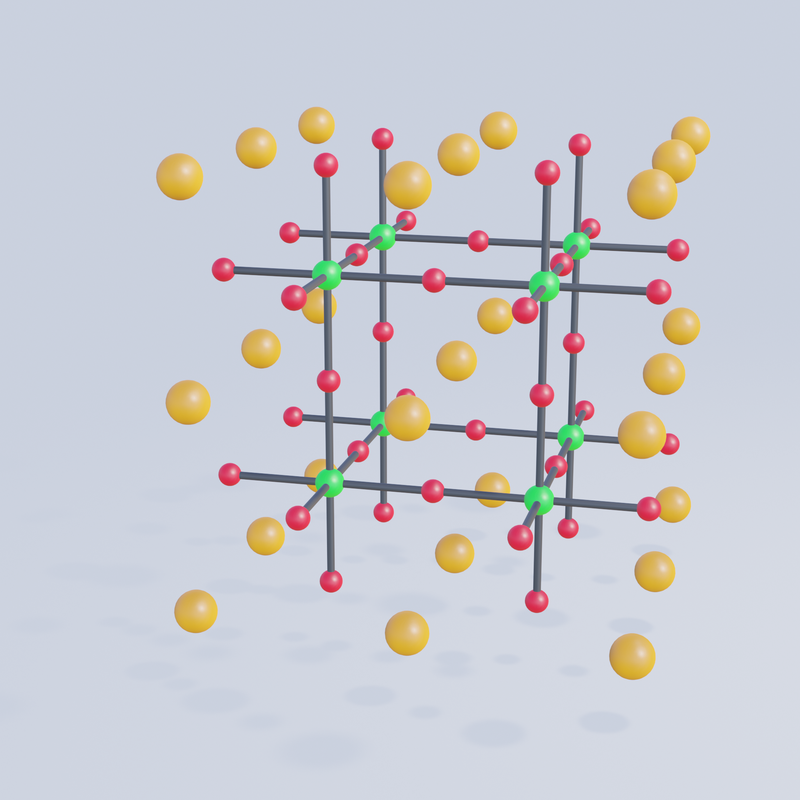}}
	\caption{\protect\subref{fig:Naturalperovskite}~Crystals of perovskites in mineral
	matrix %\textbf{(CC: what does that mean 'on matrix', not english)} 
	from Magnet Cove, Arkansas, USA.
	The cubic structure, particularly unusual for a natural compound, is
	clearly visible. \textit{(credits Wikimedia Commons)}
	\protect\subref{fig:PerovStruct}~3D Structure of a perovskite
	(\ch{ABX3}): in red the \ch{X} atoms, cations, in green the B atoms and in
	yellow the A atoms. Both A and B are metal cations, with B atoms smaller
	than the A ones.}% \textbf{(CC: problem here no?)}.}
\end{figure}

Historically ``Perovskite'' was the name given to the mineral  Calcium
Titanium Oxide (\ch{CaTiO3}), discovered in the Ural mountains in 1893. It was named after Lev Perovski, a Russian mineralogist\cite{katzPerovskitea}. This
mineral (represented in Figure~\ref{fig:Naturalperovskite}) had the peculiarity to present
a particular cubic structure. %The first description of its structure is

In the following the name Perovskite is given to a class of materials with a
structure similar to Calcium Titanium Oxide, most of which nowadays are
synthetic. This structure, generally represented as \ch{ABX3}, is indeed
peculiar to many oxides and usually takes the form \ch{ABO3}. They all present
a structure shown in Figure~\ref{fig:PerovStruct}. The X atom is an anion,
while both A and B are metal cations with A bigger than B.

In most of the perovskite natural compounds the anion X is oxygen:
particularly \ch{(Mg,Fe)SiO3} and \ch{CaSiO3}, known as Silicate Perovskite are
largely present in the lower part of the Earth mantle.

\subsection{Interest for perovskites}
The particular structure of Perovskites give them peculiar characteristics that
have been studied in different physical aspects.

It was first discovered that perovskites shows colossal magnetoresistance, that is
the capability of a material to dramatically change their resistance in
presence of a magnetic field\cite{jonker1950Ferromagnetica}.
Perovskites-like ceramics are used in superconductive applications, enabling to
reach relative high superconductive temperature. What is particularly remarkable was
the discovery in 1986 of the possibility that some of these materials have a
critical temperature above \SI{90}{K}.

From an optical point of view, perovskites have been studied for
solar-cell applications. The first report of a perovskite-based solar cell was
published in 2012\cite{chung2012Allsolidstate} using \ch{CsSnI3} and since then
they have reached remarkable efficiencies, up to
$24\%$\cite{deschler2019Perovskite}. Despite the high efficiency, the main
problem of the stability is still open, as the perovskites undergo degradation if
exposed to moisture or light. Meanwhile, other possible uses of
perovskite semiconductors have emerged: particularly the possibility to produce
perovskite nanocrystals has opened the possibility to use them as quantum dots
in quantum optics applications.

\subsection{Perovskite nanocrystals}
Quantum dot nanocrystals are interesting to study for quantum optics. They
allow  to fabricate single photon sources, working
at room temperature and they can be used deposited on a glass plate or
integrated into different photonic structures such as
nanowires\cite{geng2016Localised} or nanofibers\cite{shafi2020Occurrence}.

Perovskites nanocrystals maintain the advantage to be easy to fabricate and add
the possibility to tune their emission playing on
their shape and on their composition: this is a main advantage in comparison
with other kinds of colloidal quantum dots such as \ch{Cd/CdSe} quantum dots, 
that can also be tuned but can hardly reach ranges outside the visible one.
Their main limitation is the optical stability but with different approaches we
can obtain good improvements on this field, as I will explain later.

%TODO add? Exciton Fine Structure in Perovskite Nanocrystals

\subsubsection{Level structure in perovskite nanocrystals}
The energy level structure in perovskite nanocrystals is currently under experimental and theoretical investigations. 
The level structure of the Brillouin zone of a \ch{CsPbBr3} nanocrystal is shown in figure
\begin{figure}
    \centering
    \includegraphics[width=0.6\linewidth]{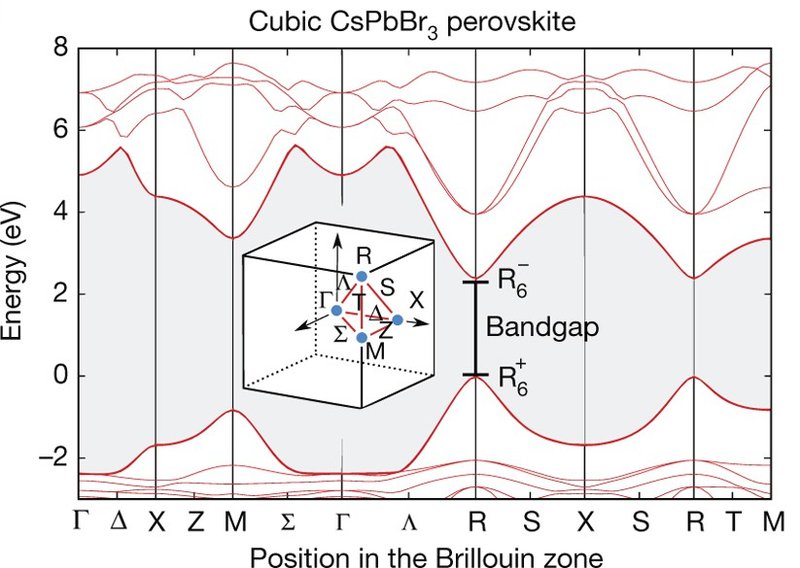}
    \caption{Band structure of \ch{CsPbBr3} perovskite, calculated by~\textcite{becker2018Brighta}. In the inset, the first Brillouin zone of a cubic crystal lattice is visible. \ccredit{becker2018Brighta}}
    \label{fig:my_label}
\end{figure}

From the most recent results obtained on this topic, the perovskite level structure seems to differ from other kinds of emitters such as \ch{CdS}/\ch{CdSe} quantum dots. Even if the fact is still debated, some studies claim that perovskites have a bright ground state~\cite{sercel2019Exciton,becker2018Brighta} which seems to be due to the Rashba Effect. Moreover, they have a momentum dependent splitting of the spin that can be derived using the perturbation theory and results in an additional term in the Hamiltonian of the form:
\begin{equation}
    H_{\mathrm{R}}=\alpha(\boldsymbol{\sigma} \times \mathbf{p}) \cdot \hat{z}
\end{equation}

where $\boldsymbol{\sigma}$ is the Pauli matrix vector and $\alpha$ is the Rashba coupling.
The Rashba effect influences only the Hamiltonian in the case of a weak confinement, while in the case of a strong confinement, it does not~\cite{sercel2019Exciton}.
Finally, the hyperfine structure is influenced by the size of the nanocrystal and by the medium in which the nanocrystal is immersed~\cite{sercel2019Exciton}.

\subsubsection{Perovskite nanocrystals as single photon emitters}
Single photon emission from perovskite nanocrystals has been reported firstly
by \textcite{park2015Room} in 2015 with an experiment at
room temperature. They describe single photon emission from perovskite
nanocubes made of \ch{CsPbBr3}, \ch{CsPbI3} and \ch{CsPbBr_xI_{(3-x)}}. In
particular, the last compound, where $x$ can be any positive rational number
smaller than~$3$, showed the flexibility of these materials.
The emission spectra are reported in Figure~\ref{fig:parkSpectra}:
\begin{figure}[tbh]
	\centering
	\subfloat[]{\label{fig:parkSpectra}\includegraphics[height=0.3\linewidth]
		{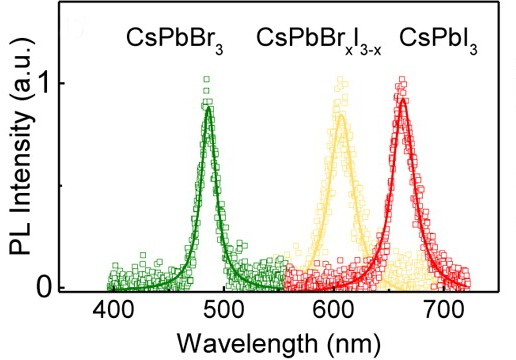}}\qquad\qquad
	\subfloat[]{\label{fig:park_g2_pulsed}\includegraphics[height=0.3\linewidth]
		{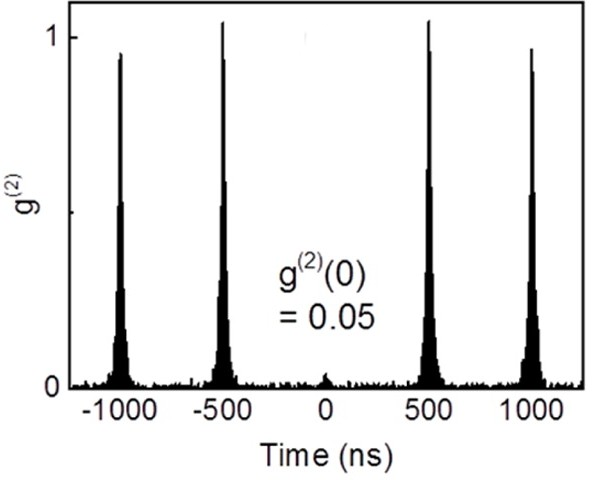}}
	\caption{\protect\subref{fig:parkSpectra}~Different emission spectra with
	different composition of perovskite nanocrystals.
	\protect\subref{fig:park_g2_pulsed}~$g^{(2)}\of{\tau}$ measurement of light
	emitted from a single perovskite nanocrystal, excited with a pulsed laser.
	\textit{(credits: \textcite{park2015Room}).}
	}
\end{figure}
it is clearly visible that changing the composition of the perovskites will modify the
emission wavelength. In Figure~\ref{fig:park_g2_pulsed}, we can
see that the autocorrelation function $g^{(2)}\of{0}$ of the emitted light is very low and thus these
nanocrystals are very good single photon emitters.

Given these facts it is not difficult to understand that the community has a big
interest in this kind of materials for single photon applications and that
multiple studies have analyzed their emission from a quantum optics point
of view in the last few years.

\subsubsection{Blinking properties}
\label{sec:blinkingPark}
\begin{figure}[tbh]
	\centering
	\includegraphics[%width=0.7\linewidth,
	height=0.3\linewidth]{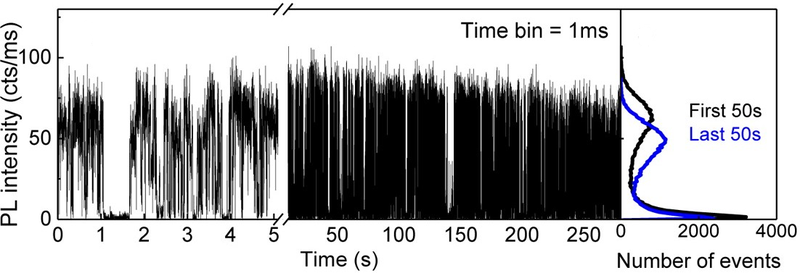}
	\caption{On the left, the luminescence trace of a single perovskite
	\ch{CsPbI3}
	nanocrystal: it is clearly possible to distinguish two different states, a
	bright one and a dark one. The binning is \SI{1}{\milli \second}. On the right
	the histogram of the first and the last \SI{50}{\second}, showing a degradation of
	the signal. \ccredit{park2015Room}
	}
	\label{fig:park_blinktrace}
\end{figure}

Like other kinds of quantum dots, these emitters do blink. From what is reported in
\textcite{park2015Room}, the blinking is the intermittency of the intensity luminescence  between a dark and a gray state: this is evident from the luminescence of a single \ch{CsPbBr3}
nanocrystal trace reported in
Figure~\ref{fig:park_blinktrace}~\cite{park2015Room}.
In this case the signal is binned every \SI{1}{\milli \second} and it is
possible to distinguish in this case between two different emission levels
normally called bright and dark states.

As already shown, the blinking is normally due to two or more different states
and can be categorised in type A and B blinking.
To have a deeper understanding of this blinking behavior, it is possible to
look at the FLID %\textbf{(CC: not defined)} 
images: here the relation between the blinking and the
intensity emission is evidenced, as explained in
section~\ref{sec:FLID} at page~\pageref{sec:FLID}.

\begin{figure}[tbh]
	\centering
	\includegraphics[width=1\linewidth]{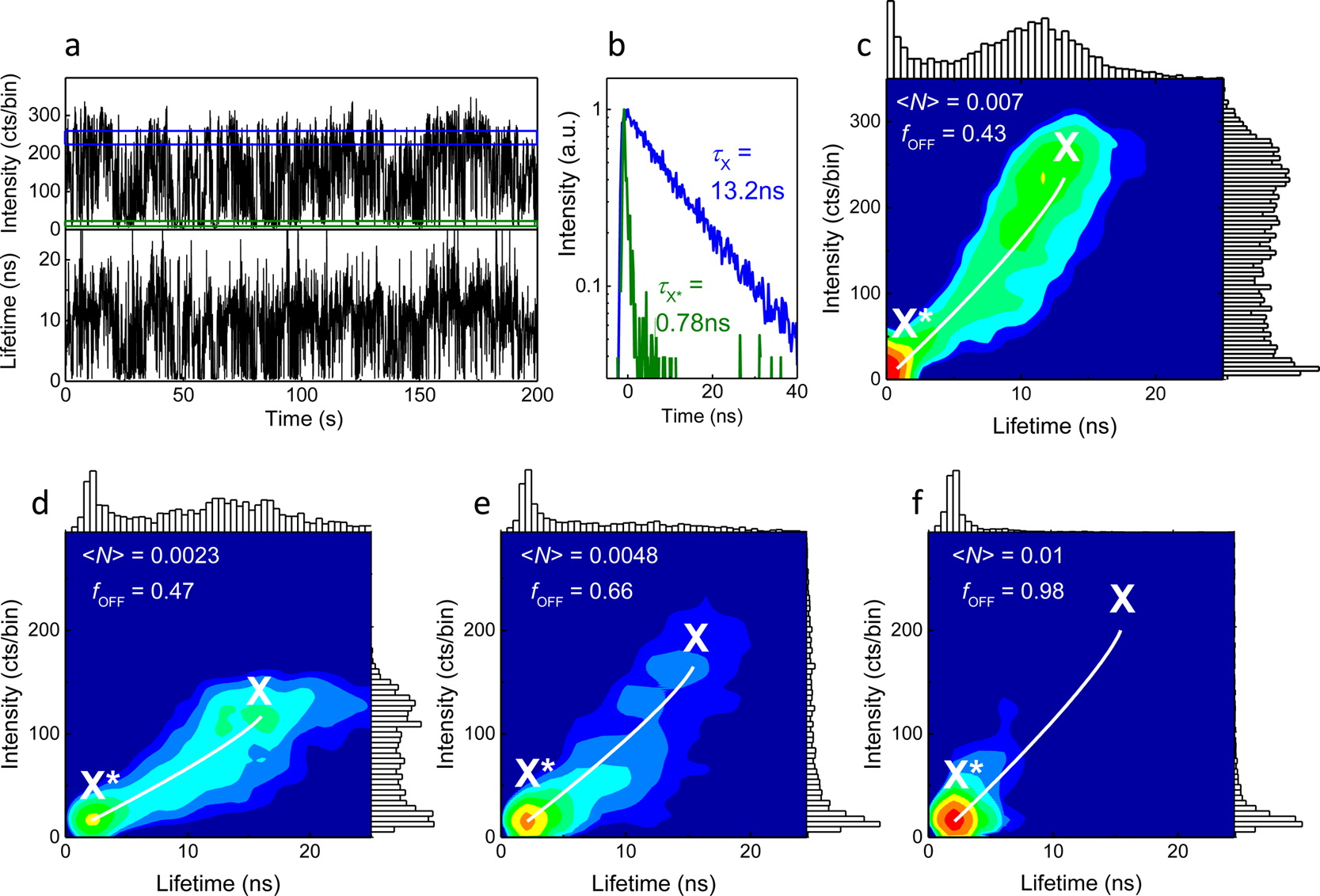}
	\caption{
		a)~Intensity trace of a single perovskite nanocrystal excited with a
		pulsed laser
		b)~Lifetime histogram of the neutral (blue) and charged (green) states
		realised using photons in bins between, respectively, the blue and green
		lines of panel~a.
		c-f)~FLID images obtained at different excitation powers, showing the
		increasing prevalence of the charged (``off'') state increasing the
		mean number of excitons created in each pulse of the laser
		$\avr{N}$.
		\ccredit{park2015Room}}
	\label{fig:parkflid}
\end{figure}

In figure~\ref{fig:parkflid} we can see a measurement of this type performed
by the authors of~\cite{park2015Room}. In the first panel
(figure~\ref{fig:parkflid}a) it is possible to see the correlation between the blinking trace (upper
part), obtained in this case with a bin size of~\SI{50}{\milli \second} and with a
pulsed laser, and the evolution of the mean lifetime (lower panel). We know the lifetime indicates the delay from the moment when the emitter is excited and the moment when a photon is emitted by spontaneous emission. The mean lifetime in this case is the arithmetical mean of the lifetimes of all the excitation-collection events registered in a certain time interval, corresponding to the bins of the histogram.
In this case, as the binning is larger than the
blinking characteristic time, we talk about flickering.
The authors supposed two different states to be present. By only taking the
photons arrived in the lower part of the intensity trace (between the green and
the blue lines respectively), they created an histogram to deduce the lifetimes
of the two different states: the neutral one, with a lifetime of
$\tau_x=\SI{13.6}{\nano \second}$, and the charged one, with a lifetime of
$\tau_{x^*}=\SI{13.6}{\nano  \second}$. This histogram is represented in
figure~\ref{fig:parkflid}b. They then reported the FLID images for different
excitation powers, using $\tau_x$ and $\tau_{x^*}$ to deduct the state that is
mostly probable as a function of the mean number of excitons created in each
pump pulse, indicated by $\avr{N}$. In Figure~\ref{fig:parkflid}c-f we can
see
that the probability for the emitter to be in the ``on'' state decreases as they
increase the excitation power, having with $\avr{N}=0.01$  $98\%$ of
probability
to find the emitter in the off state.

In addition to this analysis, looking at Figure~\ref{fig:parkflid}, we can
deduce that the emission lifetime depends on the emission intensity. This is evident from the FLID images and tell us that we are in the presence of a
type~A blinking. %TODO add reference to TypeA blinking

\subsubsection{Stability of the nanocrystals}
Unfortunately there are some limitations that make difficult to use such
emitters for practical applications. The main one is the stability as underlined for solar-cell applications. Perovskites are degraded
by moisture and light and this is true even more for nanocrystals, which have
by definition a big surface/volume ratio.

It has been observed that by exciting perovskite nanocrystals, their
structure progressively degrades, starting from the external layers. This
can be seen looking at the spectrum of the emitted light that, during the
measurement, is blue-shifted: the easiest interpretation that we can do, is
that the crystals progressively become smaller and the confinement increases,
decreasing consequently the emission wavelength.

\begin{figure}[tbh]
	\centering
	\includegraphics[width=0.6\linewidth]{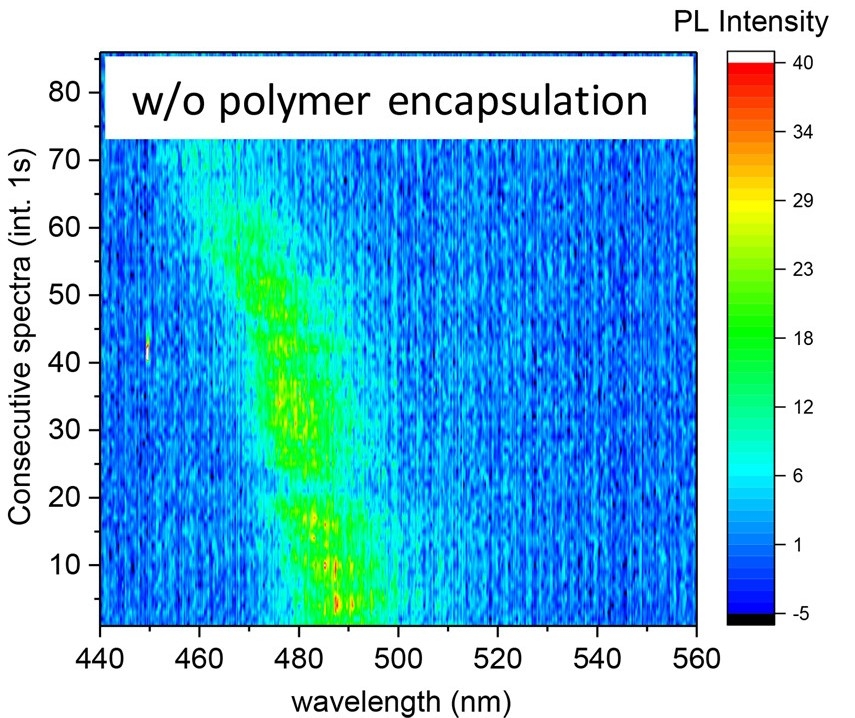}
	\caption{Degradation of the emission spectra of a single perovskite
	nanocrystal. The green trace shows the evolution of the emitted signal
	during time. We observe a blueshift due to the degradation of the crystal.
	\ccredit{raino2019Underestimated}}
	\label{fig:rainoundernopolimerspectra}
\end{figure}

This can be seen in Figure~\ref{fig:rainoundernopolimerspectra}: the central
emission wavelength of the emitter decreases by \SI{30}{\nano \meter} after
around one minute of light illumination.
It is possible to see that while the central emission wavelength decreases, the emission intensity decreases as well. At the end, the emitter becomes completely dark, as chemical reactions have irreparably modified its structure: this phenomenon of permanent loss of the emission is commonly called bleaching. 
An interesting study on this aspect has been performed
by~\textcite{raino2019Underestimated}.
\begin{figure}[p]
	\centering
	\includegraphics[width=\linewidth]{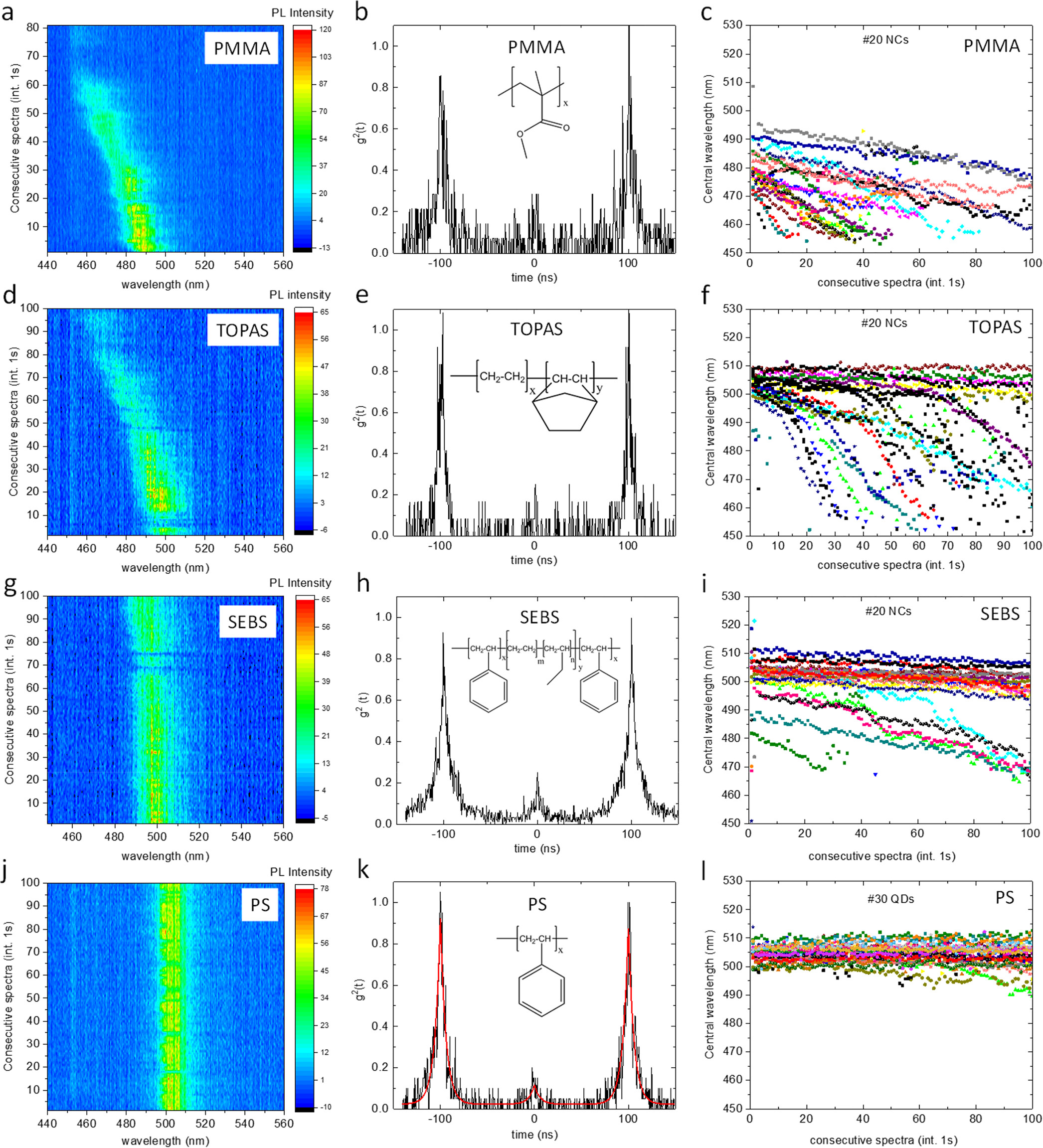}
	\caption{Resistance of single photon perovskite nanocrystals protected by
		different polymers (Poly(methyl methacrylate) (PMMA), TOPAS,
		styrene-ethylene-butylene-styrene block copolymer (SEBS), Polystyrene
		(PS)). It is possible to see that the latter shows the best performance
		in
		protecting the emitter, as the emission drift is almost completely
		removed.
		\ccredit{raino2019Underestimated}}
	\label{fig:rainounderestimated}
\end{figure}
The idea is thus to reduce the bleaching
by encapsulating the emitters in a polymer matrix, to protect them from the
combined action of moisture and light. To this end \citeauthor{raino2019Underestimated} tried different
polymers. The result of their study is shown in
figure~\ref{fig:rainounderestimated}. They selected and test five different polymers, among the mostly used polymers:
\begin{description}
	\item[\textbf{PMMA}] poly(methyl methacrylate), also known as acrylic,
	acrylic glass, or plexiglass, is a polymer with a refraction-index similar
	to the
	glass one;
	\item[\textbf{TOPAS}] a commercial cyclic olefin copolymer;
	\item[\textbf{SEBS}] styrene-ethylene-butylene-styrene block copolymer;
	\item[\textbf{PS}] polystyrene, an aromatic polymer that is one of the most
	used
	plastics.
\end{description}
PMMA polymer is know for its good performance in replacing glass and it is commonly used as matrix for nanocrystals. Despite this, polystyrene provided better performances for this application. This is due, according to the authors, to the
different binding it forms with the ligands that are at the surface of the perovskite itself.

\section{Experimental setup for nanocrystal characterisation}
To characterize our nanocrystals we prepare the samples via spin-coating
deposition over a glass plate and we use an inverted
microscope to perform the optical analysis. In this section I will discuss the experimental
procedure and describe the scheme of the setup.

\subsection{Perovskite fabrication}
All the perovskite nanocrystals I studied were fabricated by Emmanuel Lhuillier
at the ``Nanoscience Institut of Paris'' (INSP). I have studied perovskite
nanocubes fabricated by two different ways which both required only few hours. This is one of the main advantages that perovskites share with other
colloidal nanocrystals, with respect to other kind of single photons emitters, such as
\ch{SiV} defects in nanodiamonds, whose fabrication is longer and more
complex.
\begin{figure}[tb]
	\centering
	\subfloat[]{\label{fig:SEM_A}\includegraphics[height=0.42\linewidth]
		{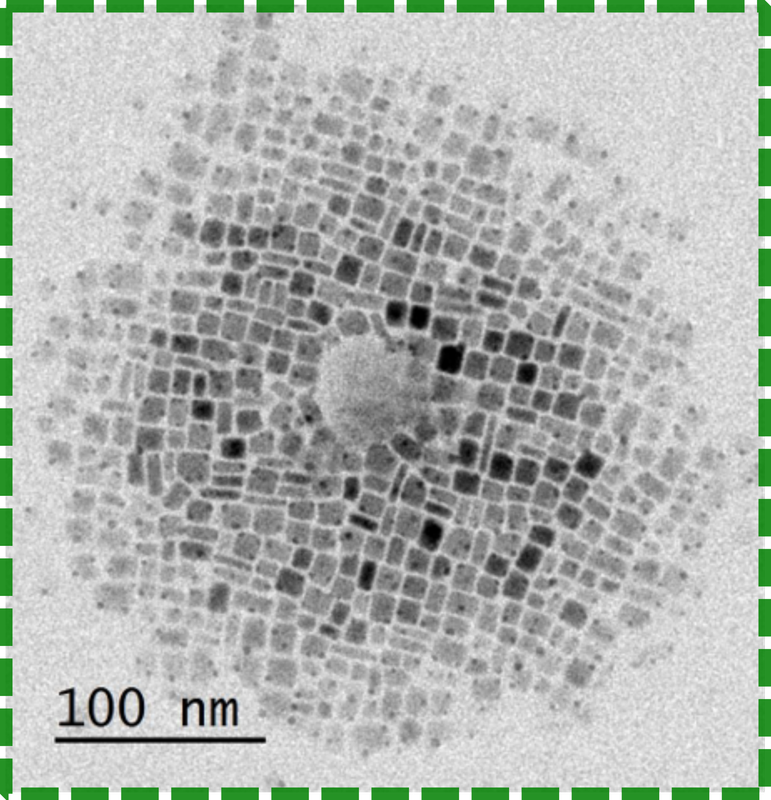}}\qquad
	\subfloat[]{\label{fig:SEM_B}\includegraphics[height=0.42\linewidth]
		{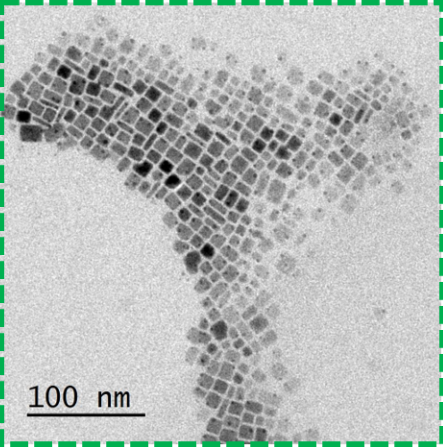}}
	\caption{\label{fig:SEM_AB} Scanning electron microscope images of for \protect\subref{fig:SEM_A}~Sample~A and
		\protect\subref{fig:SEM_B}~Sample~B. The nanocrystals present two sides of~\SI{10}{nm} while the other is shorter (between \SI{2}{nm} and \SI{6}{nm}). This difference is visible as some nanocristal appear leaning on the lateral side.
		}
\end{figure}
\subsubsection{Sample A}
The sample was prepared using the procedure provided by \textcite{protesescu2015nanocrystals}. This procedure leads to bright
\ch{CsPbBr3} nanocubes. Unfortunately the final solution is very concentrated
and it is impossible to look for a single emitter without diluting it. The
dilution procedure, even when performed with toluene (i.e. an appropriate solvent for perovskite nanocrystals), affects
the stability of the emitters. A SEM image of the nanocrystals prepared with this method is shown in figure~\ref{fig:SEM_A}. It is possible to see that they are parallelepipeds, with two sides of the crystal of about \SI{10}{nm} while the other one has a size of few nanometers.

\subsubsection{Sample B}
This procedure was initially intended to fabricate a slightly different kind of
nanoemitters, the \ch{CsPbBr3} nanosheets and was initially described
by~\textcite{weidman2016Highly}. Compared to the procedure used to
prepare the sample~A, there are three important changes:
\begin{enumerate}
	\item a smaller amount of cesium oleate is introduced  in order to favour
	the growth of the free phase of cesium;
	\item two additional ligands are introduced during the process, to allow
	the crystallization of the cesium free phase;
	\item the reaction is carried on for a longer time (from \SI{10}{\second}
	to \SI{35}{\minute}).
\end{enumerate}

With this method we obtained three different products.
\begin{description}
	\item[\ch{CsPbBr3} Nanoplatelets] are the main product of the synthesis, or
	at least the product for which this synthesis was intended for; the
	obtained nanoplatelets can be written with the form
	\ch{L2[CsPbBr3]_{n-1}PbBr4}, where L represent the ligand.
	\item[\ch{Cs}-free nanoplatelets] consist in two-dimensional
	nanoplatelets where a plane of led-bromide is sandwiched
	between two planes of ligands with C\,8 chains.
	\item[\ch{CsPbBr_3} nanocubes] are like the ones described for the sample~A.
	These cubes are the ones we are actually interested in: I will
	show how they show a better stability compared to the cubes of the
	sample~A, stability most likely due to the different ligands present in the
	solution.
\end{description}

In order to remove the first two products, we added a centrifugation step at the
end of the procedure. This procedure removes the bigger compounds leaving only the third
product in the solution. This provides us with a solution sufficiently diluted to be directly spin-coated and still allows us to find single photon emitters without any additional dilution as I will describe below. However, it was still too
concentrated to be deposited directly on a nanofiber as I will show at a later stage. A SEM image of the nanocrystals prepared with this method is shown in figure~\ref{fig:SEM_B}. As for sample~A, they have a parallelepiped shape.

\subsection{Sample preparation}
Once fabricated, a new sample can be deposited on a glass coverslip to be
observed with a microscope.
First of all, we proceed with a suitable dilution, in order to reduce the
concentration of the nanoemitters and to be able to resolve single emitters.
Then we deposit it using a spin-coater that allows to spread almost homogeneously
a thin film of the solution over the coverslip. This method avoids the unpleasant effects
that happen when the droplet is inhomogeneously dried. Indeed the mechanism of droplet drying has been largely studied. An interesting review was done by~\textcite{tarafdar2018Droplet}. What is important to notice in our application is that during the drying process, the substances in suspension are often deposited at the edges and tend to form agglomerates that we want to avoid. The spin-coater rotates the sample at high velocity (usually \SI{1500}{rounds/\minute}, distributing the solution over it and, at the same time, drying the sample.

Let us note that a different technique, often used in absence of a spin-coater, consists in using a solution of hexane-octane in the proportion of 9:1 as a solvent. This is known to naturally form a thin film over a glass plate simply by adhesion forces, without the need of spin-coating. However, to have the same concentration on the glass, the solution needs to be diluted more, with effects on the sample stability (as I will show later). For that reason, we decided to
keep the same solvent used during the fabrication (toluene) and preferred the
spin-coating method.

Once prepared, the samples need to be stored at room temperature, protected
from light and moisture. With some exceptions, they does not last longer than a
working week.

\subsection{Experimental Setup}
\label{sec:2ExperimentalSetup}
\begin{figure}[tb]
	\centering
	\includegraphics[width=\textwidth]{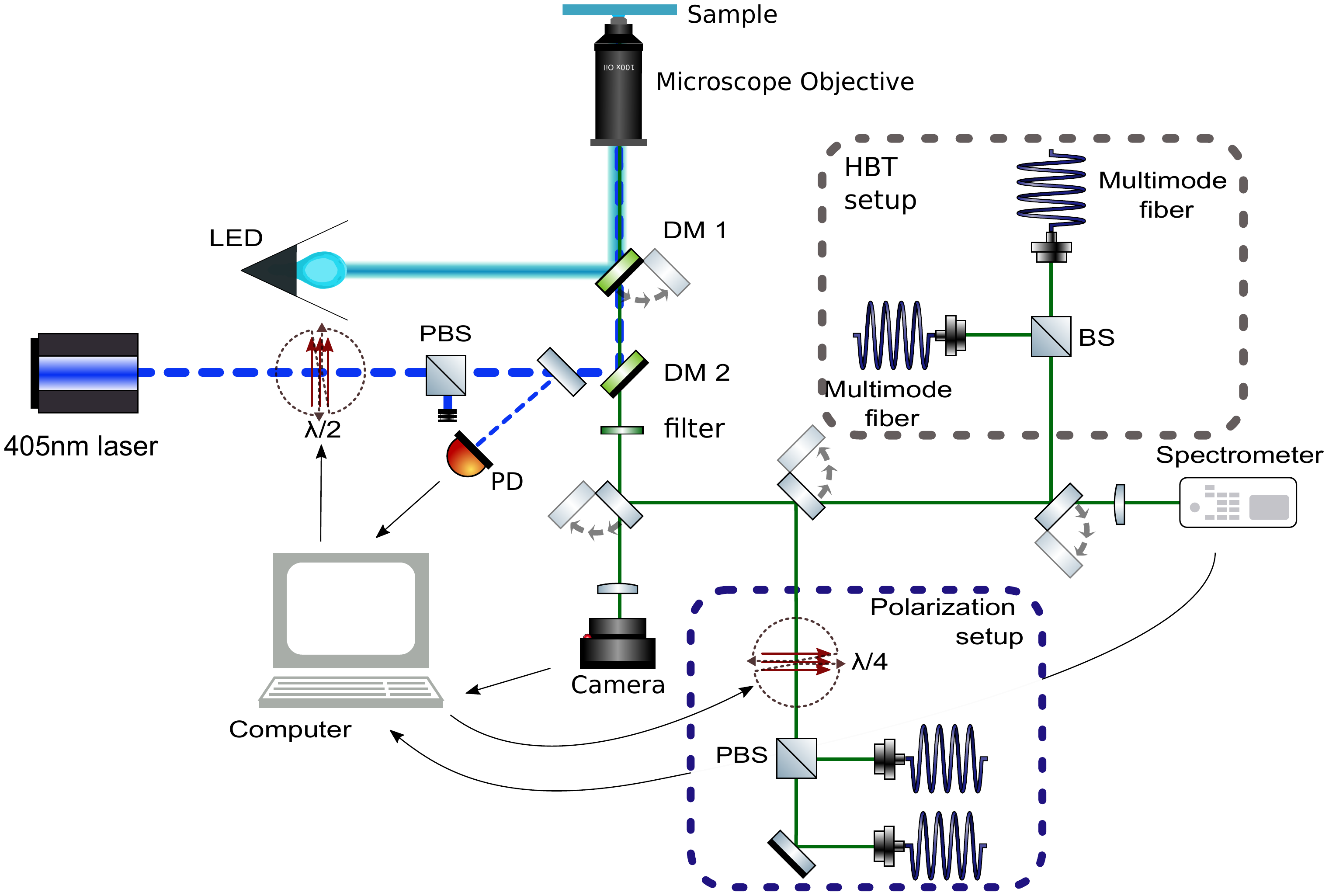}
	\caption{Experimental setup for single nanocrystal analysis. For a wide-field view, a LED light
	is reflected by a \SI{432}{\nano \meter} long pass
	dichroic beamsplitter and sent to the sample, via the microscope objective. For the excitation of a single nanocrystal, a pulsed laser light is
	reflected towards the microscope objective with a \SI{432}{\nano \meter}
	filter, while the dichroic beamsplitter used for the LED is removed. The
	collected light is filtered to remove any trace of the excitation and then
	sent optionally to the camera, to the spectrometer or to the HBT setup.
	DM:~dichroic mirror, BS:~non polarizing beamsplitter, PBS:~polarizing beamsplitter, PD:~photodiode,% \textbf{(CC: PD or PHD is better for photodiode)}, 
	$\lambda/2$: rotating half-wave plate,
	$\lambda/4$: rotating quarter-wave plate}
	\label{fig:setup}
\end{figure}

The experimental setup used to study the perovskite nanocrystal is shown in
figure~\ref{fig:setup}.
To facilitate the understanding of the measurement principles, I will give a quick overview of the setup, discussing the details later on.

\begin{figure}[tb]
	\centering
	\includegraphics[width=0.7\linewidth]{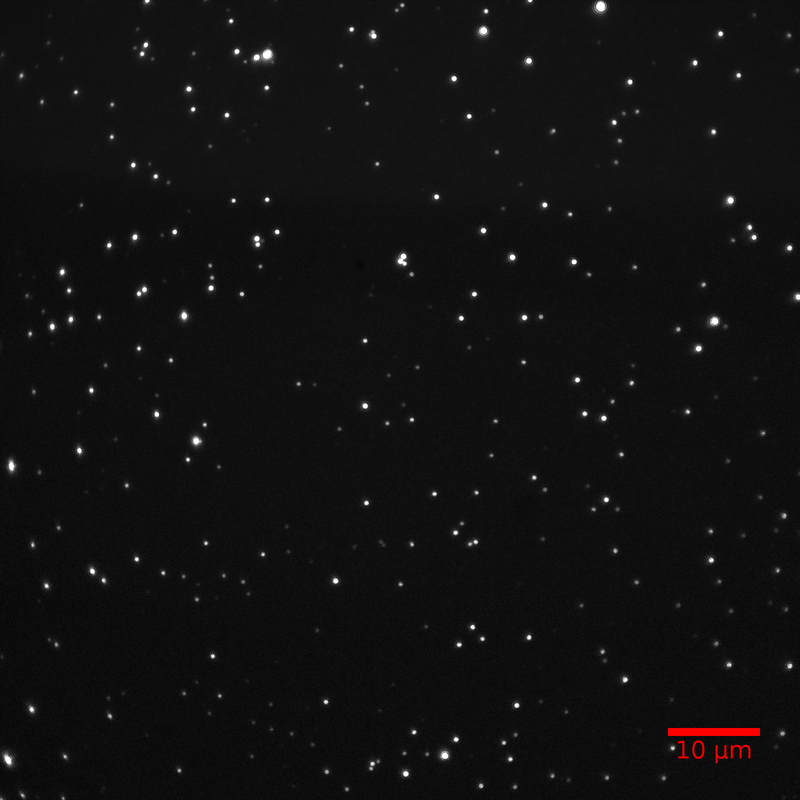}
	\caption{Example of an image obtained with the CCD camera when the sample is illuminated by the LED. The
	bright spots are the nanoemitters, while the background is dark because the excitation light is filtered out.}
	\label{fig:nopsstartscale}
\end{figure}

First of all, we spread the emitters over a glass coverslip using the spin-coater and we place it under the
microscope. %\textbf{(you mean a solution on a glass plate and the plate is placed under the microscope no?)}. 
In order to address a single emitter, we need to find it and
place it under the laser spot. Several methods are available to achieve this. A method that is often used, for example, is to create an image measuring the
luminescence spot-by-spot scanning the sample under the laser light: this
method has the inconvenience to be quite complicated to realize and slow to
perform. Another method which is the one we chose, is to focus an intense light beam (such as the one coming from an LED) into the microscope,
illuminating in this way the whole field of view. That way, we obtain a full image of the
sample. This image, after filtering the excitation wavelength, allows us to visualize
the emitters as bright spots on a dark background. A typical image obtained with this method
is reported in figure~\ref{fig:nopsstartscale}.

This technique allows to directly ``see'' the
image of the sample in a fraction of second, with all the advantages it has.

It is worth underlining that this method can be used as the intensity at which we
need to excite each emitter to obtain a detectable emission is quite low. Indeed, as opposed to the ``scanning'' method where only a
single spot is illuminated (area of the order of \SI{1}{\micro \meter ^2}), the
light of the LED is shined over a much larger area (on the order of
\SI{0.1}{mm^2}, four order of magnitude bigger) and then the applicability of this method can be limited by the difficulty to reach suitable intensities.

Once the image is performed and the emitter position identified, we move the sample via a precise translation stage to
place the emitter under the laser illumination. The laser is aligned so that it focuses at the center of the image, so we know accurately its position.
Except for the case of stability measurement that I will present later and
where the whole image is recorded, at this point we switch from the LED to the
laser illumination. For that, the first dichroic mirror (DM1) is removed and the laser is sent through the objective.
The collected light cleaned from the excitation (or in other words the emitted
light), can be analyzed with different instruments.
\begin{description}
	\item[CMOS Camera:] used to measure the emitted intensity, thanks to its
	linear response with the incident light;
	\item[spectrometer:] to perform the spectrum of the emitted light, useful to know the wavelength of the emission and the spectral width;
	\item[HBT setup:] used to perform the $g^{(2)}\of{0}$ measurement. Thanks to the time-tagged time resolved technique, this setup allows us to know the
	arrival time of each photon allowing, at the same time,
	to measure the fluorescence lifetime of the emission, the fluorescence trace and to perform interesting blinking analysis;
	\item[polarization setup:] used to perform polarization measurement. Two single photon detectors are used to correct the signal from the blinking effect, as shown below.
\end{description}

After the description of the main elements of the setup, it is important to detail
the characteristics of some of the instruments used.
\begin{description}
	\item[LED:] Pizmatich 400nm LED, controlled with a digital interface that
	allows to tune the emitter power, it is collimated inside the objective to illuminate the full field of view.

	\item[405 nm LASER:] Picoquant PDL800 (model LDH 405B) pulsed laser with pulse width smaller than\SI{50}{\pico \second}, repetition rate from
	\SIrange{2.5}{40}{\mega \hertz}.

	\item[rotating waveplate:] controlled via a servomotor and combined with the
	following polarizing beam-splitter permits to control the excitation power
	sent by the laser inside the microscope with the computer.

	\item[photodiode:] home made, receives a small fraction of the intensity,
	and through an initial calibration, permits to know the excitation power.

	\item[dichroic mirrors:] both DM1 and DM2, are longpass dichroic mirrors
	with a cutoff wavelength of \SI{432}{\nano\meter}. They reflect the
	excitation light inside the objective and transmit the emitted light to the
	detection part of the setup.

	\item[microscope objective:] an oil microscope objective with a numerical
	aperture of \num{1.4} and a magnification of \SI{100}{\times}. In this case
	the sample with the emitters lies on the opposite side with respect to the
	microscope. We used the oil objective for all the measurements, except to perform polarization
	measurements, indeed in this case an objective with a so large numerical aperture
	does not allow to measure the polarization correctly. Indeed, when collecting the emission with an high numerical aperture objective, we collect the emission into different directions, each of them having a different polarization: this results in a sum of different polarizations in the collected beam and thus in a lower degree of polarization measured. This point has been studied in detail by~\textcite{lethiec2014Measurement}.
	To reduce this problem, we used a
	dry objective with the same magnification and numerical aperture of~
	\num{0.95}. This is, in our experiment, a good compromise between the need to use a lower numerical aperture objective and the need of collect enough signal.

	\item[filter:] high quality long pass filter with a cutoff wavelength at
	\SI{420}{\nano \meter}, to remove the residual excitation light.

	\item[CCD camera:] Hamamatsu Orca, with a quantum efficiency between
	\SIrange{70}{80}{\percent} at our detection wavelengths and squared pixels
	of \SI{6.5}{\micro\meter} size.

	\item[spectrometer:] Princeton Instruments Spectrum Analyzer, allows to
	record the spectrum of the light emitted by the perovskite nanocrystals.

	\item[mirrors:] All the mirrors in the exciting path are dielectric mirror,
	while in the detection, when needed for polarization measurement, we have
	used silver mirrors to avoid changes in polarization.
\end{description}

\section{Measurements}
\subsection{Perovskite Spectra}
\begin{figure}[tb]
	\centering
	\includegraphics[width=0.4\linewidth]
	{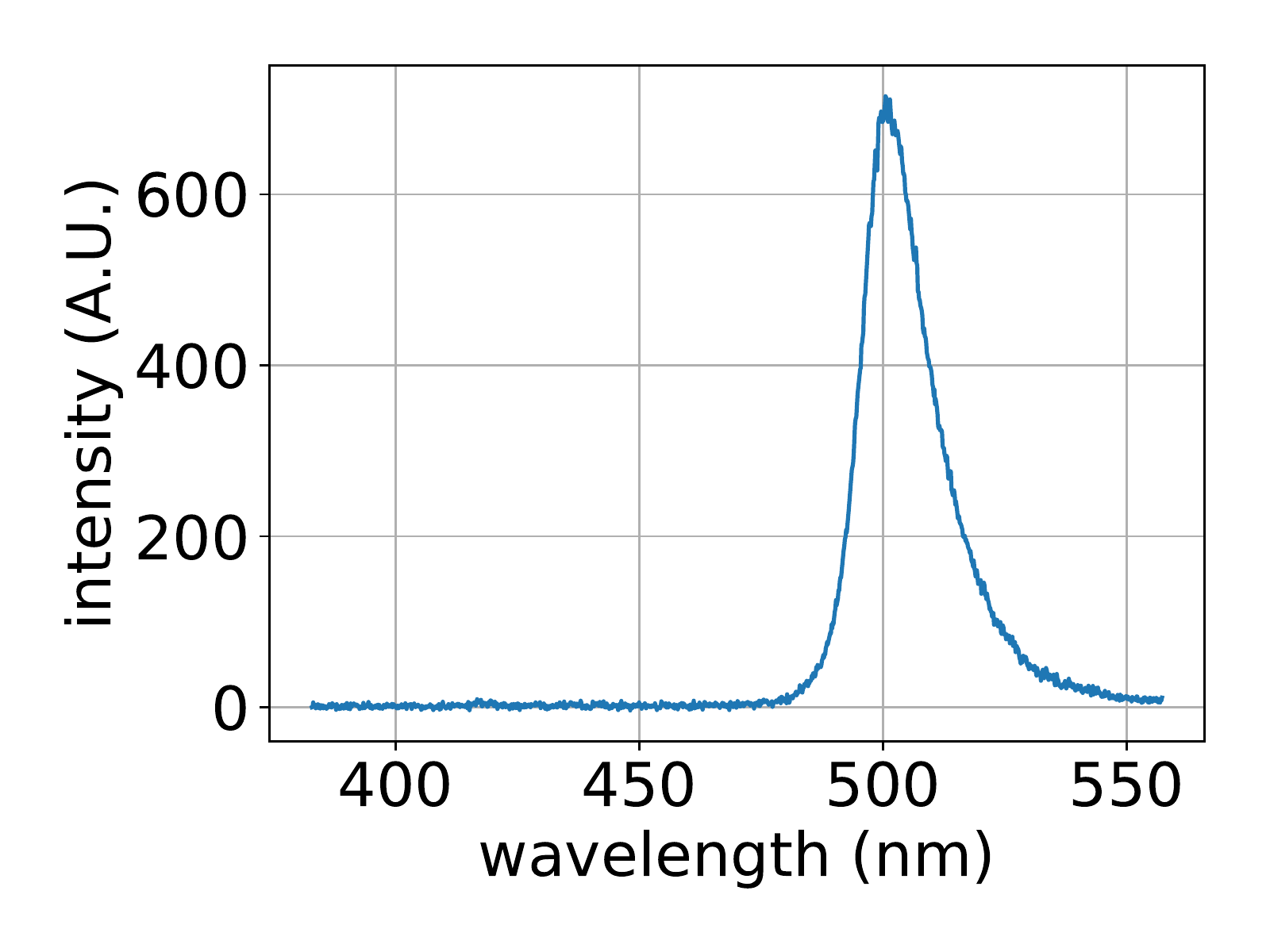}
	\caption{Emission spectrum of a single perovskite nanocrystal. In this
	example we can see that the peak is at \SI{501}{\nano \meter} and have a FWHM of \SI{15}{\nano \meter}. %\textbf{(CC: can we be more precise? It is not 500 nm. Also, can we the FWHM?)}
	}
	%data of file /home/pierinis/Documenti/Dottorato/DatiParigi/LightField/PL4/2019-11-21 14_09_35.spe
	\label{fig:spettroemissione}
\end{figure}

\begin{figure}[tb]
	\centering
	\subfloat[]{\label{fig:wave_vs_FWHM_A}\includegraphics[width=0.45\linewidth]
		{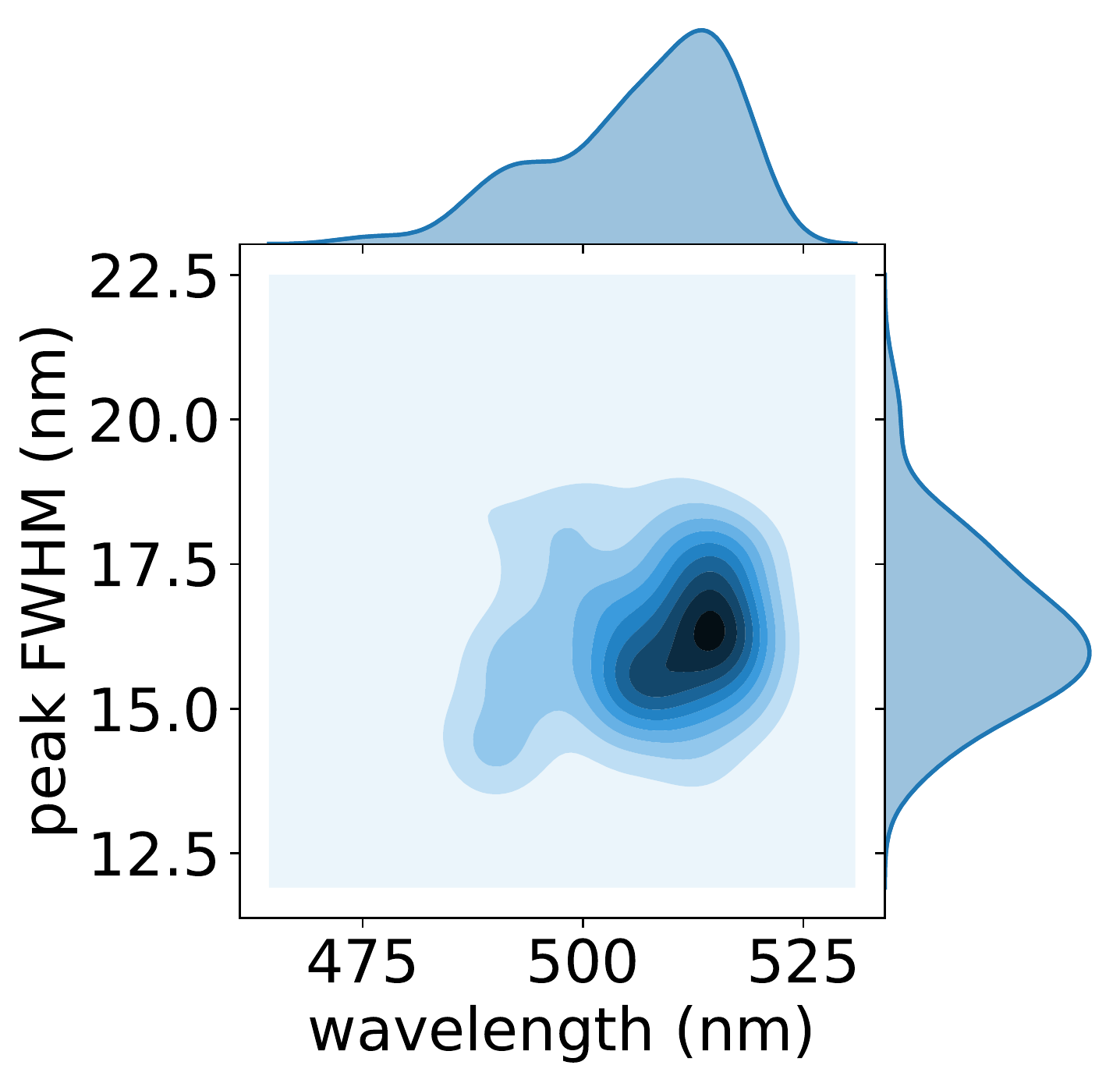}}\qquad
	\subfloat[]{\label{fig:wave_vs_FWHM_B}\includegraphics[width=0.45\linewidth]
		{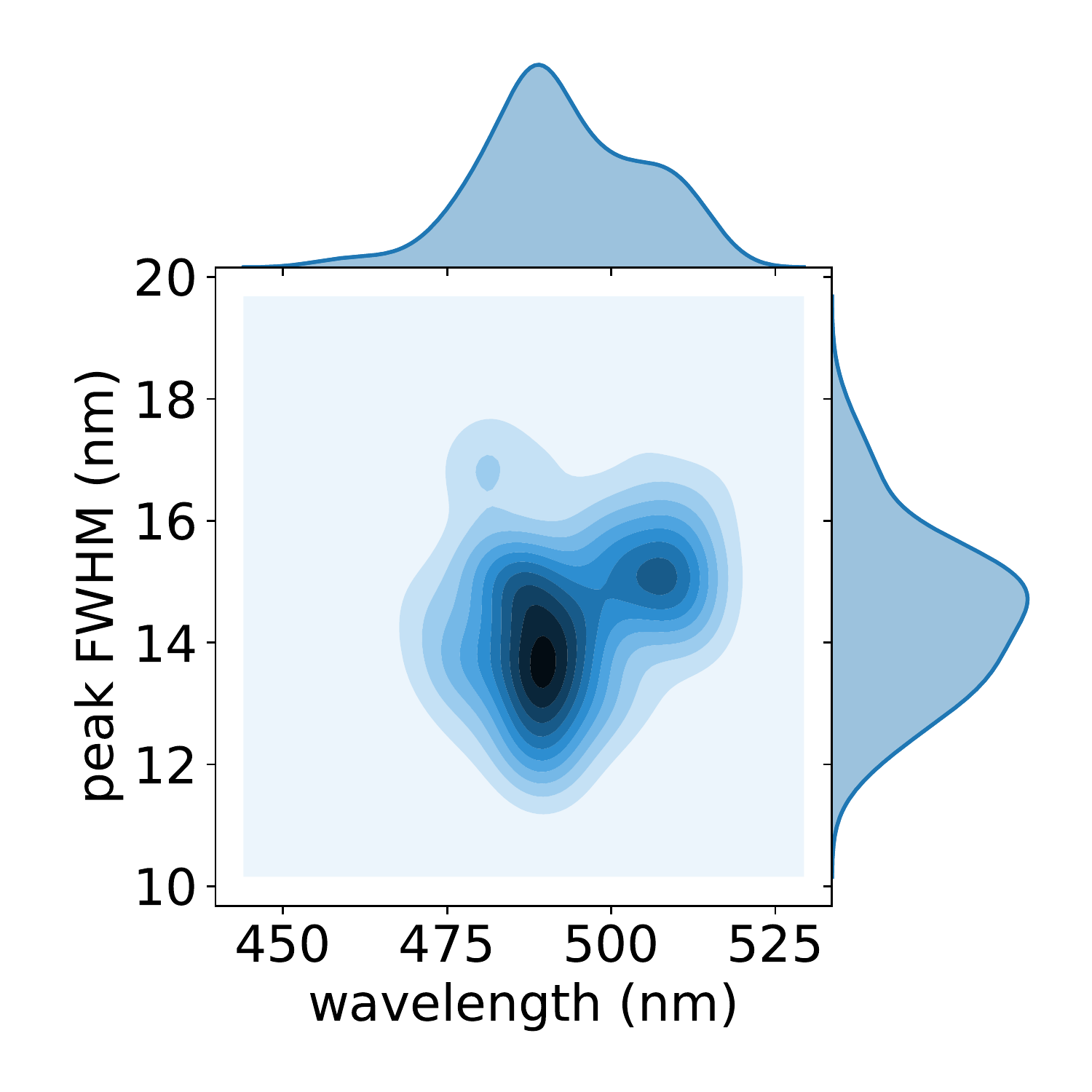}}
	\caption{\label{fig:wave_vs_FWHM} Distribution of central wavelength
	emission and the peak Full Width Half Maximum (FWHM) for the samples
		A~\protect\subref{fig:wave_vs_FWHM_A} (calculated on
		\SI{135}{emitters}) and
		B~\protect\subref{fig:wave_vs_FWHM_B} (calculated on \SI{74}{emitters}).
		}
\end{figure}

From the emission spectra, we can extract different information on the
nanoemitters such as the central emission wavelength and the width of their
emission. At room temperature, the fine structure of the emission with the different energy levels cannot be resolved in the spectrum. It is still very useful to have a quick characterization of the emitter we
are measuring. An example of a single perovskite emission spectrum is shown in
figure~\ref{fig:spettroemissione}.
Another interesting information can be obtained studying the distribution of the spectra of the two different samples. This analysis is reported in
figure~\ref{fig:wave_vs_FWHM_A}. %\textbf{(CC: in general you should only refer to figures and not have the figures below the text like you have for equations)}

We can observe that sample~A has both a broad FWHM and a higher central
emission wavelength compared to sample~B. These data can be correlated to the quantum confinement: 
%\textbf{(CC: what confinement?)}
we can thus expect a stronger confinement in
sample~B due to its lower wavelength of emission.

\subsection{Saturation behaviour}
The measurement of the saturation intensity also provides valuable information
about the sample and each emitter. Such kind of measurement
is usually performed by exciting the sample with different excitation powers and collecting
its emission. This allows us to correlate the emitted power with the exciting one and
plot a saturation curve. However, in presence of blinking or flickering, the
measurement needs to be slightly adapted to minimize the influence of the
emission fluctuations on the saturation curve.

Our emitters have two drawbacks that we need to consider to correctly interpret the results of the performed measurements: first of all, they bleach and this implies that we should excite them for the shortest possible time and with the lowest possible intensity. Secondly, they blink which affects the measurement of the emitted
power by inducing strong fluctuations of the signal.
To overcome these issues, we chose to automatize the acquisition process which allows us to perform the measurement quickly and collect more data points to reduce the effect of the blinking on the measurement.
Via a \Python{} script, a computer controls the multiple instruments for the measurement, namely: a shutter, to block the excitation laser when not needed; a
rotation wave-plate to tune the laser power, a photodiode and a camera to measure both input and output intensities.
The photodiode (represented with \textit{PD} in figure~\ref{fig:setup}) returns an electrical tension proportional to the light power impinging on it, that is also proportional to the light power shined on the emitter.
Before the measurement, we calibrate it  in order to be able to convert the  measured voltage in the power effectively shined on the emitter. % \textbf{(CC: don't understand what you meant?)}.
The measurement steps are the following: % \textbf{(CC: you like these lists but they have no numbering)}:
\begin{itemize}
	\item First, the maximum power obtained from the laser, as well as the
	minimum one and the corresponding rotation position of the wave-plate are
	measured with a power-meter and recorded by the control program.
	\item The emitter is placed under the laser spot, we double check
	its position, firstly verifying with the LED that the emitter is in the position where the laser is, and then shining the laser to verify it really excites the selected emitter. We also verify that no-other emitter is excited by the laser. % \textbf{(CC: last sentence is no english)}.
	\item The region where the emitter appears in the camera is set in the
	code, permitting to exclude all the pixels that are not concerned and thus
	to increase the signal/noise ratio.
	\item When the measurement starts, with the shutter off, the wave-plate
	goes in the position of minimum intensity. At each cycle, the wave-plates moves by 	some degrees (depending on the settings) up to reach the maximum. At each
	step multiple operations are performed.
	\begin{itemize}
		\item With the shutter closed, the camera is read to measure the
		background.
		\item The excitation power measured by the photodiode is recorded
		\item The shutter is opened and the emitted intensity on the camera is
		measured several times, keeping only the highest measured value to
		reduce the effect of the blinking. This allows to select the measure
		%\textbf{(CC: not english)}
		in which the state has been ``on'' for most of the time, minimizing
		the probability that the emitter is in the ``off'' state for a significant time duration.
	\end{itemize}
	\item At the end of the measure, we plot the emission power as a function
	of the excitation intensity, in order to obtain the saturation curve.
\end{itemize}

An example of one of the measurements is presented in
figure~\ref{fig:satintensity}:
\begin{figure}
	\centering
	\includegraphics[width=0.5\linewidth]{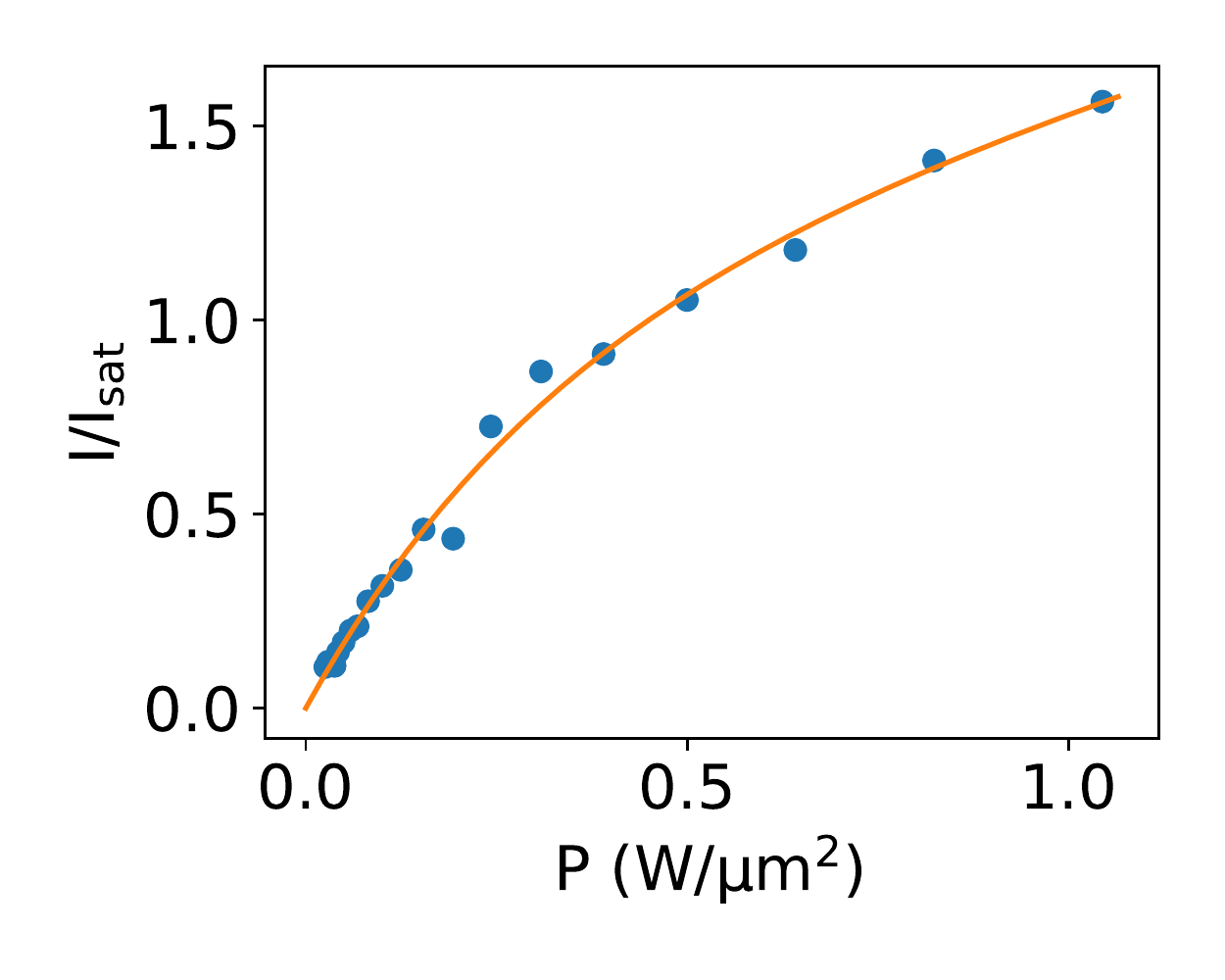}
	\caption{Example of saturation measurement}
	\label{fig:satintensity}
\end{figure}
the experimental points are represented by the blue dots. The orange curve is
the fit with the usual saturation function
\begin{equation}
	P=A \cdot \quadra{ 1-e^{-\frac{I}{I_{sat}}}} +B \cdot I
\end{equation}
Here, as seen in section~\ref{sec:sat}, %TODO add the correct reference
the first term of the sum represents the saturating part that is due to the
single exciton component, while the second term is due to the bi-exciton
emission.
This curve can be used to compare more easily the characteristics of different
emitters. Indeed exciting the sample with the same intensity ratio with respect to the
saturation power leads to the same number of excitons created, and thus a
similar excitation. To do this, anyway, the emitters need to be photo-stable enough;
in the case of sample~A, less stable than the sample~B, it was not possible to perform a preliminary measurement of the saturation
intensity, as the emitters were bleaching too fast.

\subsection{Polarization measurements}
\begin{figure}[tbp]
    \centering
    \includegraphics[width=0.5\linewidth]{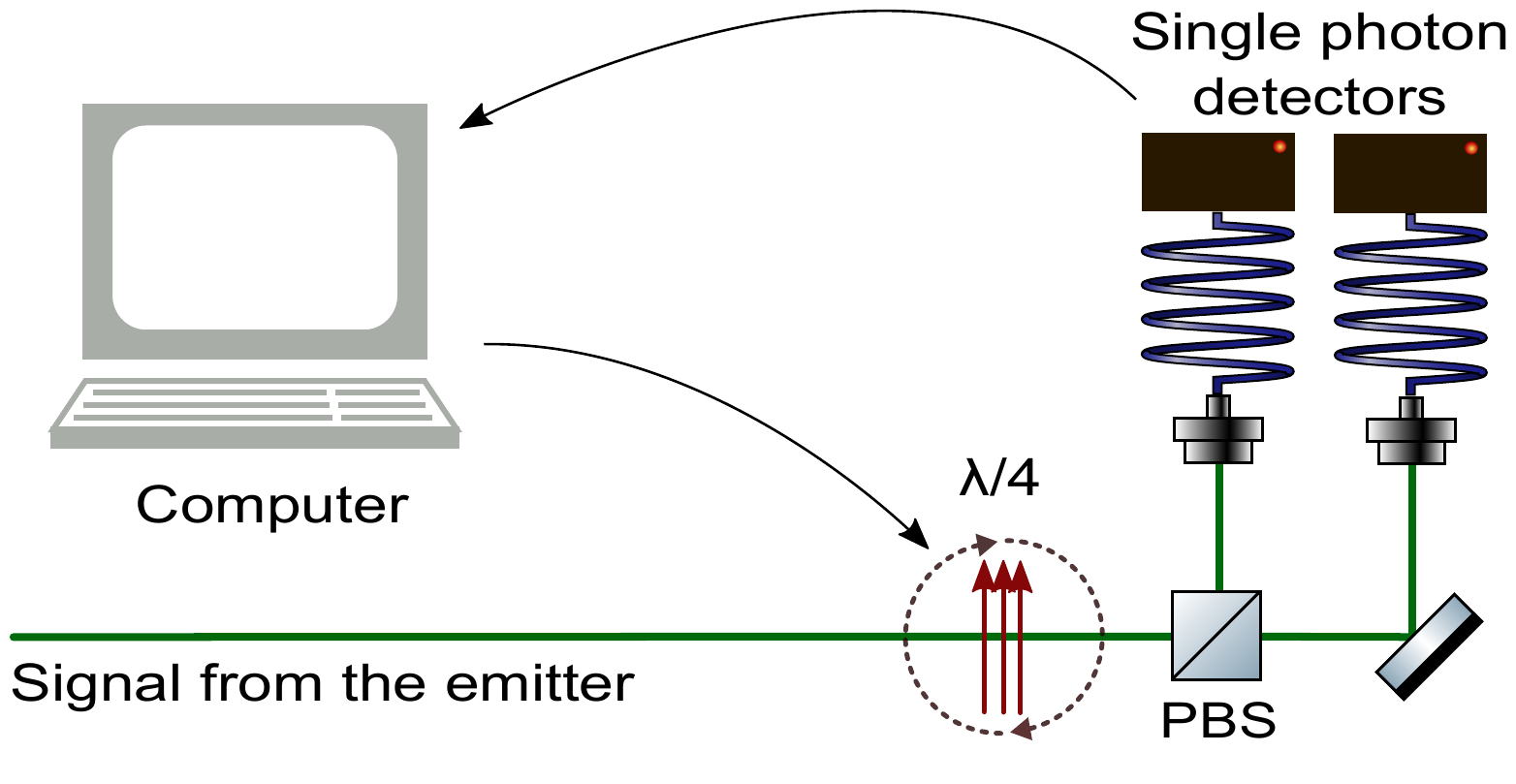}
    \caption{Detail of figure~\protect\ref{fig:setup} representing the experimental setup used to measure the polarization. The quarter-wave plate ($\lambda/4$) and the acquisition of are controlled by the computer.}
    \label{fig:polsetup}
\end{figure}
A polarized emission for a single photon emitter is an important characteristic such as dot in rod nanocrystals~\cite{pisanello2010Room}. To characterize the polarization of a given emission we can measure
the Stokes parameters.
The setup is shown in figure~\ref{fig:setup} and allows to measure all the
Stokes parameters from the light transmitted after a polarizing
beam splitter as a function of the position of a quarter-wave plate in front of it.

Let us consider only one transmitted beam in the polarization setup (represented in ~\ref{fig:polsetup}) and
imagine for the moment the reflected beam to be absorbed by a beam-stopper.
We can see the beam splitter as a fixed polarizer: when rotating the half
waveplate, we rotate the polarization of the incoming light, obtaining a
variation on the number of counts detected by the single photon detector.
Studying the obtained curve which represents the intensity collected by the
photodiode as a function of the angle of the polarizer, it is possible to
reconstruct the Stokes parameters of the light. This method is clearly reported by~\textcite{berry1977Measurement}.
\begin{figure}[tbp]
	\centering
	\includegraphics[width=0.7\linewidth]{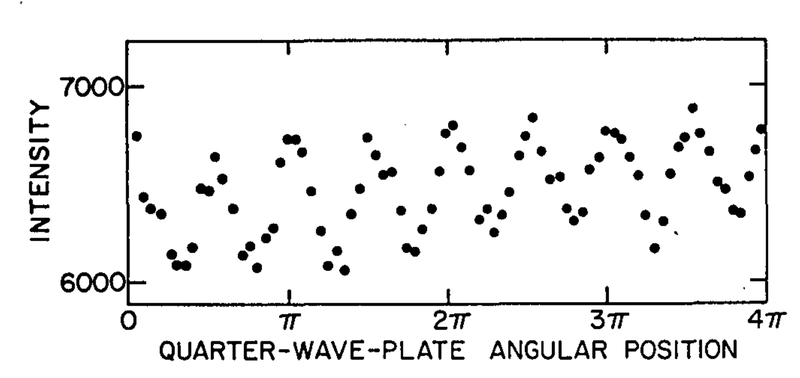}
	\caption{Measurement of the emission Stokes parameters in case of a linearly
	polarized light. The intensity collected is reported as a function of the
	rotation of the quarter-wave plate. \ccredit{berry1977Measurement}}
	\label{fig:polarization}
\end{figure}
The result of one of these measurements is reported in
figure~\ref{fig:polarization}. % \textbf{(CC: cette figure a peu d'intérêt)}.
The curve is described by the
equation~\eqref{eq:relation_Stokes_comp}:
\begin{equation}
\label{eq:relation_Stokes_comp}
\begin{split}
I_{T}(\alpha, \beta, \delta)&=\frac{1}{2}\left[S_0+\left(\frac{S_1}{2} \cos 2
\alpha+\frac{S_2}{2} \sin 2 \alpha\right)(1+\cos \delta)\right]
+\\&+\frac{1}{2}[S_3 \sin \delta \cdot \sin (2 \alpha-2 \beta)]+\\&+
\frac{1}{4}[(S_1 \cos
2
\alpha-S_2 \sin 2 \alpha) \cos 4 \beta
+\\&+(S_1 \sin 2 \alpha+S_2 \cos 2 \alpha) \sin 4 \beta](1-\cos \delta).
\end{split}
\end{equation}

where $S_0$, $S_1$, $S_2$ and $S_3$ are the four Stokes parameters of the light
as described in section~\ref{sec:stokes}, 
$\alpha$ is the rotation angle of the polarizer (in our case the polarizing
beam splitter) and $\beta$ is the rotation angle of the fast axis of the
%retardation %\textbf{(CC: what's that?)} 
wave-plate, while $\delta$ is the delay introduced by the wave-plate, in our case the quarter wave-plate.

If we choose to rotate the polarizer, meaning varying the value of $\alpha$, we need to pay
attention to the fact that all the detection optics need to be polarization
insensitive. For this reason, it is more convenient to keep the polarization fixed
and to rotate the wave-plate.
A deeper inspection on~\eqref{eq:relation_Stokes_comp} allows to notice the following behavior: let us imagine a change in $\beta$ at a frequency $\omega$; the first term does not depend on
$\beta$, while the second oscillates with a frequency $2\omega$ and the
third oscillates at a frequency $4\omega$. Therefore we can interpret this equation as a
Fourier series.

Generally, if a real-valued function $f(x)$ is integrable in an
interval $P$, its Fourier series is given by:
\begin{equation}
	\label{eq:fourier_series}
	f_{N}(x)=\frac{c_{0}}{2}+\sum_{n=1}^{N}\left(c_{n} \cos \left(\frac{2 \pi n
	x}{P}\right)+s_{n} \sin \left(\frac{2 \pi n x}{P}\right)\right).
\end{equation}

In equation~\eqref{eq:fourier_series},  $P$ is the interval and the period
of $f_N$ while $c_n$ and $s_n$ are defined as follow:
\begin{align}
\label{eq:fourier_coeff_integral}
%\begin{array}{l}
c_{n}=&\frac{2}{P} \int_{P} f(x) \cdot \cos \left(2 \pi x \frac{n}{P}\right) d
x
\\
s_{n}=&\frac{2}{P} \int_{P} f(x) \cdot \sin \left(2 \pi x \frac{n}{P}\right) d x
%\end{array}
\end{align}
This is valid for a continuous function; as we measure $N$ discrete points, the last
two equations have to be rewritten as follows ($\delta$, as before, is the delay introduced by the waveplate):
\begin{align}
\label{eq:fourier_coefficient_c}
%\begin{array}{l}
c_n=&\frac{2}{N} \dfrac{1}{1+\delta_{n0}} \sum_{i=0}^{N-1} s_i \cdot \cos\left(
2
\pi x_i \frac{n}{P}\right),
\\
\label{eq:fourier_coefficient_s}
s_n=&\frac{2}{N} \dfrac{1}{1+\delta_{n0}} \sum_{i=0}^{N-1} s_i \cdot \sin\left(
2
\pi x_i \frac{n}{P}\right).
%\end{array}
\end{align}
%\textbf{(CC:define delta and N)} 
We can thus rewrite equation~\eqref{eq:relation_Stokes_comp} as a function of
$\beta$, with the substitution $\beta \rightarrow \beta_0 + \beta$: 
\begin{equation}
\label{eq:I_fourier}
I_{T}(\beta)=c_{0}+c_{2} \cos 2 \beta+c_{4} \cos 4 \beta+s_{2} \sin 2
\beta+s_{4} \sin 4 \beta
\end{equation}
where $\beta$ varies from $0$ to $2\pi$. This is a finite Fourier series and
its non-zero coefficients $c_{0}$,  $c_{2}$, $c_{4}$, $s_{2}$, $s_{4}$ can be
computed using
equations~\eqref{eq:fourier_coefficient_c}
and~\eqref{eq:fourier_coefficient_s}.
Clearly, as in any experiment, the number of measured points cannot be
infinite. In order to estimate the number of points necessary, from the Nyquist–Shannon sampling theorem~\cite{shannon1949Communication}, we know that to
correctly sample a signal we need to sample it at least twice as much as its maximal
frequency. In other words, we need to measure at least
8 points in our curve: if we choose to use an even number of samples $N=2L$ and
$L=4$ we need to slightly modify the equations~\eqref{eq:fourier_coefficient_c}
and~\eqref{eq:fourier_coefficient_s} in the following
way \cite{berry1977Measurement}:
\begin{align}
\label{eq:fourier_coefficient_c_delta}
c_n=&\frac{2}{N} \dfrac{1}{1+\delta_{n0}+\delta_{nL}} \sum_{i=0}^{N-1} s_i
\cdot \cos\left(
2
\pi x_i \frac{n}{P}\right)
\\
\label{eq:fourier_coefficient_s_delta}
s_n=&\frac{2}{N} \dfrac{1}{1+\delta_{n0}+\delta_{nL}} \sum_{i=0}^{N-1} s_i
\cdot
\sin\left(
2
\pi x_i \frac{n}{P}\right)
%\end{array}
\end{align}

It can be experimentally more convenient to vary $\beta$ from
$0$ to $\pi$ as the signal is periodic with a period of $\pi$. In this case,
we can make the formal substitution $\beta'=2\beta$ in the
equation~\eqref{eq:I_fourier} and obtain:
\begin{equation}
\label{eq:I_fourier'}
I_{T}(\beta)=c_{0}+c'_{1} \cos \beta'+c'_{2} \cos 2 \beta'+s'_{1} \sin
\beta'+s'_{2} \sin 2 \beta'
\end{equation}
where, we have
\begin{equation*}
	c'_{1}=c_{2},\qquad c'_{2}=c_4, \qquad s'_{1}=s_2,\qquad s'_{2}=s_4.
\end{equation*}
Once again, the equation~\eqref{eq:I_fourier'} is a Fourier series (with respect to
the angle $\beta'$), and we can thus calculate its Fourier coefficients. We can
notice that in this case, using the fact that the measurement is periodic, we
divided by two the maximal frequency. Thus, once again, the
minimum number of points necessary to obtain the non-zero coefficients of the
equation~\eqref{eq:I_fourier'} is given by the Nyquist–Shannon sampling theorem: four points are sufficient. If four is the minimum number to correctly reconstruct the signal, experimentally it is always better to collect a bigger number of points. 

Now we can compare equations~\eqref{eq:I_fourier}
and~\eqref{eq:I_fourier'}, in order to obtain the Stokes
coefficients~\cite{berry1977Measurement}:
\begin{subequations}
\begin{gather}
 S_0=c_{0}- \dfrac{1+\cos \delta}{1-\cos \delta} \cdot\left[c_{4} \cos \left(4
\alpha+4 \beta_{0}\right)+s_{4} \sin \left(4 \alpha+4 \beta_{0}\right)\right]\\
 S_1=\dfrac{2}{1-\cos \delta}\left[c_{4} \cos \left(2 \alpha+4
\beta_{0}\right)+s_{4} \sin \left(2 \alpha+4 \beta_{0}\right)\right] \\
 S_2=\dfrac{2}{1-\cos \delta}\left[s_{4} \cos \left(2 \alpha+4
\beta_{0}\right)-c_{4} \sin \left(2 \alpha+4 \beta_{0}\right)\right] \\
 S_3=\dfrac{c_{2}}{\sin \delta \sin \left(2 \alpha+4
\beta_{0}\right)}=\dfrac{-s_{2}}{\sin \delta \cos \left(2 \alpha+4
\beta_{0}\right)} \\
\abs{S_3}=\dfrac{\left(c_{2}^{2}+s_{2}^{2}\right)^{\frac{1}{2}} }{ \sin ^{2}
\delta}
\end{gather}
\end{subequations}
We can choose the reference system in which $\alpha=0$, $\beta_0$ and $\delta$
can be obtained by calibrating the system as explained
by~\textcite{berry1977Measurement}.
$S_0$ represents the intensity and we can renormalize the other Stokes
parameters to it.

In our analysis we did not take into account the effect of the blinking: it has
been shown~\cite{andreev2017Selfcalibrating} that the intensity fluctuations have a
significant effect on the measurement. For this reason it is important to implement a
normalization scheme: this is implemented by adding a second detection channel. This way, all the emitted light, transmitted and reflected, is collected. The signal
transmitted by the beamsplitter is thus normalized by the entire signal
collected, reducing the possible influence of the blinking.

\begin{figure}[tb]
	\centering
	\subfloat[]{\label{fig:polplot_measure}\includegraphics[width=0.45\linewidth]
		{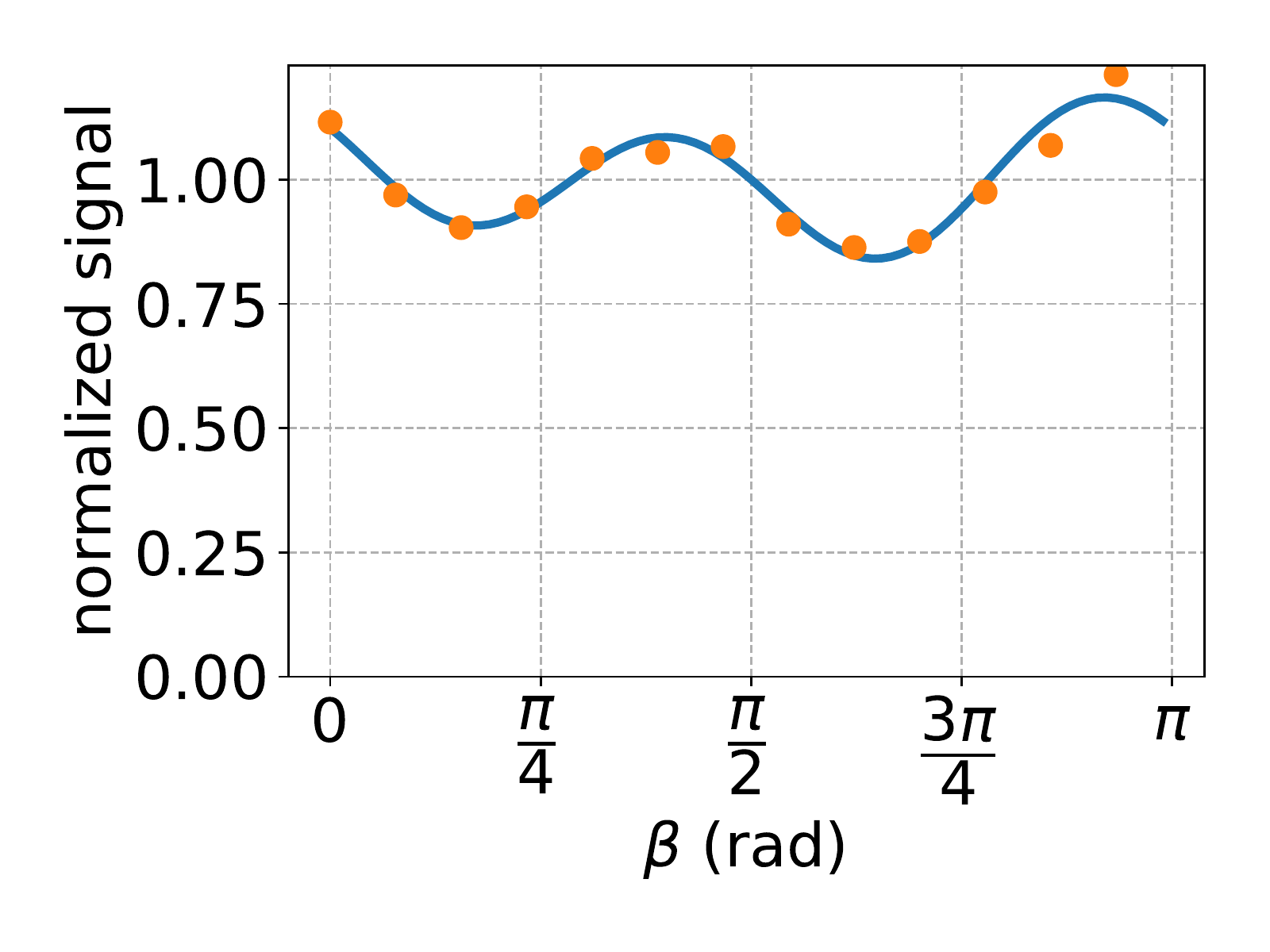}}\qquad
	\subfloat[]{\label{fig:polplot_ell}\includegraphics[width=0.45\linewidth]
		{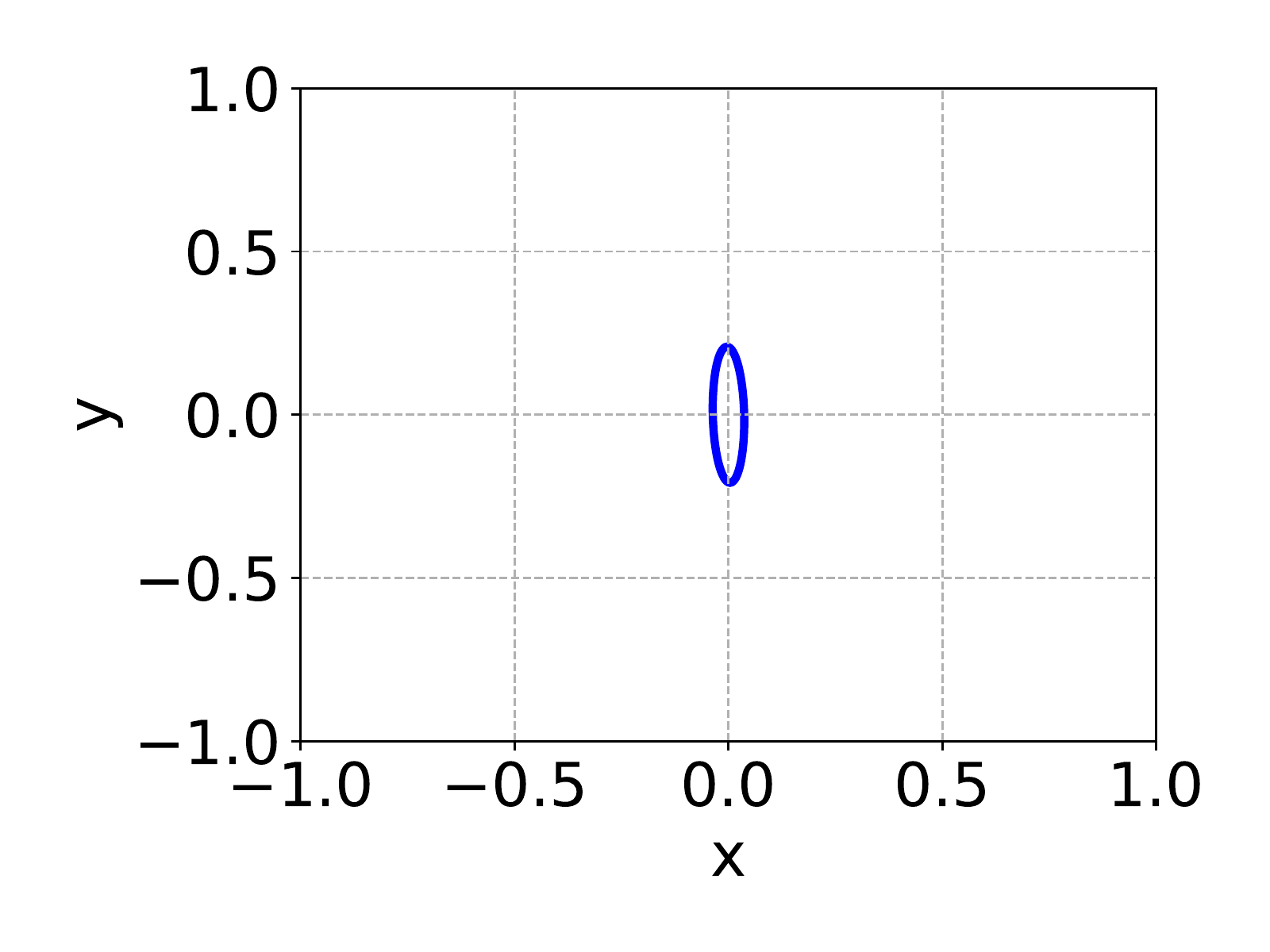}}
	\caption{\protect\subref{fig:polplot_measure}~Example of measure of Stokes
	parameters on a perovskite
	nanocrystal. The red dots are the experimental points while the blue curve
	is obtained via the described procedure.
	\protect\subref{fig:polplot_ell}~Polarization ellipse. The degree of
	polarization is~0.18 and the degree of linear polarization is~0.17, showing
	an almost non polarized emission.}
	\label{fig:polplot}
\end{figure}

An example of these measurements is shown in figure~\ref{fig:polplot}. The red
dots are the experimental points while the red curve is obtained via the described
Fourier analysis. Unfortunately, the light emitted by our emitters happened to be non-polarized. We repeated the measurement for several different emitters, with the same conclusion.  %\textbf{(CC: how manye emitters did you try? Say it.)}.

This behavior was expected due to the cubic symmetrical structure and shape which do not favor any preferential polarization. However, this finding does not exclude that
future studies of asymmetrical perovskites could give polarized photons.

\subsection{\gd{} measurements}
\label{sec:g2_measurement}
I've shown in the Introduction how important it is, from the quantum optics point
of view, to perform \gd measurements. I detail here how the measurement is
performed and how the data are treated to obtain the \gd autocorrelation function. %This setup is a slight modification of the setup used by~\textcite{berry1977Measurement}.
\subsubsection{Acquisition}
\begin{figure}[tb]
	\centering
	\includegraphics[width=0.7\linewidth]{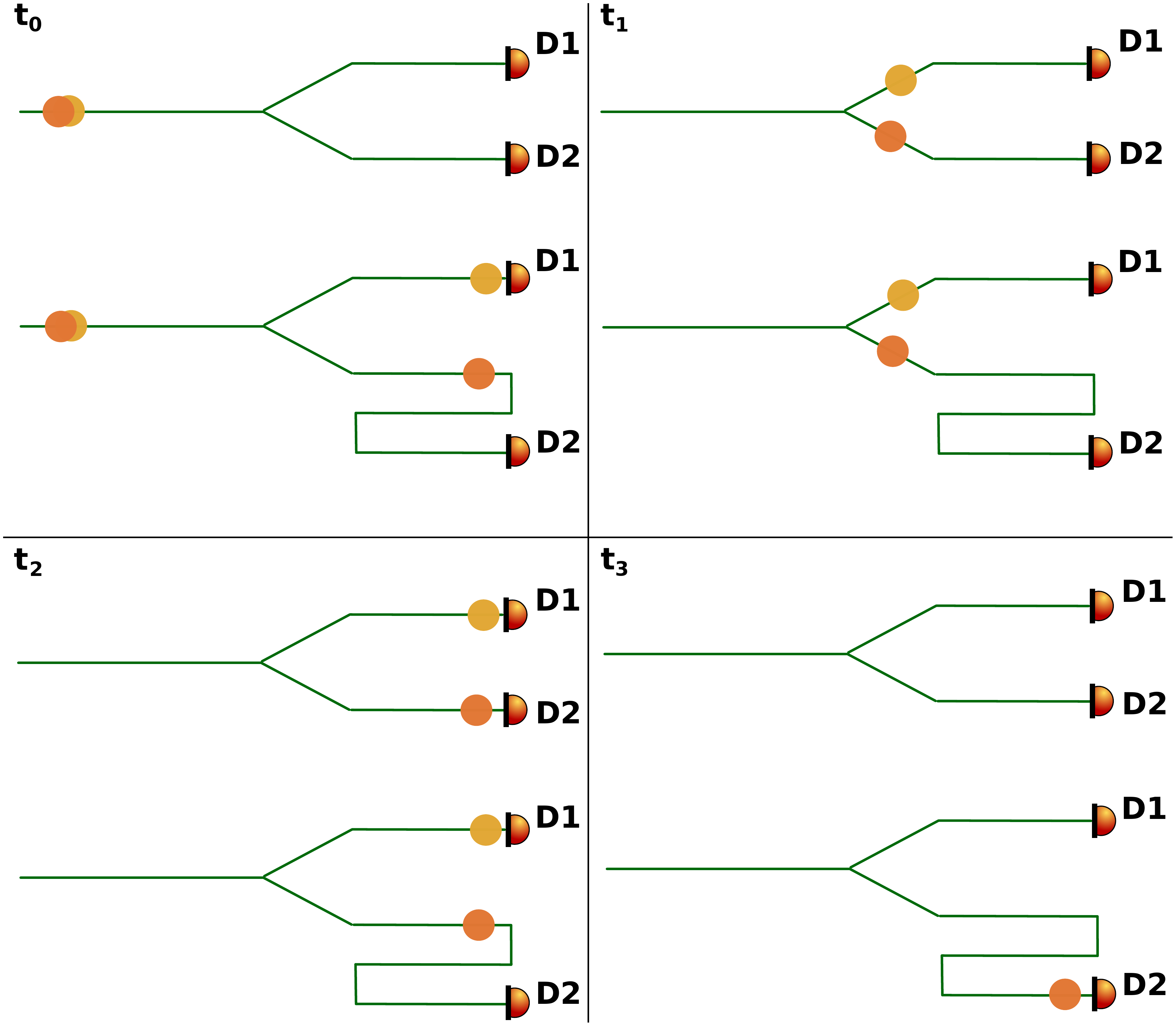}
	\caption{Example of the delay line effect on photon detection. In each
		panel, the situation without (upper path) and with (lower part) the
		delay line is represented. $t_0$)~Two photons are emitted
		simultaneously and $t_1$) separated on different paths by the
		beamsplitter. $t_2$) in case of no
		delay line, the two photons arrive at the two detectors and only one
		is 	recorded. In the other case, only one photon arrives on the
		detector, while the other arrival is delayed. $t_3$) In case of the
		presence of the delay line, the second photon arrives on the detector:
		its true arrival time with respect to the first one can be obtained
		by subtracting $\delta=t_3-t_2$.}
	\label{fig:photonsdelay}
\end{figure}
\begin{figure}
    \centering
    \includegraphics[width=0.9\linewidth]{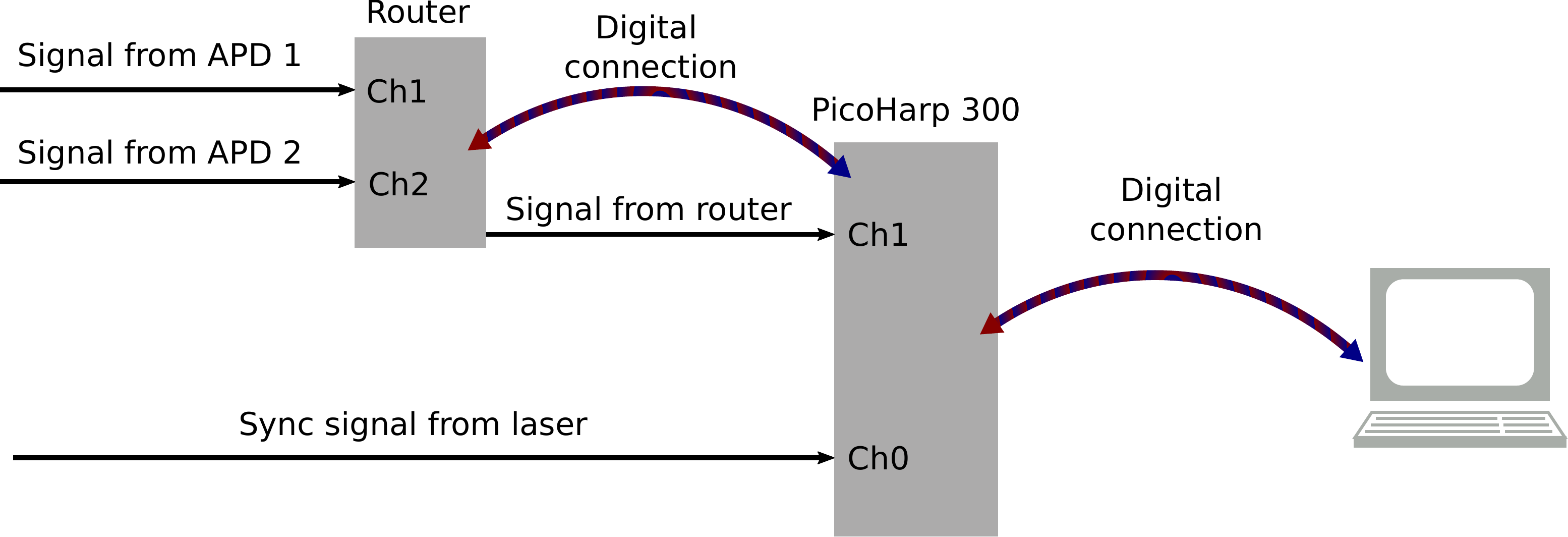}
    \caption{Scheme of the detection apparatus: the router transmits the detected signal to the PicoHarp card, passing the information on which channel the signal arrived from. The PicoHarp card measures the arrival time of the sync signal from the laser, given by the channel~0.The detection signal from the router is received in channel~1. After the stop signal is received, the card has a dead time on the order of hundreds of nanoseconds before being able to detect another signal on the same channel.}
    \label{fig:SchemePicoHARP} 
\end{figure}

To perform an autocorrelation measurement, also known as \gd, we use the setup
evidenced by a gray dashed line in figure~\ref{fig:setup}. The signal is
split in two parts and sent to two different avalanche photodiode single
photon detectors (Excelitas SPCM-AQRH-14-FC) via two multimode fibers.
To record the signal, a Picoquant PicoHarp~300 TCSPC card is used, together
with a Picoquant PHR~800 router. In practice the ``sync'' channel of the laser,
after an appropriate voltage conversion (the laser and the card use a different standard, thus the signal needs to be attenuated and inverted as explained in the manual), is sent to one channel of the
PicoHarp card (usually channel 0) while the other channel receives the signal
from the router. The router collects the signal from the two single photon
detectors and is sent it to the card, recording in which channel the photon has
arrived. The scheme of this mechanism is shown in figure~\ref{fig:SchemePicoHARP}.
If we use this system with the Time Tagged Time-Resolved mode, we can record a
huge amount of information. For each detected photon, we record:
\begin{itemize}
	\item the laser pulse that generated it, as an integer number that starts
	from 1 (first laser pulse sent in the measurement) and increases at each
	pulse (even when no photons are collected);
	\item the delay between the collected photon and the laser pulse;
	\item on which channel, and thus on which APD the photon was detected.
\end{itemize}

A disadvantage of this measurement technique is that the signal of both APDs is redirected from the router on the same channel on the card (channel~1 in figure~\ref{fig:SchemePicoHARP}).
The channel~1
of the card that records all the photons arrived is the same for both APDs.
After each photon detected, there is a dead time $t_{\mathrm{dead}}$ for the channel during which it is
not able to collect any other signal. As we want to measure \gdz{} this behaviour can be a huge problem, as
the dead time prevents to measure delays $t_\delta$ shorter than $t_{\mathrm{dead}}$. We can write this limitation in a more formal way by saying that we cannot measure $\tau$ if the following condition is satisfied:
\begin{equation}
\label{eq:delayLine}
    \abs{\tau}\leq t_\delta
\end{equation}
This can be resolved with a workaround, we can delay the arrival of the photons
of one of the two channels using a delay line: if we call this delay $T_D$ we can rewrite the condition~\eqref{eq:delayLine} as
\begin{equation}
\label{eq:delayLine2}
    T_D- t_\delta \leq \tau \leq T_D+t_\delta
\end{equation}
With a $T_D$ long enough, the interval fixed by equation~\eqref{eq:delayLine2} does not contain $\tau=0$.

To clarify the effect of the delay line in single photon detection it is useful
to refer to the image~\ref{fig:photonsdelay}. An ideal experiment is presented with
and without the addition of a delay line and is shown at different times. Let us
imagine two photons emitted almost simultaneously at the time $t_0$. They have
the same trajectory until (time $t_1$) when they encounter the beamsplitter. They then get separated in two different paths. Clearly, the photons can also continue on the
same path, but we will consider the case in which they propagate to different paths.
In the case where no delay line is present, both signals will arrive almost at
the same time $t_2$ onto different detectors, with the result that the card
will be able to record only one of them. In the case where a delay line is
present, the orange photon will arrive later, at time $t_3$; calling $\delta$
the fixed time delay introduced by the delay line. We have $t_3 = t_2 +
\delta$ and it is sufficient to subtract $\delta$ from the time measured by the
second detector to reconstruct the original signal.

\subsubsection{Signal cleaning}
\begin{figure}[tbh]
	\centering
	\subfloat[]{\label{fig:g2raw}\includegraphics[width=0.45\linewidth]
		{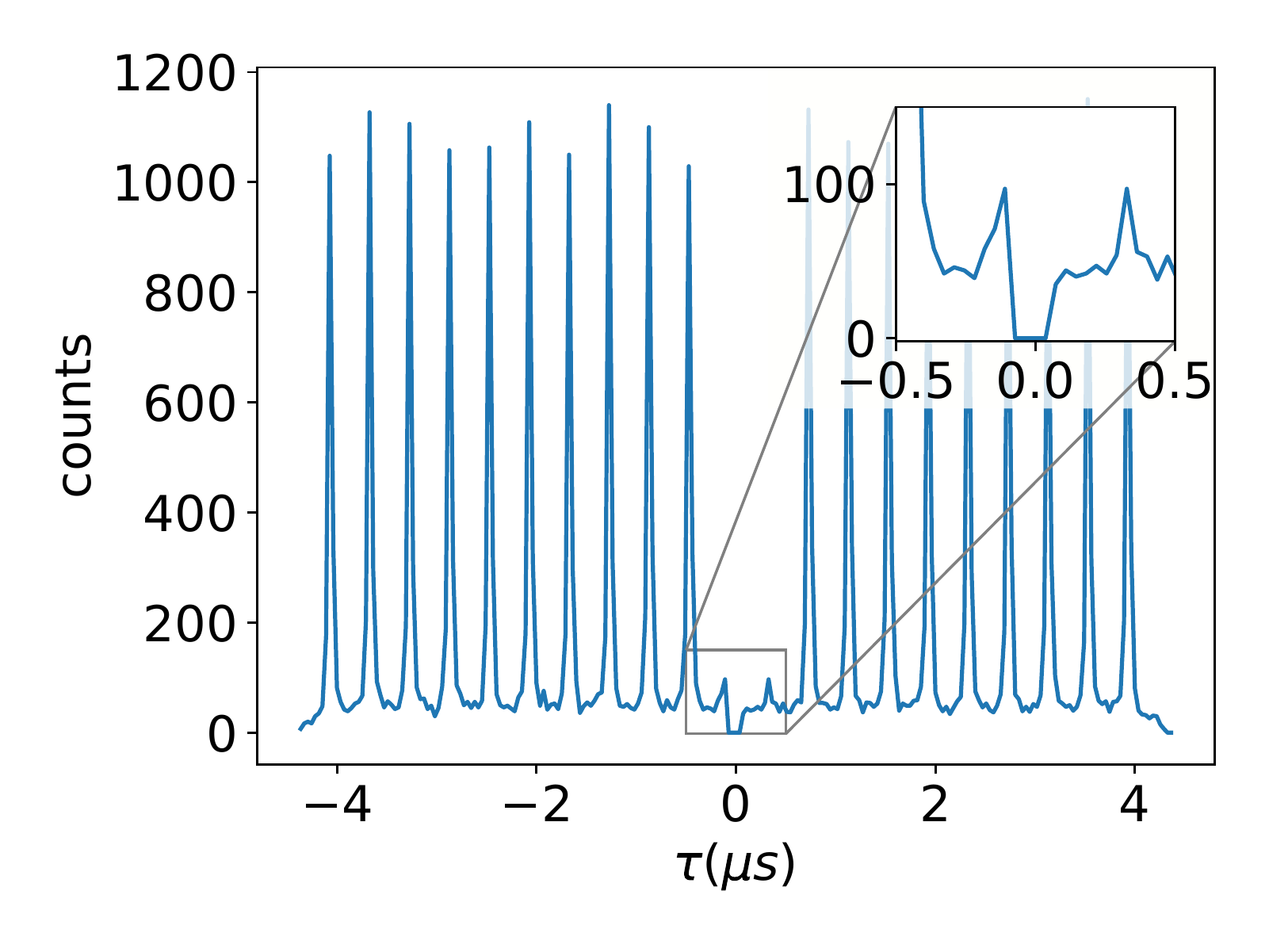}}\qquad
	\subfloat[]{\label{fig:g2clean}\includegraphics[width=0.45\linewidth]
		{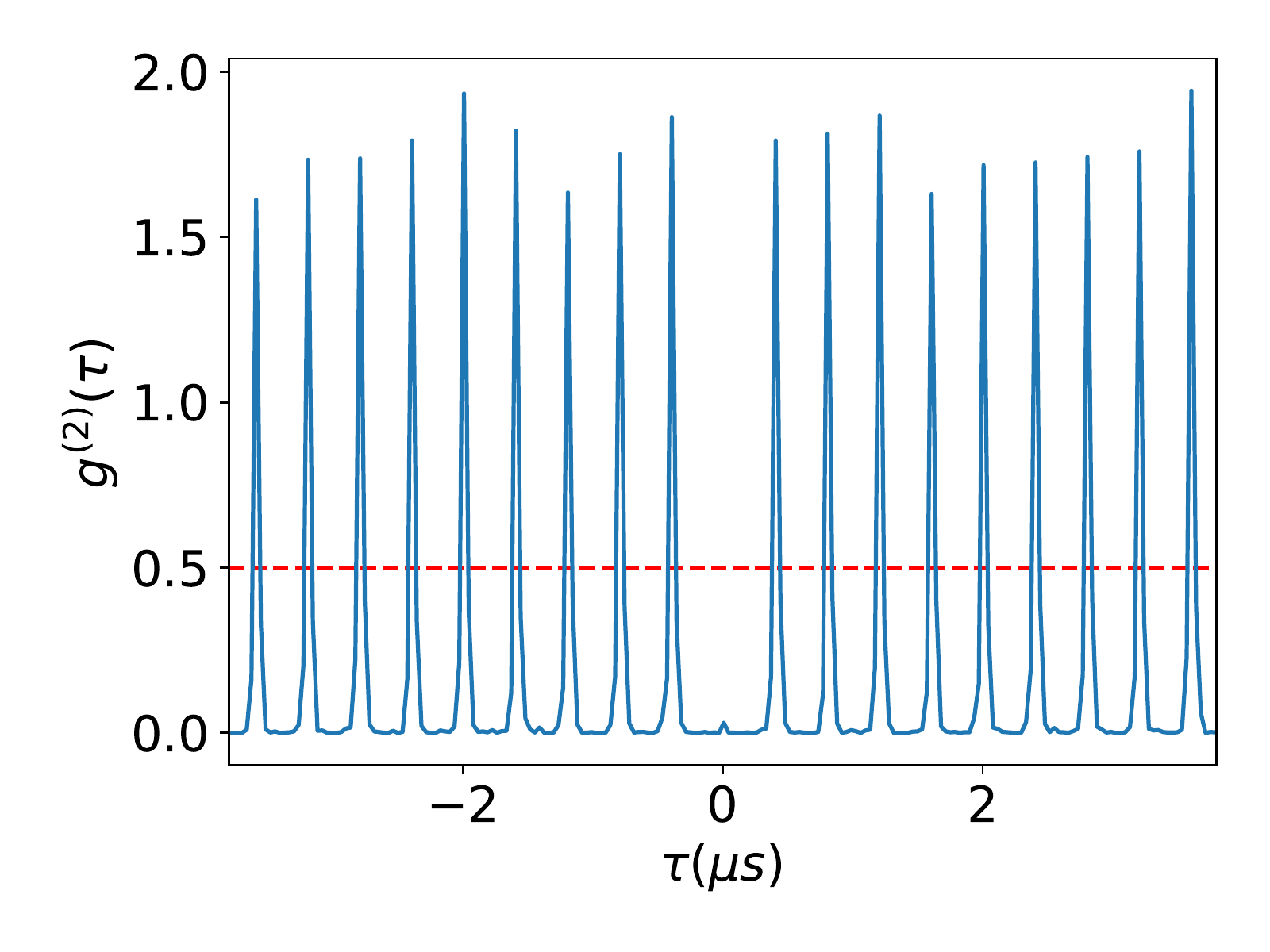}}
	\caption{%\label{fig:g2_example}
		\protect\subref{fig:g2raw}~\gdt{} measurement without any treatment, in
		the inset is clearly visible the dead time of the instrument
		\protect\subref{fig:g2clean}~\gdt{} elaborated and cleaned. The
		emitters shows a $ \textrm{g}^{(2)}\of{0}\approx0$, thus a very good
		single photon emission.
	}
\end{figure}

An example of the signal acquired is shown in figure~\ref{fig:g2raw}.
The inset clearly shows the dead time (as explained) while the
real zero delay of the \gd{} is moved to about \SI{300}{\nano \second} from the
origin of the time axis. On the y-axis is reported the number of photon
coincidences.
This gives some useful information but it needs to be
'clean' in order to obtain the final signal.

To do this we need to:
\begin{enumerate}
	\item move the zero of the time axis in order to remove the effect of the
	delay line,

	\item normalize the data to have the \gd{} function instead of the number
	of
	coincidences,

	\item remove the effects caused by the background of the measurement.

	\item remove the confusing zeros due to the dead-time
\end{enumerate}
If the first point is straightforward the other two are not.
For point~$2$, an approach is to look at the \gd{} measurement at long
delays
compared to the blinking time. There, the \gd{} function should converge to~1, due to its normalization.  %\textbf{(CC: why?)}.
In our case, using the pulsed laser, it means the height of the peaks should converge
to~1. We can therefore compute the \gdt{} for $\tau\approx\SI{10}{\milli
\second}$ to find the mean height of the peaks and normalize it to one.

Concerning point~3, we can make some considerations on how the
coincidences are counted. We call $M(\tau)$ the number of counts measured at a
given delay $\tau$. Each count can be generated either by a start from the
signal or from the background and by a stop from the signal or from the
background. Calling $s(\tau)$ the probability to have a start (or stop)
generated by the signal and $b(\tau)$ the probability to have the start (or
stop) generated by the background, we can write (when both $b(\tau)$ and
$c(\tau) \ll 1$):
\begin{equation}
M(\tau)=C\tonda{b(\tau)+s(\tau)}\tonda{b(\tau)+s(\tau)}
\end{equation}
where C is a constant of proportionality. We define $M'=\frac{M}{C}$.
We consider $\tau_b$ in between two consecutive peaks, we know that there is no
signal there and $s(\tau_b)=0$.
\begin{equation}
M'(\tau_b)=b^2(\tau)
\end{equation}
To find $M_{c}(\tau)=Cs^2(\tau)$ we solve the system of these two equations
and  we find that
\begin{equation}
M_{c}(\tau)=M(\tau)+M(\tau_b)-2\sqrt{M(\tau)}\sqrt{M(\tau_b)}
\end{equation}
With this formula, we obtain the $g^{(2)}(\tau)$ histogram cleaned from the
background counts.

The last step allows us to remove the ``dead-time''. This is easily done by removing a
number of points equivalent to one peak and we can see that it does not perturb
the result.

The resulting signal is shown in figure~\ref{fig:g2clean}. We can notice that
the peaks around $\tau=0$ (except for the one at $\tau=0$), reach a value higher than one. This is a well known effect of the blinking~\cite{manceau2018CdSe}. We can understand this behaviour by taking into account the fact that the probability to have an emission is higher when a photon has recently been detected. In this case, indeed, the state is known to be ``on''. On the opposite, at far delays from the first detected photon, the state could be both in ``on'' and ``off'' states. It is thus normal, fixing the normalization of the \gdt{} at 1 for large values of $\tau$, that the \gdt for small values of $\tau$ will reach values bigger than 1. 

\section{Stability}
As already explained, one of the main problems with perovskite nanocrystals is
their stability under light illumination. To address this problem we initially
chose to follow the approach of~\textcite{raino2019Underestimated}, and see if
the positive effect of polystyrene was confirmed with our setup.
To this end, we tried to add polystyrene to the Sample~A before the deposition, mixing a solution of this polymer to the solution containing the nanocrystals (both of them where dissolved in toluene),
to see if we were able to measure an increased stability.

\subsection{Role of the polystyrene}
\label{sec:rolepolymer}
\begin{figure}[tbh]
	\centering
	\subfloat[]{\label{fig:bleach_withpsornoti}\includegraphics[height=0.40\linewidth]
		{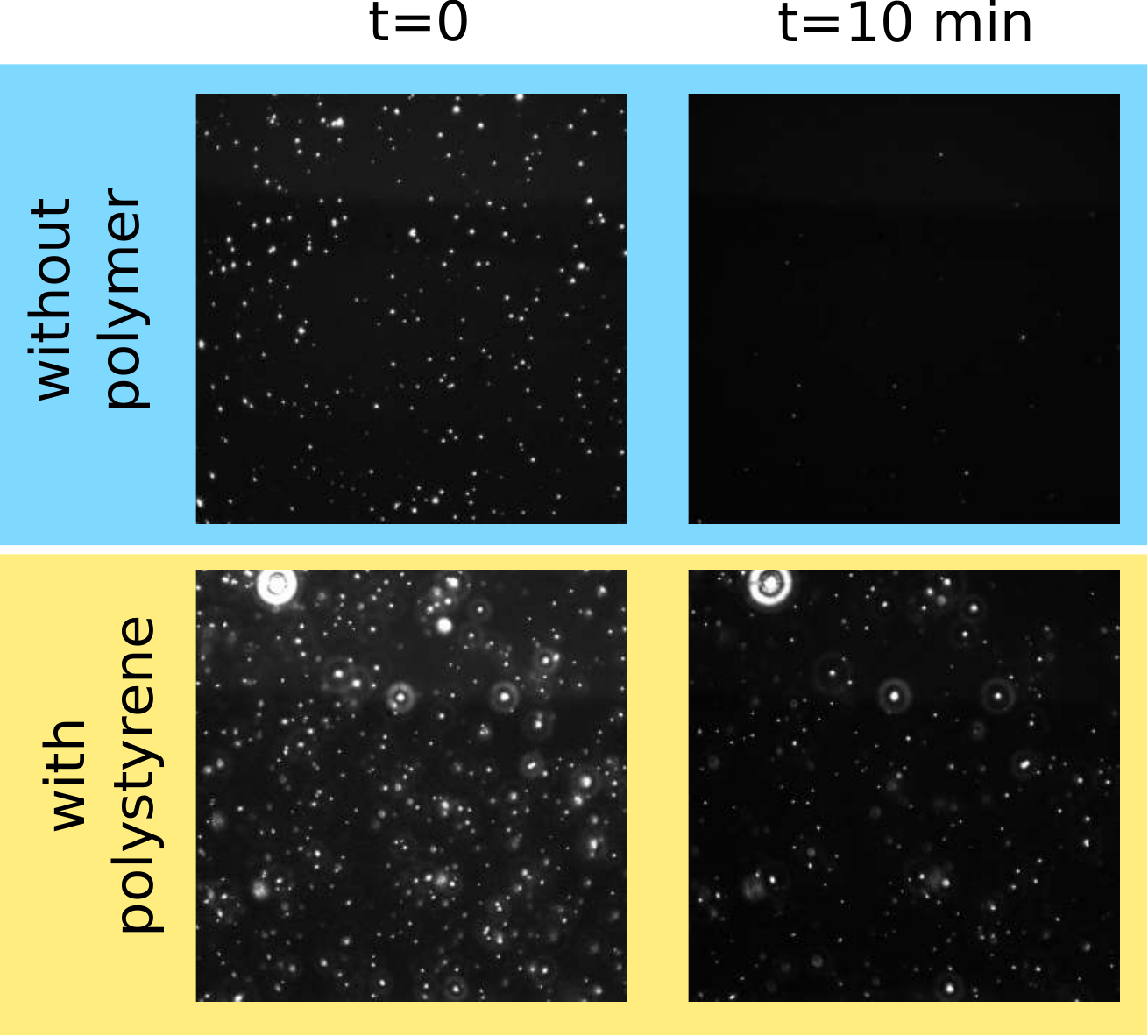}} \qquad
	\subfloat[]{\label{fig:bleach_withpsornott}\includegraphics[height=0.40\linewidth]
		{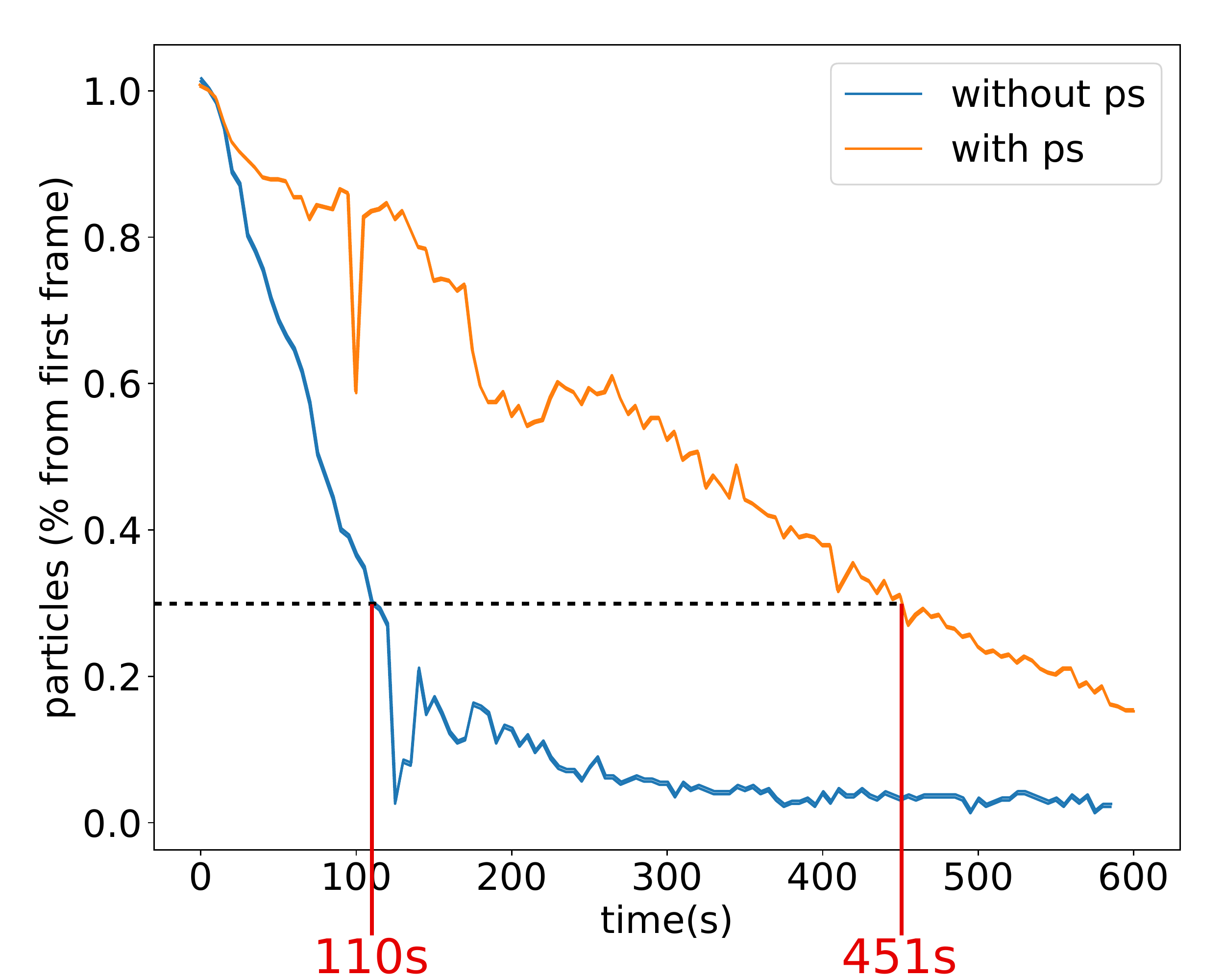}}
	\caption{\protect\subref{fig:bleach_withpsornoti}~Images of the samples at
	the beginning of the measurement ($t=0$) and after \SI{10}{\minute}
	($t=\SI{10}{min}$) under strong LED illumination, without and with
	polystyrene protection. \protect\subref{fig:bleach_withpsornott}~Quantitative analysis
	of the measurement: the stability is increased by a factor 4.}
	\label{fig:bleach_withpsornot}
\end{figure}

To perform the measurement, we decided to compare the sample with polystyrene and
a sample without polystyrene under the emission of the light of an LED at maximum power. Using the camera, we recorded an image of the
sample each~\SI{20}{\second} and counted the number of distinct bright emitters in each image via a
program written for this purpose.
This way, we were able to obtain a statistics of the emitters' behavior by a single set of images, directly measuring their bleaching time.
The result of this measurement is reported in figures~\ref{fig:bleach_withpsornot}.
On the left picture (\ref{fig:bleach_withpsornoti}) the images of the samples
with and without polystyrene are shown at the initial time of the measurement
and after 10~minutes of strong LED illumination. It is clearly visible that
the stability is increased by the presence of the polystyrene. Indeed, after 10~minutes, most of the emitters are still bright. This effect can be quantified
counting the number of emitters that are still bright in each frame. This is
shown in figure~\ref{fig:bleach_withpsornott}. If we measure the time at which the emission of the emitters is reduced by $\SI{30}{\percent}$ of the initial amount, we
can see that this time is multiplied by four thanks to the polystyrene protection. 

\subsection{Role of the Sample preparation}
\label{sec:rolesampleprep}
\begin{figure}[tbh]
	\centering
	\includegraphics[width=0.6\linewidth]{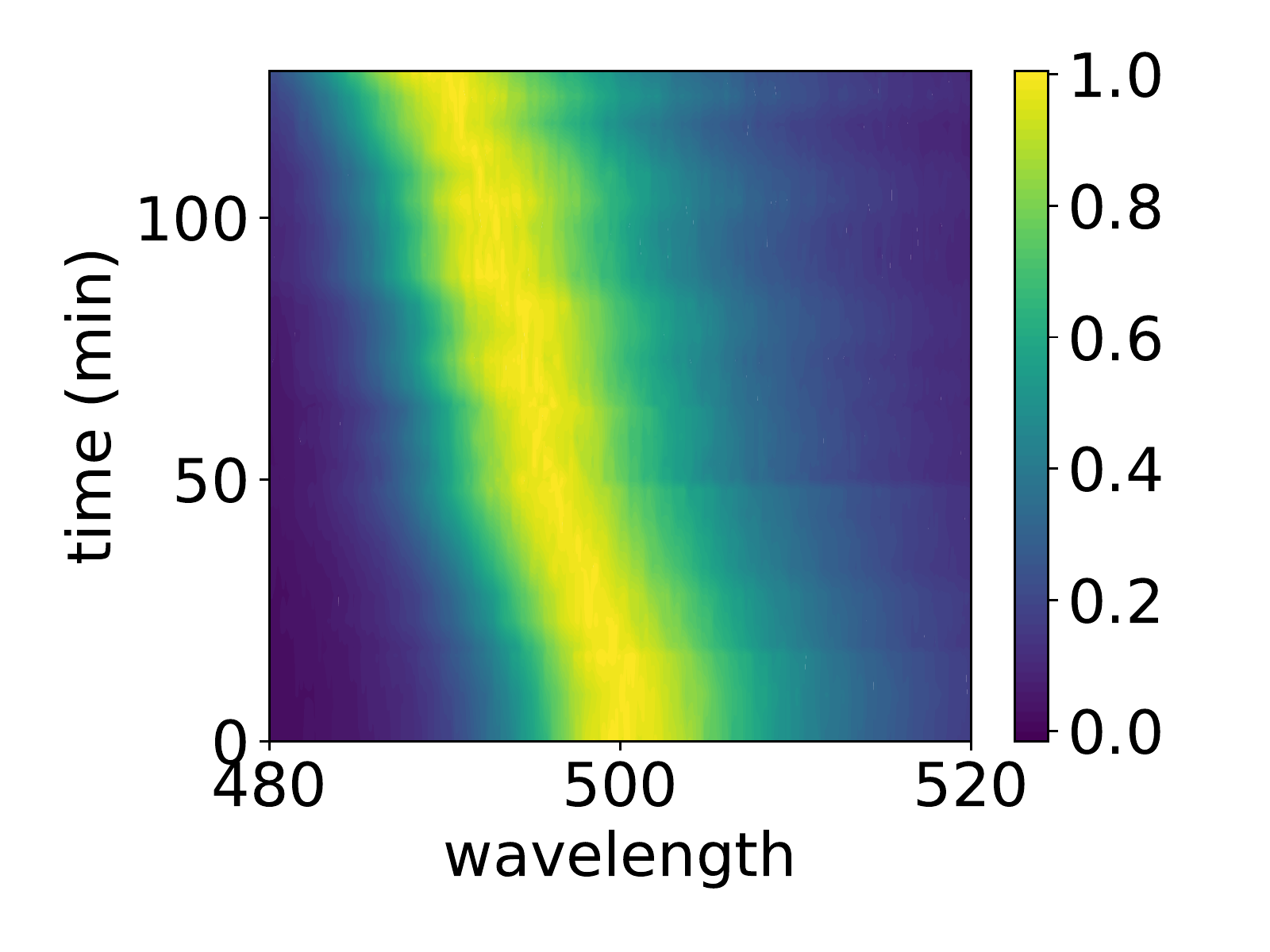}
	\caption{Evolution of the fluorescence spectrum of a single perovskite nanocube
	of sample~B as a function of time. The emitter was excited at the
	saturation intensity. In order to clear the image from the effect of the blinking, each spectrum (i.e. each row) is individually normalized. As a consequence, the image does not report the intensity evolution of the signal. A
	blue-shift of about \SI{10}{\nano \meter} after two hours is visible.}
	\label{fig:spectradegrationn}
\end{figure}

\begin{figure}[tbh]
	\centering
	\includegraphics[width=0.6\linewidth]{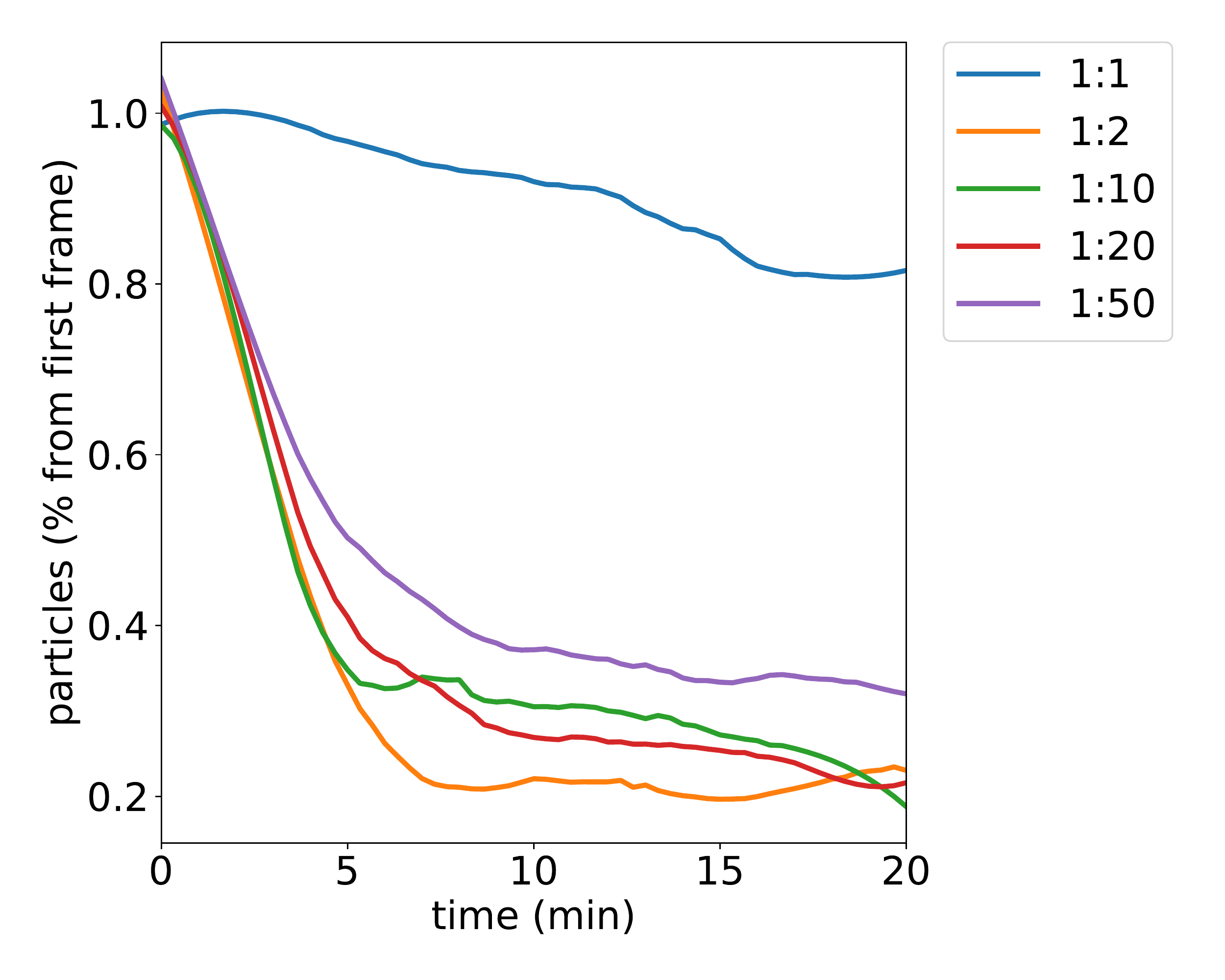}
	\caption{Measure of the robustness of perovskites nanocrystals for samples with different concentrations. The samples were obtained by diluting the original solution in toluene.}
	\label{fig:resistenza_diluizione}
\end{figure}

Protecting the perovskite emitters with a polymer is not the best option suitable for photonic
applications. For applications that need single nano-objects to be precisely
positioned, the presence of the polymer, that takes space, can perturb the near field.
For this reason we looked for alternative solutions and in particular we investigated different samples and different fabrication methods. Surprisingly, the fabrication of the sample~B that was intended to
obtain two-dimensional platelets, produces also very stable perovskite nanocubes.
To study the stability properties of these nanocubes, we performed a measurement similar to the one discussed in~\cite{raino2019Underestimated}. We illuminated a single photon emitter
at the saturation power with a pulsed laser and measured the evolution of the
spectrum as a function of time. This measure is shown in
Figure~\ref{fig:spectradegrationn}.
It is possible to see that there is a blue shift, as already observed in previous
works\cite{raino2019Underestimated} but with a very different timescale. In our
case we were able to measure the emission up to~\SI{2}{\hour} with a shift of
about~\SI{10}{\nano \meter}. We can compare our results with those shown in
figure~\ref{fig:rainoundernopolimerspectra}, where a degradation of more
than~\SI{20}{\nano \meter} is observed in some tens of seconds.
The blue shift can be attributed to the degradation of the
external layers of the emitters, that become smaller and thus are more
confined. It is interesting also to perform the same measurement shown in
section~\ref{sec:rolepolymer} for this sample.
We performed this measurement for different dilutions of the original solution,
finding that the robustness of the emitters is strongly concentration-dependent.
 This is shown in
figure~\ref{fig:resistenza_diluizione} where
the blue curve represents the number of emitters that are still alive as a
function of time for a sample prepared by spin-coating directly from the original
solution. The other curves represent the same measurement for samples prepared by diluting the
initial solution with toluene at different ratios.
It is clear that, while the concentrated sample is stable even
after~\SI{20}{\minute}, any subsequent dilution reduces the sample stability with
the result that only \SI{30}{\percent} of the emitters are still alive
after few minutes of strong excitation with the LED.
This is probably due to the presence of ligands alongside the perovskites in
the original dilution. When we increase the amount of solvent, we reduce
the ratio ligands/solvent in the solution, affecting this way the stability.

The effect of dilution on the stability can be a limit for using these emitters
for photonic applications as often very diluted samples are needed. With this study, we
have shown that the preparation of the emitters plays a crucial role on their
stability under light illumination and suggest a possible strategy  to increase the emitters stability  compensating the effect of the dilution. For example, we can add, along with toluene, the appropriate proportion of ligands needed to maintain the initial ratio ligands/solvent.

\section{Blinking properties}
It is interesting to characterize the blinking behavior of single quantum dots. Indeed, blinking is usually an undesired characteristic of nano-emitters and
its characterization helps to find the possible path to reduce it. In
addition, studying the blinking, we can access interesting information about the states contributing to the
emission.

In other kind of colloidal quantum dots, such as \ch{CdS}/\ch{CdSe} rods, this
behavior has been widely studied, as shown in the section~\ref{sec:blinking-ch1}.
%ref
Previous studies have shown that perovskite nanocrystals present
luminescence fluctuations, similar to those observed for other kinds of analogous
emitters. The first study of this behavior, described in
section~\ref{sec:blinkingPark} was
performed by~\textcite{park2015Room}. We decided to perform a similar study
with our emitters in order to understand their behavior. In order to acquire
a good statistics and to have a good signal to noise ratio, we need the emitters to stay in the bright state for a sufficient long
time: for this reason I will limit this study to the sample~B.

\begin{figure}[tbh]
	\centering
	\subfloat[]{\label{fig:blinking_a_glob}\includegraphics[width=0.45\linewidth]
		{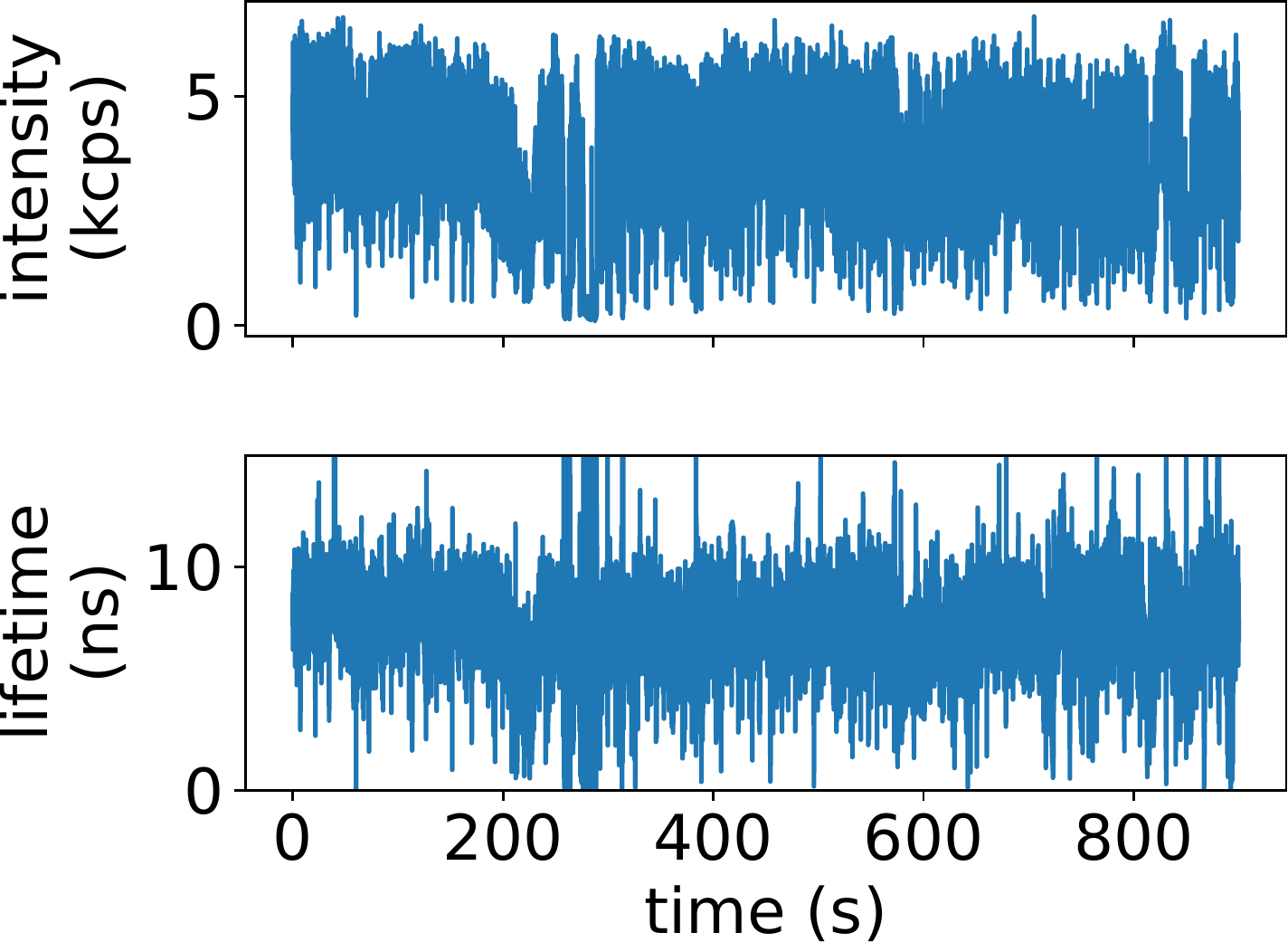}} \qquad
	\subfloat[]{\label{fig:blinking_a_zoom}\includegraphics[width=0.45\linewidth]
		{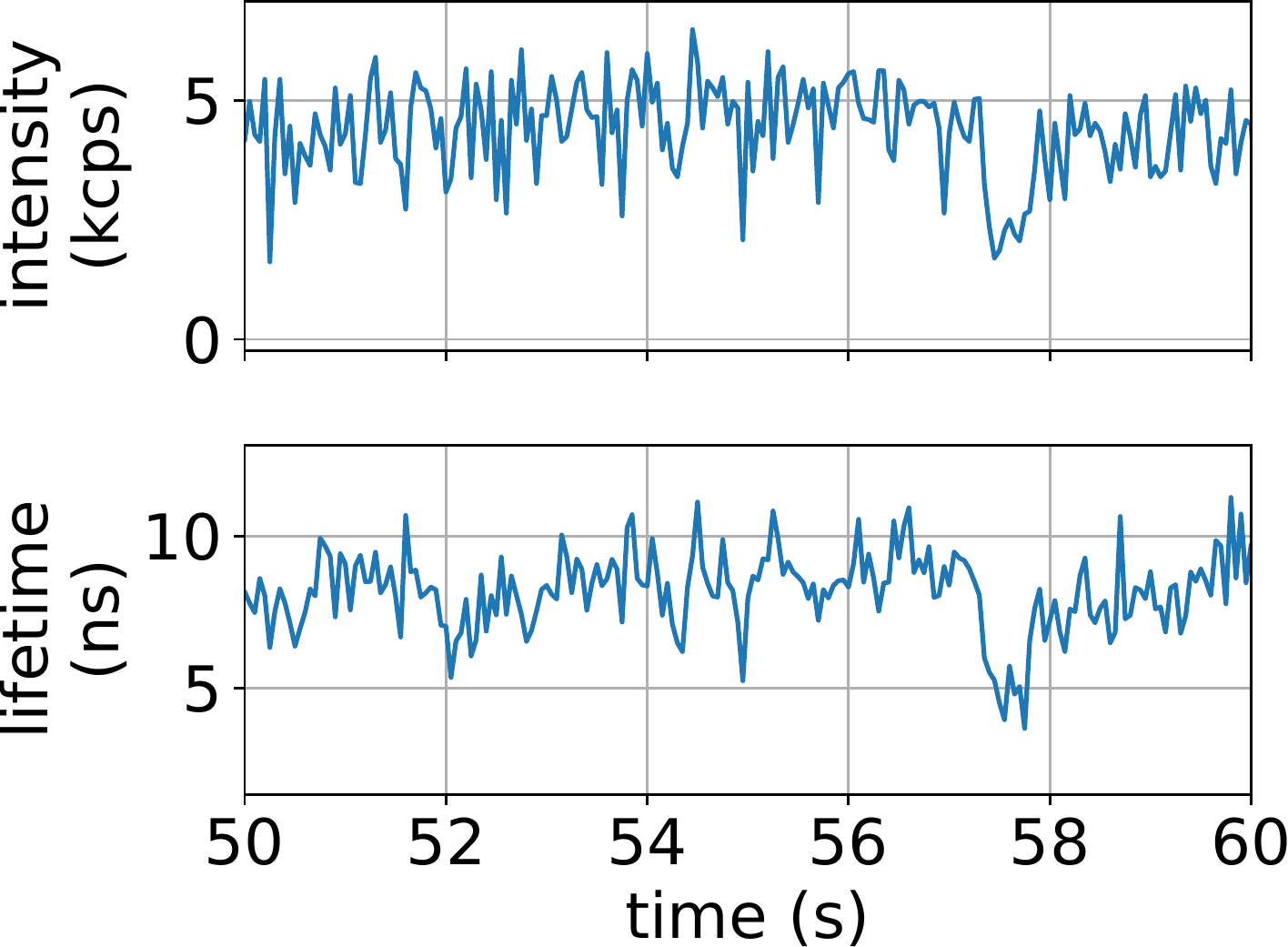}}
	\caption{Fluorescence trace acquired for a single photon emitter excited at
	half of the saturation intensity (top panels) correlated with the mean measured
	lifetime  (bottom panels).
	\protect\subref{fig:blinking_a_glob}~Full
	trace acquired. \protect\subref{fig:blinking_a_zoom}~Zoom on a part of the
	trace.}
	\label{fig:blinking_a}
\end{figure}

First of all, it is useful to observe the timetrace to have a rough idea of the
kind of blinking we are dealing with. This is reported in
figure~\ref{fig:blinking_a}
Specifically, from figure~\ref{fig:blinking_a_glob} on the top panel, the full
trace recorded for this emitter is shown. It is possible to see that there are fluctuations of the intensity, but it is not possible to identify regions completely
``dark''. This is confirmed by inspecting the zoomed curve reported in
the top panel of figure~\ref{fig:blinking_a_zoom}.
This means that the characteristic blinking time is actually shorter than our
binning time: in this case, it is better to talk about
flickering~\cite{galland2011Two}.

On the lower panels of figure~\ref{fig:blinking_a}, the evolution of the
lifetime as a function of time is reported. It is calculated by averaging the arrival time of the
photons in each bin.

Comparing this evolution with the fluorescence trace shown in the top panels, we can identify a correlation between the emission state and the
emission lifetime. This is clearer in figure~\ref{fig:blinking_a_zoom}
where, thanks to the zoom, we can clearly see that a reduction in the emission
at~$t=\SI{57}{\second}$ reflects in a shorter lifetime.
This is the signature of a type~A blinking.

\begin{figure}
	\centering
	\includegraphics[width=0.5\linewidth]{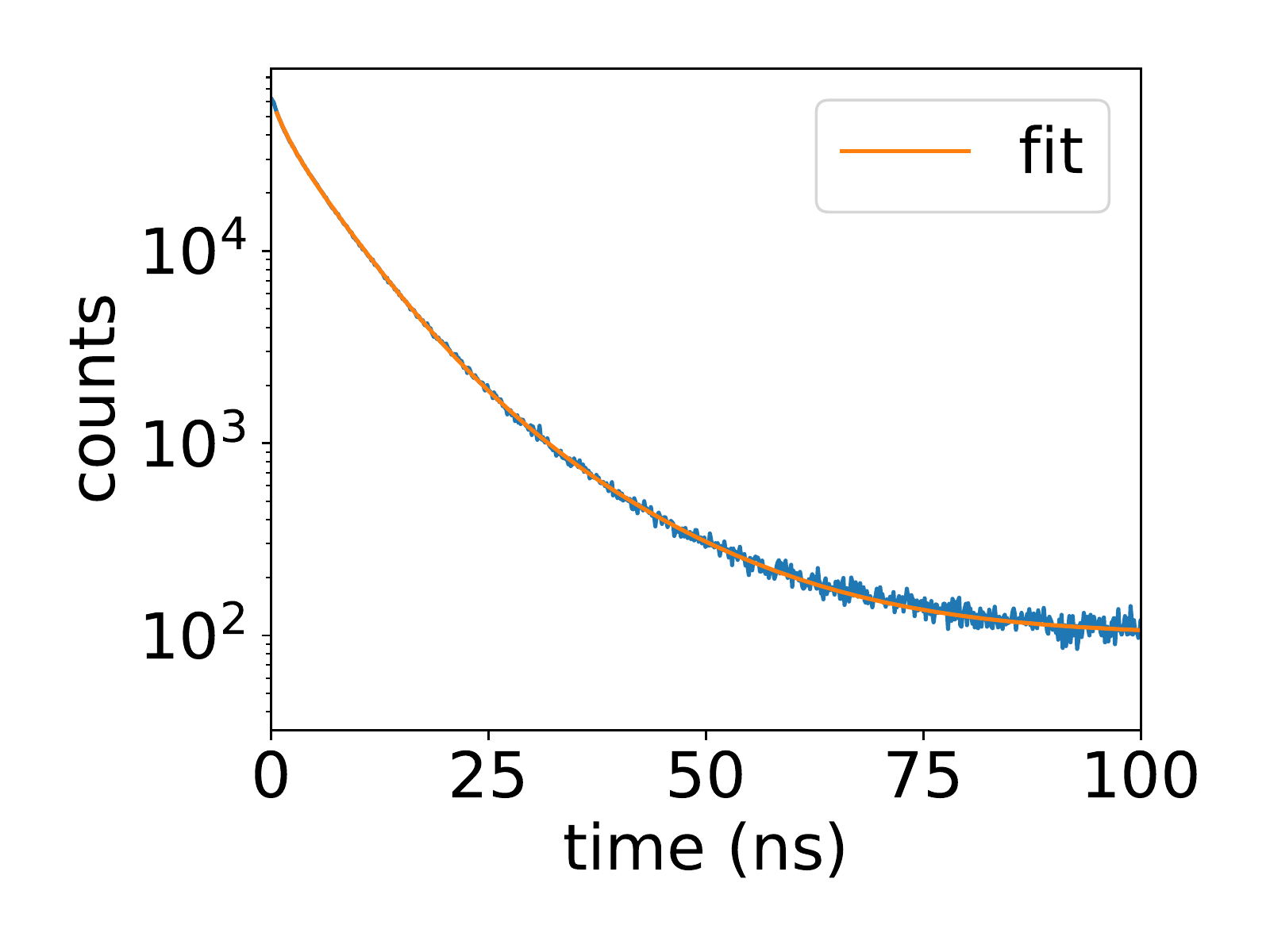}
	\caption{Lifetime of a single perovskite nanocrystal, fitted with a triple
		exponential decay model. We obtain \SI{1,4}{ns}, \SI{6,1}{ns} and
		\SI{14.7}{ns}, corresponding respectively to the lifetimes of the
		biexciton, grey and neutral emission states.}
	\label{fig:em1alifetime}
\end{figure}

Concentrating our analysis on these emitters, we can observe the normalized
lifetime
histogram shown in figure\ref{fig:em1alifetime}.
\begin{figure}
	\centering
	\subfloat[]{\label{fig:FLIDa}\includegraphics[width=0.45\linewidth]
		{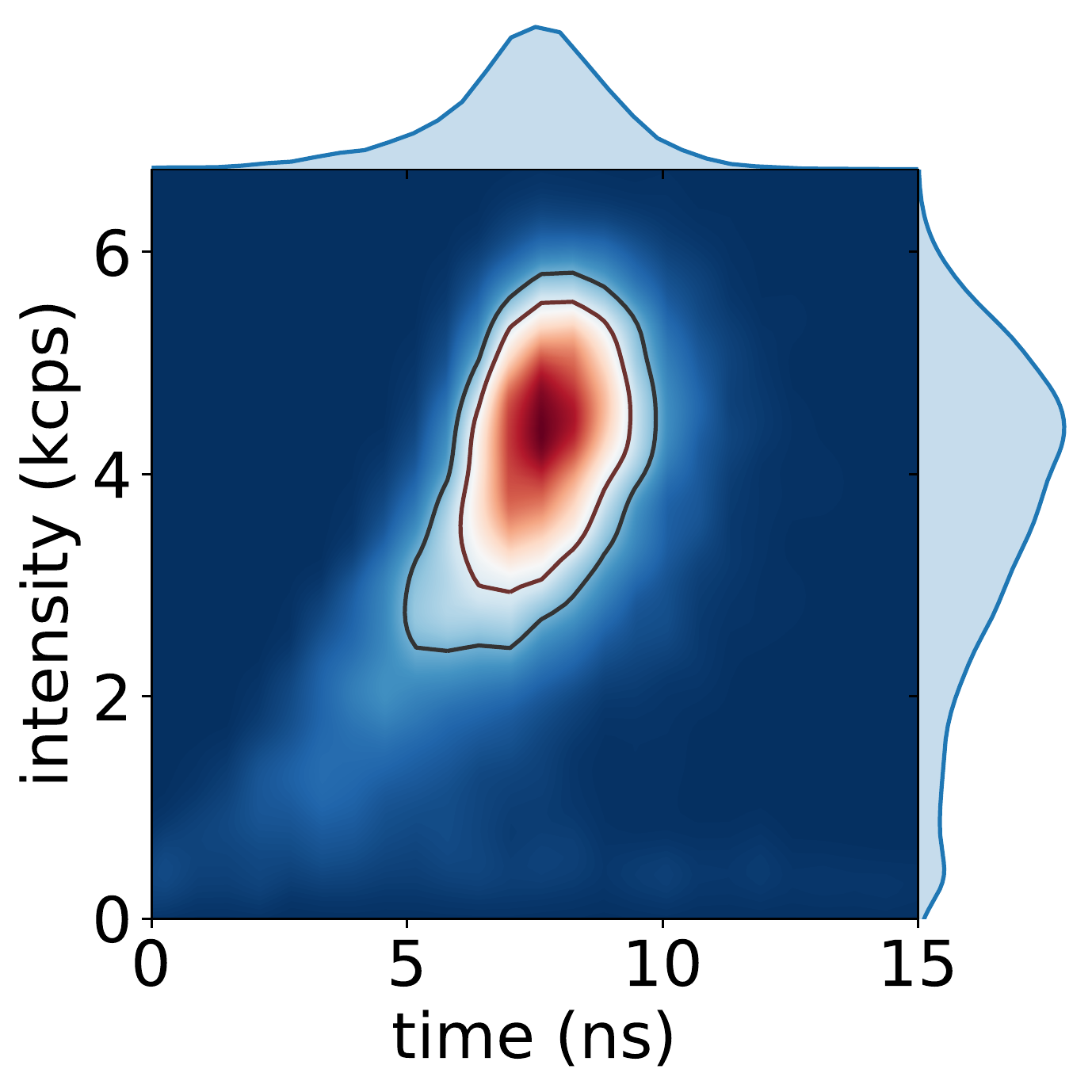}} \qquad
	\subfloat[]{\label{fig:FLIDb}\includegraphics[width=0.45\linewidth]
		{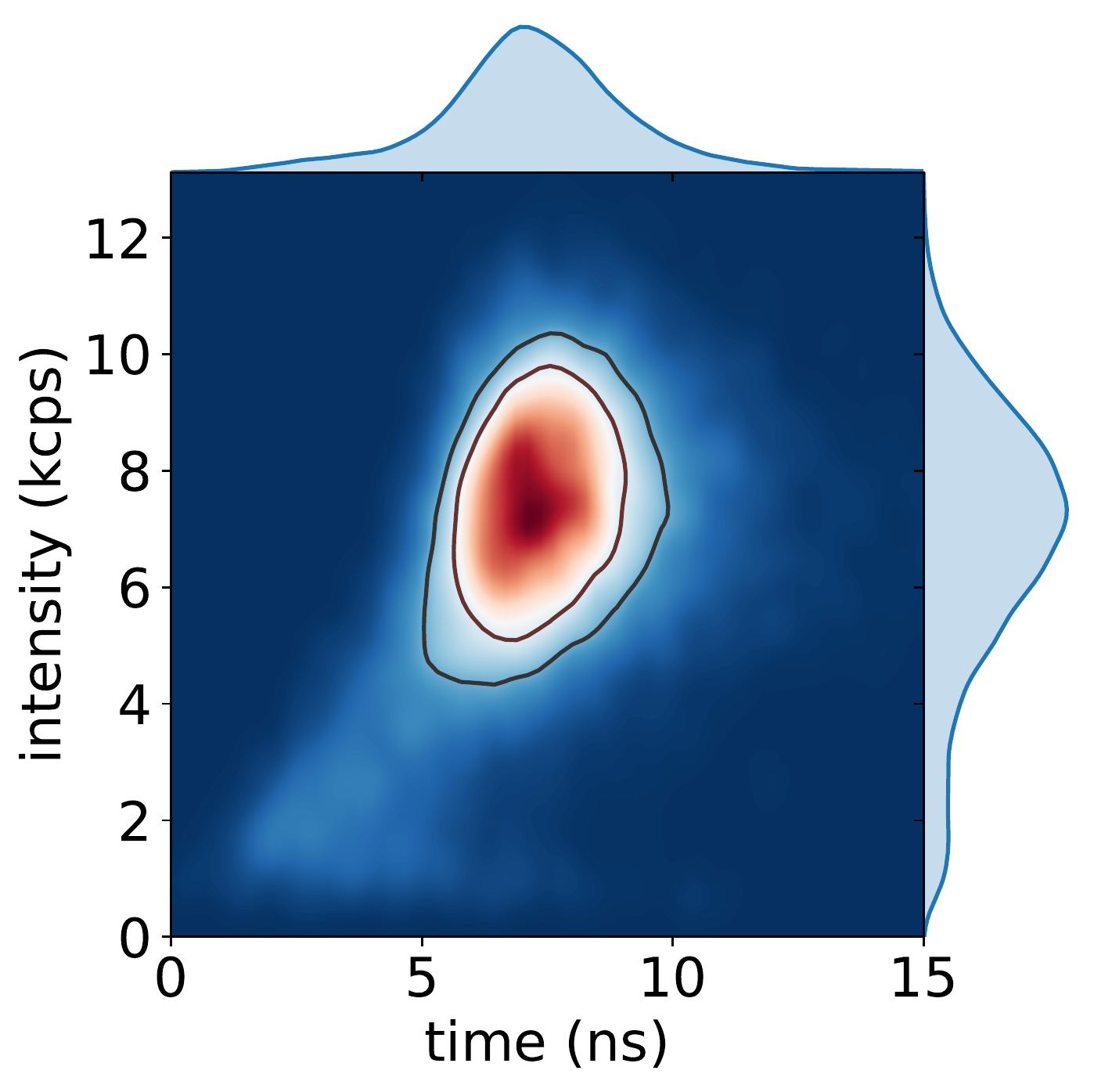}}
	\caption{fluorescence lifetime intensity distributions (FLID) for a single
	perovskite nanocrystal excited at half (\protect\subref{fig:FLIDa}) and
	twice (\protect\subref{fig:FLIDb}) the saturation intensity. The enclosed
	black curves contain respectively \SI{50}{\percent} and \SI{68}{\percent} of the
	occurrences.
		}
	\label{fig:FLID}
\end{figure}
The blue curve is the experimental points, while the orange curve is a fit
using a triple exponential decay model:
\begin{equation}
A_1\cdot e^{-\frac{t-t_0}{\tau_1}} +
A_2 \cdot e^{-\frac{t-t_0}{\tau_2}} +
A_3 \cdot e^{-\frac{t-t_0}{\tau_3}} + B
\label{eq:lifetime}
\end{equation}
where the three different lifetimes $\tau_1$, $\tau_2$ and $\tau_3$,
corresponds to the neutral, the charged and the biexciton emission
respectively~\cite{galland2011Two} %TODO add reference to theory section
(see in section~\ref{sec:lifetime-ch1}).
$A_1$, $A_2$ and $A_3$ are the respective amplitudes while $B$ represents the background counts.

We can now study the lifetime as a function of the emitted intensity with the
fluorescence lifetime intensity distributions (FLID) that are represented in
figure\ref{fig:FLID}.
The same emitter has been excited with two different intensities, at half of the
saturation intensity (figure~\ref{fig:FLIDa}) and at twice the saturation
intensity (figure~\ref{fig:FLIDb}). In both cases, the area surrounded by the
closed curves contains \SI{50}{\percent} (inner curve) and
\SI{68}{\percent} of the occurrences.
If we compare these distributions with the ones measured
by~\textcite{park2015Room} shown in figure~\ref{fig:parkflid}, we can see that, even at high excitation intensity, our emitters remain in a bright state. This
property shows again the robustness of the emitters produced with this
fabrication method. A good level of emission remains even under
non-optimal strong excitation conditions as the ones presented here.

\section{\gd{} distribution}
\begin{figure}
	\centering
	\includegraphics[width=0.6\linewidth]{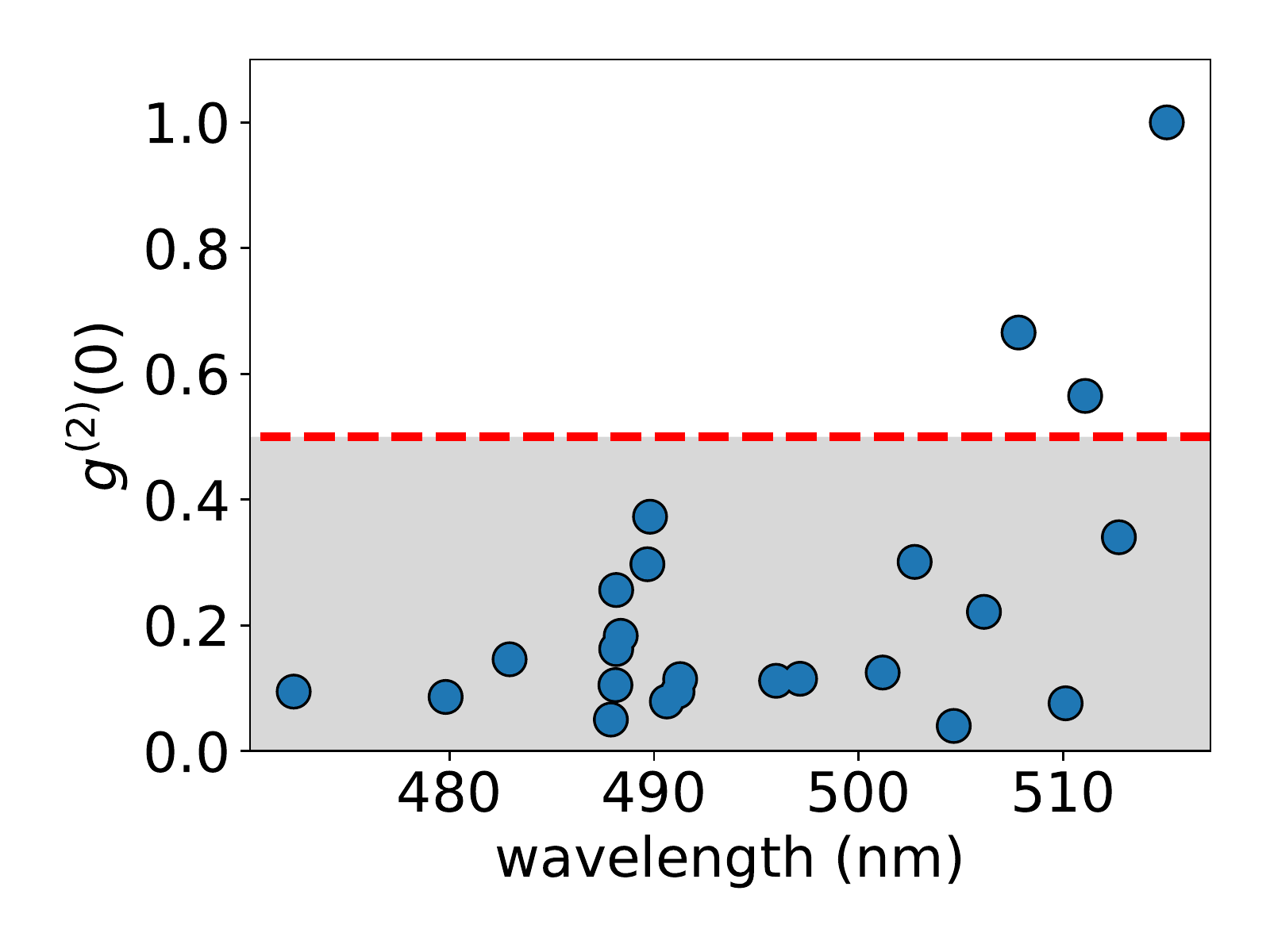}
	\caption{\gdz{} distribution as a function of the central emission wavelength
	for 33 different emitters excited at the saturation intensity.}
	\label{fig:g2vslambdapsat}
\end{figure}
We have seen before how it is possible to measure the spectrum and the \gd{} for
single photon emitters, as well as the information we can extract from these measurements. It
is also interesting to combine this information and visualize the distribution
of the \gd{} as a function of the wavelength. We performed this for sample~B and it
is reported in figure~\ref{fig:g2vslambdapsat}.
For each emitter, the \gdz{} function is shown as a function of the central emission
wavelength. The shaded region contains the emitters that can be considered
emitting single photons. We can observe a degradation of the single photon
emission when the central emission wavelength increases. 
In order to find an explanation to this phenomena it is interesting to study the size distributions of the cubes in our sample, as well a the mean central emission wavelength for larger nanocubes.

\begin{figure}
    \centering
    \includegraphics[width=\linewidth]{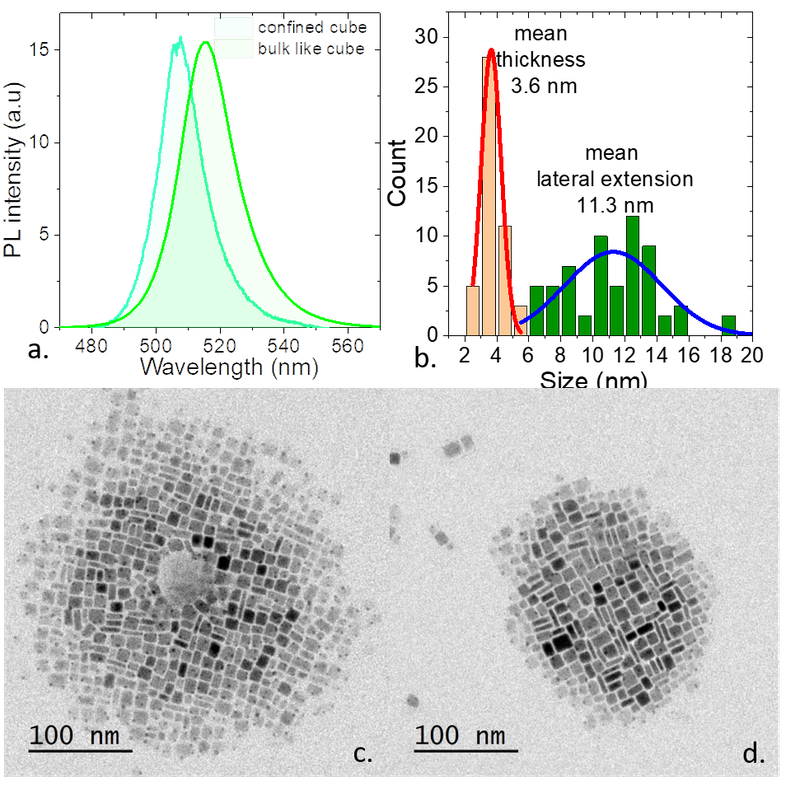}
    \caption{a)~PL spectra for confined cubes obtained in sample~A and for the bulk like \ch{CsPbBr3} cubes from the Protesescu’s method\cite{protesescu2015nanocrystals}. b)~Size histogram for the confined cubes showing two peaks which are fitted by two gaussians leading to a mean thickness of \SI{3.6}{nm} and a mean lateral extension of \SI{11.3}{nm}. c)~and d)~are TEM images of the confined cubes}
    \label{fig:figS2}
\end{figure}

In figure~\ref{fig:figS2}a the ensemble emission spectra obtained for sample~A is shown in comparison to what obtained for larger bulk-like cubes obtained using Protesescu’s method\cite{protesescu2015nanocrystals}. 
%As already observed in the statistics reported in figure~\ref{fig:wave_vs_FWHM}, the cubes present in sample~B has a lower emission wavelength and a narrower emission compared to sample~A.
From this image we can see how the size plays a role on the central emission wavelength of the cubes.
In panel~b) of figure~\ref{fig:figS2} the size distribution of nanocubes in sample~A is reported, while panels~c) and~d) reports images of the cubes in sample~A obtained with a scanning electron microscope. It is possible to see a variability in the sizes of produced cubes, the larger cubes having a lateral size of several times the smaller ones. 
A similar variability can be observed for nanocrystals of sample~B.

%\ref{fig:figS2}
It seems reasonable to attribute to size variability the different central emission wavelengths observed for nanocrystals as shown in figure~\ref{fig:wave_vs_FWHM}. In addition, knowing that the bulk perovskite does not show single photon emission, the increase of the size induces, at a certain point, the loss of the single-photon emission. For this reason we attribute the effect reported on figure~\ref{fig:g2vslambdapsat} to the loss of the confinement due to the larger size of the emitters.

\clearpage
\begin{chrecap}
	\begin{itemize}
		\item Perovskites are well known chemical structures in which the
		interest of the scientific community increases.
		\item Perovskite nanocrystals are promising single photon emitters but
		they have stability problems
		\item polymers can improve the stability of the nanocrystals protecting
		them from moisture and light, but have the inconvenience to make
		difficult the coupling of the emitters with integrated platforms
		\item different fabrication methods can have a huge effect on the emitters'
		stability
		\item I have described a fabrication method that
		\begin{itemize}
			\item improves the stability of the emitters, making them last for
			more than \SI{1}{\hour}
			\item produces nanocrystals with a shorter characteristic blinking time and they remain in a
			bright state even at high excitation power
		\end{itemize}
		\item I have shown the relation between the central wavelength
		emission of the emitters and their \gdz, showing a loss of single
		photon emission for higher emission wavelengths
	\end{itemize}
\end{chrecap}
\chapter{Coupling nanoemitters to nanofibers}
\label{chap:fiber}
\minitoc
\section{Optical fiber}
\label{sec:fiber}
By optical fiber, we mean a transparent and flexible cylinder made by a
dielectric material that is able to guide light using total internal reflection.

\begin{figure}[tbh]
	\centering
	\subfloat[]{\label{fig:TIR_scheme}\includegraphics[height=0.5\linewidth]
		{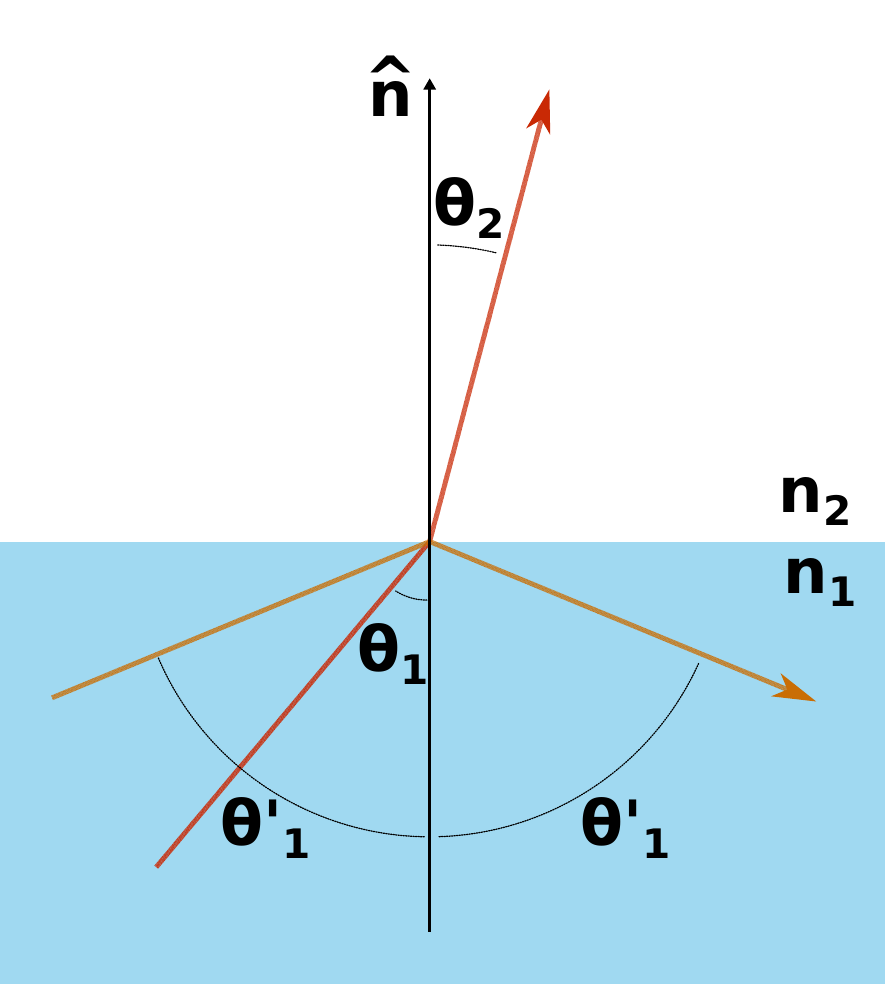}}\qquad\qquad
	\subfloat[]{\label{fig:TIR_Colladon}\includegraphics[height=0.5\linewidth]
		{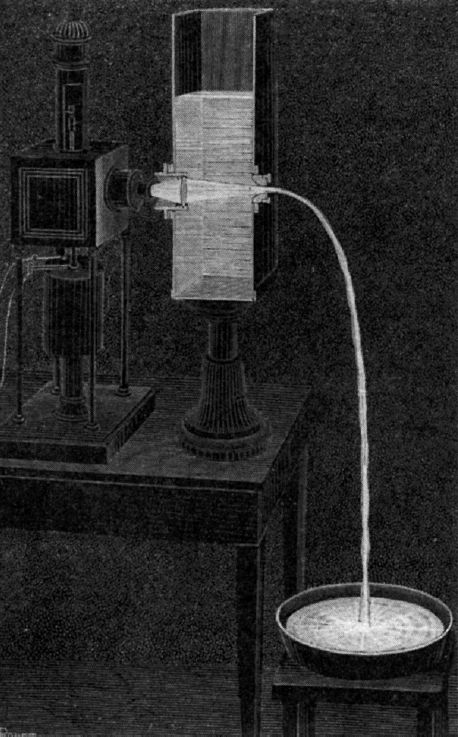}}
	\caption{\protect\subref{fig:TIR_scheme}~Low of refraction: the ray in red
	hits the surface with an angle $\theta_1<\theta_l$ with respect to the normal and it
	is refracted with an angle $\theta_2$; the ray in orange hits the
	surface with an angle $\theta'_1>\theta_l$ and it is totally reflected.
	\protect\subref{fig:TIR_Colladon}~Jean-Daniel Colladon experiment: A tank
	contains water that falls down through a hole in open air.
	The light from the lamp on the left side is guided by the water. This image
	was published in the ``La Nature'' review in 1884.
		}
\end{figure}

This is a fundamental principle in optics and its discovery is usually attributed to Johannes
Kepler\cite{mach2013Principles}. He did not derive the right laws of
refraction but he was the first scientist to observe that, while the angle of light incident in water
is varied between $0$ and \SI{90}{\deg}, the angle of the light propagating inside the water is at maximum for
$42\deg$. After this first observation and thanks to the contribution of René Descartes, Christian Huygens and Isaac Newton (amongst others), it became clear that light,
when going from a more refractive to a less refractive medium, can be refracted
only if the incidence angle is small enough. When the angle is larger than a
certain threshold, light is completely reflected.
This phenomenon, called total internal reflection, is described by the Snell law which is given by:
\begin{equation}
	\dfrac{\sin \theta_2}{\sin \theta_1}=\dfrac{n_1}{n_2}
\end{equation}
where $\theta_1$ and $\theta_2$ are the incidence angle and the refraction
angle respectively, $n_1$ and $n_2$ are the refraction index of the first and the second medium respectively
(figure~\ref{fig:TIR_scheme}).
It is seen that when $n_1>n_2$, the equation does not have a solution for any
$\theta_1$. Indeed, we can multiply by $\sin \theta_1$ both sides of the
equation
to obtain
\begin{equation}
{\sin \theta_2}=\dfrac{n_1}{n_2} \sin \theta_1
\end{equation}
For this equation to have a solution, we need to have:
\begin{equation}
\dfrac{n_1}{n_2} \sin \theta_1 \leq 1
\Longrightarrow
\theta_1 \leq \arcsin \tonda{\dfrac{n_2}{n_1}}
\end{equation}
The angle for which the equality is true is known as limit angle, $\theta_l$. We
can thus reformulate the description of total internal refraction saying that
when the incidence angle is larger than the limit angle, light is completely
reflected at the incidence plane.
With this principle and neglecting the roughness of the surface, it is possible
to guide light. Today this seems trivial, but this possibility has been
demonstrated for the first time only in early 1840
by~\textcite{colladon1842reflections}.
The first practical applications of this discovery were used to guide light to
illuminate body cavities (as surgical lamp). The scientists tried earlier on to use optical guides to transport information. The first
attempt was done with the photophone invented by Bell that used the sun light
focused with a lens and a vibrating mirror to translate the vibration of the
mirror in optical signal. The vibration on the ray was converted back in electrical signal at
the arrival. This system had unfortunately the problem to be unusable in cloudy
days, limits that forced Bell to abandon it.

The technique of  transmission of information via optical fibers was then used in the second half of the XXth century
to capture images from inside the human body and by NASA in cameras that were sent
to the moon. For long distance communications, unfortunately, the main problem
was the losses too high. At the beginning, the origin of the losses was not understood.
The discovery that the losses were caused mainly by glass impurities, (effect that go the Nobel prize to Charles Kuen Kao in 2009) opened the path to low loss fibers that nowadays are broadly used in communication technologies providing  a repeater each \SI{150}{\kilo \meter} is used for very long distance communications.

A fiber is usually made of two collinear cylinders: the core and the
cladding, the second one with a refraction index that is slightly lower than the
first one. Depending on the cladding size and thus on its ability to
transport one ore more optical modes, fibers can be classified into single mode
and multi-mode fibers for a given wavelength.

Single mode fibers have usually a core of few
micrometers in diameter and a cladding of \SI{125}{\micro \meter}.

In the following I will first explain in detail the mechanism of guiding light inside  an optical fiber and the peculiarity of nanofibers before describing the nanofiber fabrication method.

\section{Light propagation inside optical fibers}

\subsection{Propagation in a planar guide}

\begin{figure}
    \centering
    \includegraphics[width=0.6\linewidth]{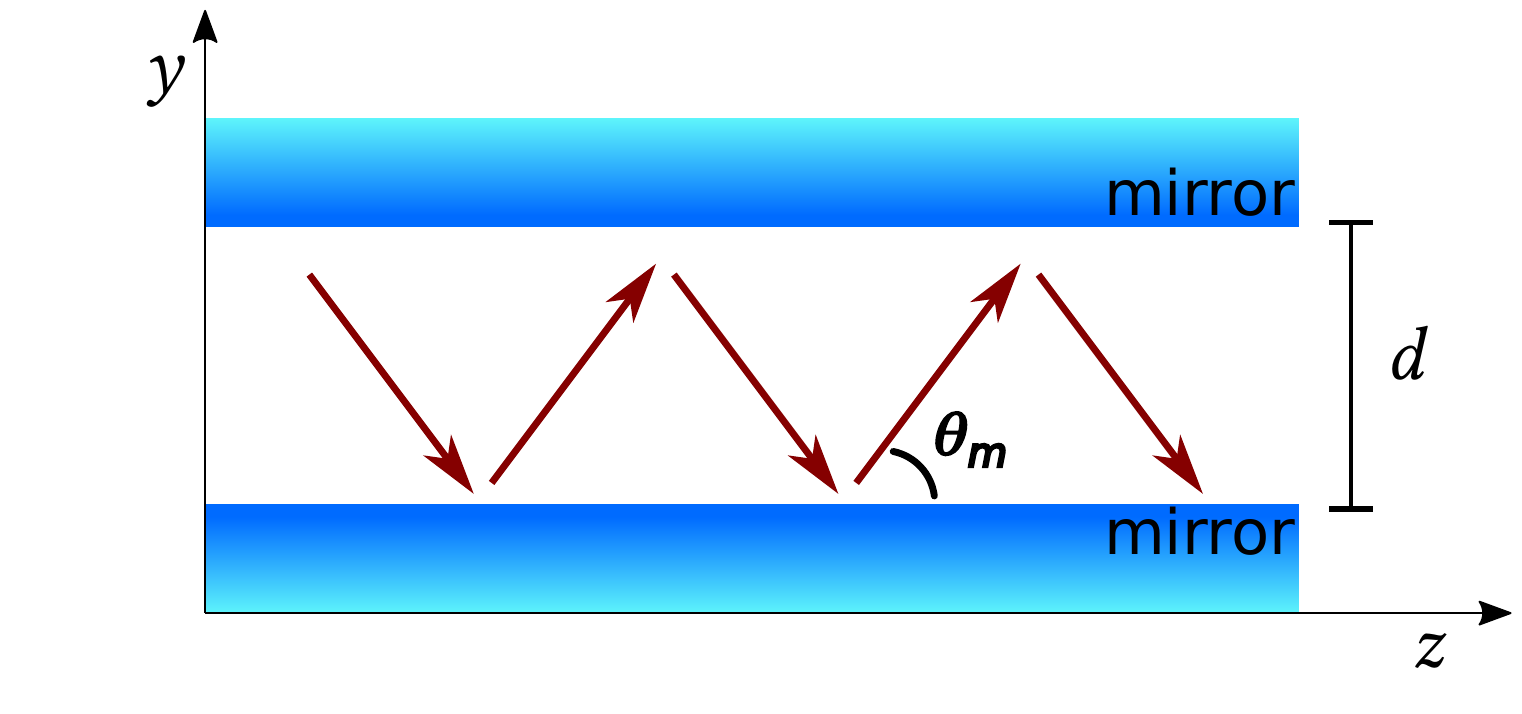}
    \caption{Scheme of the geometrical model for a plane wave propagation between two mirrors. In blue are represented the mirrors, spaced by a distance $d$. The light propagates along the path indicated in red. }
    \label{fig:PlaneWave}
\end{figure}

The intuitive explanation given in the previous section is not sufficient
to explain all the underlying physics. To go a step further, let us start with the simple
example of the light guided between two planar mirrors and consider a
transverse electric plane wave propagating inside them. This is  illustrated  in figure~\ref{fig:PlaneWave}. With transverse electric (TE) we indicate an electromagnetic wave with no electric field in the direction of the propagation. In analogy, transverse magnetic (TM) waves are electromagnetic waves with no magnetic field in the direction of the propagation.
Considering the mirrors extending indefinitely in both directions, in order to have constructive interferences,  we need to impose the so called ``self consistent''
condition: all the waves propagating downwards must be in phase, in order to not destructively interfere (this is also valid for all the waves propagating upwards).

A wave that satisfies this condition is called \textit{Mode}: in other words the modes are specific distributions of the electromagnetic field that maintain the same transverse shape (i.e. on the surface orthogonal to the waveguide) and
polarization at all the locations along the waveguide
axis~\cite{saleh2007Fundamentals}. If the incident angle for a given mode is
$\theta_m$, to be self consistent, the optical path between two different
reflections has to be an integer multiple of $2\pi$. In this case we have.
\begin{equation}
	\label{eq:modemirror}
	\sin \theta_m=m \dfrac{\lambda}{2d}
\end{equation}
where %$n$ is the refraction index of the medium,
$\lambda$ is the wavelength of the
radiation in the medium, $d$ the distance between the planes and $m$ a positive
integer, called order of the mode. When this condition is fulfilled, the upward
and downward components interfere destructively and only the horizontal
component is present. We recall that $\lambda=\lambda_0 / n$, where $\lambda_0$ is the wavelength of the radiation in void and $n$ the refraction index of the medium.
%its $k$ number can be written as
%\begin{equation}
%	k_{ym}=m\dfrac{\pi}{d}
%\end{equation}
%differents modes propagates along different paths and thus have a different
%group velocity
The propagation constant $\beta$ is given by the modulus of the projection
of $\vec{k}$-vector over the y-axis:
\begin{equation}
\label{eq:betamDef}
	\beta_m=k \cos \theta_m
\end{equation}
using the equation~\eqref{eq:modemirror} and we recall that $\sin^2 \theta
+\cos^2 \theta=1$, we can write:
\begin{equation}
	\beta^2_m=k^2-\dfrac{m^2 \pi^2}{d^2}
\end{equation}
This means that modes of higher order travel with a smaller $\beta$ constant. The amplitude of the electric field $E_x\of{y,z}$ can be written as~\cite{saleh2007Fundamentals}:
\begin{equation}
    E_{x}(y, z)={a}_{m} u_{m}\of{y} e^{-i \beta_{m} z}
\end{equation}
where $a_m$ is the amplitude of the mode $m$, $i$ is the imaginary unit and the $u_m$ are given by equation~\eqref{eq:um}.
\begin{equation}
\label{eq:um}
u_{m}(y)=\left\{\begin{array}{ll}
\sqrt{\dfrac{2}{d}} \cos \left(m \pi \dfrac{y}{d}\right), & m=1,3,5, \ldots \\
\sqrt{\dfrac{2}{d}} \sin \left(m \pi \dfrac{y}{d}\right), & m=2,4,6, \ldots
\end{array}\right.
\end{equation}

From equation~\eqref{eq:modemirror}, we can also extract the maximal number
of modes $M$ that a mirror guide can accept, given the fact that $\sin
\theta_m\leq1$
\begin{equation}
	m \dfrac{\lambda}{2d} \leq 1 \qquad \Longrightarrow \qquad M=
	\dfrac{2d}{\lambda}
\end{equation}
the maximal number of modes thus decreases as the ratio between
the distance of mirrors and the wavelength of the propagating light decreases.

We can also note that when $\lambda<2d$, no mode is allowed to propagate in the waveguide and for this reason $\lambda<2d$ is called  the cut-off wavelength. A guide that can
guide only one mode (i.e. with $\lambda_c\leq \lambda < 4d$) is called
a single-mode waveguide.
It is also important to note that, as each mode has a different $\beta_m$, it
propagates with a different group velocity. The group velocity dispersion is a peculiarity that makes
difficult the transmission of information over long distances with multimode
waveguides (and fibers).

To conclude the description of the planar waveguide example, we can note that if we want to
consider TM
modes instead of the TE ones, the mathematical treatment will be similar: an extensive  analysis of the physics governing the light propagation in planar waveguides can be found
in the book of~\textcite{saleh2007Fundamentals}.

%\textbf{(CC: qq remarques: I think an expression of the E field with beta would help to understand what it is. Recall what group velocity is and give and example with an index depending on lambda. FInally, you mention TE and TM for the first time here. You should recall what they are before no? When you describe the waveguide)}

\subsection{Propagation in optical fibers}
To study light propagation inside an optical fiber, we need to take care
of two main differences with respect to the waveguide geometry discussed in the previous paragraph: the
cylindrical geometry and the presence of a dielectric core and cladding. In optical
fibers, as well as in the planar waveguide seen before, the light propagates
in the form of modes of the electromagnetic field : if the core of the fiber is small enough, only one mode
can be guided and the fiber is called a single mode fiber, if it is bigger, thus more
than one mode can be guided and the fiber is called a multimode fiber.
In this second case, as evidenced in the case of planar guide, the optical
modes have different group velocities: this
normally limits the repetition rate of a pulse that can be transmitted in a
long fiber avoiding the overlap with the following pulse. To reduce this problem,
graded index multimode fibers can be used, where instead of an abrupt jump between
the values of the refraction indices of the core and the cladding, the index decreases continuously from the center to
the edge.

In the following, I will first recall the main properties of the step index fibers, with a
focus on single mode fibers and I will then describe single mode nanofibers, which will be coupled to a single perovskite nanocube towards building up a hybrid photonic  device. A detailed description of these systems can be found in the ninth chapter
of~\citetitle{saleh2007Fundamentals} by \textcite{saleh2007Fundamentals}.

To obtain the propagation modes for an optical fiber, we need to consider
Maxwell equations in a dielectric medium:
\begin{equation}
\label{eq:Maxwell}
	\begin{aligned}
	\nabla \times \mathcal{H} &=\frac{\partial \mathcal{D}}{\partial t} \\
	\nabla \times \mathcal{E} &=-\frac{\partial \mathcal{B}}{\partial t} \\
	\nabla \cdot \mathcal{D} &=0 \\
	\nabla \cdot \mathcal{B} &=0
	\end{aligned}
%	\begin{aligned}
%	\nabla \times \mathbf{H} &=\dfrac{\partial \mathbf{D}}{\partial t} \\
%	\nabla \times \mathbf{E} &=-\dfrac{\partial \mathbf{B}}{\partial t} \\
%	\nabla \cdot \mathbf{D} &=0 \\
%	\nabla \cdot \mathbf{B} &=0
%	\end{aligned}
\end{equation}
%Now let's consider a monochromatic wave, and let's write the electric and
%magnetic fields as:
%\begin{equation}
%\begin{aligned}
%\mathcal{E}\of{\mathbf{r},t}&=\Re\graffa{
%	{\mathbf{E}
%		\of{\mathbf{r} e^{i\omega t}}
%	}
%}\\
%\mathcal{H}\of{\mathbf{r},t}&=\Re\graffa{
%	{\mathbf{H}
%		\of{\mathbf{r} e^{i\omega t}}
%	}
%}\\
%\mathcal{B}\of{\mathbf{r},t}&=\Re\graffa{
%	{\mathbf{B}
%		\of{\mathbf{r} e^{i\omega t}}
%	}
%}\\
%\mathcal{D}\of{\mathbf{r},t}&=\Re\graffa{
%	{\mathbf{D}
%		\of{\mathbf{r} e^{i\omega t}}
%	}
%}
%\end{aligned}
%\end{equation}
%where $\mathbf{E}$ and $\mathbf{H}$ are the complex amplitudes vectors of the
%electric and magnetic field, respectively, and $\Re$ indicates the real part
%operator.
 For simplicity, we consider our medium to be:
\begin{description}
	\item[linear] -- the polarization density vector $\mathcal{P}$
	and the electrical field $\mathcal{E}$ are linearly related;
	\item[non dispersive] -- meaning that, at a certain time $t$,
	$\mathcal{P}(t)$ is determined by  $\mathcal{E}(t)$ but not by
	$\mathcal{E}(t')$ for any time $t'\neq t$
\end{description}
We can thus write $\mathbf{D}=\epsilon\of{\mathbf{r}} \mathbf{E}$ and
$\mathbf{B}=\mu
\mathbf{H}$. Now we can distinguish two cases, the case in which the medium is
homogeneous and the case in which it is not.

 \paragraph{Homogeneous medium} \mbox{}\\
 In this first case, the medium is homogeneous i.e.
 $\epsilon\of{\textbf{r}}=\epsilon$.
 Let us consider a monochromatic wave and let us write the electric and
 magnetic fields as:
 \begin{equation}
 \begin{aligned}
 \label{eq:monocromatic}
 \mathcal{E}\of{\mathbf{r},t}&=\Re\graffa{
 	{\mathbf{E}
 		\of{\mathbf{r} e^{i\omega t}}
 	}
 }\\
 \mathcal{H}\of{\mathbf{r},t}&=\Re\graffa{
 	{\mathbf{H}
 		\of{\mathbf{r} e^{i\omega t}}
 	}
 }\\
 \mathcal{B}\of{\mathbf{r},t}&=\Re\graffa{
 	{\mathbf{B}
 		\of{\mathbf{r} e^{i\omega t}}
 	}
 }\\
 \mathcal{D}\of{\mathbf{r},t}&=\Re\graffa{
 	{\mathbf{D}
 		\of{\mathbf{r} e^{i\omega t}}
 	}
 }
 \end{aligned}
 \end{equation}
 where $\mathbf{E}$ and $\mathbf{H}$ are the complex amplitudes vectors of the
 electric and magnetic fields, respectively, and $\Re$ indicates the real part
 operator.
 Maxwell equations~\eqref{eq:Maxwell} can thus be simplified
as
\begin{equation}
\label{eq:Maxwell_om}
\begin{aligned}
	\nabla \times \mathbf{H} &=i \omega \epsilon \dfrac{\partial
	\mathbf{E}}{\partial t} \\
	\nabla \times \mathbf{E} &=-i \omega \mu \dfrac{\partial
	\mathbf{H}}{\partial t} \\
	\nabla \cdot \mathbf{D} &=0 \\
	\nabla \cdot \mathbf{B} &=0.
\end{aligned}
\end{equation}
%
%From these equations it is possible to distiguish two cases.
By applying the curl operator to the first and
the second equations of~\eqref{eq:Maxwell_om}, we can derive the wave equation to be satisfied by each complex component $u$ of the electric and
the
magnetic fields:
\begin{equation}
%\label{eq:3.11}
	\nabla^2 u+ \dfrac{1}{c^2} \dfrac{\partial^2 u}{\partial t^2}=0
\end{equation}
where $c$ is the velocity of light in the medium $c=1/\sqrt{\mu \epsilon}$.
This equation, using the time dependace expressed by the equations~\eqref{eq:monocromatic} can be rewritten as
\begin{equation}
\label{eq:3.11}
\nabla^2 u+ k^2 u=0
\end{equation}
where $k=nk_0=\omega\sqrt{\epsilon \mu}$ with $n=\sqrt{\epsilon \mu /\tonda{
\epsilon_{0} \mu_0} }$, $k=\omega/c_0 $ and $c=c_0/n$, as usual.

\paragraph{Inhomogeneous medium} \mbox{}\\
In this second case, more general, we do not
make the homogeneity hypothesis. By following the same mathematical procedure and applying
the curl operator to the first and the second equations in~\eqref{eq:Maxwell}, we obtain different equations for the magnetic and electric fields:
\begin{subequations}
	\begin{alignat}{2}
	\label{eq:wave_el}
	\frac{\epsilon_{0}}{\epsilon} \nabla \times(\nabla \times
	\mathcal{E})&=-\frac{1}{c_{0}^{2}} \dfrac{\partial^{2}
	\mathcal{E}}{\partial
	t^{2}}\\
	\nabla \times\left(\dfrac{\epsilon_{0}}{\epsilon} \nabla \times
	\mathcal{H}\right)&=-\frac{1}{c_{0}^{2}} \frac{\partial^{2}
	\mathcal{H}}{\partial t^{2}}
	\end{alignat}
\end{subequations}

Equation~\eqref{eq:wave_el} can be rewritten in a different form, more
useful. First of all, we can use the relation (valid in a system of linear
coordinates):
\begin{equation}
	\nabla \times \tonda{ \nabla \times \mathcal{E} } = \nabla \tonda{\nabla
	\cdot \mathcal{E} } - \nabla^2 \mathcal{E};
\end{equation}
substituting it in \eqref{eq:wave_el} we obtain:
\begin{equation}
\dfrac{\epsilon_{o}}{\epsilon}
\quadra{\nabla \tonda{\nabla \cdot \mathcal{E} } - \nabla^2 \mathcal{E}
}=-\dfrac{1}{c_{0}^{2}}
	\dfrac{\partial^{2} 	\mathcal{E}}
	{\partial t^{2}}.
\end{equation}
Now, we multiply both terms by $\epsilon/\epsilon_{0}$ and we use the fact
that $c_0=1/\sqrt{\epsilon_0 \mu_0}$ to write:
\begin{equation}
\label{eq:3.15}
\nabla \tonda{\nabla \cdot\mathcal{E} } - \nabla^2 \mathcal{E}
=-\epsilon \mu_0
\dfrac{\partial^{2} 	\mathcal{E}}
{\partial t^{2}}.
\end{equation}
From Maxwell equations~\eqref{eq:Maxwell} we know that $\nabla \cdot
\mathcal{D}=0 $ and thus $\nabla \cdot \tonda{ \epsilon \mathcal{E}}=0 $, so it
is true that:
\begin{equation}
\label{eq:3.16}
0=
\nabla \cdot \tonda{ \epsilon \mathcal{E}}=
\tonda{\nabla  \epsilon } \cdot \mathcal{E} +
\epsilon \tonda{\nabla \cdot \mathcal{E}} \Rightarrow
\nabla \cdot \mathcal{E}=-\dfrac{
\tonda{\nabla  \epsilon } \cdot \mathcal{E}
}{
\epsilon
}
\end{equation}
Using the result of equation~\eqref{eq:3.16} in equation~\eqref{eq:3.15}
and moving all terms to the left we obtain:
\begin{equation}
\label{eq:3.17}
\nabla^2 \mathcal{E}
-\epsilon \mu_0
\dfrac{\partial^{2} 	\mathcal{E}}
{\partial t^{2}}
+\nabla \tonda{
	\dfrac{\nabla \epsilon }{ \epsilon} \cdot \mathcal{E}
}
=0
\end{equation}
We notice here that the last term of this equation is the one that mixes
the vector components. In other words, when this therm is zero, the first two terms give rise to 3 scalar equations, one for each component, that are independent. When the last term is not zero on the other hand, the 3 equations are coupled as any component appears in any equation. When the last term is zero, each component satisfies an
equation analogous to~\eqref{eq:3.11}.

It is thus important to know the amplitude of these terms to see if the last
one can be neglected and when. In particular, the order of magnitude of the different
terms is the following:
\begin{subequations}
	\label{eq:3.18}
	\begin{alignat}{1}
	\nabla^2 \mathcal{E} & \approx \frac{E}{\lambda^2}
	\\
	\epsilon \mu_0 \dfrac{\partial^{2} 	\mathcal{E}}
	{\partial t^{2}} &\approx \frac{E}{\lambda^2}
	\\
	\nabla \tonda{	\dfrac{\nabla \epsilon }{ \epsilon} \cdot \mathcal{E}}
	&\approx
	\dfrac{E }{\lambda^2 }  \dfrac{n_1^2 -n_2^2 }{n_2^2 }
	\label{eq:third_element}
	\end{alignat}
\end{subequations}
where $E$ is the modulus of the electric field and $\lambda$ the radiation
wavelength. When $n_1^2-n_2^2 \ll 1$ we refer to weak guiding and this is the
case for standard fibers. In the opposite case, when this is not true, the third
term cannot be neglected. This is the case in tapered optical nanofibers where, as I will show in the following, $n_1\approx1.4$ and $n_2=1$. In this case, we need to solve the general equation~\eqref{eq:helm_core_cladding} and the solution is a bit trickier.

\paragraph{Step index fibers} \mbox{}\\
Let us now consider a monochromatic wave in the
form described by equations~\eqref{eq:monocromatic}, propagating inside a step index fiber. We
consider, for simplicity, a core surrounded by a cladding and that the cladding
extends indefinitely in space. This is a good approximation when the core is much smaller than
the cladding, such as in single mode fibers. The third term in
equation~\eqref{eq:3.18} is thus zero both in the core and the cladding and
the equation becomes the Helmholtz equation. When written
in radial coordinates, it reads as:
\begin{equation}
\label{eq:Helm_fiber}
	\frac{\partial^{2} U\of{r,\phi,z} }{\partial r^{2}}+\frac{1}{r}
	\frac{\partial
	U\of{r,\phi,z}}{\partial r}+\frac{1}{r^{2}} \frac{\partial^{2}
	U\of{r,\phi,z}}{\partial
	\phi^{2}}+\frac{\partial^{2} U\of{r,\phi,z}}{\partial z^{2}}+n^{2}
	k_{0}^{2} U\of{r,\phi,z}=0
\end{equation}
with the modes traveling in the $z$ direction.
%and the propagation
% constant $\beta$  defined as in the case for the planar guide (equation~\eqref{eq:betamDef}) by the equation~\eqref{eq:betaDef}:
% \begin{equation}
%     \label{eq:betaDef}
% 	\beta=k \cos \theta
% \end{equation}
% where $\theta$ is the angle that the propagation direction of the wave forms with the $z$-axis.
As done before for the planar waveguide, we chose the $z$-axis laying in the same direction of the fiber axis.
In addition, $U\of{r,\phi,z}$
is periodic with respect to the angle $\phi$ with a period of $2\pi$, thus we can write:
\begin{equation}
\label{eq:fromUtou}
	U\of{r,\phi,z}=u(r)e^{-jl\phi} e^{-j\beta z}
\end{equation}
Equation~\eqref{eq:fromUtou} implicitly defines $l$ and $\beta$: we will see that $l$ corresponds to the order of the solution, while $\beta$ is called the propagation constant, and has the same physical meaning described for planar waveguides (cfr. equation~\eqref{eq:betamDef}).
Substituting equation~\eqref{eq:fromUtou} in equation~\eqref{eq:Helm_fiber} we obtain
\begin{equation}
\label{eq:Helm_upiccolo}
\frac{d^{2} u}{d r^{2}}+\frac{1}{r} \frac{d u}{d r}+\left(n^{2}(r)
k_{0}^{2}-\beta^{2}-\frac{l^{2}}{r^{2}}\right) u=0
\end{equation}.

In order for a mode to be guided, the following conditions must be
valid $n_2k_0<\beta<n_1k_0$, where $n_2$ is the refraction index of the cladding and $n_1$ the refraction index of the core.
It is convenient to define $k_T$ and $\gamma$ as:
\begin{equation}
	\label{eq:kt_gamma_def}
	k_T^2=n_1^2k_0^2-\beta^2, \qquad \qquad \gamma^2 = \beta^2 - n_2^2 k_0^2.
\end{equation}
where $k_T$ and $\gamma$ are real quantities for guided waves. We can use them to rewrite equation~\eqref{eq:Helm_upiccolo} in two separate expressions for the core and the
cladding:
\begin{subequations}
	\label{eq:helm_core_cladding}
	 \begin{eqnarray}
		 \frac{d^{2} u}{d r^{2}}+\frac{1}{r} \frac{d u}{d
		 r}+\left(k_{T}^{2}-\frac{l^{2}}{r^{2}}\right) u=0, & r<a  & \text {
		 (core) } \label{eq:helm_core_claddinga}
		 \\
		 \frac{d^{2} u}{d r^{2}}+\frac{1}{r} \frac{d u}{d
		 r}-\left(\gamma^{2}+\frac{l^{2}}{r^{2}}\right) u=0, & r>a & \text {
		 (cladding)
		 } \label{eq:helm_core_claddingb}
	 \end{eqnarray}
\end{subequations}

The differential equations~\eqref{eq:helm_core_cladding} have well known
solutions given by the Bessel functions of the first kind $J_l\of{k_Tr}$ for
equation~\eqref{eq:helm_core_claddinga} and the modified Bessel functions
of the second kind $K_l\of{\gamma r}$ for any order $l$.

It is useful to note that from equations~\eqref{eq:kt_gamma_def}, we can deduce that
$k_T^2+\gamma^2$ is constant and equal to $\tonda{n_1^2-n_2^2} k_0^2$.

With some geometrical considerations \cite{saleh2007Fundamentals}, it is possible
to show that:
\begin{equation}
	\sqrt{n_1^2-n_2^2}=\NA
\end{equation}
where the numerical aperture $\NA$ is defined as the sine of the maximal
acceptance angle, i.e. the maximal angle $\theta_a$ with respect to the longitudinal
axis the light, previously propagating in air, can have to be guided: $\NA=\sin \theta_a$.
Thus we can state that:
\begin{equation}
k_T^2+\gamma^2=(\NA )^2 k_0^2.
\end{equation}
An important parameter in fibers is the so called fiber parameter or
$\mathsf{V}$-parameter, defined as:
\begin{equation}
\mathsf{V} \defeq \NA\, k_0 a = 2\pi \frac{a}{\lambda_0} \NA
\end{equation}
where $a$ is the fiber radius. The $\mathsf{V}$-parameter, also called fiber parameter, determines the number of allowed modes in the fiber, as I will show in the following and their propagation constant.
Additionally if we define $\mathsf{X}=k_T a$ and $\mathsf{Y}=\gamma a$ as the
normalized
parameters, we can write:
\begin{equation}
\mathsf{X}^2+\mathsf{Y}^2=\mathsf{V}^2
\end{equation}.

If now we consider the case of a weakly guiding fiber, as most of the fibers
are, most of guided rays are paraxial, as $n_1^2-n_2^2 \ll 1$. In this case the
term~\eqref{eq:third_element} is negligible and equation~\eqref{eq:3.17}
becomes analogous to equation~\eqref{eq:3.11}.
By imposing the continuity of the function $u(r)$
and of its derivative at $r=a$, we obtain \cite{saleh2007Fundamentals} the following equation, known as
characteristic equation:
\begin{equation}
\label{eq:characteristic_equation}
\mathsf{X} \frac{J_{l \pm 1}(\mathsf{X})}{J_{l}(\mathsf{X})}=\pm \mathsf{Y}
\frac{K_{l \pm 1}(\mathsf{Y})}{K_{l}(\mathsf{Y})}
\end{equation}
Using the fact that $\mathsf{X}^2+\mathsf{Y}^2=\mathsf{V}^2$ we can rewrite
$\mathsf{Y}$ in function of $\mathsf{X}$
\begin{equation}
\mathsf{X} \frac{J_{l \pm 1}(\mathsf{X})}{J_{l}(\mathsf{X})}=\pm
\sqrt{\mathsf{V}^2-\mathsf{X}^2} \frac{K_{l \pm
1}(\sqrt{\mathsf{V}^2-\mathsf{X}^2})}{K_{l}(\sqrt{\mathsf{V}^2-\mathsf{X}^2})}
\end{equation}.
This equation can be solved graphically, plotting the left and the right side
terms as a function of X and looking for their intersections.
\begin{figure}[tbh]
	\centering
	\includegraphics[width=0.7\linewidth]{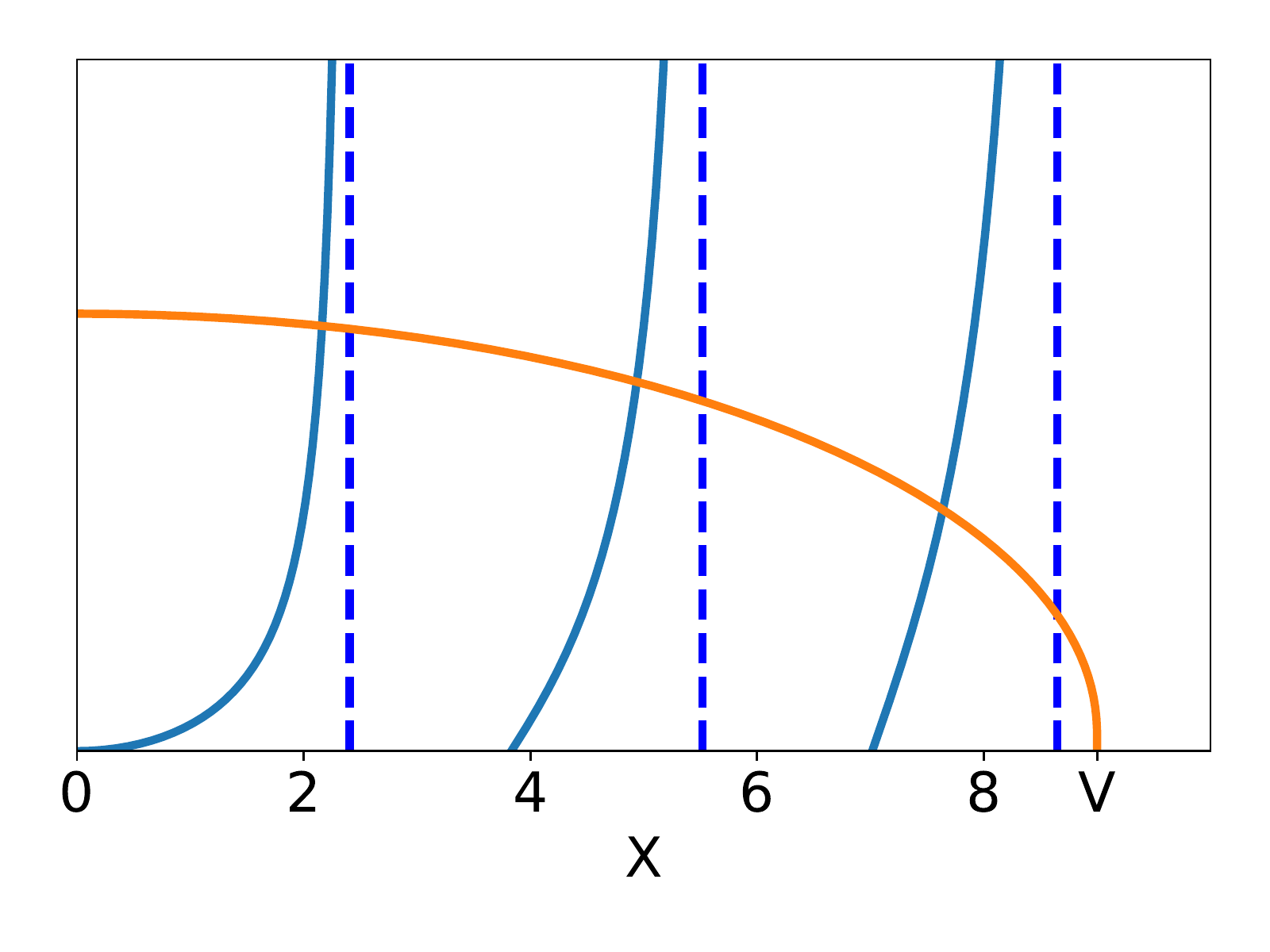}
	\caption{Graphical solutions of the equation
	\protect\eqref{eq:characteristic_equation} for $l=0$. In blue, the term on
	the left side, independent from $\mathsf{V}$, in orange the term on the
	right side.
	The dashed lines marks the asymptotes where the equation is not defined. }
	\label{fig:bessel}
\end{figure}
This is represented in figure~\ref{fig:bessel}: in blue is represented the term on
the right side, while in orange is represented the term on the left side. The solutions are
represented by the intersections between the two curves.
Any intersection represents a possible solution, i.e. a possible guided mode.
As the first term is independent from $\mathsf{V}$, it is clear that the number of
intersections can be determined looking at the second term. In particular, when
$V< 2.405$ there will be only one intersection between the blue and the orange
curves, meaning that only the mode $\mathrm{LP}_{01}$ is a solution and is thus allowed. In this case, the fiber is a single mode fiber.
We can rewrite this condition recalling the definition of $\mathsf{V}$ in the
form:
\begin{equation}
	\label{eq:single_mode}
	\frac{2\pi a}{\lambda_0} \NA <2.405 \qquad \Rightarrow \qquad a< 2.405
	\frac{\lambda_0}{2
	\pi \sqrt{n_1^2 -n_2^2 } }
\end{equation}
Equation~\eqref{eq:single_mode} represents a very important condition in
order to obtain single mode fibers in practice.

\subsection{Modes of a tapered optical nanofiber}
We described in the previous section the main characteristics of a standard optical fiber. In the following, we detail the case of a nanofiber. A nanofiber is also an optical waveguide but with a diameter dimension comparable or even smaller than the wavelength of the guided light.
In this case, we cannot consider anymore that the light is traveling totally inside the
nanofiber. We will see that, in this case, there is a strong evanescent field extending outside the nanofiber.
The light is guided between the nanofiber and the air surrounding it, thus in this case the
approximation that $n_1^2 -n_2^2 \ll 1$ is not valid anymore. It is still possible to
use equations~\eqref{eq:helm_core_cladding} and their solutions are given by the solution of a generalized characteristic equation:
\begin{equation}
	\label{eq:charac}
	\left[\frac{J_{v}^{\prime}(\mathsf{X})}{\mathsf{X}
	J_{v}(\mathsf{X})}+\frac{K_{v}^{\prime}(\mathsf{Y})}{\mathsf{Y}
	K_{v}(\mathsf{Y})}\right]
	\left[n_{1}^{2} \frac{J_{v}^{\prime}(\mathsf{X})}{\mathsf{X}
	J_{v}(\mathsf{X})}+n_{2}^{2} \frac{K_{v}^{\prime}(\mathsf{Y})}{\mathsf{Y}
	K_{v}(\mathsf{Y})}\right]
	=
	\beta^{2} \frac{l^{2}}{k^{2}}
	\left(
	\frac{1}{\mathsf{X}^{2}}+\frac{1}{\mathsf{Y}^{2}}
	\right)^{2}
\end{equation}
where $\mathsf{X}$, $\mathsf{Y}$, and $\mathsf{V}$ are the same quantities defined in
the previous section.

Equation~\eqref{eq:charac} gives origin to different families of modes:
$HE_{vm}$ and $EH_{vm}$,
where $v$ and $m$ characterize the azimuthal and the radial
distributions respectively.
In the special case in which $v=0$, the $z$ component of the electric field
and the magnetic fields are zero, giving origin to the transverse electric and
transverse magnetic modes $TE$ and $TM$.
We thus obtain~\cite{tong2004Singlemode}:
\begin{subequations}
	\label{eq:modesnf}
	\begin{alignat}{2}
		\left\{\frac{J_{v}^{\prime}(\mathsf{X})}{\mathsf{X}
		J_{v}(\mathsf{X})}+\frac{K_{v}^{\prime}(\mathsf{Y})}{\mathsf{Y}
		K_{v}(\mathsf{Y})}\right\}\left\{\frac{J_{v}^{\prime}(\mathsf{X})}
		{\mathsf{X}
		J_{v}(\mathsf{X})}+\frac{n_{2}^{2}
		K_{v}^{\prime}(\mathsf{Y})}{n_{1}^{2} \mathsf{Y}
		K_{v}(\mathsf{Y})}\right\}&=\left(\frac{v \beta}{k
		n_{1}}\right)^{2}\left(\frac{\mathsf{V}}{\mathsf{X}
		\mathsf{Y}}\right)^{4} &
		\qquad &
		\hfil
			\boxed{
				\genfrac{}{}{0pt}{}{HE_{\nu m}}{EH_{\nu m}}
			}
		\\
		\frac{J_{1}(\mathsf{X})}{\mathsf{X}
		J_{0}(\mathsf{X})}+\frac{K_{1}(\mathsf{Y})}{\mathsf{Y} K_{0}(Y)}&=0&&
		\hfil
		\boxed{
			TE_{0m} }\\
		\frac{n_{1}^{2} J_{1}(\mathsf{X})}{\mathsf{X}
		J_{0}(\mathsf{X})}+\frac{n_{2}^{2} K_{1}(\mathsf{Y})}{Y
		K_{0}(\mathsf{Y})}&=0. && \hfil
	\boxed{
	TM_{0m}
	}
	\end{alignat}
\end{subequations}

\begin{figure}[tb]
	\centering
	\includegraphics[width=0.7\linewidth]{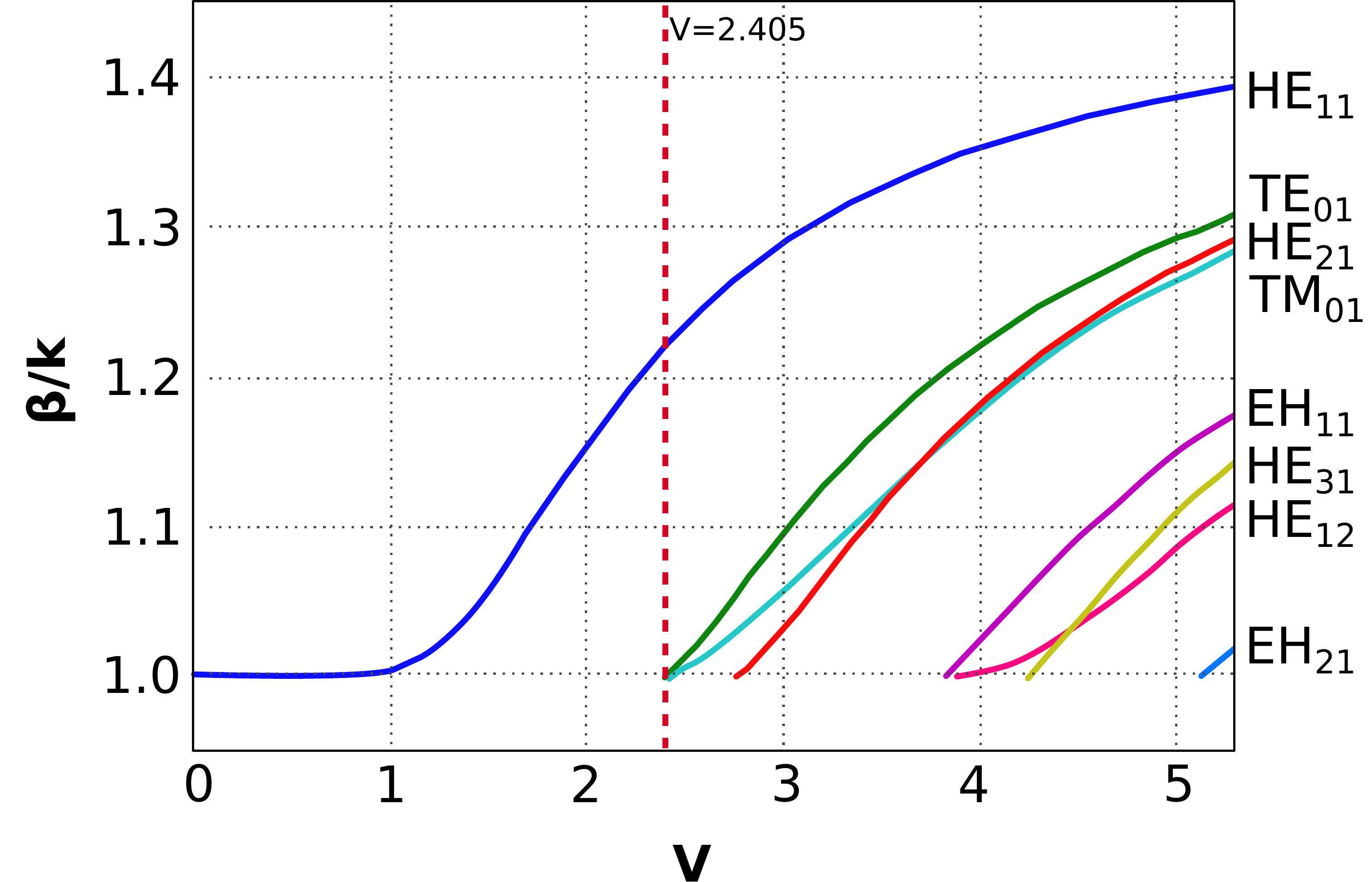}
	\caption{Guided modes of a nanofiber, allowed values of $\beta/k$ as a
	function of $\mathsf{V}$. The red dashed line indicates the threshold of
	$V$ under
	which the nanofiber is single mode where the $\text{HE}_{11}$ mode only is allowed. The graphs are obtained by solving numerically equations~\protect\eqref{eq:modesnf} for a silica glass waveguide with a constant refraction index $n_1=1.46$.}
	\label{fig:graficoclean}
\end{figure}

It is possible to solve numerically these equations and to find
the mode solutions. The first solutions of these equations for a glass nanofiber
are reported in
figure~\ref{fig:graficoclean}.
 When $V<2.405$, only one solution, the mode $HE_{11}$, is admitted while
the others are not.  This means that in the case of $V<2.405$ the nanofiber can
be considered a single mode fiber and for this reason the mode $HE_{11}$ is called the
fundamental mode of the fiber. In our experiment, we choose the diameter of the nanofiber so that we only have
the fundamental mode guided by the nanofiber.
Looking at figure~\ref{fig:graficoclean}, we can also notice that in all the solutions we have $n_2 k<\beta<n_1 k$. For this reason,
we can define an effective refractive index of the nanofiber as
\begin{equation}
	n_{eff}=\frac{\beta}{k}
\end{equation}
and consider at it as the index propagation of the light that a certain mode actually ``sees''. From this point of view, it is clear that it has to be in
between $n_2$ and $n_1$.

Considering now the fundamental mode, it is useful to calculate the field distribution of this mode. Indeed, in order to optimize the near field coupling, the knowledge of
the spatial distribution of electrical field is important.
The characteristic equation for this mode becomes:
\begin{equation}
\frac{J_{0}(\mathsf{X})}{\mathsf{X}
J_{1}(\mathsf{X})}=-\frac{n_{1}^{2}+n_{2}^{2}}{2 n_{1}^{2}}
\frac{K_{1}^{\prime}(\mathsf{Y})}{\mathsf{Y}
K_{1}(\mathsf{Y})}+\frac{1}{\mathsf{X}^{2}}
-\left\{\left[\frac{n_{1}^{2}-n_{2}^{2}}{2 n_{1}^{2}}
\frac{K_{1}^{\prime}(\mathsf{Y})}{\mathsf{Y}
K_{1}(\mathsf{Y})}\right]^{2}\right.
\left.+\frac{\beta^{2}}{n_{1}^{2}
k^{2}}\left(\frac{1}{\mathsf{X}^2}+\frac{1}{\mathsf{Y}^2}\right)^{2}\right\}^{\!1
 / 2}
\end{equation}

It is useful to define the parameter $s$ as follows:
\begin{equation}
s=\frac{
	\ddown{\mathsf{Y}^{2}}+\ddown{\mathsf{X}^{2}}
}
{J_{1}^{\prime}(\mathsf{X}) /\left(\mathsf{X}
J_{1}(\mathsf{X})\right)+K_{1}^{\prime}(\mathsf{Y}) /\left(\mathsf{Y}
K_{1}(\mathsf{Y})\right)}
\end{equation}

The electric field can be written as follows for $r<a$:
\begin{subequations}
\begin{alignat}{2}
E_{r}(r, \varphi, z, t)&=i A \frac{\beta}{2k_T}
%\frac{K_{1}(\mathsf{Y})}{J_{1}(\mathsf{X})}
\left[(1-s) J_{0}\of{k_T r}-(1+s)
J_{2}(k_T r)\right] e^{i(\omega t -\beta z)}\\
E_{\varphi}(r, \varphi, z, t)&=-p A \frac{\beta}{2k_T}
%\frac{q}{h} \frac{K_{1}(\mathsf{Y})}{J_{1}(\mathsf{X})}
\left[(1-s) J_{0}(k_T r)+(1+s)
J_{2}(k_T r)\right] e^{i(\omega t -\beta z)}\\
E_{z}(r, \varphi, z, t)&=f A
%\frac{2 q}{\beta}  \frac{K_{1}(\mathsf{Y})}{J_{1}(\mathsf{X})}
 J_{1}(k_T r) e^{i(\omega t -\beta z)}
\end{alignat}
\end{subequations}

where $A$ is a normalization constant that can be determined \textit{a
posteriori}, $p$ depends on the polarization ($p=1$ in case of a clockwise polarization and
$p=-1$ in case of an anticlockwise polarization) and $f$ depends
on the propagation direction ($f=1$ for a forward propagating mode and $f=-1$
for a backward propagating mode).
Similarly, we can write these equations for $r>a$:
\begin{subequations}
	\begin{alignat}{2}
E_{r}(r, \varphi, z, t) &=-i A \frac{\beta}{2 \gamma}
\frac{J_{1}(\mathsf{X})}{K_{1}(\mathsf{Y})}\left[(1-s) K_{0}(\gamma r)+(1+s)
K_{2}(\gamma r)\right] e^{i(\omega t-\beta z)} \\
E_{\varphi}(r, \varphi, z, t) &=-p A \frac{\beta}{2 \gamma}
\frac{J_{1}(\mathsf{X})}{K_{1}(\mathsf{Y})}\left[(1-s) K_{0}(\gamma r)-(1+s)
K_{2}(\gamma r)\right] e^{i(\omega t-\beta z)} \\
E_{z}(r, \varphi, z, t) &=f A \frac{J_{1}(\mathsf{X})}{K_{1}(\mathsf{Y})}
K_{1}(\gamma r)
e^{i(\omega
t-\beta z)}
\end{alignat}
\end{subequations}

\begin{figure}[tb]
	\centering
	\includegraphics[width=\linewidth]{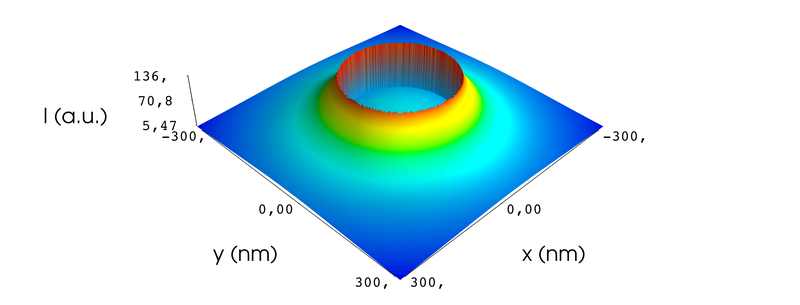}
	\caption{Quasi circular modes: calculated intensity distribution for the modes $HE^+_{11}$ and $HE^-_{11}$ guided by
		a nanofiber with a radius $a=\SI{125}{nm}$. The light used in the
		simulation has a wavelength $\lambda=\SI{500}{nm}$.}
	\label{fig:ifiber}
\end{figure}

In this case, by analogy with the weakly guiding case~\cite{okamoto2006Fundamentals,gloge1971Weakly}, has been referred in literature as~\textit{quasi-circular} modes~\cite{petersen2014Chirala,mitsch2014Quantuma}. Anyway, this does not mean that the polarization is truly circular.
The field intensity distribution for these modes is represented in
figure~\ref{fig:ifiber}.
\begin{figure}[tbh]
	\centering
	\label{fig:e_fiber}
	\subfloat[]{\label{fig:e_r}\includegraphics[width=0.33\linewidth]
		{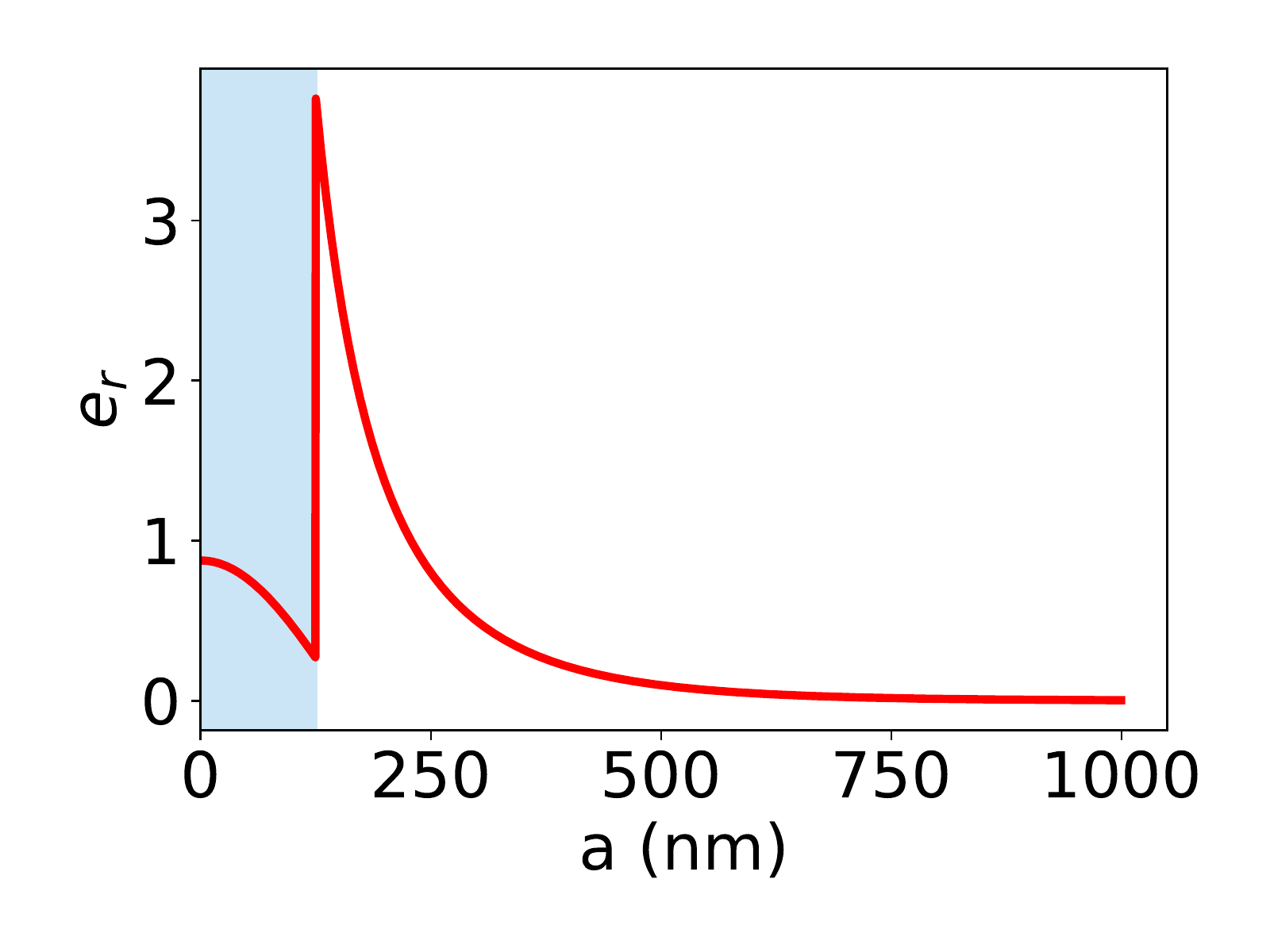}}
	\subfloat[]{\label{fig:e_phi}\includegraphics[width=0.33\linewidth]
		{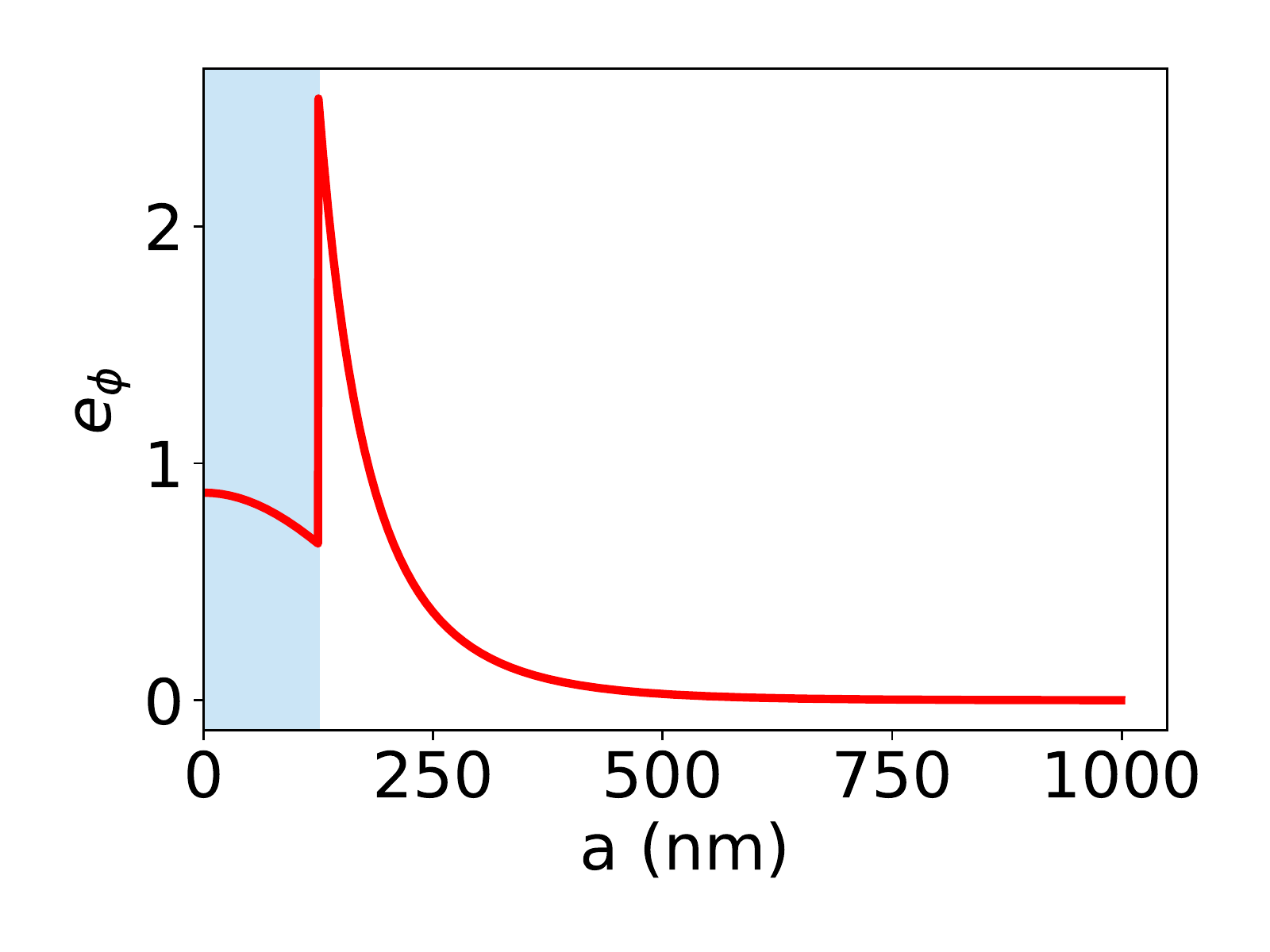}}%\qquad
	\subfloat[]{\label{fig:e_z}\includegraphics[width=0.33\linewidth]
		{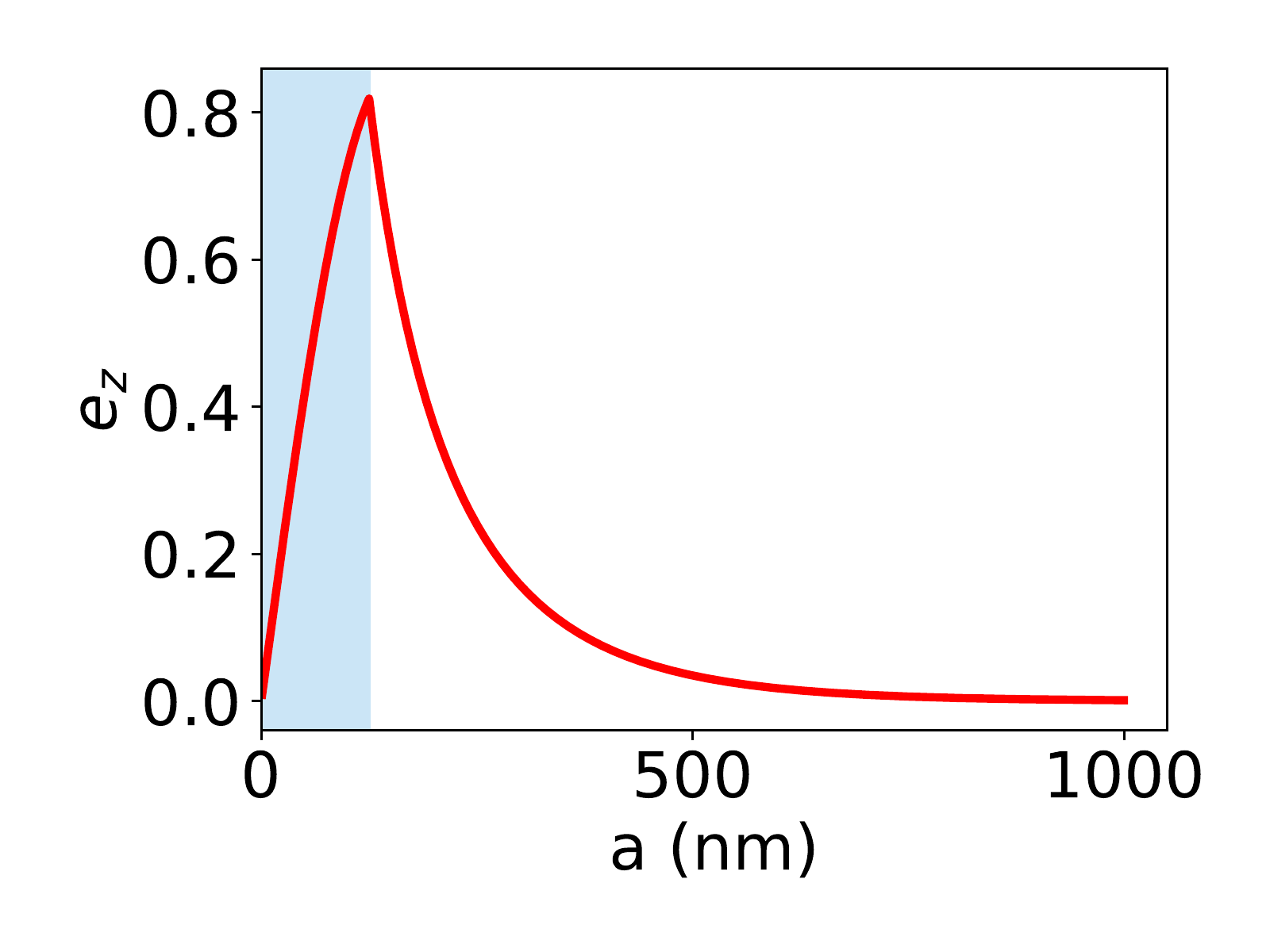}}
	\caption{%\label{fig:g2_example}
		\protect\subref{fig:e_r}~Modulus of the electrical field $E_{r}$ of the
		mode
		$HE_{11}$ of the fiber, calculated for $t=0$ and $z=0$,
		\protect\subref{fig:e_phi}~Modulus of the electrical field
		$E_{\varphi}$ of the mode $HE_{11}$ of the fiber, calculated for $t=0$
		and $z=0$,
		\protect\subref{fig:e_z}~Modulus of the electrical field $E_z$
		of the mode $HE_{11}$ of the fiber, calculated for $t=0$ and $z=0$.
		}
\end{figure}
It is important to note that any linear superposition of any two orthogonal modes is still a solution. If we call $HE^+_{11}$ and $HE^-_{11}$ the modes with $p=\pm1$, we
can create a new pair of orthogonal modes $\graffa{HE^+_{11} \pm HE^-_{11}}$.
It is useful to express them in cartesian coordinates, for $r>a$ we obtain:
\begin{subequations}
	\begin{alignat}{2}
		E_{x}(r, \varphi, z, t) &=-i A \frac{\beta}{2 q}
		\frac{J_{1}(\mathsf{X})}{K_{1}(\mathsf{Y})}\left[(1-s) K_{0}(\gamma r)
		\cos \left(\varphi_{0}\right)+(1+s) K_{2}(\gamma r) \cos \left(2
		\varphi-\varphi_{0}\right)\right] e^{i(\omega t-\beta z)} \\
		E_{y}(r, \varphi, z, t) &=-i A \frac{\beta}{2 \gamma}
		\frac{J_{1}(\mathsf{X})}{K_{1}(\mathsf{Y})}\left[(1-s) K_{0}(\gamma r)
		\sin \left(\varphi_{0}\right)+(1+s) K_{2}(\gamma r) \sin \left(2
		\varphi-\varphi_{0}\right)\right] e^{i(\omega t-\beta z)} \\
		E_{z}(r, \varphi, z, t) &=A \frac{J_{1}(\mathsf{X})}{K_{1}(\mathsf{Y})}
		K_{1}(\gamma r) \cos \left(\varphi-\varphi_{0}\right) e^{i(\omega
		t-\beta z)}
	\end{alignat}
\end{subequations}
with $\varphi_{0}$ defining two orthogonal polarizations, usually
$\varphi_{0}=\graffa{0,\pi/2}$.
The intensity distribution for these modes is represented in
figure~\ref{fig:ifiberlinear}.
\begin{figure}[tb]
	\centering
	\includegraphics[width=\linewidth]{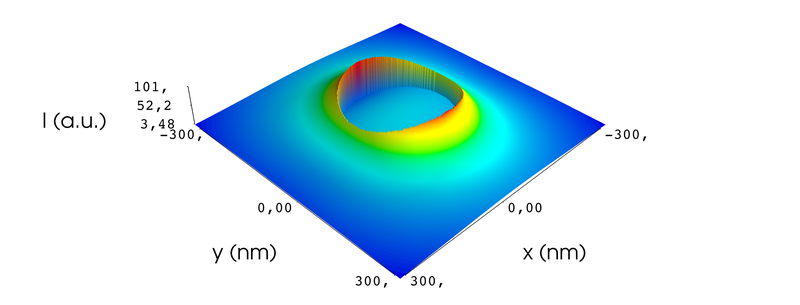}
	\caption{Quasi-linear modes:  calculated intensity distribution for the modes $\graffa{HE^+_{11} \pm HE^-_{11}}$ guided by
		a nanofiber with a radius $a=\SI{125}{nm}$. The light used in the
		simulation has a wavelength $\lambda=\SI{500}{nm}$. The reported mode is the quasi-polarized one along the $y$-axis, the other is similar but rotated by \SI{90}{\degree} and thus quasi-polarized along the $x$-axis.}
	\label{fig:ifiberlinear}
\end{figure}
As we can see, in this case the axial cylindrical symmetry is
broken. These modes, by analogy with free space, are called quasi linear modes.

\begin{figure}
    \centering
    \includegraphics{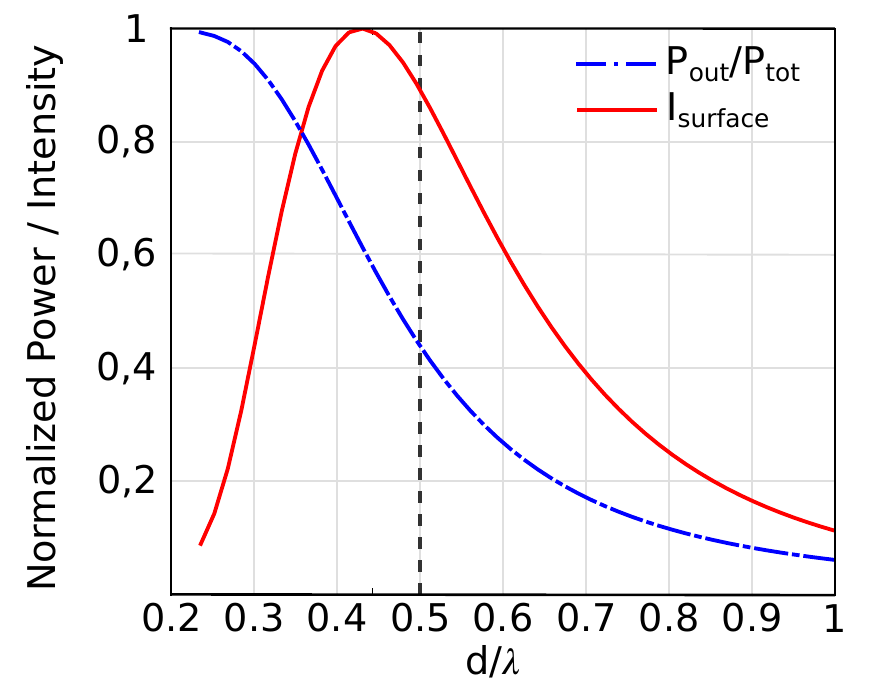}
    \caption{In the figure, we show the comparison between the power and the intensity of the fundamental mode. The blue curve represents the ratio between the power that is guided inside and outside the fiber, while the red curve represents the intensity of the field at the surface, both of them are normalized. Both the curves are normalized to their respective maximum. These curves are calculated for $n_1=1.46$ and $n_2=1$. The black dashed line represents the diameter we choose for the pulling (fabrication of the nanofiber). Smaller diameters present problems of mechanical stability.}
    \label{fig:power_ratio}
\end{figure}

From the field distribution it is clear that the guided field in our example is
strongest outside the nanofiber, in its proximity. This, as expected, depends on the radius of the fiber: increasing the radius will decrease the
evanescent field and the field is more and more guided inside the
nanofiber. Vice-versa, decreasing the nanofiber radius will increase the evanescent field while the field contained inside the nanofiber will decrease. This can be understood thinking in terms of different limit cases: when the fiber radius is much bigger than
the wavelength we have a standard fiber where the light is fully guided inside the
fiber; in the opposite case, when the radius is much smaller than the
wavelength the light is almost completely outside the nanofiber. In this second
case, in the ideal case, with the radius equal to zero, there is no
fiber anymore and the light is not guided anymore, we have the propagation of a plane wave. This means that there is an
optimal radius in order to have a strong evanescent field together with a
guided mode and a nanofiber sufficiently mechanically stable. The optimum turns out to be $r\approx \lambda/4 = \SI{125}{nm}$. This can be seen in figure~\ref{fig:power_ratio},
where the comparison between the intensity at the surface and the ratio of the power guided outside
the fiber is shown. The dashed line represents our fiber diameter: a slightly smaller diameter can be done to increase the intensity at the surface of about \SI{10}{\percent} but it comes with more stability problems.
% TODO:eventually add a graph to explain this fact.

As we will see in the following sections, this allowed us to obtain a transmission in the fiber up to $98\%$ while preserving a good mechanical stability of the nanofiber which could be fabricated and manipulated without breaking it.

\section{Nanofiber fabrication}
I will describe in this section the experimental setup we use in the laboratory to fabricate the nanofibers. Beside the desired final diameter, we need to be able to control also the shape of the nanofiber in order to adiabatically guide the light from the standard fiber
to the nanofiber and vice-versa as I will describe.

\subsection{Choosing the profile of the fiber}
A possible way, which is probably the simplest solution, to obtain a nanofiber
from a standard fiber without loosing the transmission is to slowly reduce the
radius up to reaching the targeted diameter. This procedure can
theoretically work and if the radius variation is slow enough, in principle it is possible to
adiabatically guide the mode of the original fiber towards the one of the nanofiber.
Unfortunately this procedure necessitates very
long transition regions, that presents numerous practical inconveniences. One of
the main problem is the mechanical stability: when the tapered and thin part of the fiber is very long, the fiber is much more sensitive to the mechanical vibrations. Therefore, when the transition region
becomes too large, moving
the fiber from a setup to another is a complex operation with a high risk of breaking the fiber. In addition, the fabrication setup is usually limited in space and it does not allow to fabricate long tapered regions. In our setup, it is possible to obtain tapered regions of few centimeters.

For all of these reasons, it is preferable to produce a transition region that
is as short as possible, compatible with our requirements about the transmission.
A good method to calculate the ideal profile has been described by~\textcite{nagai2014Ultralowloss}. In practice the tapered angle needs to be in any point small enough such that the power coupling from the fundamental mode to the higher-order modes is negligible. To quantify this, it is useful to define the delineation angle~\textcite{nagai2014Ultralowloss, love1991Tapered} $\Omega\of{r}$ such that:
\begin{equation}
\Omega(r)=\frac{r}{2 \pi}\left(\beta_{1}(r)-\beta_{2}(r)\right).
\end{equation}
This angle provides an approximate delineation
between approximately adiabatic and lossy tapers~\cite{love1991Tapered}.
In the previous equation, $\beta_{1}(r)$ and $\beta_{2}(r)$ are the propagation constants of the first and second order modes, respectively. In addition, we can define the adiabaticity factor as:
\begin{equation}
\theta(r)<F \Omega(r).
\end{equation}
The smaller $F$ is, the best transmission we can expect from the nanofiber. In particular when keeping $F<0.4$, it is possible to obtain fibers with a transmission over \SI{98}{\percent}\cite{nagai2014Ultralowloss}.

In our platform, we use a Matlab routine to calculate and simulate the fiber profile before the pulling, routing that generates the instructions for the motors used during the pulling itself as I explain in the following.
\subsection{Pulling mechanism}
\begin{figure}[ht]
	\centering
	\includegraphics[width=0.4\linewidth]{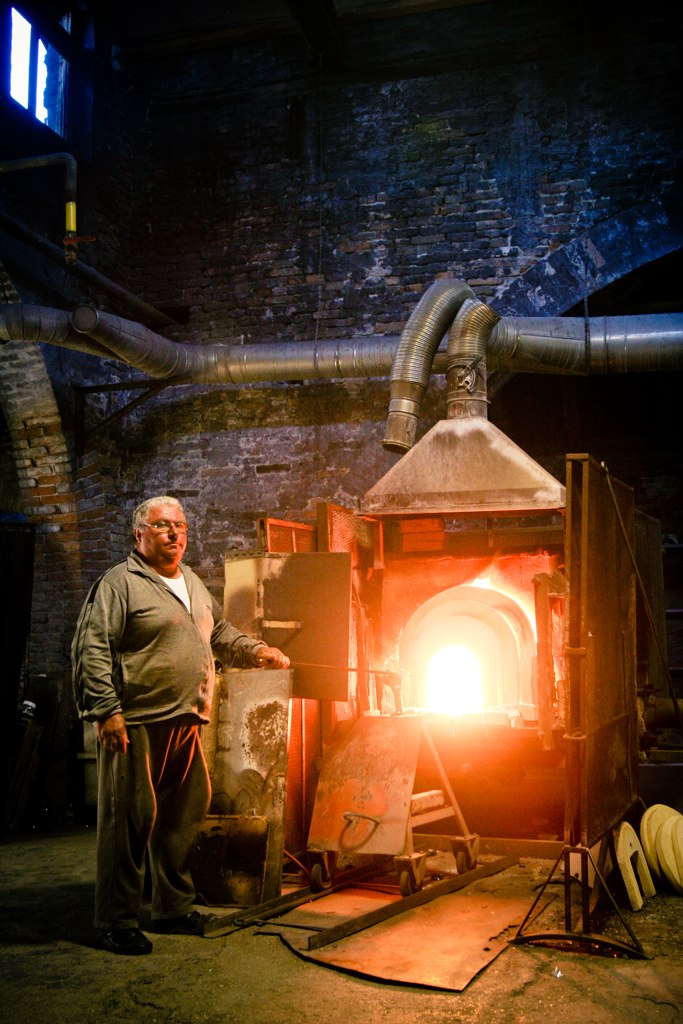}
	\caption{A man heating up glass in order to obtain the desired shape in the
	Island of Murano, Italy. \credit{Zanetti Murano Studio}}
	\label{fig:vetraiomurano}
\end{figure}

\begin{figure}[tb]
	\centering
	\includegraphics[width=0.7\linewidth]{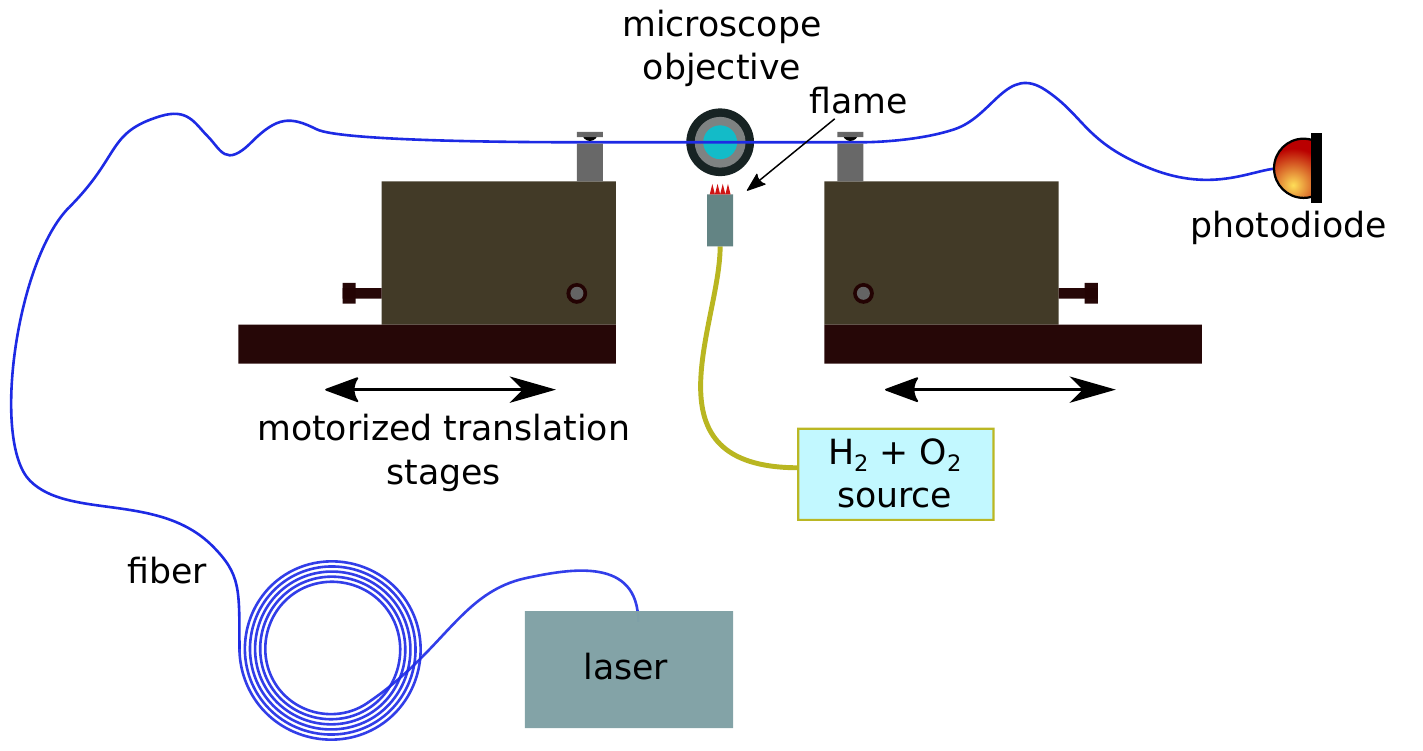}
	\caption{Experimental setup for pulling nanofibers. The laser is used to
	constantly monitor the transmission during the process.}
	\label{fig:fiberpulling}
\end{figure}

One of the most reliable ways of producing a nanofiber is to start with a
standard fiber, heat the glass and pull it in order to reduce its diameter.
This is conceptually not very different from the standard way to shape glass. It relies on the fact that the glass is an amorphous material that can be
easily shaped by heating it to change its viscosity coefficient. In the case of standard glass, this is usually heated at \SI{800}{\celsius} inside an oven. A
man heating the glass for this purpose in the Murano island is shown in
figure~\ref{fig:vetraiomurano}.
In order to pull the nanofiber, we use the setup shown in
figure~\ref{fig:fiberpulling}.
The fiber is clamped between two translation stages and pulled over a clean flame of oxygen and hydrogen. The procedure is detailed in section~\ref{sec:pulldet}. Before the practical description of the experiment, i is useful to understand the physical model on which it relies.

In the case of fiber pulling, the procedure has been described in details
by~\textcite{birks1992Shape}. The variation of the viscosity of the fiber is
generally described by fluid-mechanics~\cite{dewynne1989mathematical}, but it can
be approximated as a cylindrical melting zone of a length $L_0$ also called in
the following effective flame diameter.
If we pull the fiber without moving the flame we obtain a waist with the
diameter described by the following equation:
\begin{equation}
	r\of{x}=r_0 \exp{-\dfrac{z_e}{2L_0}}
\end{equation}
where $r_0$ is the initial fiber radius, $r{x}$ the radius after extending the
fiber of a length $z_e$, while in the transition area the profile is given by
\begin{equation}
		r\of{z}=r_0 \exp{-\dfrac{z}{2L_0}}
\end{equation}
where we placed the origin in the last unprocessed point of the fiber and it
increases towards the waist.
In this way, by simply pulling the fiber over the flame, only one shape can be produced and for this reason a more versatile technique is needed, to be able to realize the appropriate profile.

\begin{figure}
	\centering
	\includegraphics[width=0.7\linewidth]{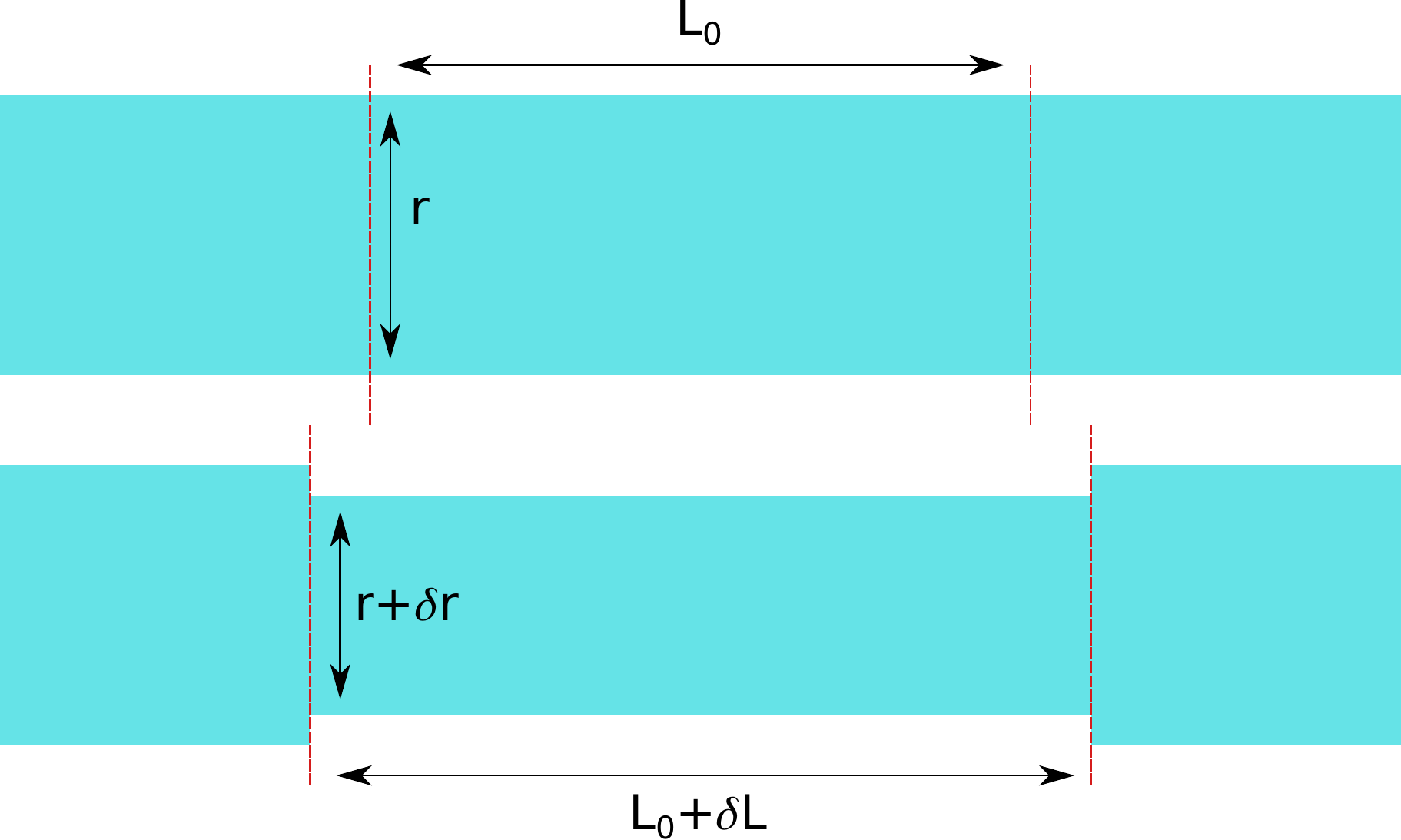}
	\caption{Scheme of the fiber pulling. When the fiber is pulled by a
	small length $\delta L$, the region heated by the flame increases its length
	from $L_0$ to $L_0+\delta L$. At the same time, the radius of the region changes due to the volume conservation and becomes $r+\delta r$ with $\delta r <0$.
	This scheme is valid when $\delta L$ is small enough to allow the
	transition region to be neglected.}
	\label{fig:fiberpullingparams}
\end{figure}

To understand better the physics behind it, we need to make two simple assumptions:
\begin{itemize}
	\item the glass in the heated region is soft enough to be stretchable but
	is hard enough not be stretched by its own weight and the other parts of the fiber, outside the heated region, are solid and non stretchable,
	\item the volume of the glass does not change significantly due to the
	heating and the conservation of the volume comes from the conservation of
	the mass.
\end{itemize}
From these assumptions, we can use the mass conservation to state that the volume of the
heated cylinder at the time $t$ and the volume after the pulling at the time
$t+\delta t$ must be the same leading to equation~\eqref{eq:3-42}:
\begin{equation}
\label{eq:3-42}
\pi\left(r+\delta r\right)^{2}(L_0+\delta L)=\pi r^{2} L
\end{equation}
where $\delta r$ is the variation of the radius (and thus a negative quantity), L is
the length of the taper and $\delta L$ is the length of the fiber which has been pulled. The
meaning of the parameters is clarified in figure~\ref{fig:fiberpullingparams}.

 In the limit $\delta L \rightarrow 0$, we can derive  equation~\eqref{eq:3-volume} from equation~\eqref{eq:3-42}~\cite{birks1992Shape}:
\begin{equation}
\label{eq:3-volume}
\frac{d r}{d L}=-\frac{r}{2 L}
\end{equation}

We thus found that, by controlling the pulling, we can control the final radius of the heated zone. We can deduce that with multiple pulling steps and $L$ small enough so that we can obtain an arbitrary profile.
To obtain from equation~\eqref{eq:3-volume} the pulling procedure necessary to obtain a given
profile, there is no analytical solution and it is necessary to proceed numerically (see ~\textcite{birks1992Shape}).
In conclusion, it is possible to find a procedure in order to obtain
the desired shape for the transition region. This involves the ability to pull while
translating the flame under the nanofiber in order to be able to choose $\delta L$ and the heated region. For practical reasons, it is more suitable to keep the flame fixed and to move the fiber back and forth instead, resulting in a
translation movement superposed to the pulling one.
Usually, the pulling procedure is constituted by
different consecutive pulling steps that reduces progressively the radius,
approximating at the best the calculated optimal desired profile .

To perform these calculations, in the lab we use a matlab code, that generates a
file with the instructions in terms of movements of the stages in order to produce the
correct profile.

The heating can be performed using a flame or a focused \ch{CO2} laser. In
our case, a flame of hydrogen and oxygen was chosen as it presents less
alignment inconvenience than a laser spot. In different situations,
especially when spatial constraints are present, the laser can be the best option.

\subsection{Experimental details}
\label{sec:pulldet}
\begin{figure}[tb]
	\centering
	\includegraphics[width=0.7\linewidth]{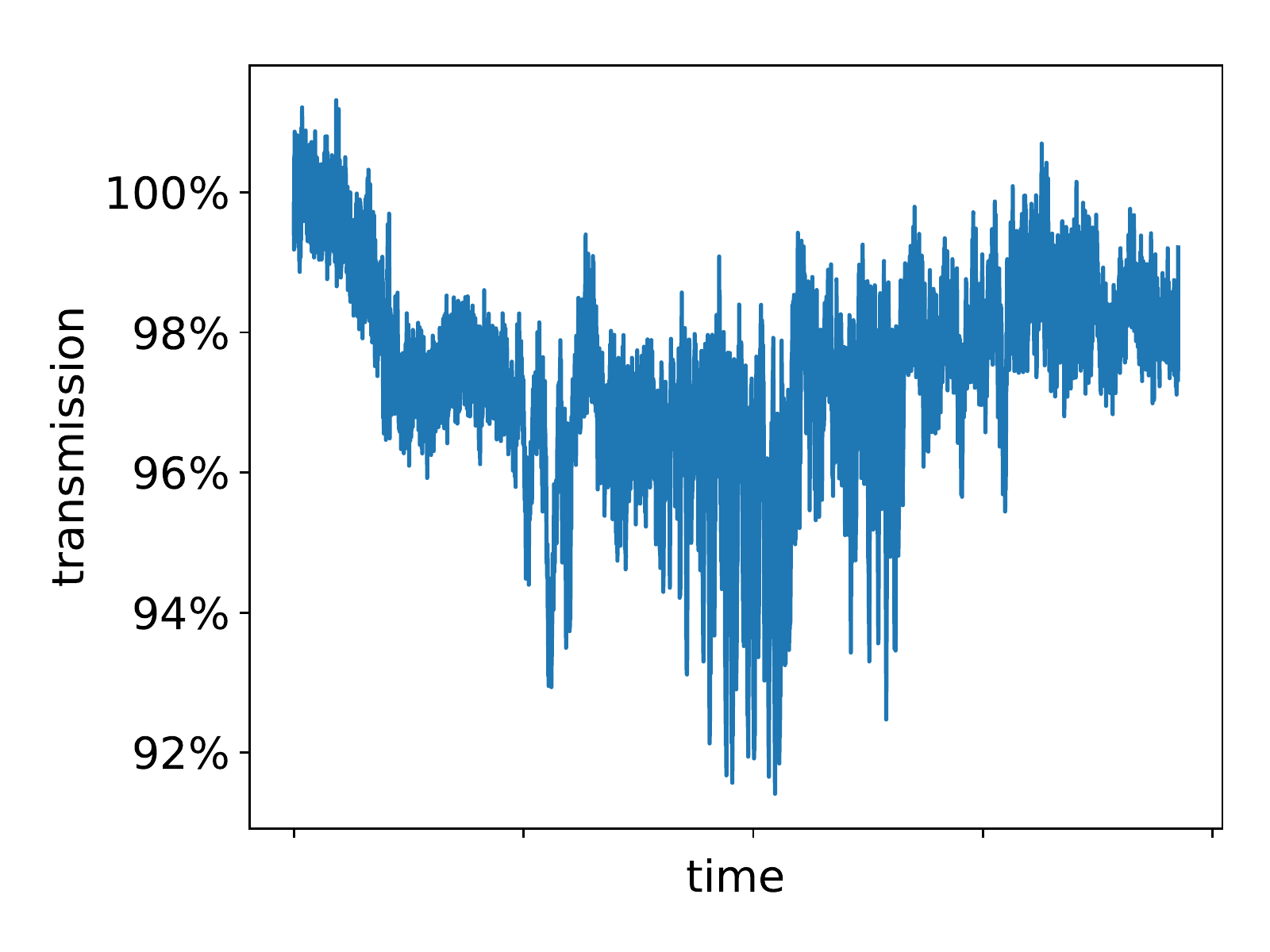}
	\caption{Intensity trace measured during the fiber pulling process. The trace
	starts at the beginning of the pulling phase and it is recorded until its end. At the
	beginning, the transmission is \SI{100}{\percent}, during the pulling we
	loose part of the transmission (down to \SI{92}{\percent}) while the profile is not optimal, but it is partially recovered when the nanofiber reaches its final shape.}
	\label{fig:transmission}
\end{figure}

The fiber is placed over a pure flame of hydrogen and oxygen. The amount of gas is controlled by two gas mass flow controllers that allows a fine control over the gas flow and, consequently, over the flame size.
When the pulling starts, a motor moves the flame near the fiber.
At the same time, the fiber is translated and pulled via two precise
translation stages following the instructions written in a text file. The whole
procedure is controlled by the computer: at the end of the pulling phase, the flame is
moved away from the fiber.
During the process, we constantly monitor the transmission of the fiber in
order to have information on the quality of the pulling. The transmission is monitored measuring the transmitted light with a photodiode as a function of the time normalized to the power transmitted at the beginning of the pulling. This is accurate as long as all the other parameters of the set-up are unchanged (laser emitted power, coupling of the laser, etc).  An example of the
intensity trace measured is reported in figure~\ref{fig:transmission}.

\begin{figure}[tb]
    \centering
    \includegraphics[width=\linewidth]{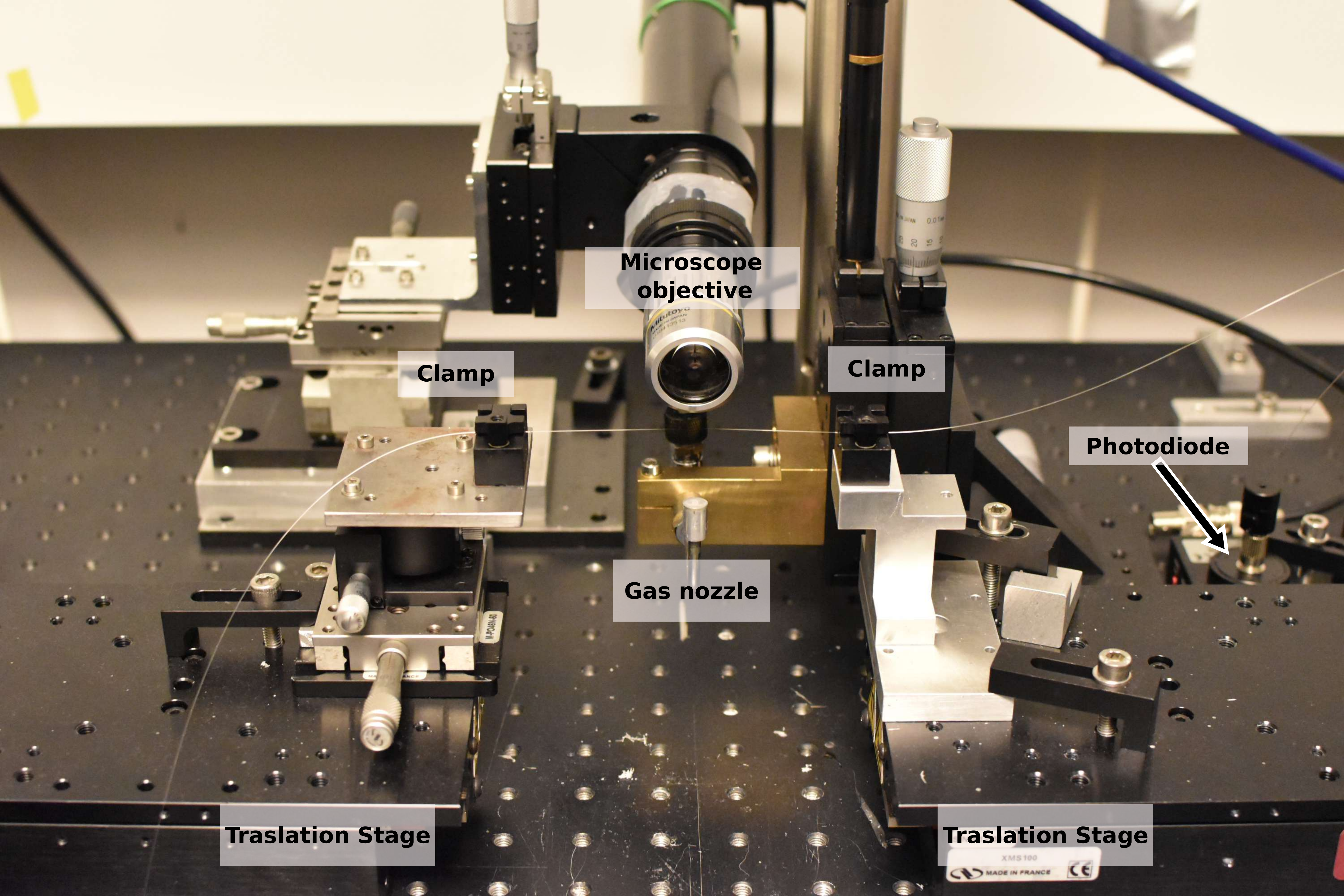}
    \caption{Photo of our pulling setup: the fiber has been pulled and it is clamped between two supports. At the center is visible the microscope objective, while under the fiber, it is possible to see the gas nozzle where the flame burns. At the left and right bottom of the picture, the two translation stages that control the pulling are visible.}
    \label{fig:photoPS}
\end{figure}

A microscope objective with a large working distance is used to monitor the fiber
position and to correctly place the flame under the fiber. A picture of the pulling setup is reported in figure~\ref{fig:photoPS}

The procedure for the fiber pulling is composed of several steps, all of them are crucial for obtaining an optimal result.

\begin{enumerate}
	\item We start from a standard fiber. In this particular case of the
	experiment, I start with a Thorlabs SM450 fiber. The first step
	is to couple the laser inside the fiber and send the output light from the fiber to the photodiode in order to
	monitor its transmission
	\item We switch on the flame opening the flux of oxygen and using an electrical gas lighter. We then open the hydrogen flux, verifying that the
	flame changes color and gets from red to blue, meaning that the correct temperature is reached (the burning temperature of hydrogen and oxygen is \SI{2660}{\celsius}).
	\item Commercial fibers are covered by an acrylate jacket to give them a better mechanical resistance. It is important to remove it in the part that
	we want to pull (few centimeters are normally enough), using a fiber stripper.
	\item A special care needs to be done for cleaning the fiber in order to
	remove any dust. Any residual dust will burn under the flame and reduce the final achieved
	transmission. The first step is to wipe the
	fiber with isopropanol to remove any trace of grease present on it as well as the powder left by the jacket removal. Secondly, we use acetone on
	the uncovered part to dissolve residual traces of the jacket that are still
	in place. We finally clean again with isopropanol.
	\item The fiber is placed on the holder and blocked with two mechanical clamps. The whole setup is under a clean laminar air-flow, to avoid the
	presence of dust and to reduce the oscillation of the fiber.
	\item The pulling starts with the profile previously calculated. The computer controls the movement of the translation stages (left and right)
	and of the flame (up and down). During the pulling, it is necessary to adjust
	the flame position as the fiber is slightly pushed up by the lifting hot
	air around the flame.
	\item Once the pulling is successfully achieved the flame is moved away
	from the fiber by the computer. It is important to automate this task as timing is crucial in the success of the pulling procedure, if the flame is not removed when the pulling  stops, the fiber can break.
	\item It is useful to slightly increase the tension of the nanofiber to reduce
	its vibrations. This is done by monitoring the fiber position with the microscope
	and moving the translation stages (usually few micrometers is enough). In this
	situation, the nanofiber dimension is below the microscope resolution but it is
	possible to see its luminescence if we keep the room in the dark.
	\item The fiber is now ready and we can switch off the flame.
\end{enumerate}
Following this protocol we are usually able to produce nanofibers with a transmission of \SI{98}{\percent} which is enough for our experiments.
Better transmissions have been reported in the literature. With accurate optimization of the setup and of the fiber profile,
\textcite{nagai2014Ultralowloss} claim to be able to produce nanofibers with
more than \SI{99.7}{\percent} of transmission.
The pulling phase requires few minutes, but if we consider the whole procedure, we can estimate that two hours is necessary to fabricate a nanofiber. The success rate depends largely on the ability of the experimenter but we can consider one success for two trials.

Once the pulling is completed, we need to move the fiber to the experiment. This is how it is done:

\begin{enumerate}
	\item In order to move the fiber, we need to fix it to
	its final support. We asked the mechanical service of the lab to produce
	some U-shaped support that can maintain the fiber fixed during the
	experiment. To fix the fiber onto it we:
	\begin{itemize}
		\item use a translation stage to place the holder directly
		under the nanofiber, until it gently touches it;
		\item we glue the fiber on the holder using a UV glue. This step
		needs a particular care as the nanofiber is very fragile.
	\end{itemize}
	\item We place the nanofiber in a clean and hermetic plastic box to avoid the deposition of dust over the nanofiber. We made two small notches in the box
	to allow the fiber to exit from it.
	\item Once the box is closed, we can disconnect the fiber from the
	photodiode and stop monitoring the transmission. We cut the fiber in order to have about \SI{50}{cm} at each ends and we carefully move the box with the nanofiber from the pulling setup to the experiment.
	\item We weld the two extremities of the tapered fiber to the fibers of the
	experiment using a splicing machine in order to minimize losses. Usually, losses in the connection are below the instrument sensitivity
	(i.e. \SI{0.1}{\decibel}) and thus small enough.
\end{enumerate}

\section{Experimental Setup}
\begin{figure}[tb]
	\centering
	\includegraphics[width=0.8\linewidth]{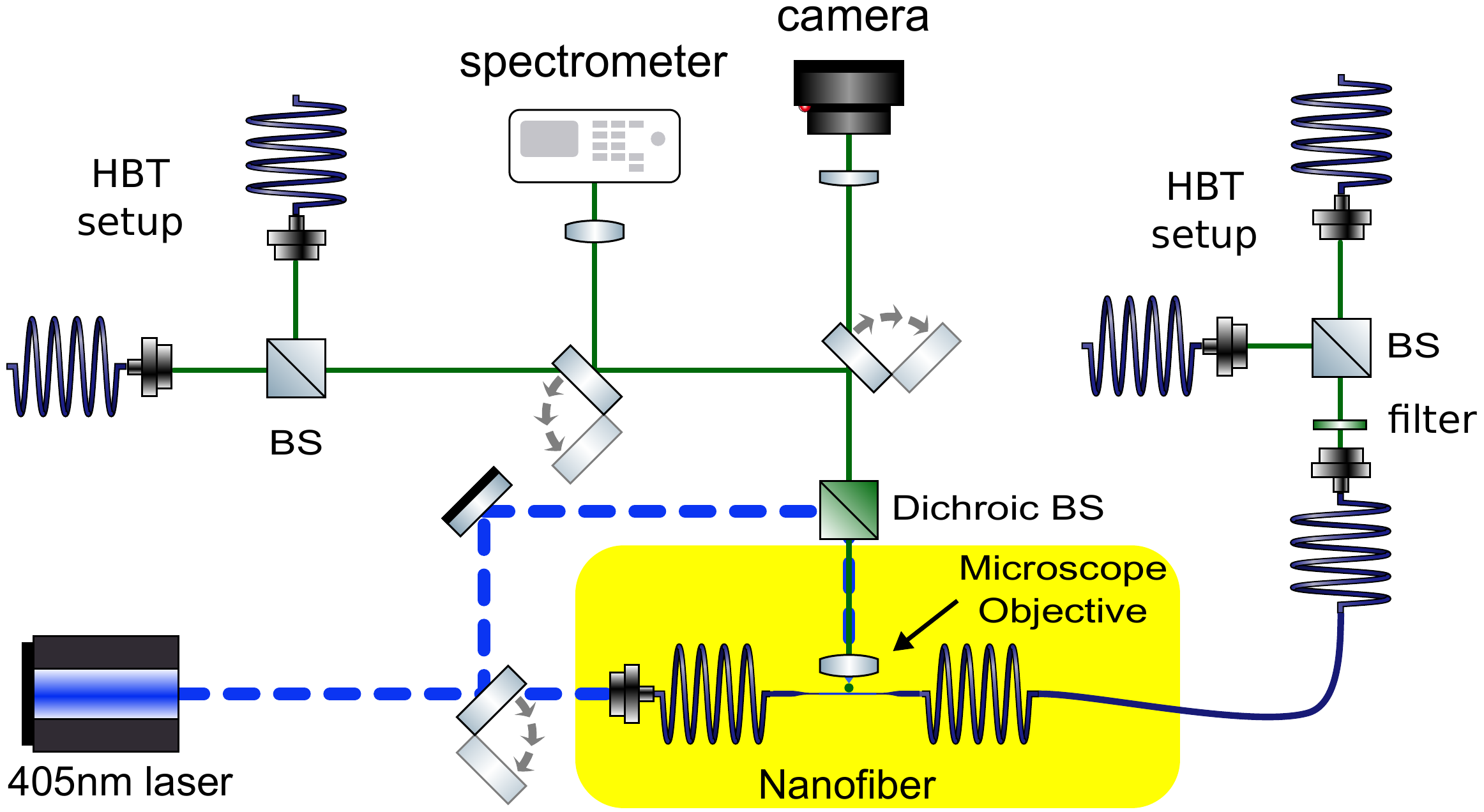}
	\caption{Setup for the nanofiber experiment: a blue pulsed laser can
	excite the emitter both from the nanofiber and from freespace. The
	image and the spectra of the nanocrystals can be recorded using a camera. Ultimately, it is possible to perform the \gd{} measurement in both
	configurations. The part surrounded by in yellow is placed in a box under a
	clean laminar air-flow. BS: non polarizing beamsplitter, HBT: Hanbury Brown
	and Twiss.}
	\label{fig:setupfibra}
\end{figure}

\begin{figure}
    \centering
    \subfloat[]{\label{fig:setupF_all}\includegraphics[width=0.9\linewidth]
		{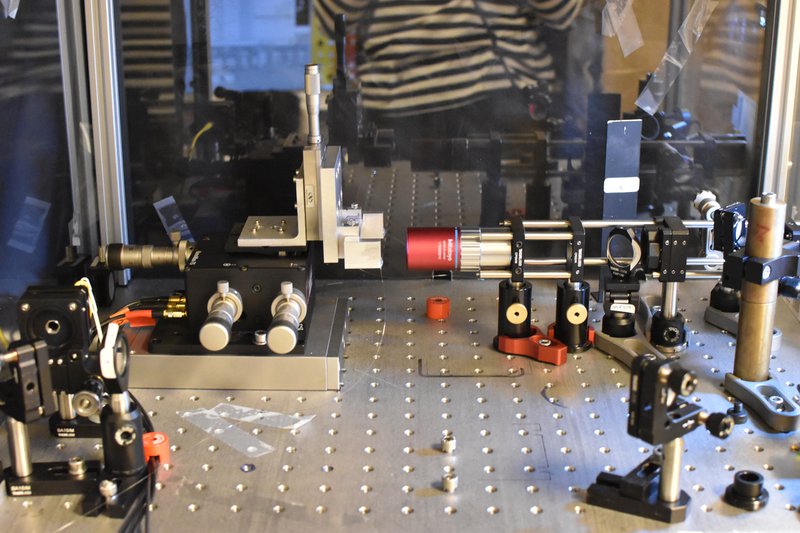}} \\
	\subfloat[]{\label{fig:setupF_small}\includegraphics[width=0.9\linewidth]
		{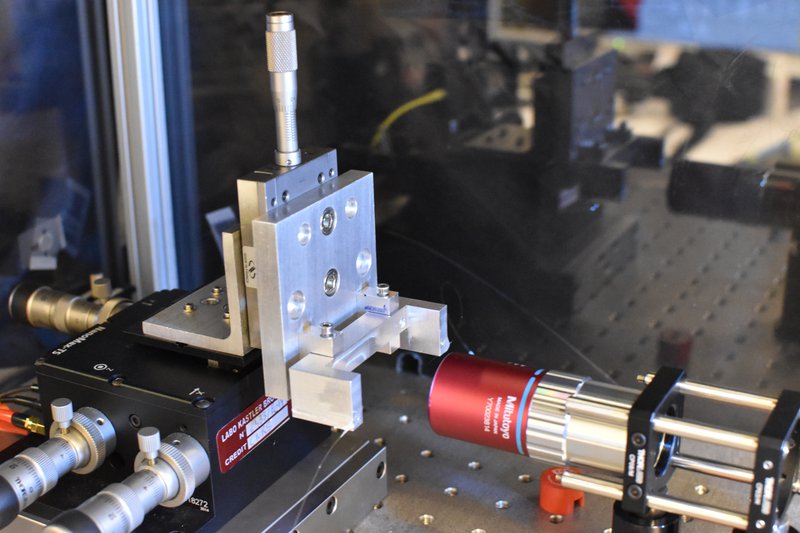}}
    \caption{Photo of the experimental setup for observing single photon emitters on the nanofiber from different prospective. The U-support to which the fiber is glued is visible.}
    \label{fig:setupF}
\end{figure}

The setup used for our experiment is shown in figure~\ref{fig:setupfibra}. A photo of the setup is reported in figure~\ref{fig:setupF}.
The setup allows to excite the emitter on the fiber both from the nanofiber and
from the free space thanks to a home-made confocal microscope. By the same token, the luminescence of the emitter can be
collected by the microscope objective, analyzed with the camera or the
spectrometer or sent to the HBT setup to perform the \gd{} measurement. It is
also possible to connect directly the fiber to the \gd{} measurement, in order to study the light that is coupled to the nanofiber via the near field coupling.

It is important to stress the fragility of the nanofiber in any experimental
procedure. In particular we have shown that most of the mode propagates
outside the nanofiber: this makes it possible to
couple the light emitted by the nanocrystal directly into the fiber through the near field coupling but it implies also that any kind of dust lying
over the nanofiber can disturb the light propagation and reduce the
transmission. For this reason, it is essential to protect the fiber from dust
using a clean air laminar glow. This is obtained by surrounding the part of the
experiment in yellow in figure~\ref{fig:setupfibra} with a box and sending a
laminar flow inside it.
This has also the advantage of protecting the fiber from any accidental shock and
to protect the experimentalist from accidentally touching the nanofiber. Indeed, the diameter of the nanofiber is smaller than the size of the
pores of the skin (usually \SI{50}{\micro \meter}) and for this reason can be
dangerous to touch it.

In the following, I will explain how we deposit the nanocrystals on the
nanofiber and I will discuss the experiment we performed, concluding with the results we
obtained.

\section{Nanocrystal deposition and antibunching measurement on the nanofiber}
%added part
The coupling of the light emitted by the nanocrystals with the modes of the fiber happens via the strong evanescent field present at the nanofiber surface. This technique has been used in the past to couple with nanofibers different emitters, such as single atoms\cite{nayak2008Single,nayak2009Antibunching}, nanodiamonds containing \ch{NV} centers\cite{schroder2012Nanodiamondtapered,liebermeister2014Tapered} and \ch{CdS}/\ch{CdSe} colloidal quantum dots\cite{fujiwara2015Ultrathin, yalla2012Fluorescence}.
However, to the best of our knowledge no references about perovskite nanocrystals coupling with nanofibers nor with other integrated photonic structures are available in literature. The poor photostability of these emitters made indeed particularly challenging the integration with this kind of structures.
Different techniques to place the emitters in the proximity of the nanofiber have been used. In case of atoms, this is obtained placing the nanofiber under vacuum and trapping atoms near it using optical trapping. Differently, in case of solid emitters, deterministic deposition can be achieved using and AFM tip to place the emitter at the desired position making use of another nanofiber to deposit the emitter. In the case of our perovskites, we decided to use a technique for non-deterministic deposition consisting in touching the fiber with a droplet of solution containing the nanocrystals. Compared to more complex techniques, this method, detailed in the following, allows a faster and simpler deposition, important in case of fragile emitters. When further optimization of perovskites stability will be performed, the other techniques could be applied to deterministically place them on the fiber.
%end added

\begin{figure}
	\centering
	\includegraphics[width=1\linewidth]{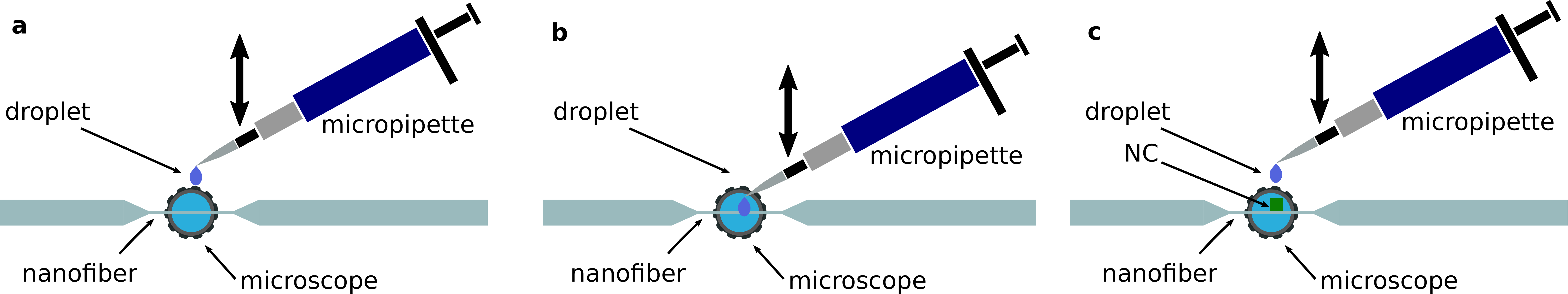}
	\caption{Scheme of the procedure for deposition of single nanocrystals over
	a nanofiber. A droplet of \SI{20}{\micro \liter} is created (a), and the
	fiber is touched with it (b) using a precise translation stage. The droplet
	is removed and 	this results in one or more NCs deposited on the fiber (c).}
	\label{fig:schemefiberdepositionall}
\end{figure}

To deposit the nanocrystals on a nanofiber, we first take a droplet of  \SI{20}{\micro \liter} with the
solution containing the nanocrystals with a micropipette. Then, with the droplet at the extremity of the micropipette, we move the droplet towards the fiber until we touch it, leaving some nanocrystals over the fiber. In order to make a sturdy movement, we use a 3-axis translation stage to control the position of the micro-pipette.
The scheme of the procedure is shown in
figure~\ref{fig:schemefiberdepositionall}.

% After the fabrication of the nanofiber, a particular care is needed for the NCs
% deposition: the
% nanofiber, as explained before, is very sensitive to any mechanical
% solicitation and it is very easy to break it.
% In addition, any dust deposited over it can perturb the surrounding evanescent
% field and degrade the nanofiber transmission: the whole procedure needs thus to
% be performed under a clean laminar air flux, to avoid any possible
% contamination.

% We create a droplet of \SI{20}{\micro \liter} with the solution containing the
% NCs at the point of a micropipette, and gently approach the fiber using a
% translation stage. A scheme of this procedure is shown in
% figure~\ref{fig:schemefiberdepositionall}.

We monitor the procedure with the microscope objective shown in
figure~\ref{fig:setupfibra}. Once the droplet is in contact with the fiber, we gently move the droplet away from it.
This procedure often results in one or more emitters deposited over the fiber. This can easily
be verified by shining the excitation laser inside the fiber
and filtering out the excitation
beam to collect only the perovskite nanocube
emission wavelength.
\begin{figure}
	\centering
	\subfloat[]{\label{fig:beforedepo}\includegraphics[height=0.4\linewidth]
		{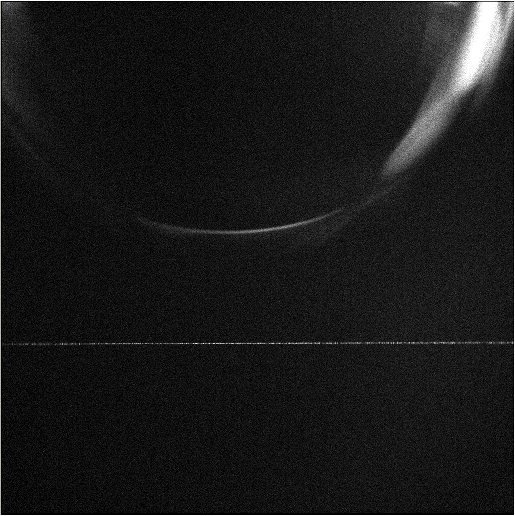}}\qquad\qquad
	\subfloat[]{\label{fig:afterdepo}\includegraphics[height=0.4\linewidth]
		{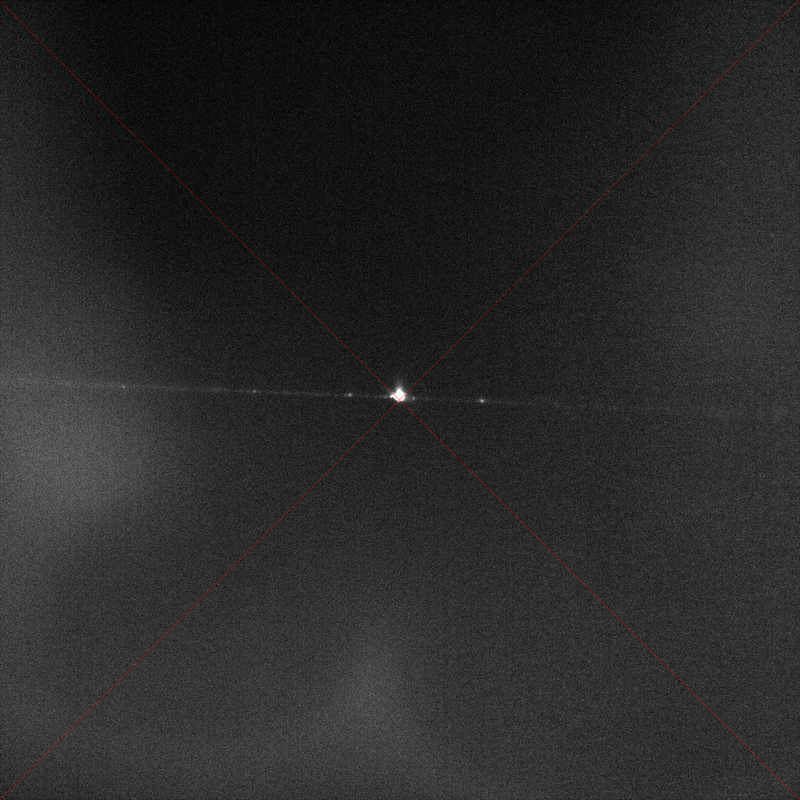}}
	\caption{Image of the nanofiber detected from the camera. A laser is shined trough the fiber.
	\protect\subref{fig:beforedepo}~Before the deposition, the nanofiber is visible on the camera thanks to the weak diffused light from small imperfections at on the nanofiber surface;
		\protect\subref{fig:afterdepo}~after the nanoparticle deposition, the nanocrystals scatter light and they are visible as a strong bright spot on the fiber viewed by the
		camera. Both of the images were taken without filtering the excitation
		light. \label{fig:fiberdeposition}
	}
\end{figure}
\begin{figure}
	\centering
	\includegraphics[width=0.5\linewidth]{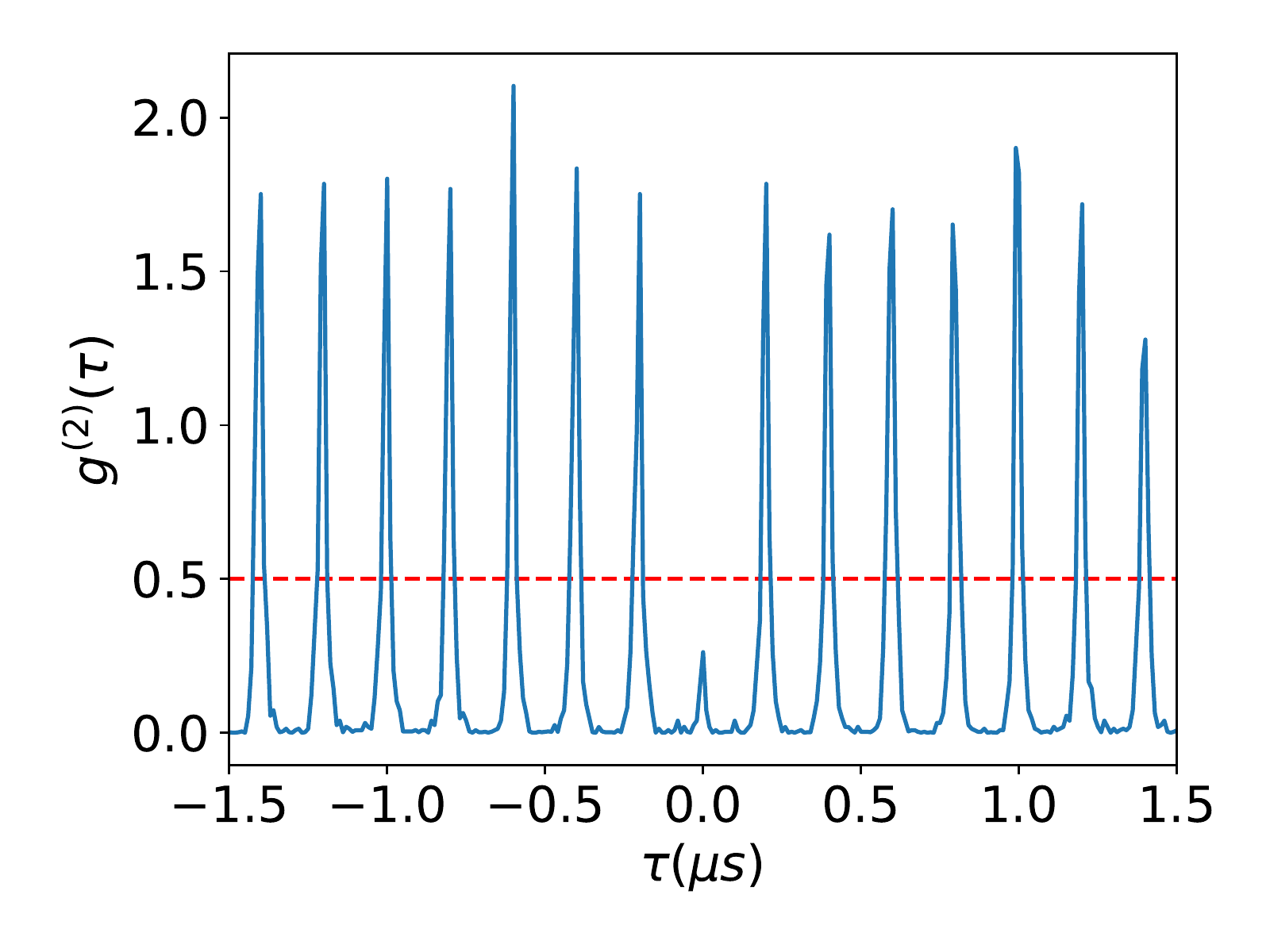}
	\caption{\gd{} function of a single perovskite nanocrystal placed onto a
	nanofiber: the photons are emitted directly throw the nanofiber and collected
	via the fiber output. We obtain a $ g^{(2)}\of{0}=0.24$.}
	\label{fig:fig9pltem3norm}
\end{figure}
Figure~\ref{fig:fiberdeposition} shows how we can verify when we deposit the emitter
over the nanofiber. During the procedure a laser is shined trough the fiber. First of all, we see the fiber, thanks to the light diffused by small imperfections on the nanofiber surface,
and the approaching droplet (both of them visible in
figure~\ref{fig:beforedepo}). At the end of the procedure, an emitter is stuck
to the nanofiber and scatters light resulting in a bright spot on the camera
(as shown in figure~\ref{fig:afterdepo}).

%\section{Experimental results}
In order to deposit the emitters sufficiently spaced from one to another, we need to reduce the nanocrystal concentration. To do this, we were forced
to dilute the initial solution by \SI{100}{times}. As I have shown in
section~\ref{sec:rolesampleprep}, this results in perovskites that bleach in a
very short time: this means we have to illuminate the emitters for very short times and we cannot perform long statistical measurements.

We illuminate the nanocrystals via the nanofiber in order to find them and place them under the laser light coming from the microscope objective. We then use the fiber to collect the photons, sending the collected light to the antibunching setup and use the laser through the microscope objective to excite the emitter.

Even by having a short measurement time, we were able to measure a good single photon emission from the signal
collected via the nanofiber. The result of this measurement is shown in figure~\ref{fig:fig9pltem3norm}.
With this proof of principle measurement we demonstrate the possibility to couple a single perovskite nanocrystal with a tapered nanofiber. In order to perform a
deeper study on this hybrid photonic device, a strong effort needs to be done to improve the nanocrystal
stability to be able to use them for a long time after having reached the correct dilution.

\clearpage
\begin{chrecap}
	\begin{itemize}
		\item Fiber optics allow to guide light in very efficient way.
		\item I have reported a brief introduction on the fiber development and
		the theory of the light propagation in fibers.
		\item Depending on the fiber $\mathsf{V}$ parameter, the fiber can allow either the propagation of one single mode or of several modes, which propagate with different group
		velocities.
		\item When the fiber diameter is smaller than the  wavelength of the guided light, it propagates mostly outside the fiber: it is the case for a so-called nanofiber.
		\item It is possible to fabricate an optical nanofiber by tapering a standard optical fiber.
		\item It is important to have an accurate control on the transition
		region between the standard fiber and the nanofiber in order to reduce the losses.
		\item It is possible to use the evanescent field to directly excite a
		nanocrystal that is placed on a nanofiber and vice-versa, to collect the light emitted  by the nanocrystal via the nanofiber.
		\item I have shown that it is possible to  couple the single
		photon emission of a perovskite nanocrystal directly inside the nanofiber
	\end{itemize}
\end{chrecap}

\chapter{Outlook and future perspectives}
\label{chap:outlook}
\minitoc
\section{Introduction}
Despite the versatility of the nanofiber platform integrated with perovskite
nanocrystals there are some limitations that need to be overcome in order to
be able to practically use it.
In this chapter I will discuss some possible paths to address the main problems
of
this platform, keeping in mind the final goal: to produce integrated single
photons sources for quantum applications.
First of all, I will concentrate on the emitters, describing how perovskites
nanocrystals could be improved and presenting also different kinds of emitters that could
solve the main limitations of perovskites in the future, such as single defects in nanodiamonds.

In the second part of the chapter, I will describe another integrated  platform I had the opportunity
to work with, that could improve the nanofiber limitations, mainly its
fragility, while keeping its advantages: this platform is the \IEW{} platform.

\section{Perovskite optimization}

Perovskite nanocrystals have shown to be high quality emitters, with a narrow
emission at room temperature and easy fabrication method. The versatility of
the chemical structure is so wide that almost any interesting wavelength can
be reached: at the moment, emission in visible and near infrared ranges
has been shown.

In addition, recent studies on perovkite nanoplatelets have shown that the emission wavelength depends on the number of layers: with the improvement of the growth techniques, this can be an advantage to produce several identical emitters~\cite{bekenstein2015Highly,akkerman2016Solution, protesescu2015nanocrystals}.
%todo: add reference

The main drawback of perovskite nanocrystals is their photo-stability: this is the
direction on which it is important to focus the research to start with. Our study~\cite{pierini2020Hybrid} has shown that one of the factors is the role of the ligands. That is actually not completely understood but that could offer in the near future the key to obtain stable perovskite nanocrystals.

A recent study in this direction has demonstrated that it is possible to stabilize
nano\-crystals replacing the oleic
acid with a more appropriate ligand, the 2-hexyldecanoic acid~\cite{yan2019Ultrastablea}. That could be the way out for such emitters.

\section{Single defects in nanodiamonds}
Defects in diamonds are interesting sources of single photons, as they are very stable.

Crystallographic defects in diamonds results from crystal irregularities, as well as from
substitution or interstitial impurities. When present in huge quantity, they
have effects in the crystal color and electrical conductivity. In jewelry, they are used to produce colored diamonds. More than \num{100} luminescent defects are known
in diamond, and a large part of them has been studied in
detail~\cite{davies1994Properties}.
Diamond luminescent defects are of great interest for single photon
generation, due to their great stability. During my PhD, I studied two different
single-photon emitting defects in diamonds, Nitrogen-Vancancy and Silicon-Vacancy, whose main characteristics are
detailed below. At L2n laboratory, Germanium-Vacancy color centers are also currently studied~\cite{nahra2020Single}, but I did not have the opportunity to work with them and, for this reason, I will not treat them here.

\subsection{Nitrogen-Vacancy color centers}
One of the most important defects in diamonds for quantum photonics is the
Nitrogen-Vacancy/\ch{NV} defect. It is very abundant due to the presence of nitrogen in atmosphere which makes it highly probable to find them even in a clean production environment.

\begin{figure}[tb]
	\centering
	\includegraphics[width=0.35\linewidth]{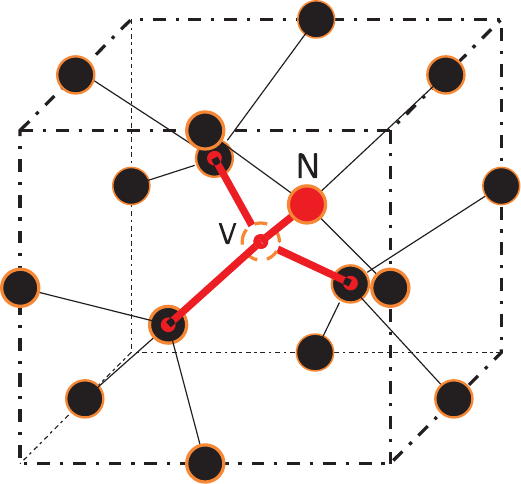}
	\caption{Representation of an \ch{NV} center defect in a diamond crystal.
	\ccredit{trupke2011Enhancing}}
	\label{fig:nvstruct}
\end{figure}

\begin{figure}[tb]
	\centering
		\subfloat[]{\label{fig:NV-spectr}\includegraphics[width=0.45\linewidth]
		{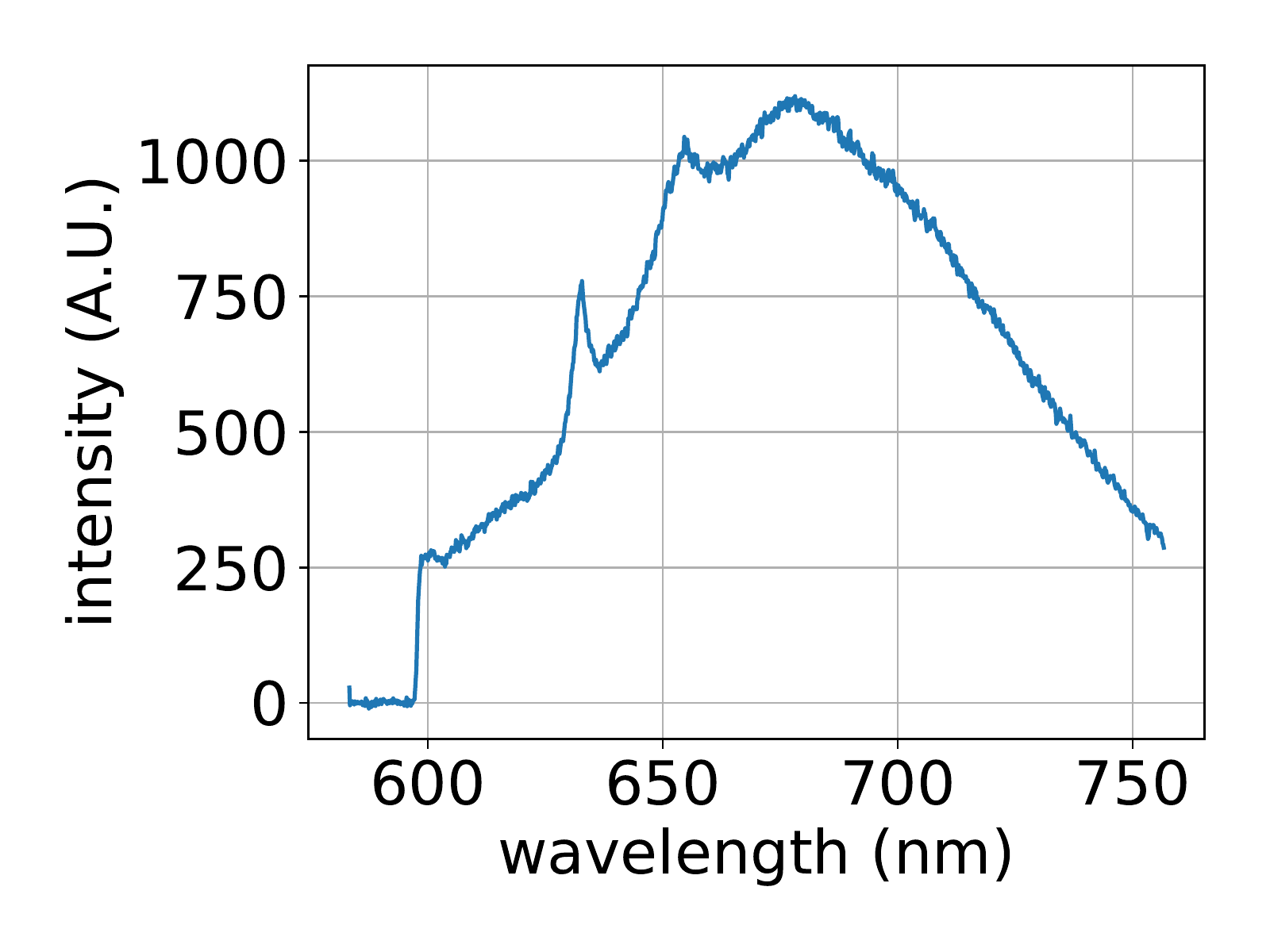} }\quad
	\subfloat[]{\label{fig:NV-diag}\includegraphics[width=0.48\linewidth]
		{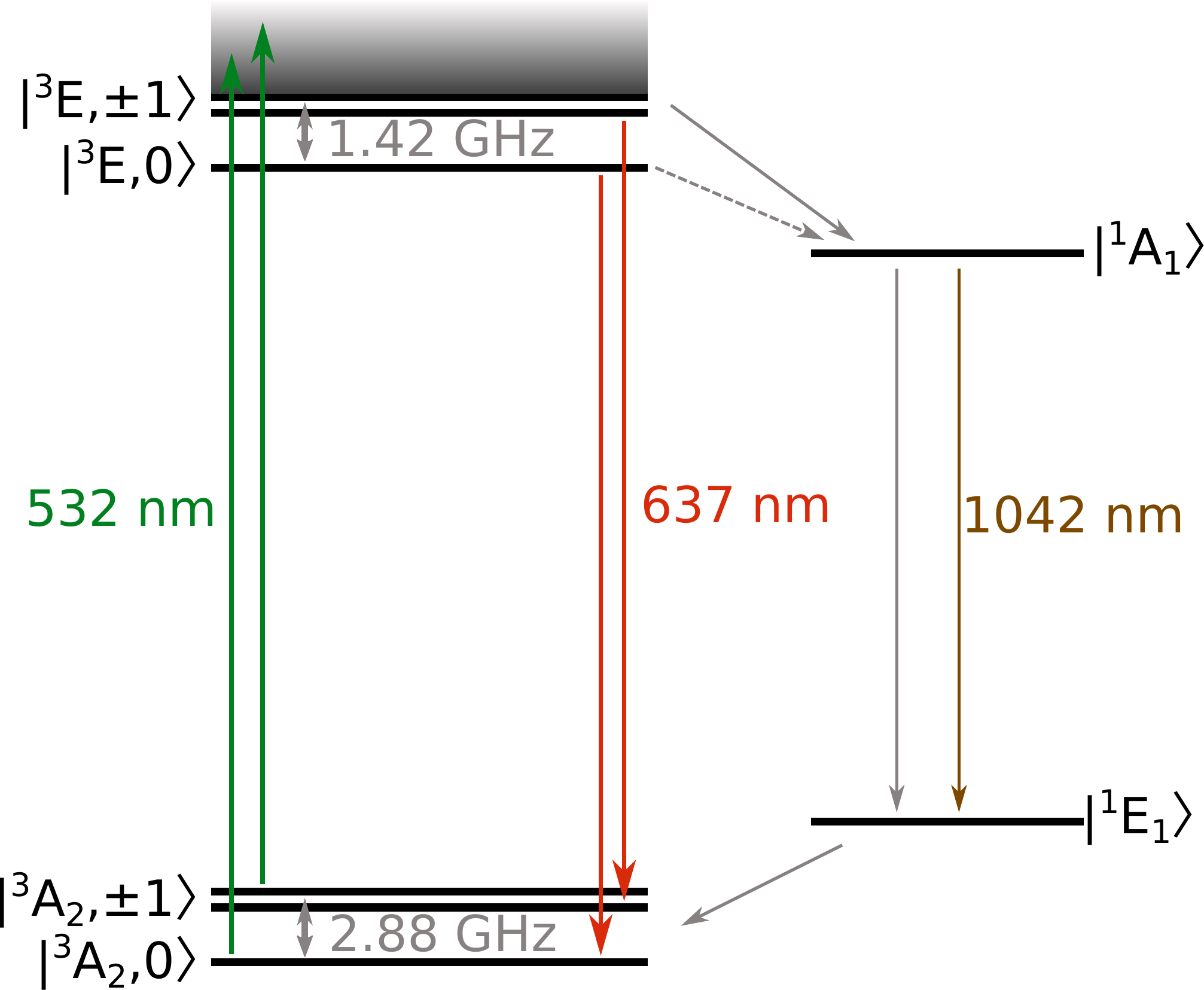}}

	\caption{\protect\subref{fig:NV-spectr}~Fluorescence emission of a single NV
	center in a nanodiamond. The
	fluorescence is cut by a \SI{600}{\nano \meter} long-pass filter. The first
	peak on the left, at about \SI{634}{\nano \meter} is the zero phonon line:
	in single defects in bulk it is at \SI{637}{\nano \meter} but reticular
	strains can slightly modify its wavelength.
	\protect\subref{fig:NV-diag}~Level structure of an NV defect. The green line
	represent the usual excitation light (out of resonance), while the red
	represent the emission wavelength. The gray lines represents non-radiative
	relaxation process.}
	\label{fig:nvspectrum}
\end{figure}

The structure of \ch{NV} defect is reported in figure~\ref{fig:nvstruct}.
% TODO: \usepackage{graphicx} required
%todo  add diamond NV- structure figure.
This defect is present in both the neutral \ch{NV^0} state and in the charged state
\ch{NV-}. I focused in my thesis on the \ch{NV-} defect which is the most studied for single-photon
generation as its fluorescence is easy to detect, compared to the neutral one. As
other defects in diamonds, the \ch{NV-}
does not present stability issues so much. A very strong laser illumination can
transform a \ch{NV-} center in a \ch{NV^0} one, but this problem is not
present with standard excitation power.
As opposed to perovskite quantum dots, \ch{NV-} nanodiamonds have a broad
spectrum at room temperature, due to the thermal broadening, i. e. due to coupling with phonons from the diamond crystal structure. As an example, a
spectrum of a \ch{NV-} defects in a nanodiamond is reported in
figure~\ref{fig:NV-spectr}, while in figure~\ref{fig:NV-diag} is sketched the
level diagram with a focus on the emission wavelengths. The setup used to measure the spectrum is the same as described for perovskites at page~\pageref{sec:2ExperimentalSetup}, using a \SI{532}{\nano \meter} continuous wave laser and appropriate dichroic and long-pass filters.
Here, it is interesting to note that the usual excitation, out of resonance and
typically with a wavelength of~\SI{532}{\nano \meter} does not mix the spin
states $\ket{0}$ and $\ket{\pm1}$. In addition the non radiative decay rate
from $\ket{^3E,0}$ to $\ket{^1A_1,0}$ is much lower than the one from
$\ket{^3E,\pm1}$: as a result, by exciting repeatedly the emitter, it is possible to
bring the spin of the NV center to the $m_s=0$ level; such phenomenon is called spin
polarization~\cite{robledo2011spin} and allows to easily reset the spin state for applications that use the spin to encode the qubit~\cite{holzgrafe2019Error, knowles2014Observing}.
% /home/pierinis/Documenti/Dottorato/DatiParigi/NV centers/histogram/2018_04_26
At cryogenic temperatures, most of the emission is in the zero-phonon-line and the
linewidth is reduced to a width of few megahertz.

In conclusion, \ch{NV-} color centers in diamond are interesting sources of
single photons that show, with respect to the
perovskite nanocrystals, the advantage of the stability  and the possibility to control and use their spin state. The main limitation is that the broad spectrum makes it
difficult to obtain two photons at the same frequency at room temperature. The coupling of \ch{NV-} nanodiamonds with nanofibers have been successfully achieved by different groups both at room and at cryogenic temperatures~\cite{fujiwara2015Ultrathin,liebermeister2014Tapered, schroder2012Nanodiamondtapered}.

\subsection{Silicon-Vacancy color centers}

\begin{figure}[tbp]
	\centering
	\subfloat[]{\label{fig:SiVStruct}\includegraphics[height=0.33\linewidth]
		{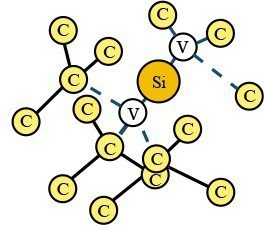} }\quad
	\subfloat[]{\label{fig:sivemission}\includegraphics[height=0.33\linewidth]
		{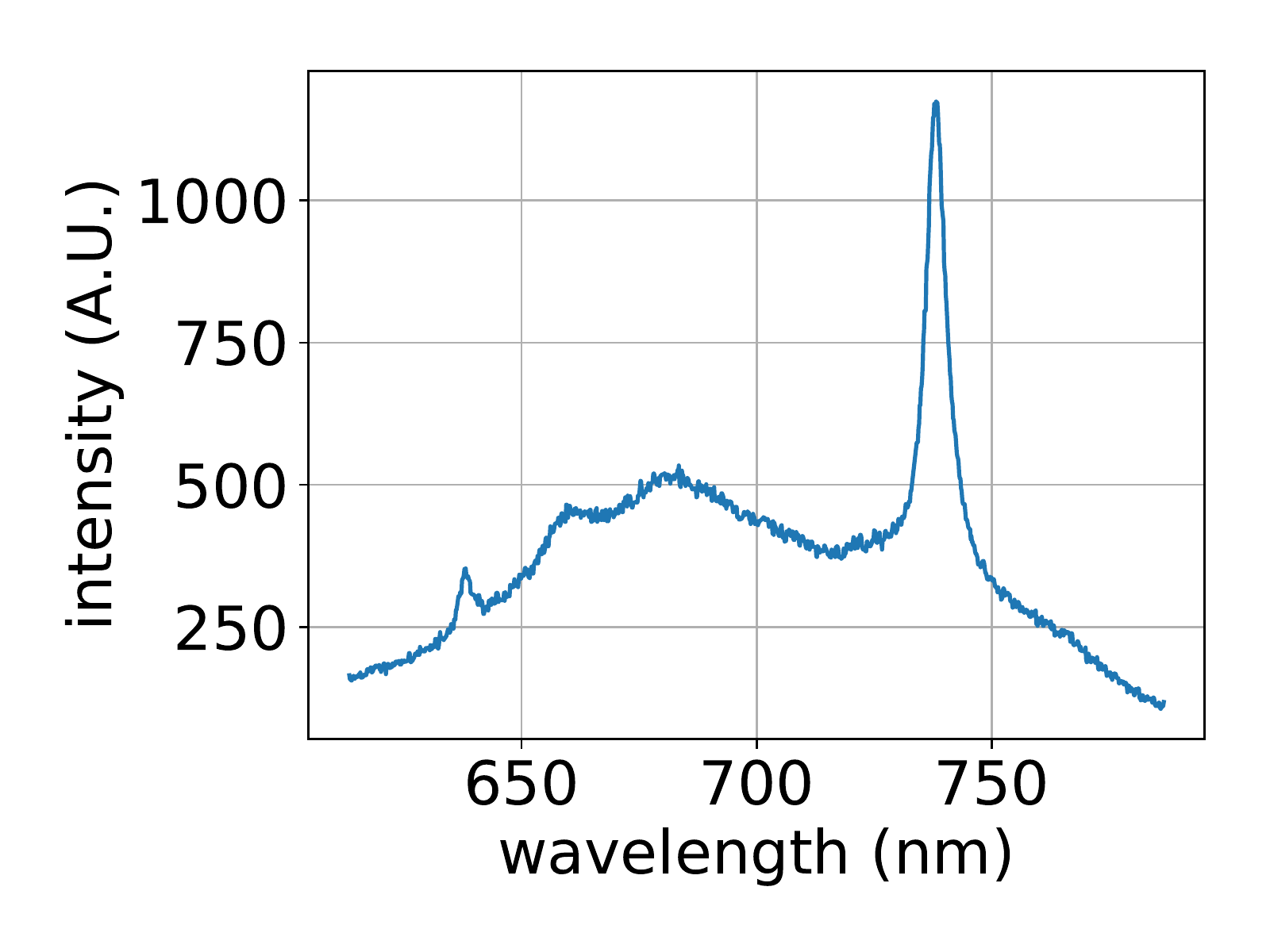}}
	\caption{\protect\subref{fig:SiVStruct}~Structure of a SiV color center in a
		nanodiamond. The \ch{Si} atom
		is in between two vacancies.\ccredit{zeleneev2020Nanodiamonds}
		\protect\subref{fig:sivemission}~Measured emission of \ch{SiV} defects
		in a
		nanodiamond at~\SI{737}{\nano \meter}. It is clearly visible that the emission of \ch{NV} defect is also present with the \ch{SiV}
		emission.}
	%\label{fig:SiVStruct}
\end{figure}
\begin{figure}[tbp]
	\centering
	\subfloat[]{\label{fig:siv-lev}\includegraphics[height=0.6\linewidth]
		{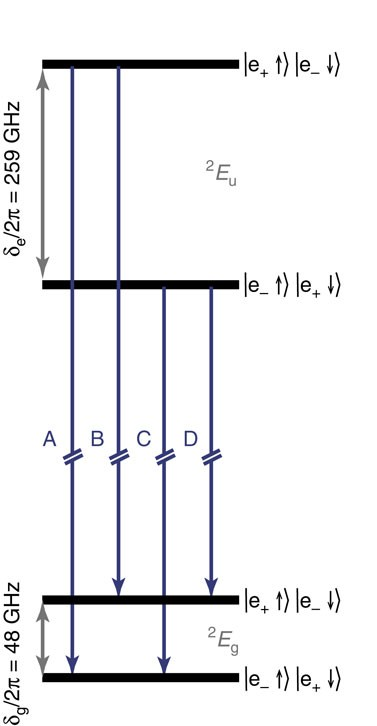} }\qquad\qquad
	\subfloat[]{\label{fig:siv-crio}\includegraphics[height=0.58\linewidth]
		{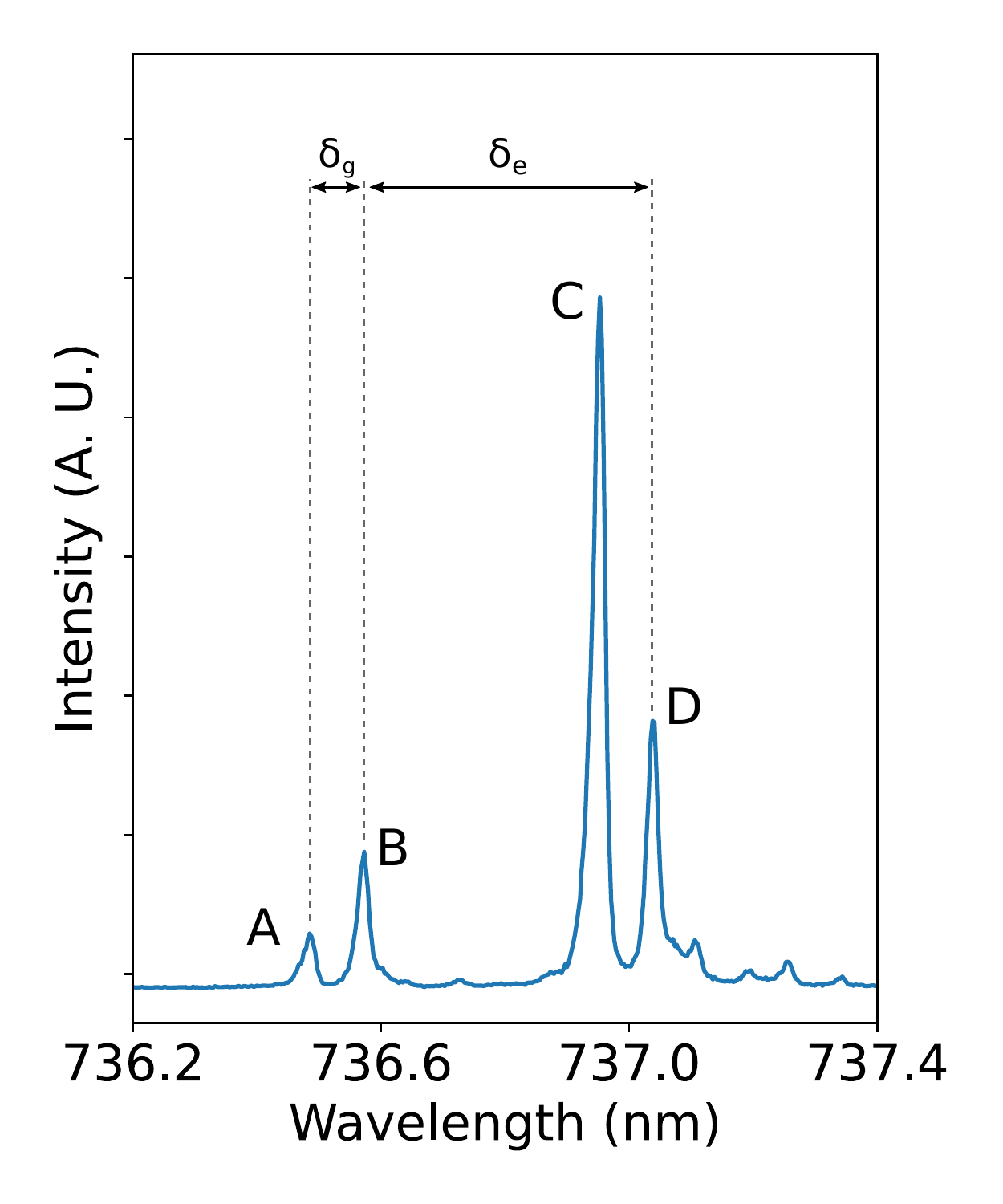}}
	\caption{\protect\subref{fig:siv-lev}~Energy levels of an \ch{SiV} color
		defect in diamond \ccreditnp{becker2016Ultrafast} and \protect\subref{fig:siv-crio}~emission
		spectrum
		at~\SI{4}{\kelvin}
		of the zero phonon line transition of the \ch{SiV} color defect \textit{(measured by M. Nahra at L2n)}.}
	\label{fig:siv-levstruct}
\end{figure}

Less abundant than the \ch{NV-} color centers, the silicon-vacancy defects are
gaining interest as single photon emitters: indeed, they have most of their
emission in the zero phonon
line even at room temperature~\cite{marseglia2018Bright,neu2012Photophysics}.
The structure of this defect is more symmetrical, as the \ch{Si} atom is in
between the places left by the two missing carbon atoms as shown in
figure~\ref{fig:SiVStruct}. Its spectrum is shown in
figure~\ref{fig:sivemission}.

The fluorescence wavelength of the zero phonon line is in the near-infrared
range, at about \SI{737}{nm} and at room temperature \SI{70}{\percent} of the
light is emitted in it~\cite{marseglia2018Bright, haussler2017Photoluminescence}.
%neu2011Single,neu2012Photophysics}
This is a very large amount if compared with other kind
of nanodiamonds. In figure~\ref{fig:siv-levstruct} the energy levels of an \ch{SiV} color defect are shown, as well as a low-temperature zero phonon
line emission: four transitions are present, all of them are included in the
larger emission obtained at room-temperature~\cite{becker2016Ultrafast} (figure~\ref{fig:sivemission}).

\begin{figure}[tbp]
	\centering
	\subfloat[]{\label{fig:SiVCEW}\includegraphics[width=0.4\linewidth]
		{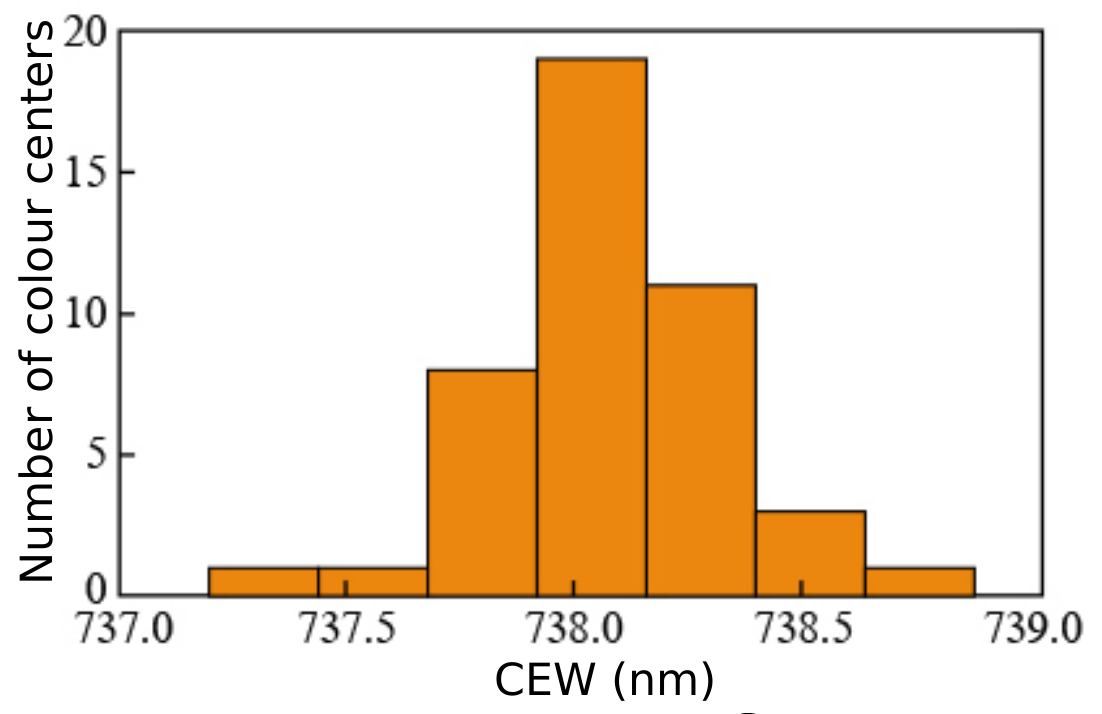}}\qquad\qquad
	\subfloat[]{\label{fig:SiVFWHM}\includegraphics[width=0.4\linewidth]
		{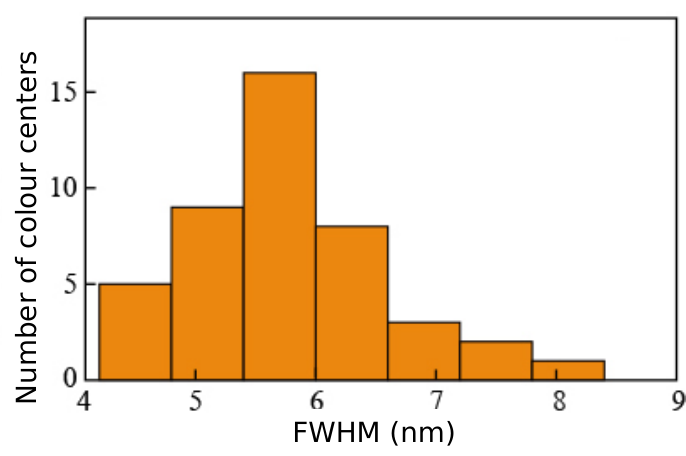}}
	\caption{Characteristics of SiV nanodiamonds obtained
	by~\textcite{zeleneev2020Nanodiamonds} \protect\subref{fig:SiVCEW}~Central emission wavelength of the peak (CEW) at room temperature
		\protect\subref{fig:SiVFWHM}~Full width at half-maximum of the peak at room temperature.\ccredit{zeleneev2020Nanodiamonds}}
	\label{fig:SiV-CEW-FWHM}
\end{figure}

As the silicon is less abundant with respect to nitrogen, this kind of defect is naturally less abundant, it is thus more difficult to fabricate nanodiamonds with
a single \ch{SiV} center and without any other defect. One interesting way to
produce high quality nanodiamonds containing \ch{SiV} color centers is to
create them directly in nanodiamond form, avoiding mechanisms that can
induce reticular strain inside them: this is of importance as the reticular strain
can affect the emission spectrum; the absence of reticular strains is indeed the main advantage of the direct fabrication of nanodiamonds with respect to the method based on nanodiamond explosion~\cite{boudou2013Fluorescent,schirhagl2014NitrogenVacancy}. Recently, a fabrication procedure
to obtain single high-quality nanodiamonds with a size of few nanometers has
been described by~\textcite{zeleneev2020Nanodiamonds}: with this fabrication
technique the authors obtained the remarkable wavelength distribution and full
width at half maximum of the fluorescence peak shown in
figure~\ref{fig:SiV-CEW-FWHM}.
This results show that these emitters are good candidates for future
integration on nanofibers: we started for this reason a collaboration with Prof. Viatcheslav Agafonov from the university of Tours in order to study the application of his nanodiamonds in our platform.

\begin{figure}
    \centering
    \includegraphics[width=0.45\linewidth]{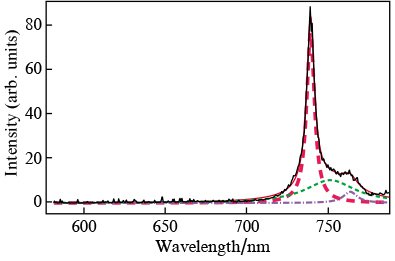}
    \caption{Typical spectrum of \ch{SiV} color center fluorescence from a nanodiamond fabricated via the procedure described in~\textcite{zeleneev2020Nanodiamonds}. The red dashed line represents the fit of the zero phonon line with a sum of three Lorenz functions, while the others represent the fit of the phonon wings. \ccredit{zeleneev2020Nanodiamonds}}
    \label{fig:SiV_spectra}
\end{figure}

In the test we made, it was not possible to integrate them, as the first samples were not small enough and contained, together with \ch{SiV}
color centers, also some \ch{NV} color centers, as clearly visible in the
spectrum reported in figure~\ref{fig:sivemission}. Recent improvements within the fabrication shown in~\textcite{zeleneev2020Nanodiamonds} seems to point towards solving this issue. The spectrum observed is shown in figure~\ref{fig:SiV_spectra}.

\section{Ion-exchange glass waveguides}
\begin{figure}[tb]
	\centering
	\includegraphics[width=0.7\linewidth]{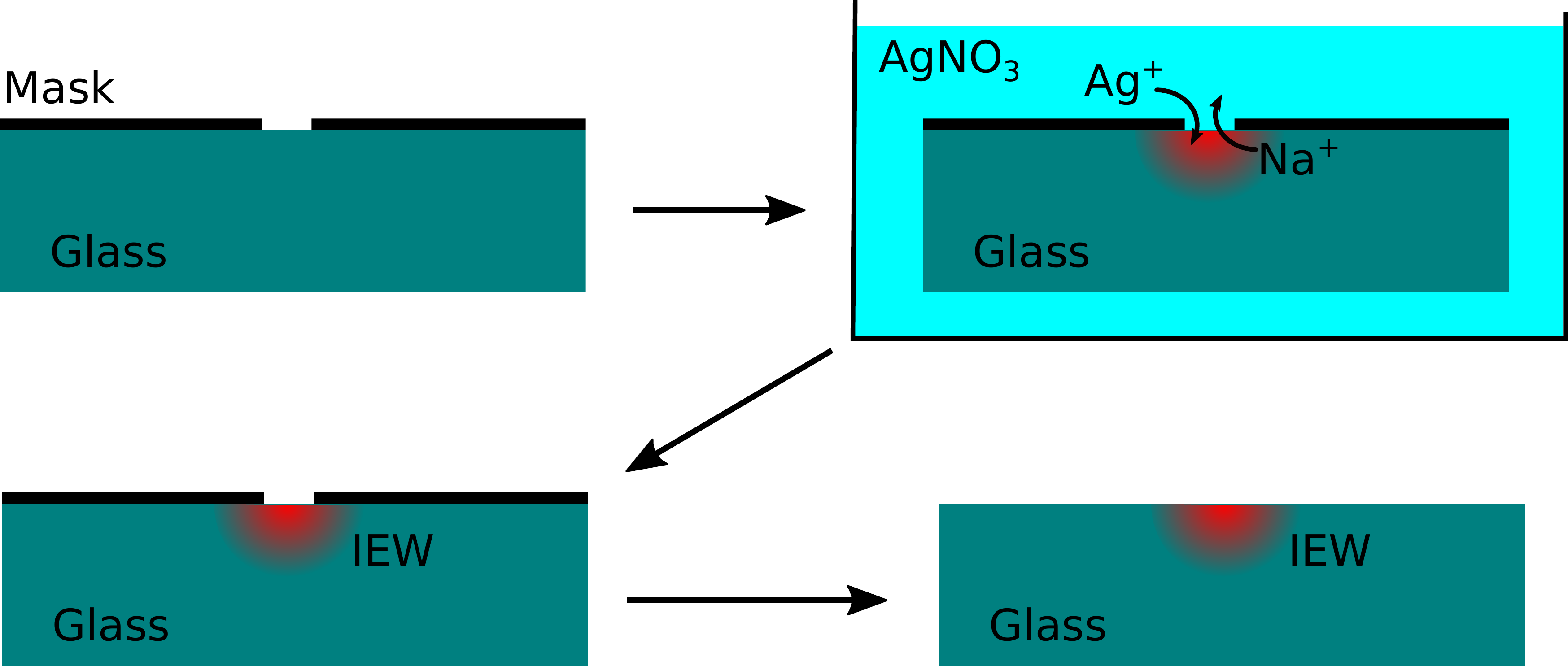}
	\caption{Fabrication procedure of \protect\IEW{}: first of all, a mask is
	created on the surface of a glass plate; then the glass is immersed in a solution
	containing \ch{Ag+} ions, as for example a solution of \ch{AgNO3}. Some of the \ch{Ag+}
	ions replace some of the \ch{Na+} of the glass, modifying its refraction
	index. Eventually the support is extracted and the mask is
	removed.}
	\label{fig:iew}
\end{figure}

The ion-exchange glass waveguide \iEW{} is an alternative platform that can overcome some fragility problem
of nanofibers, while keeping most of their advantages.
In order to produce an \iEW{}, an appropriate amount of \ch{Ag+} ions are diffused in glass, creating
a gradient in the  refracting index that can weakly guide the
light~\cite{tervonen2011Ionexchanged}.
A representation of this kind of waveguide is reported in
figure~\ref{fig:iew}. By using simple thermal diffusion, the \ch{Ag+} ions
diffuse inside the glass: their concentration is smaller the farthest we go
from the glass surface towards the inner region of the guide, creating a gradient of concentration (and consequently of the refraction
index) that is indicated in red in the image. A mask is used to ensure the ions
diffuse only in the desired region; it can be a metallic mask (made of titanium
or aluminum), or, better~\cite{walker1983Integrated,weiss1995Determination}, a
dielectric mask
(aluminum oxide or silicon
oxide). The guide can then
be left in place (this was the case for the guides I used) or can be moved deeper
in the glass by applying an electrical field. These guides are produced by the company TeemPhotonics in Meylan as part of an on-going collaboration between the L2n-UTT and TeemPhotonics for many years now.
By using the near field interaction, as explained for the nanofibers in the same way, it is possible to couple light from emitters deposited over the waveguide directly
inside it.
However, as opposed to the nanofiber, the near field of an \IEW{} is weak and
the guided mode is mostly contained inside the guide. As a solution to this
problem, it has been proposed to use a \ch{TiO2} layer in order to ``pull'' the mode
outside the waveguide and use it to couple the
emitters~\cite{beltranmadrigal2016Hybrid}. Simulations and experiments
conducted at the L2n-UTT by Josslyn Beltram-Madrigal showed that this mechanism works well~\cite{beltranmadrigal2016Hybrid}.

\subsection{Deterministic positioning of emitters on top of a waveguide}
The deposition procedure described for the nanofibers is not applicable to the waveguides, as the support is now two-dimensional and an emitter could get stuck far from the guide: we need a different way to deposit the emitters on the
waveguides. The most simple one is to spin-coat a given amount of solution
containing the emitters, hoping that some of them will be in the right place
over the guide; however this method is completely non-deterministic and it is completely based on randomness. For this reason, it is not adequate for more complex experiments, like the ones involving creating plasmonic antennas over the guide.
A more deterministic approach can consist in positioning the emitter at the right place by using
an AFM tip: this approach allows a fine control of the position of the
nanoemitter~\cite{schell2011Scanning} but requires time and is difficult to imagine a possible integration for industry processes.
An alternative promising approach to address this problem is the one we proposed in the work
with~\textcite{lio2019Integration}: the idea is to use a polymer to encapsulate the emitters
at the desired place. Photopolymerization is a process in which a polymerization
reaction is induced by light: the region of the liquid irradiated by light reacts
and changes state becoming solid, while the non-illuminated regions remain in the liquid state and
can be washed out by using the appropriate solvent. If the emitters are dispersed in the polymer, there is a significant probability that one (or more) of them will
be trapped in the solidified part, while all the others will be eliminated during the washing procedure.
As any light-induced process, usually the smallest size reachable is
diffraction limited. However different techniques can be used to overcome this
limit, as for example by using two
photon
polymerization techniques\cite{maruo1997Threedimensional,cumpston1999Twophoton}.
One of the most promising techniques that was
recently
investigated is polymerization by evanescent waves~\PEW{}\cite{ecoffet1998Photopolymerization,duocastella2017Improving}.

This technique consists in using the near field of the guided mode of the waveguide to induce  the polymerization of the photo-resin. This method allows obtaining  a layer of the polymer with a thickness of few nanometers, trapping in the contained quantum dots/emitters on the waveguide
surface. We performed the experiment in two different configurations showed in
figure~\ref{fig:waveguideLio}:
\begin{figure}[tb]
	\centering
	\subfloat[]{\label{fig:waveguideLioA}\includegraphics[width=0.4\linewidth]
		{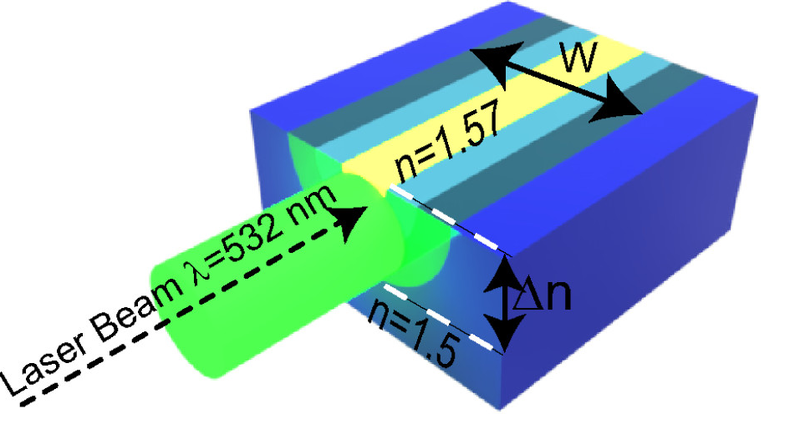}}\qquad\qquad
	\subfloat[]{\label{fig:waveguideLioB}\includegraphics[width=0.4\linewidth]
		{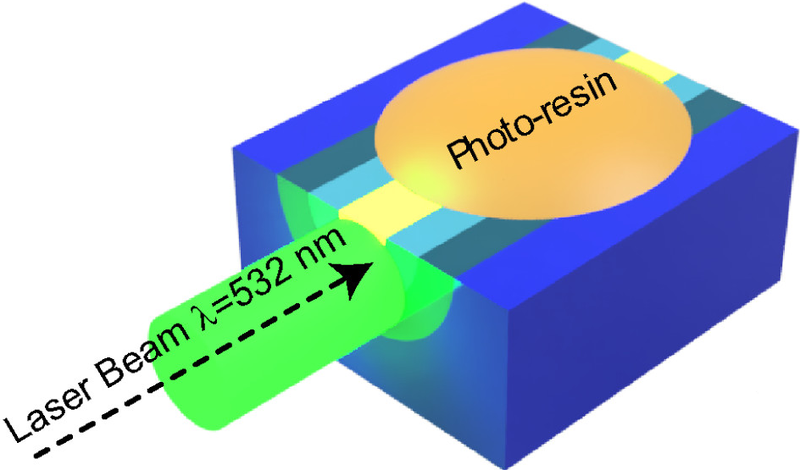}}\\
	\subfloat[]{\label{fig:waveguideLioC}\includegraphics[width=0.4\linewidth]
		{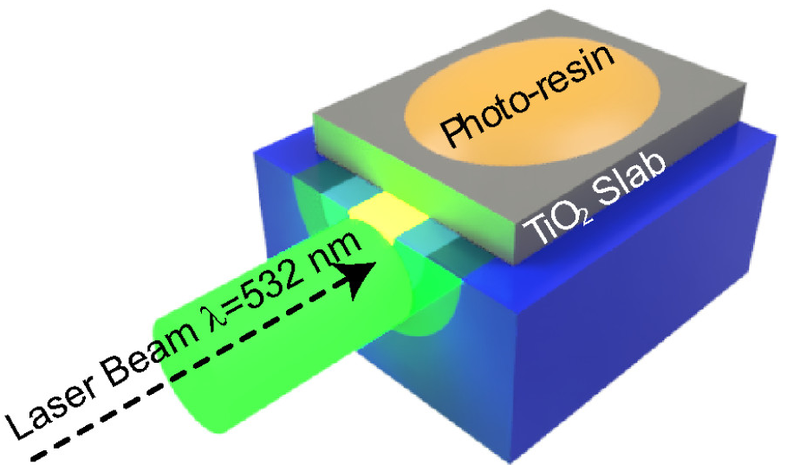}}
	\caption{
		\protect\subref{fig:waveguideLioA}~The \IEW{} studied, the
		refraction index changes from \num{1.5} in the blue region (glass
		substrate) to \num{1.57} in the inner zone of the guide, represented
		in yellow. The green cylinder represents the curing laser used to
		obtain the photopolymerization reaction. In the first configuration
		considered \protect\subref{fig:waveguideLioB} the photo-resin is
		directly placed over the \IEW{}. In the other,
		\protect\subref{fig:waveguideLioC} the waveguide
		is firstly covered by a \ch{TiO2} layer in order to create a double
		dielectric waveguide and increase the near field coupling. The resin is
		placed over the \ch{TiO2} layer. \ccredit{lio2019Integration}}
	\label{fig:waveguideLio}
\end{figure}
in one case we deposit the photo-resin directly over the waveguide, while in the other case, we deposit the photo-resin over a \ch{TiO2} layer present on top of the guide. As explained before, the \ch{TiO2} layer has the effect of creating a double
dielectric waveguide and increases the intensity of the near field compare to having the \IEW{} alone.

In the experiment, we used two different resins: one containing  
\ch{CdSe}/\ch{
ZnS} nanocrystals
while the other not. The details on the composition are reported in
table~\ref{tab:resins}.
\begin{table}[tb]
\begin{tabular}[]{|l|l|}
	\hline
	\textbf{resin 1} & \textbf{resin 2}\\
	\hline
	pentaerythritol triacrylate (PTEA)& pentaerythritol triacrylate\\
	\SI{4}{\percent} methyl diethanolamine (MDEA) & \SI{4}{\percent} methyl
	diethanolamine\\
	\SI{0.5}{\percent} eosin Y & \SI{0.5}{\percent} eosin Y\\
	& \SI{1}{\percent} \ch{CdSe}/\ch{ZnS} nanocrystals\\
	\hline
\end{tabular}
\caption{Composition of photoresins used in the experiment: both of them are
based on pentaerythritol triacrylate (PETA) with the addition of other
components (the percentage indicates the molar concentration). The
second resin contained the nanocrystals.}
\label{tab:resins}
\end{table}
 
\subsubsection{Single waveguide dose characterization}
The first interesting experiment to perform to develop this technique is to characterize
the minimal dose needed to polymerize the resin.
\begin{figure}[p]
	\includegraphics[width=0.6\linewidth]{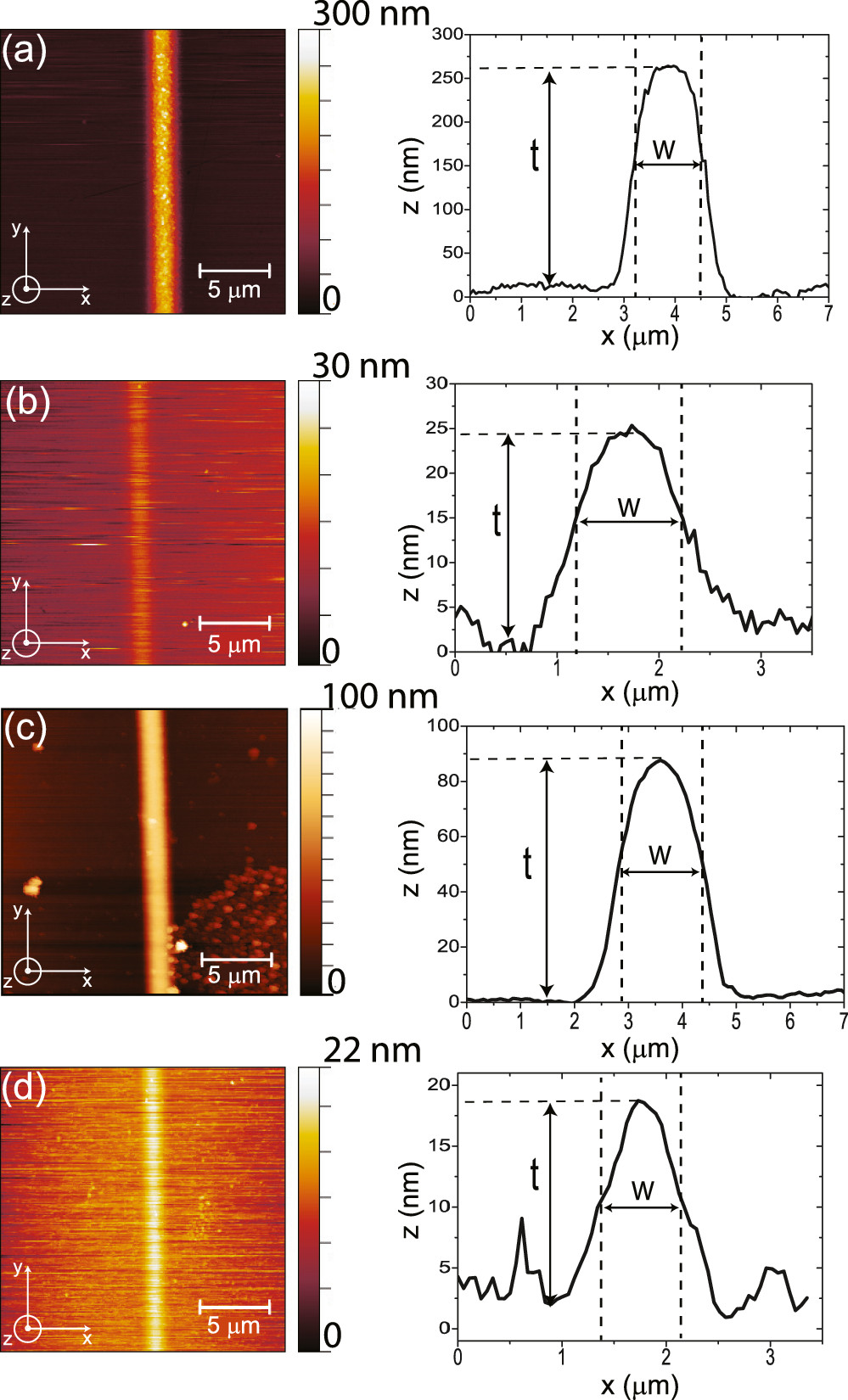}
	\caption{On the left hand-side, AFM images of photopolymerized resins. On the right hand-side the profile of the polymer section. (a) and (b) are obtained
		using the resin without the nanocrystals, respectively with a power of
		\SI{0.56}{\milli \watt} and an exposure
		time $t_e=\SI{300}{\second}$ and $t_e=\SI{60}{\second}$; (c) and (d)
		are	obtained using the resin with the nanocrystals, respectively with a
		power of \SI{0.70}{\milli \watt} and an exposure time
		$t_e=\SI{300}{\second}$	and	$t_e=\SI{60}{\second}$. In presence of
		nanocrystals the power is higher due to the greater absorption of the
		polymer. \ccredit{lio2019Integration}}
	\label{fig:lioafm}
\end{figure}
A standard time of \SI{60}{\second} was used to cure the resin with
different laser powers.
The characterization of the thickness of the polymer layer deposited over the waveguide was
performed using an atomic force microscope (AFM). It Was found that the minimal power to trigger the photopolymerization process was \SI{0.56}{\milli \watt} for the
resin without the quantum dots/nanocrystals and \SI{0.70}{\milli \watt} for the resin
containing the quantum dots. This difference can be explained by the
greater absorption of the mixed resin due to the quantum dots.
Once we defined the minimal threshold, it is interesting to have information on how
the thickness evolves by increasing the energy: this was studied using an exposure
time five times longer for each polymer. The AFM images and their section
profiles are shown in figure~\ref{fig:lioafm}.
The obtained thicknesses are as low as $(18\pm 2)~\si{\nano \meter}$ in case of
resin~1 and $(24\pm 2)~\si{\nano \meter}$ in case of resin~2. These values are
of
the same order of the diameter of the involved nanoemitters, which is about
\SI{12}{\nano \meter}.
This type of thin film containing nanoemitters (in this case nanocrystals) could be used in the future to obtain deterministic positioning of nanoemitters on plasmonic
nanostructures.

An interesting application of near-field two-photons induced polarization has been developed in our group by~\textcite{ge2020Hybrid} with the development of a single-photon switchable hybrid nano-emitter. The near-field is used to polymerize a  photo-resin containing quantum-dots near a gold nanocube, obtaining a single photon-emitter with an excitation efficiency strongly dependent of the polarization of the excitation light.

\subsubsection{Fluorescence emission spectra}
\begin{figure}[tb]
	\centering
	\includegraphics[width=0.6\linewidth]{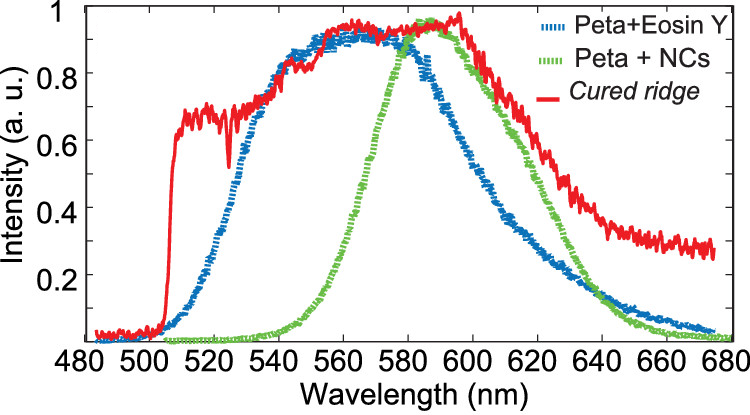}
	\caption{In red, emission spectrum of the fluorescence of the
	photopolymerized ridge on the \IEW{}. In blue, emission spectrum of the fluorescence of PETA~+~\SI{0.5}{\percent}~eosin~Y while in green, we have the fluorescence of a drop of PETA containing the nanocrystals used in the
	experiment. \ccredit{lio2019Integration}}
	\label{fig:liospectra}
\end{figure}

The emission spectra were measured with an home-made confocal microscope at the L2n-UTT,
similar to the one used for the measurements of the perovskite optical properties. The emitters were excited  with a
blue laser ($\lambda=\SI{405}{\nano \meter}$) which was removed from the detected
light with a long-pass filter with a cutoff wavelength of \SI{500}{\nano
\meter}.
The result is shown in figure~\ref{fig:liospectra} where
the blue and the green curves represent respectively the emission of the two
drops of PETA~+~\SI{0.5}{\percent}~eosin~Y and PETA~+NCs. The red one is the
emission of the cured polymer. It is possible to see that the red curve is the composition of the two others, as expected. The first part of the red curve,
that has not a correspondence in the other two, is due to Methyl diethanolamine (MDEA)
emission, the central part is due to the PETA~+~\SI{0.5}{\percent}~eosin~Y
component while the part at higher wavelengths is due to the nanocrystals.

\subsubsection{Double waveguiding behavior}
As already mentioned before, adding a \ch{TiO2} layer over the \IEW{} has the
effect to create a double dielectric waveguide: the light is now guided by both waveguides. A thickness of~\SI{85}{\nano \meter} was chosen for the \ch{TiO2} layer as it allows an optimal working range for the visible light (from
\SIrange{400}{800}{\nano \meter}) where the guide is a single mode one in this range.

\begin{figure}[tb]
	\centering
	\includegraphics[width=0.5\linewidth]{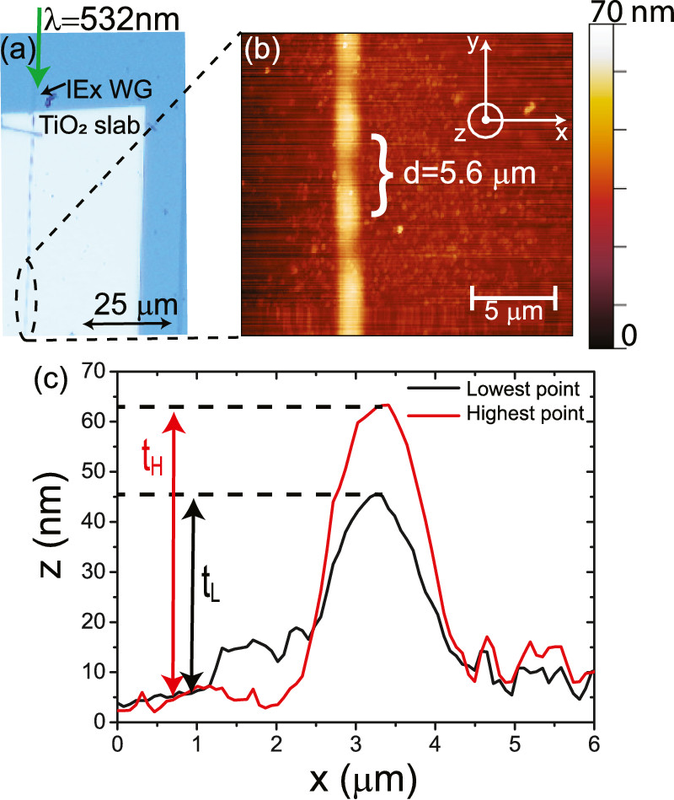}
	\caption{(a) Image of the sample, where the \ch{TiO2} slab is present,
	(b) obtained atomic force microscope image, showing a modulation of the
	thickness of the resin with a periodicity of \SI{5.6}{\micro \meter}; (c)
	profile of the fabricated ridge; the red one is taken where the thickness of the ridge is maximal, while the dark one was taken were the thickness of the ridge is
	minimal. \ccredit{lio2019Integration}}
	\label{fig:liodoublewaveguide}
\end{figure}

The procedure to photopolymerize the resin is similar to what was described before,
with the only difference that, in this case, it is not spin-coated directly on the glass
but on the \ch{TiO2} layer. Only resin~2 (the one with nanocrystals) was used
for this experiment and  we chose to use an exposure time
$t_e=\SI{300}{\second}$ and an input power of \SI{0.7}{\milli \watt} (the same
values used in the case of a single waveguide). The result is shown in
figure~\ref{fig:liodoublewaveguide}.

What is new here, is that the polymer presents a periodic modulation clearly visible
in the AFM measurements (figure~\ref{fig:liodoublewaveguide}b). We can explain
this modulation as an effect of a modulation of the near field intensity during the curing
of the resin: this modulation is due to a periodical beating between the two waveguides. The beats is present between the \IEW{} and the \ch{TiO2} layer.

This beating was observed also in previous experiments, performed in our lab by~\textcite{beltranmadrigal2016Hybrid}, in which a
SNOM~analysis on a double waveguide was performed with the goal of measuring the evanescent field present during the light propagation in the double
waveguide. The results of the simulations with
a finite element method validated this
hypothesis~\cite{lio2019Integration}.

\clearpage
\begin{chrecap}
	In this chapter I presented an insight of possible options and different approaches that can improve
	in the near future our system.
	From the emitters side:
	\begin{itemize}
		\item I explained the potential improvements that are possible with
		perovskite nanocrystals;
		\item I showed the potential improvements that the study of defects in
		nanodiamonds can offer in general;
		\item I detailed the main properties and advantages of the two kinds
		of single defects in nanodiamonds I studied during my PhD.
	\end{itemize}
	From the photonic platform side, I presented a promising alternative approach, the \IEW{} platform.
	\begin{itemize}
		\item The \IEW{} has the potential to be more stable with respect to the
		nanofiber, as it is more robust and less affected by dust,
		\item I detailed the methods of nanoemitters deposition and differences with respect to the nanofiber case.
		\item I detailed the results of an original study in which I was involved in order to use photo-polymerization as a strategy of deposition of
		quantum dots over a waveguide.
	\end{itemize}

\end{chrecap}
\chapter*{Conclusions}
\addcontentsline{toc}{chapter}{Conclusions}
\markboth{Conclusions}{Conclusions}

In this manuscript I presented the results obtained during my PhD thesis as
well as the theory required to understand them.

Regarding perovskite nanocrystals, I explained their properties and their main
limitation, that is the stability under laser excitation. This limitation, due to the degradation under moisture and light, can be reduced with
different approaches:
\begin{itemize}
	\item It is possible to protect the emitters using a polymer. This
	technique allows an improvement on the stability: our measurements show
	that, using a polymer, the degradation happens in a time-scale four time
	larger. However, this approach is not suitable for depositing emitters on the
	nanofiber, which was my final goal during my thesis.
	\item The preparation of the emitters plays an important role for their
	stability: with a slight different preparation process we were able to
	obtain emitters whose stability can be measured in hours instead than in
	minutes.
\end{itemize}
I have shown, in addition, the role of the dilution in the stability. This is
important for applications that need a diluted solution, but is also an indication
of the role of ligands in the stability. This can opens new research paths to
increase the stability of these perovskites.

The performed measurements showed that our perovskites are not only more
stable from an emission point of view, but also the spectra stability is
increased, with a drift of the spectrum less than~\SI{20}{nm} in more than
two hours.

The increased stability allowed to study the blinking properties of our
emitters: in particular the analysis of the fluorescence-lifetime intensity
distributions had shown that our emitters stay in an excited state for the most
of the time.

Thanks to the improvement on fabrication it was possible to couple single
photons emitted by the nanocrystals with a tapered nanofiber. Tapered nanofiber
are obtained by pulling a standard optical fiber over a pure flame of hydrogen
and oxygen: in this way it is possible to adiabatically change the the taper profile in
order to couple almost all the light coming from fiber in the nanofiber and vice-versa, by obtaining a nearly unitary transmission that can be over~\SI{98}{\percent}. In the nanofiber part,
a strong near-field is present around the fiber: this allows to couple the
light from a single photon emitter deposited on the nanofiber directly inside it.
This was demonstrated by performing an antibunching measurement on the photons collected at the output of the
nanofiber. Such measurement showed a $\gdz=0.24$, proving the presence of single
photons in the nanofiber.

In addition, I also had the possibility to work on a different platform, that
can in future be used instead of nanofibers: the ion-exchange glass waveguide
platform. In particular we have shown a technique for nanocrystal deposition on
waveguides. We dispersed the emitters in a photo-resin, and we used a two
photon polymerization technique in order to deposit on the glass waveguide a few nanometers thick ridge of resin containing the emitters. %We were able to prove that
%the ridge is still containing working emitters.
With future developments this technique, combined with plasmonic nanoantennas,
could bring to deterministic positioning of colloidal quantum dots on the
waveguide.

The work accomplished in this thesis gave origin to three
publications on peer-review journals and an invited oral presentation at the ``Smart Nanomaterials, Advanced Innovations and Applications'' international conference in Paris.
\appendix
\chapter{Analysis software for TTTR3 files}
\label{chap:program}
\minitoc
\section{Introduction}
During the  time-tagged time resolved measurement (TTTR) used for the
characterization of single photon emitters a large amount of data is produced.
Indeed, any detected photon arrival is recorded with its arrival time and the
channel where the event was detected. If we consider a standard \SI{3E4}{cps},
we have more than \SI{5E7}{counts} in half-an-hour of measurement. For this
reasons data are stored in a binary format, that allow to save them in a file
size smaller than the one allowed by a plain-text file.
PicoQuant decided to give to this file the extension \verb|.ptu| and to release
demo codes in different programming languages in order to allow to access it.
Among them I chose \CC, as it presents two main advantages.
\begin{itemize}
	\item It is a stable open source language, which ensures the
	availability of the written code for the lab and guarantees it will be
	usable in the future without the need of license.
	\item It is a compiled language which is faster to run on a computer.
\end{itemize}
This was a judicious choice as the computation of the \gd{} is
resource and time consuming.

In addition, \CC{} has a wide variety of tools to create stable graphic
interfaces in order to make the program user friendly.

%\section{The algorithm}
The first version of the program was a command line code generating
the \gd{} graph and writing the values in a file format readable with gnuplot. This approach was working, but in the long term the
increased needs of analysis, and thus of options, made it not
practical and its use quite complex.
For this reason I wrote a second version of the program,
using \CC alongside with Qt framework that allows the creation of a graphical
interface. Here I will describe only the actual version of the software. However, in order to understand some choices, it can be useful to take in account that it has been written firstly for a command line use and
then it has been adapted to the Qt framework to add a graphical user
interface.

I this appendix I will describe the usage of the software, without entering in
the details of the underlying code, then I will describe the tools I used to write it
and the algorithms on which it is based.

\section{The software usage}
\begin{figure}[tb]
	\centering
	\includegraphics[width=1\linewidth]{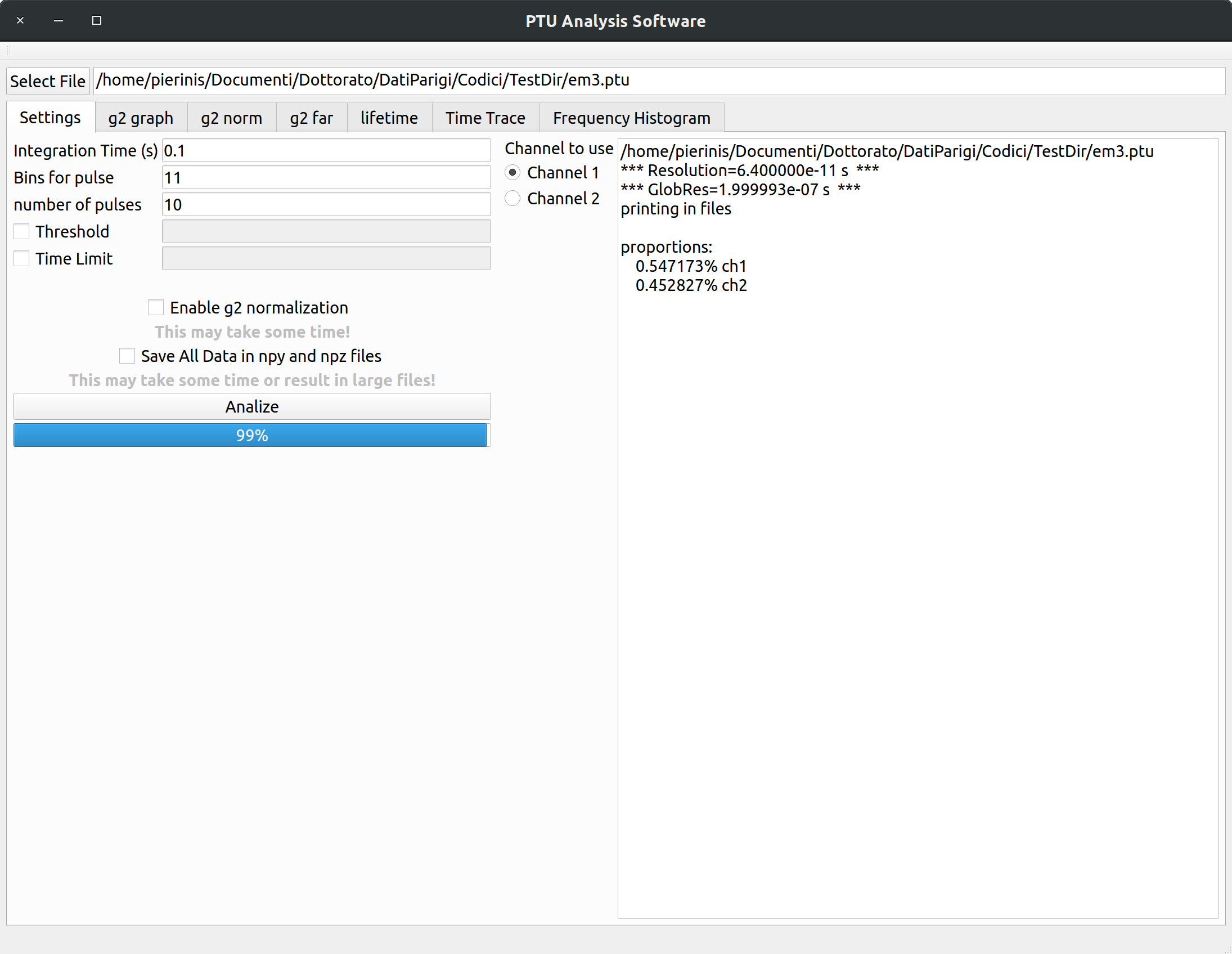}
	\caption{PTU Analysis Software: the main window that is opened when the
	program is started.}
	\label{fig:Amainscreen}
\end{figure}
When the software is launched, appears the window reported in
figure~\ref{fig:Amainscreen}.
In the first line is possible to paste the path to the interested file,
or alternatively to select the file using the \textit{Select File} button. The
name of the output files is automatically chosen by the software using the
original name and appending a specific suffix for each file, depending on
the content. This line is visible in any page of the application.
Multiple panels are available under it, the first one (\textit{Settings}) is
the one used to set the analysis parameters and perform the analysis while the
others are used to display the analyzed data. In the following I will explain
in detail these panels.
\subsection{\textit{Settings} panel}
\begin{figure}
	\centering
	\includegraphics[width=0.6\linewidth]{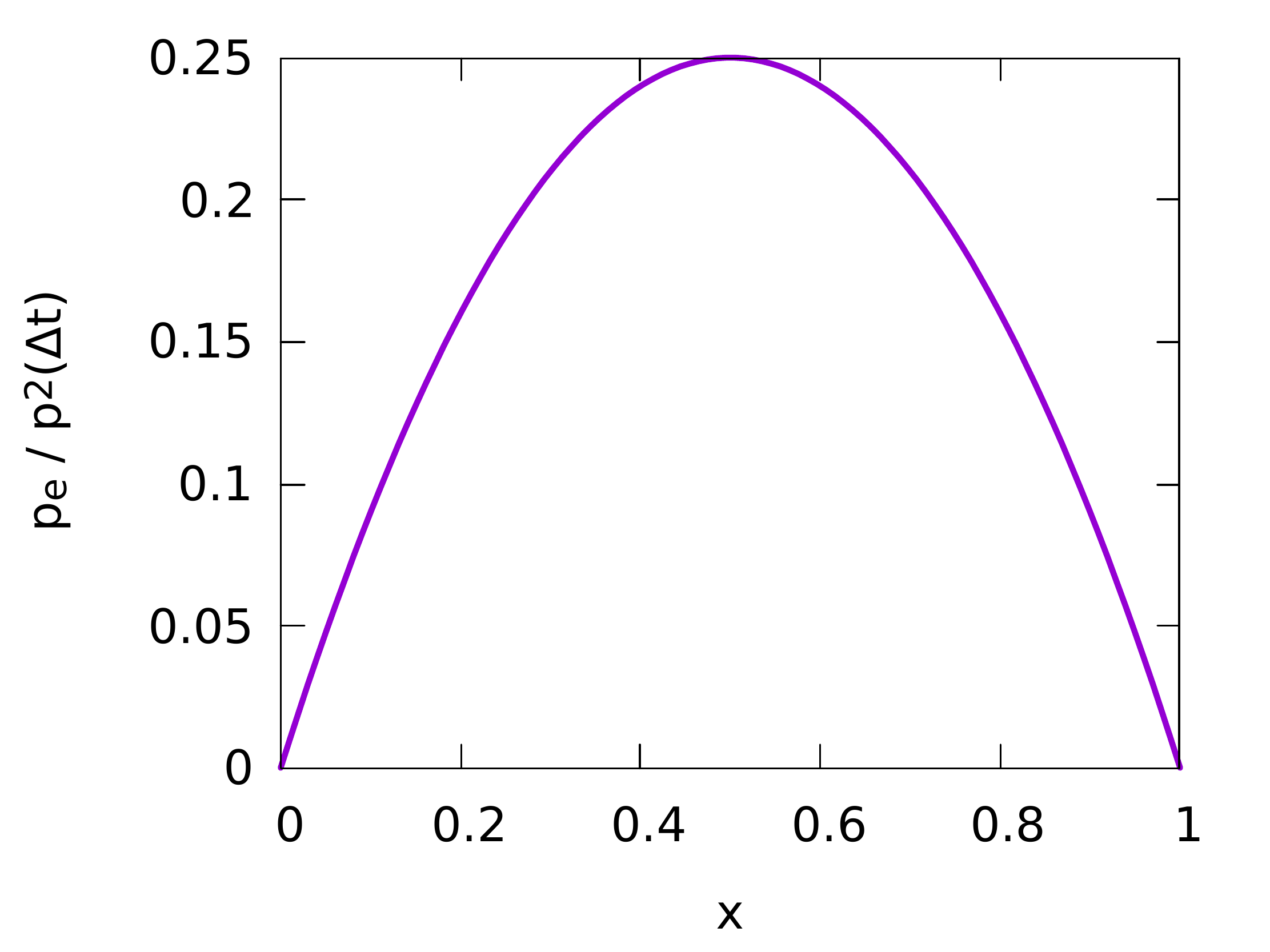}
	\caption{Graphical representation of the parabola in
	equation~\protect\eqref{fig:graphperchch}. The maximum is located in
	$x=0.5$.}
	\label{fig:graphperchch}
\end{figure}

Multiple settings are available on the left of the panel to correctly perform
the Analysis (see figure~\ref{fig:Amainscreen}). In the
following their behavior and function are explained one by one.
\begin{itemize}
\item The \textit{Integration Time (s)} line, contains, in seconds, the time
used to
create the time trace histogram and to possibly define the thresholds.
\item The \textit{Bin for Pulse} line contains the number of bins for each
laser impulsion used in the \gd{} histogram. In other words, if the laser sends
a pulse every \SI{200}{\nano \second} and we set $11$ bins for pulse, the bin
size will be $\SI{200}{\nano \second}/11$. In this way, each laser impulsion
contains exactly a entire number of bins, which reduces the artifacts in the
histogram. In my experience, an odd number of bin per pulse is better than an
even one, as in this case the maximum falls in the middle of the bin and not on the border of it.
\item The \textit{Threshold}, when set, introduces a threshold that the signal has to
reach in order to
consider the emitter ``ON''. In practice the time trace is traced, and only the
photons arrived in bins that contains more than a certain number of counts
are considered. This can be useful, if the emitter blinks between a bright
and a gray state. Indeed it allows to analyse separately the bright state excluding the gray
one.
\item The \textit{Time Limit} field, when set, introduce a time limit on the
analysed signal. This is useful if we realize that the emitter died after a
certain time and we know that after that time only noise has been recorded.
\item Option \textit{Enable \gd{} normalization} is useful to normalize the
\gd{} histogram as explained in section~\ref{sec:g2_measurement}.
of chapter~\ref{chap:perovskites}.
\item The \textit{Channel to use} settings ask which channel the user wants to
use to calculate the TimeTrace and the lifetime histograms. The choice is between
one of the two channels.
\end{itemize}
On the right of the panel, a non-editable text panel is available: here the
processing details and some information on the analysis are written during
the process. This is useful to know the progress status of the analysis. In addition, the last two lines printed on the right panel tell us which proportion of photons is
arrived in each channel: this is very useful, as to efficiently measure the
\gd{} this must be the nearest possible to the \SI{50}{\percent} in each
channel. This can be seen with a simpler calculation: first of all, let
$p(\Delta t)$ be the probability of getting a photon in a time interval of
$\Delta t$, corresponding the the size of the bin in the \gd{} histogram. Now,
the probability to obtain it in the first channel will be $p_1(\Delta t)$ and
the probability to obtain it in the second channel will be $p_2(\Delta t)$, and
suppose it to be constant in any time interval with the same width. Now, we can
say that the following equation is valid (for definition of $p(\Delta
t)$):
\begin{equation}
	p(\Delta t)= p_1(\Delta t)+p_2(\Delta t).
\end{equation}
If now we define $x$ as $p_1(\Delta t)/p(\Delta t)$ we can write:
\begin{equation}
	\label{eq:A2}
	p_1(\Delta t)=x p(\Delta t) \qquad \textrm{and} \qquad p_2(\Delta t)=x
	p(\Delta t).
%	= p_1(\Delta t)+p_2(\Delta t)
\end{equation}
The probability to register a coincidence event in the \gd{} graph $p_e$ will
be given by to the product of $p_1$ and $p_2$ as it happens only when a
photon is detected in each of the two channels. Thus we write:
\begin{equation}
	\begin{split}
		p_e &= p_1(\Delta t) p_2(\Delta t)\\
		p_e	&= x (1-x) p^2(\Delta t)
	\end{split}
\end{equation}
the last equation is the equation of a parabola, with a maximum on $x=0.5$. Its
graph is reported in figure~\ref{fig:graphperchch}.
Is interesting to note that when $0.4\leq x \leq 0.6$, $p_e(x)\geq 0.96
p_e(0.5)$: in other words we can tolerate a slightly non-homogeneous
distribution of the counts in the two channels conserving more than the
\SI{95}{\percent} of counts.

\subsection{Visualization panels}
\begin{figure}[p]
	\centering
	\subfloat[]{\label{fig:A-g2graph}\includegraphics[width=\linewidth]
		{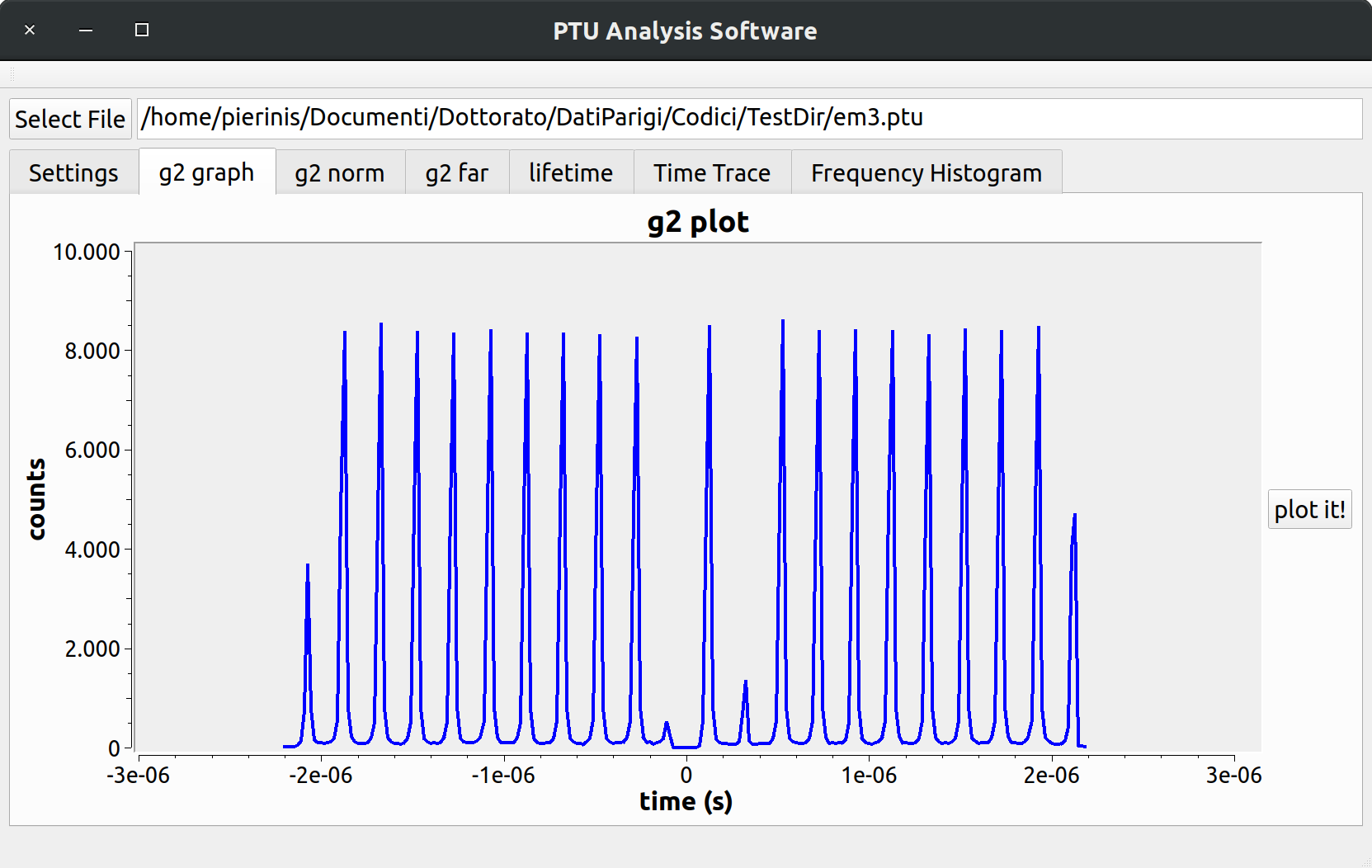}}\\
	\subfloat[]{\label{fig:A-g2norm}\includegraphics[width=\linewidth]
		{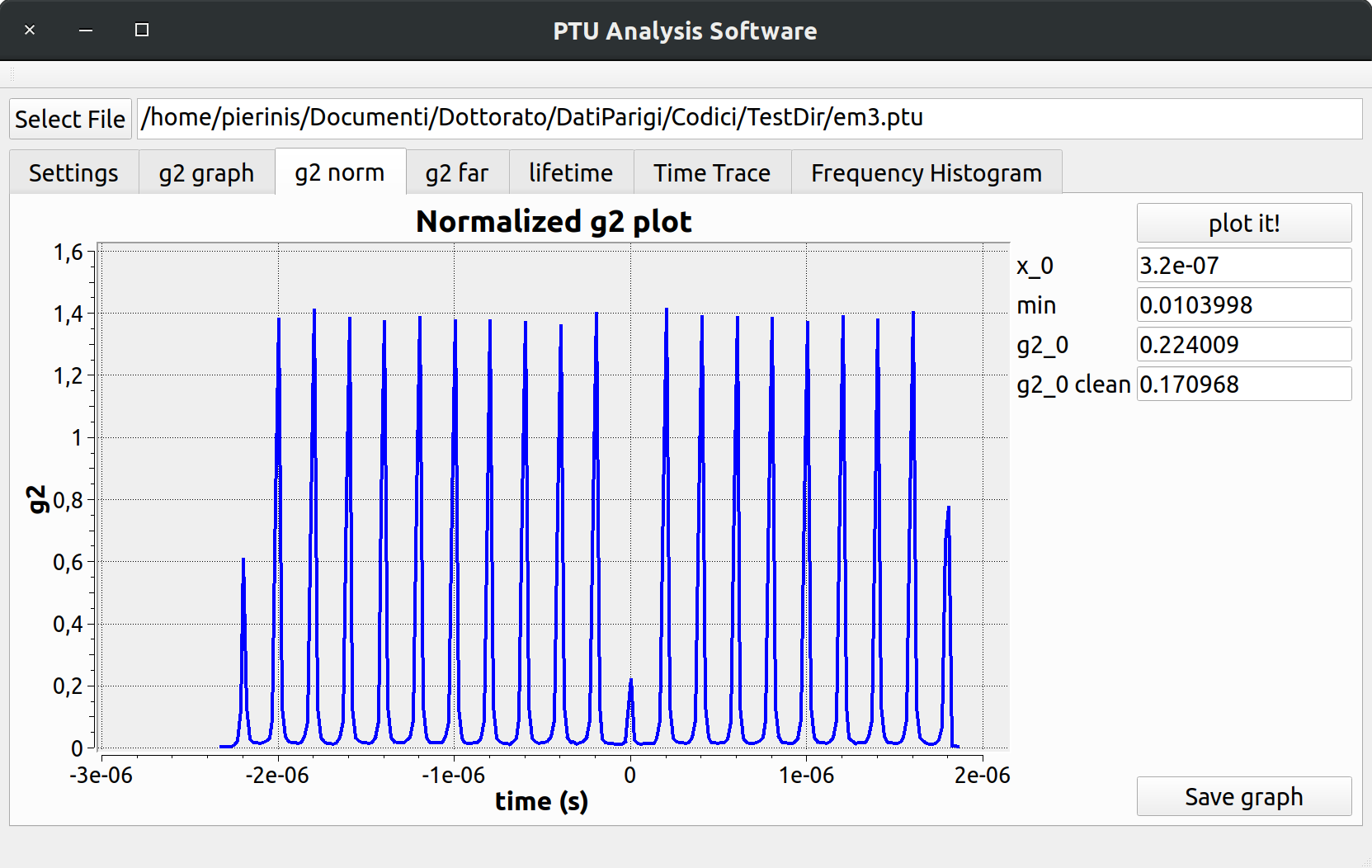}}
	\caption{Example screenshot of \textit{\gd{}
	graph}~\protect\ref{fig:A-g2graph} and \textit{\gd{}
	norm}~\protect\ref{fig:A-g2norm} panels.}
	\label{fig:A-g2pan}
\end{figure}

Three different panels are available to visualize the \gd{} histogram once the
file has been analyzed. The first one (figure~\ref{fig:A-g2graph}) reports the
raw \gd{} graph as it is measured by the instrument. Here is clearly visible,
around $t=\SI{0}{s}$, the effect of the dead time of the instrument described
in chapter~\ref{chap:perovskites}. In this graph the time axis refers to the
delay in arrival of the signal to the instrument; we are instead interested in
the delay in the arrival of the photons in the APDs. These is calculated in the
following panel, the \textit{\gd{} norm} panel, reported in
figure~\ref{fig:A-g2norm}.
\begin{figure}[p]
	\subfloat[]{\label{fig:A-g2far}\includegraphics[width=\linewidth]
		{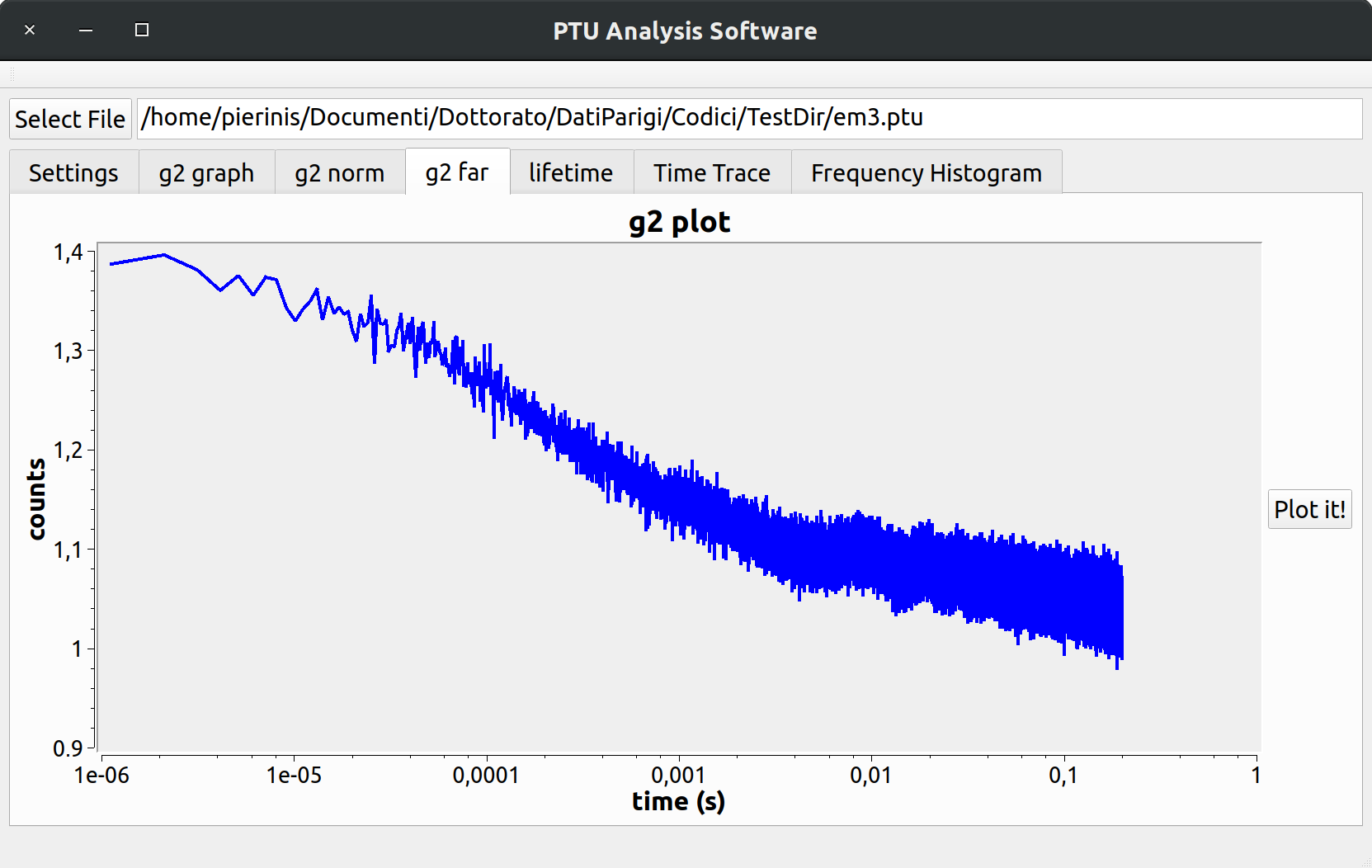}}\\
	\subfloat[]{\label{fig:A-lifetime}\includegraphics[width=\linewidth]
		{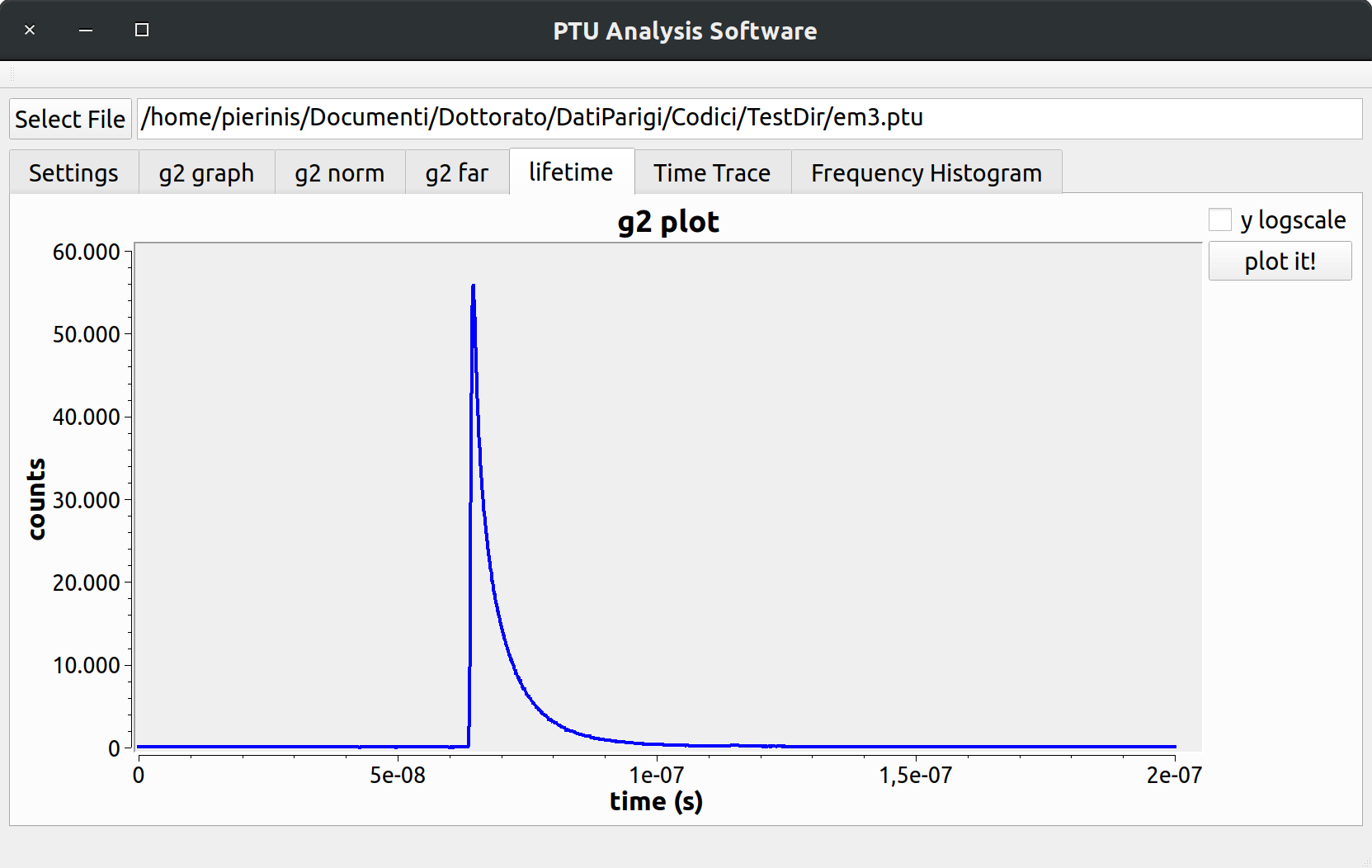}}
		\caption{Example screenshot of \textit{\gd{}
		far}~\protect\ref{fig:A-g2far} and
		\textit{lifetime}~\protect\ref{fig:A-lifetime} panels.}
	\label{fig:A-g2farandlife}
\end{figure}
\begin{figure}[p]
	\subfloat[]{\label{fig:A-TimeTrace}\includegraphics[width=\linewidth]
		{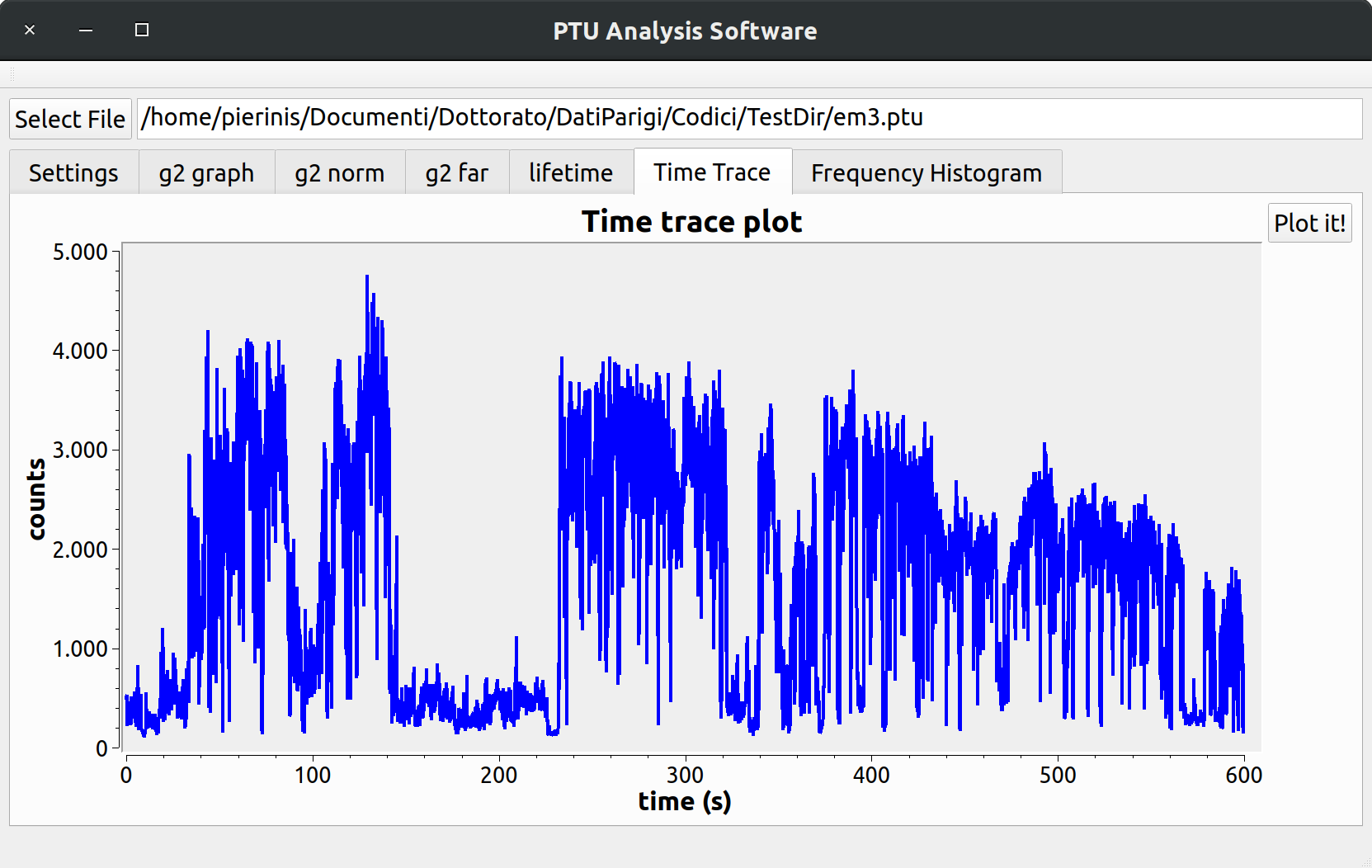}}\\
	\subfloat[]{\label{fig:A-Freq}\includegraphics[width=\linewidth]
		{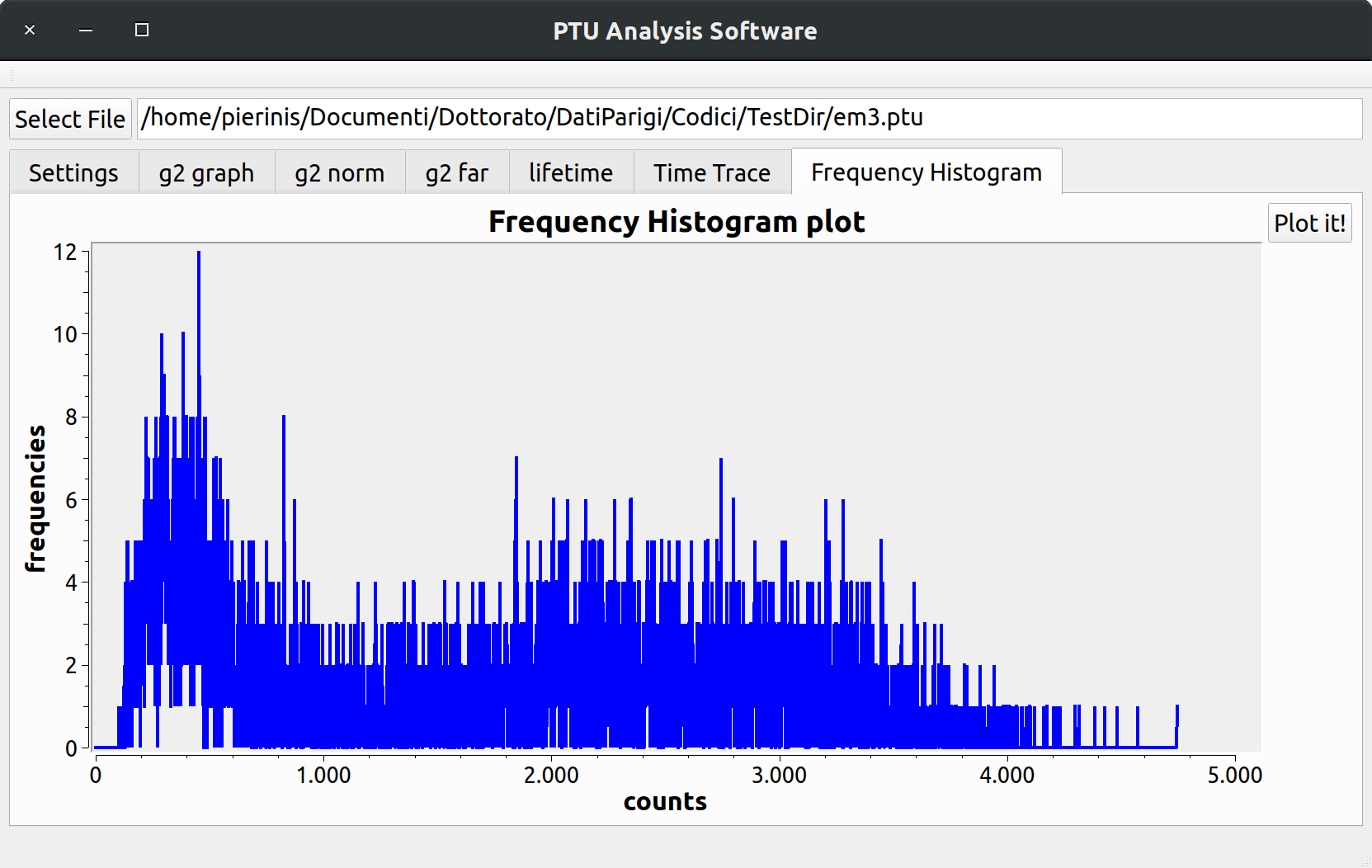}}
		\caption{Example screenshot of \textit{Time
		Trace}~\protect\ref{fig:A-TimeTrace} and
	\textit{Frequency Histogram}~\protect\ref{fig:A-Freq} panels.}
	\label{fig:A-TTandFreq}
\end{figure}
The first thing we note here is the presence of multiple text field. The first
one, marked with \textit{x\_0}, is editable: here it is important to insert the
length of the delay line, while the others are calculated by the program and
cannot be modified.
Once \textit{x\_0} value set, if the analysis has been performed with the
\textit{Enable \gd{} normalization} option enabled, it is possible to click on
the \textit{plot it!} button to visualize the plot. This time, the zero delay
is what we are looking for, while the dead time has been artificially removed
to make the graph more readable. It it also important to note that the height
of the last peaks is lower due to the fact that the last peak falls on the border of the chosen window: this is
not a physical effect, and they not need to be considered in the analysis.

The panel \textit{\gd{} far} is shown in figure~\ref{fig:A-g2far} and reports
the maximal values of the \gd{} histogram at long delays. The graph is created
only when the analysis is performed with the \textit{Enable \gd{}
normalization}. These is useful in case of blinking studies.

In panel \textit{lifetime} reported in figure~\ref{fig:A-lifetime} is possible
to plot the lifetime histogram on the selected channel. It is often useful to plot
it in semi-log scale. For these purpose, the \textit{y logscale} option can
be selected.

In panel \textit{Time Trace} reported in figure~\ref{fig:A-TimeTrace} is
possible to plot the intensity trace of the emitter along the measurement.

In panel \textit{Frequency Histogram} panel reported in figure~\ref{fig:A-Freq}
it is possible plot the frequencies histogram: it reports the occurrence of a
certain value in the intensity trace. On the $y$ axis the number of occurrences
is reported (labeled as \textit{frequencies}) in the application, while on the $x$
axis the number of counts is reported, corresponding to the $y$ axes of the
intensity trace plot. A file containing a matrix of a lifetime histogram for
each bins in the frequency histogram plot is also created. This can be used to
study the lifetime dependence of the blinking.\label{sec:A-lifeM}

Any of this plot is made reading the data previously analyzed with the
\textit{Settings} panel: the analysis saves the analyzed data in multiple files
that are read by the application when clicking on \textit{plot it} of each
panel. The data are saved in text format, allowing an easier opening with
different application to perform further analysis as for example the fits.

\subsection{Possible future improvements}
I spent time to make the software stable: now events like ``the file does not
exists'' when clicking on the \textit{plot it!} button does not make the
software crash, but raise an error.

I also added various functions, as the experience has shown that they are
important. For example, the progress bar showing the progress of the analysis
is useful to know that the program is not frozen but it is still running and to
estimate the time needed for the operation.

Anyway, several improvements can still be done. Some suggestions are listed below:
\begin{itemize}
	\item Adding a panel or an option to perform batch analysis could be useful
	as it could allow to leave the code performing the analysis of multiple
	files alone. This could be a great point as each analysis can take several minutes.
	Ideally this batch analysis could use multiple processes, in order to run
	efficiently on multi-core CPUs.
	\item An estimation of the ending time could be shown during the analysis.
	\item The ability to analyze the plotted data with the most common analysis
	could be added. Here the main problem is the difficulty to perform a fit
	using \CC. Indeed, there is a good and stable library that was written for
	the C programming language that could be used: the gnu scientific library.
	Unfortunately its usage is not straightforward (e.g. it requires to
	manually calculate and declare the Hessian matrix of the function that we
	want to fit) and requires time to be implemented. The advantage is that it
	is a low level library and, once implemented, the fitting process will be
	fast.
	\item The possibility to save the graphs in pdf format has
	been added only for the ``\gd{} far'' function, as it is the one that one
	would
	more likely want to save. Anyway, this could be added for any plot with
	minor effort.
	\item Finally, a cleaner saving of data and metadata could be realized. In the
	actual option any analysis of the same file overwrite the previous one:
	it could be useful to save the output files of each analysis in different
	directories and create a file with the history of the analysis that has
	been performed.
\end{itemize}
Of course there are other improvements that can be performed, but in my opinion
these are the most important ones.

\section{Tools used}
\subsection{\CCT{} programming language}
The C programming language is one of the most diffused languages. It was
created by Dennis Ritchie bewtween 1969 and 1973. In 1972 was born the
first Unix operating system entirely written in C: modern operating systems
derived from it are MacOS and GNU/Linux, the first one is used in personal
computers while the second one is largely diffused on servers and on mobile
phones.

The strength of the Unix operating system and its evolutions shows the power of
the C programming language, that is now the reference in most of the modern
hardware platform. As it is strongly related to Unix, it has an implementation
on almost any new operating systems.

The \CC{} programming language initially was an extension of the C programming
language in order to add classes and increase the level of abstraction of the language to make
the programming easier.  Its development was cured by Bjarne Stroustrup and
started in 1979. The advantage of the \CC{} is that it is object oriented,
differently from the C that is procedural: it allows a modern approach while
keeping most of the advantages of C. As for  the C language, there are open source compilers,
that ensure the code to be accessible free-of-charge also in the future. This
is the main advantage with respect to other languages, like Matlab, that strongly depends on
the license of the producer.

If we compare it with Python, another largely diffused programming language in science,
also based on C~libraries, \CC{} is probably more difficult to
write but is faster in the execution. Thus, it is convenient for applications
that manipulate large data files. Anyway, for
applications that do not require a huge effort from the computer side,
such as experimental routines, Python is often easier to write and to maintain.

\subsection{Qt}
``Qt is not a programming language on its own. It as a framework written in
C++''\cite{Qt}. Instead, Qt is a cross-platform framework to develop
applications in \CC. This means that writing a program in Qt it is possible
to deploy it on different operating systems (Linux, OS X,Windows) with minor
changes.

\begin{figure}
	\centering
	\includegraphics[width=\linewidth]{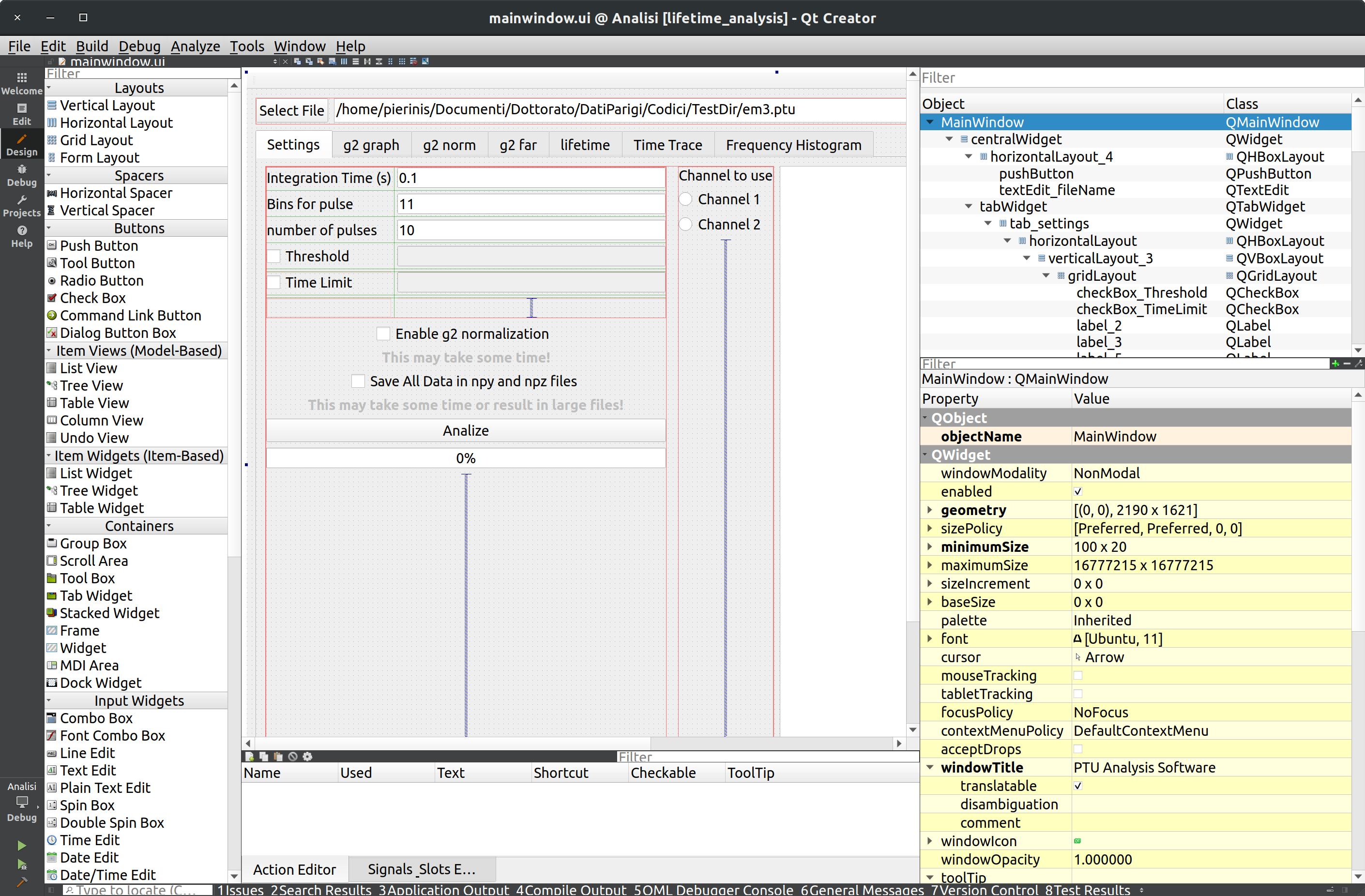}
	\caption{Qt~Creator window in the part that enable creation and editing of
	the graphical interface.}
	\label{fig:qtcreator}
\end{figure}

It is available with two licenses: one that is useful for companies that want
to sell their programs but also the GPL license, one of the most important
licenses for open source software. The last one is not adapted for producing
open source codes. It allows the creation of graphic user interfaces (GUI) in an
easy way, simply dragging and dropping the various elements and manually
designing it.
To show the graph I used the Qwt plugin to the Qt software, that makes this
quite easy.\cite{Qwt} In figure~\ref{fig:qtcreator} an image of the Qt~Creator
program that enables to create the graphical interface is shown.

\section{Algorithm and code details}
Complete code listings are not in the scope of this Appendix, and can be
requested to me if needed. However a description of the structure of the code could
be useful to future users.

The easiest way to start a new qt project is to use \textit{Qt Creator}
program, different kinds of blank projects can be selected: in my case I created
a desktop application, but mobile applications can also be created.
Any Qt project is divided in multiple files, knowing them is the first step to
get oriented in the code. I describe in the following the main file types
and then proceed with a description of the main points of the analysis
algorithm.
\subsection{Principal files of the code}
\begin{itemize}
	\item One file has a \verb|.pro| extension. It is the \textbf{project}
	file that
	configures the code and contains the list of the other files of the project, the
	list of the libraries and where to find them, the version of \CC{} used (in
	my case \CC{}11). It is usually created by \textit{Qt Creator} but it needs
	some time to be manually edited. A screenshot of this file is reported in
	figure~\ref{fig:filepro};
\begin{figure}[tbp]
	\centering
	\includegraphics[width=1\linewidth]{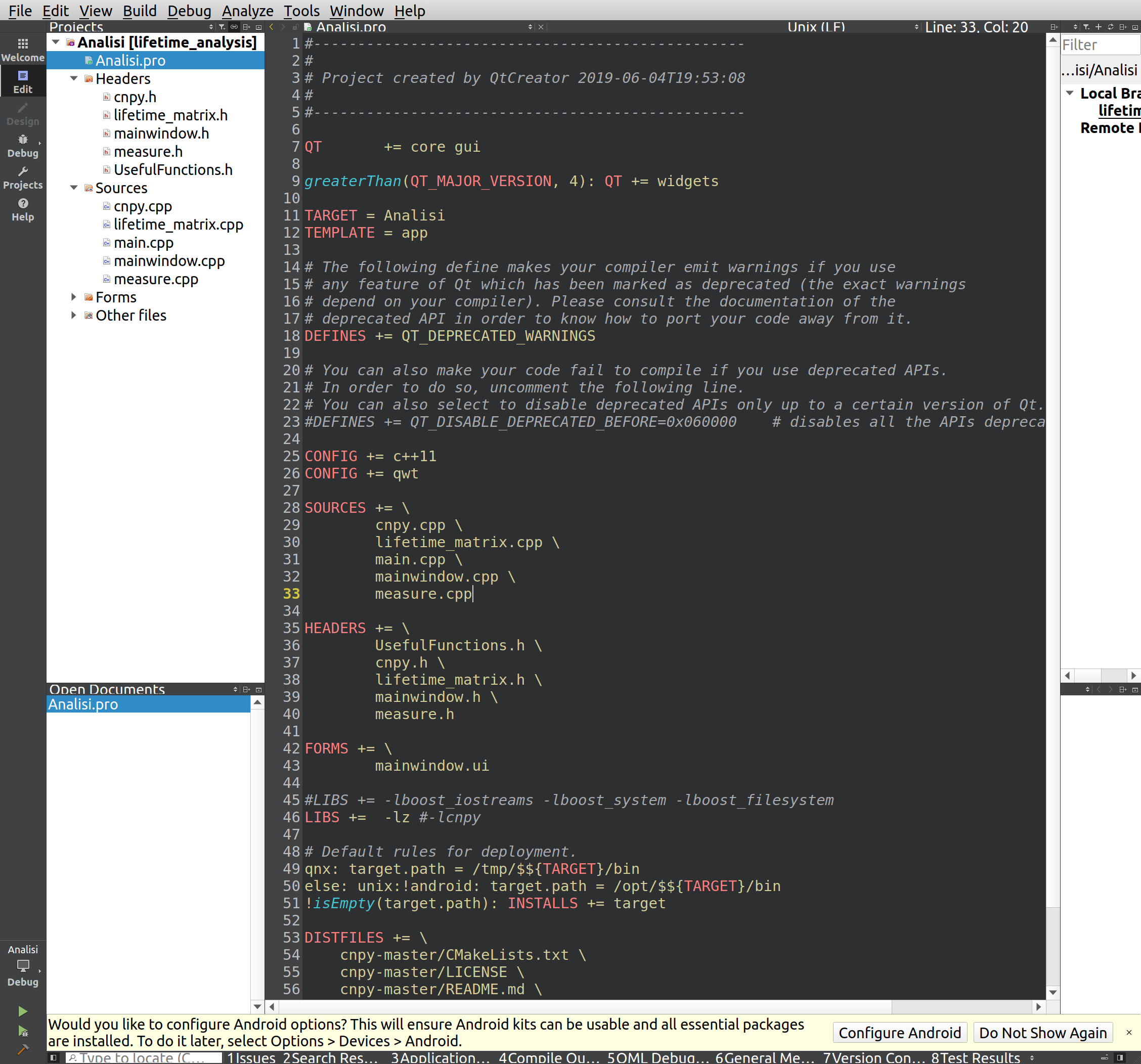}
	\cprotect\caption{Screenshot of \textit{Qt Creator} software showing the
	\verb+.pro+ file of my project. It is visible the list of the 	project
	files as described in the text and various included
	libraries, the most important of which is the Qwt library. This is the
	version running on linux; slight modification are needed on windows.}
	\label{fig:filepro}
\end{figure}
	following it, it is easier to analyze the other file types
	\item The \textbf{sources} files contain the classes used by the
	application. In objective programming, a Class is the description of the
	characteristics of a certain object. In the code, the class can be called
	to create an instance of the object. In my case four classes are present:
	\begin{itemize}
		\item the \textit{main} class is called on program starting, creates
		the application and opens the main window
		\item \textit{mainwindow} contains the instruction for the creation of
		the main window of the graphical user interface
		\item \textit{measure} here the most of the analysis code is present,
		the object is created when launching an analysis
		\item \textit{lifetime\_matrix} is used to create a matrix containing a
		lifetime histogram for each intensity recorded in the Frequency
		Histogram plot, as explained in section~\ref{sec:A-lifeM}
		\item \textit{cnpy} contains the code of an open source library used to
		write data in a format (\verb|.npy|, \verb|.npz|) easily readable with
		the \textit{numpy} library of python; \textit{numpy} is largely used in data analysis and
		the data in this format are useful to be elaborated in further analysis.
	\end{itemize}
	\item The \textbf{headers} corresponds to the class file. They contain the
	variable definitions and their initial value as well as the paradigms. A paradigm
	contains the name of a method of the class and the type of method inputs and
	outputs.
	\item the \textbf{forms} are files that contain the description of the
	user interface. They are generated by the user interface designer of
	\textit{Qt Creator} and they have the extension \verb|.ui|.
\end{itemize}

\subsection{Algorithm main points}
The code contained in the \textit{measure.cpp} file is more than
\SI{1000}{lines} and it is not convenient to report it here entirely. The code has
been written starting from the \textit{demo} code provided by Picoquant, that
extracts from the file the measurement parameter known to the instrument and the
lines where each photon arrival time is recorded. The goal was to analyze it
without the occupation of too much memory, in order to avoid
\textit{segmentation fault} errors and in the most time efficient way.
The strategy I choose was to analyze each lines on the fly during their
collection process.

For each photon in the file the function \textit{got photon} is called. The
information of the line is recorded in a structure, called  \textit{fotone}.
The structure \textit{fotone} is sent to a function that has the scope to
measure what was the intensity when the photon was collected: for this reason
it retains the structure in an array until all the collected photon of a given
bin of the intensity trace are read from the \verb+.ptu+ file. Once the number of
photons collected in the bin is known, this value is stored in each structure \textit{fotone} collected up to now. The mean lifetime in the bin is
also calculated; then a line is added to the TimeTrace file, containing the
time to which the bin corresponds, the number of photons arrived in that bin and
the
calculated mean arrival time.

At this point each of the  \textit{fotone} structures collected up to now is
sent to a function that has the goal to create the \gd{} histogram and, if
required by the user, to create the \textit{far}~\gd{}. They are created
separately.
Here two separate arrays with the last arrived photons are created large enough
to contain the interval chosen by the user. The photon is assigned to the
correct array and the delay between it and the photons already arrived in the
other arrays is evaluated, adding eventually bins to the histogram.
Here is the most time-consuming function, as it verifies for any arrived
photons all the delay with previously arrived photons until the maximal delay
desired is reached. This is the reason why the calculation of the
\textit{far}~\gd histogram require time. In this case, in addition, only the
peaks are recorded, to avoid to store too much data points that make the graph
difficult to plot and to visualize.

If the option to record all data in python files is selected, the information
on each \textit{fotone} are written there.

Finally, a function writes all the data in appropriate named files. The time required to calculate the \gd{} function depends on the number of counts recorded and on the computer processor. In my experience, in order to calculate the \gd{} far, with a quite new 8th generation i7 Intel processor under Ubuntu~18.04, it can takes half an hour for largest files. On the other hand, with lower quality processors under Windows we needed half a day of analysis. If the processor is multi-core, multiple instances needs to be launched in order to analyze multiple emitters at the same time and save time: as told before, ideally the application could be improved to natively support multi-processing without the need to launch multiple instances.

\begin{otherlanguage}{french}
%	\doublespacing
	\renewcommand{\thechapter}{R}
	\renewcommand{\thefigure}{R.\arabic{figure}}
	% !TeX encoding = UTF-8
\chapter{Résumé en français}
%\addcontentsline{toc}{chapter}{\numberline{R}Résumé en français}
\label{chap-resume}
\minitoc
\section{Introduction}
Les technologies quantiques utilisent
pour l'élaboration et la transmission de l'information les principes
fondamentaux de la mécanique quantique. Dans ce type de technologies, l'unité
de base de l'information, appelée bit dans un appareil classique, est
constituée d'un qu-bit. Idéalement, un qubit est un système quantique à deux
niveaux, notés $\ket{1}$ et $\ket{0}$ en analogie à la
notation qu'on utilise classiquement. Dans la théorie de l'information quantique, les réalisations 
physiques d'un qu-bit peuvent être variées, cependant elles possèdent toutes une caractéristique commune: chaque qu-bit peut se trouver non seulement
dans un des deux états ($\ket{\psi}=\ket{1}$, $\ket{\psi}=\ket{0}$ ) mais aussi dans
la
superposition de deux:
\begin{equation}
	\ket{\psi}=\dfrac{1}{\sqrt{\,\abs{\alpha}^2+\abs{\beta}^2}} \tonda{\alpha
		\ket{0}+
		\beta
		\ket{1}},
\end{equation}
où $\alpha$ et $\beta$ sont deux nombres complexes.
L'ensemble des états possibles pour un qubit est représenté avec la sphère de Bloch, reportée en figure~\ref{fig:ch1_blochsphereF}.
\begin{figure}[tbh]
	\centering
	\includegraphics[width=0.35\linewidth]{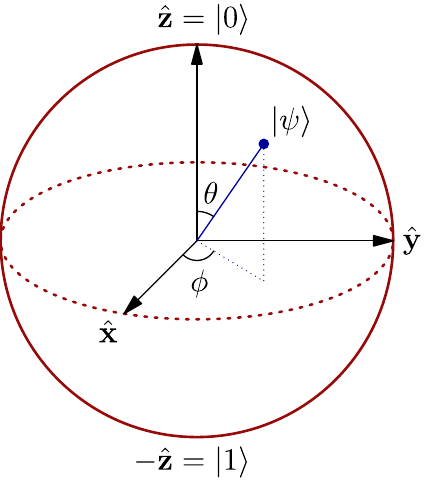}
	\caption{Représentation sur la sphère de Bloch d'un qubit $\ket{\psi}$. Chaque point sur la sphère correspond à un état $\ket{\psi}$
	possible du qubit. $\theta=0$ est en correspondance avec l'état $\ket{0}$,
		$\theta=\pi$ est en correspondance avec l'état $\ket{1}$.
	}
	\label{fig:ch1_blochsphereF}
\end{figure}
Chaque point de la sphère correspond à un vecteur d'état possible; en
particulier les points situés aux pôles de la sphère correspondent respectivement aux états
\ket{0} et \ket{1}.

L'intérêt pour les technologies quantiques a augmenté de plus en plus dans les
dernières années ; pour ce type d'applications et pour leur développement, il est
important d'avoir des systèmes reproductibles et de pouvoir disposer d'un nombre suffisant de qu-bits dans des systèmes les plus 
compacts possibles.
Comme support physique pour un qubit, les photons présentent
plusieurs avantages, en particulier du point de vue de la facilité de manipulation et de
transport. C'est pour cela que l'optique quantique, qui traite de l'étude de
la lumière et des interactions lumière-matière du point de vue de la
mécanique quantique, a connu un grand intérêt de la part de la communauté
scientifique. En optique quantique, la lumière est considérée comme
un ensemble de photons, les excitation fondamentales du champ électromagnétique.
\subsection{Émetteurs à photon unique}
Un émetteur à photon unique est défini comme un objet qui n'émet pas plus d'un
seul photon à la fois.
En pratique, on peut faire la distinction entre sources de photons uniques
stochastiques, par exemple lorsqu'un laser est atténué de telle sorte que la probabilité d'avoir deux
photons en sortie soit faible, et sources de photons uniques déterministes. Dans ce
dernier cas, la source n'émet pas plus d'un photon à la fois grâce à la nature des
processus physiques qui sont à la base de la génération du photon.

Plusieurs réalisations pratiques sont possibles pour une source de photons uniques
déterministe, comme des  atomes uniques, des ions piégés, des boîtes quantiques ou des
centres colorés dans les nanodiamants. Chacun de ces émetteurs présentent des
avantages et des inconvénients. En particulier, dans ma thèse je me suis
concentré sur les deux derniers types d'émetteurs, en étudiant les propriétés
des boîtes quantiques de pérovskite et de centres colorés dans les nanodiamants.

Je détaille brièvement ci dessous quelques unes des caratéristiques les plus intéressantes de ce genre d'émetteurs.

\subsubsection{Mécanisme d'émission}
Il s'agit d'émetteurs  avec une structure de bandes similaire à
celle d'un semi-conducteur. Leur excitation est généralement assurée par une source laser hors
résonance, qui va créer un ou plusieurs excitons. Les excitons peuvent se
recombiner de façon radiative (avec émission d'un photon) ou non radiative.
L'émission de photons uniques dans ce type d'émetteurs est principalment due à l'effet de recombinaison Auger,
qui impose qu'un seul exciton, le dernier, peut se recombiner de façon radiative, tandis que tous
les autres ont un canal privilégié de recombinaison non radiative où l'énergie est cédée à un autre exciton. Ce mécanisme est illustré en
figure~\ref{fig:AugerR-F}.
\begin{figure}[tb]
	\centering
	\subfloat[]{\label{fig:AugerR1-F}\includegraphics[width=0.3\linewidth]
		{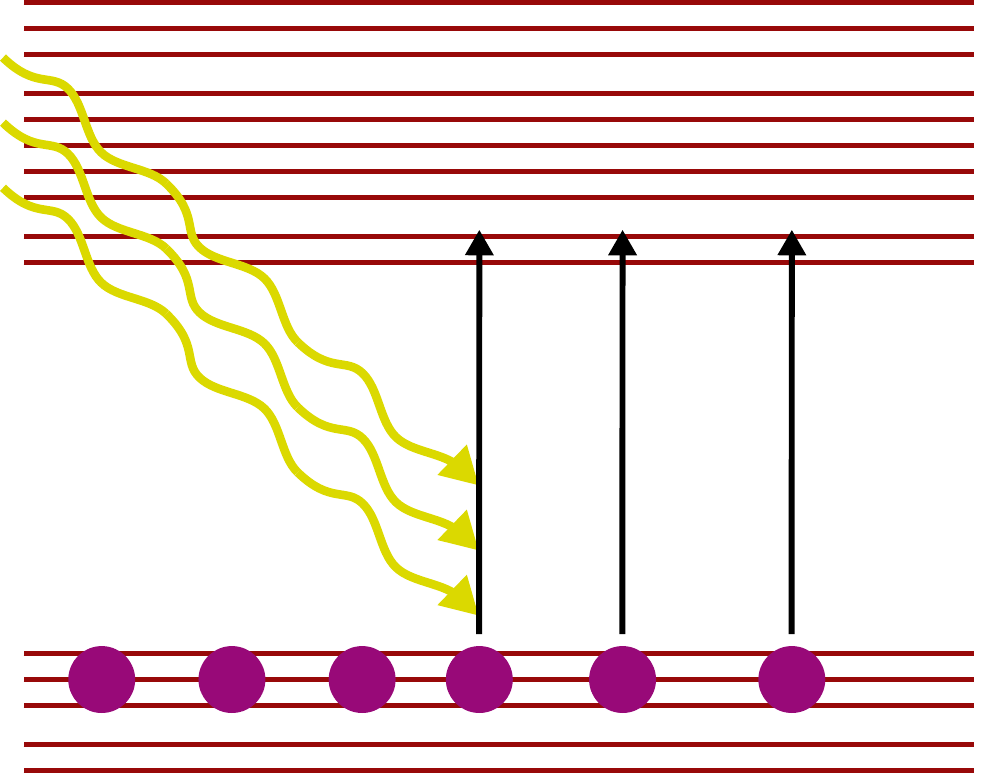}}\qquad
	\subfloat[]{\label{fig:AugerR2-F}\includegraphics[width=0.3\linewidth]
		{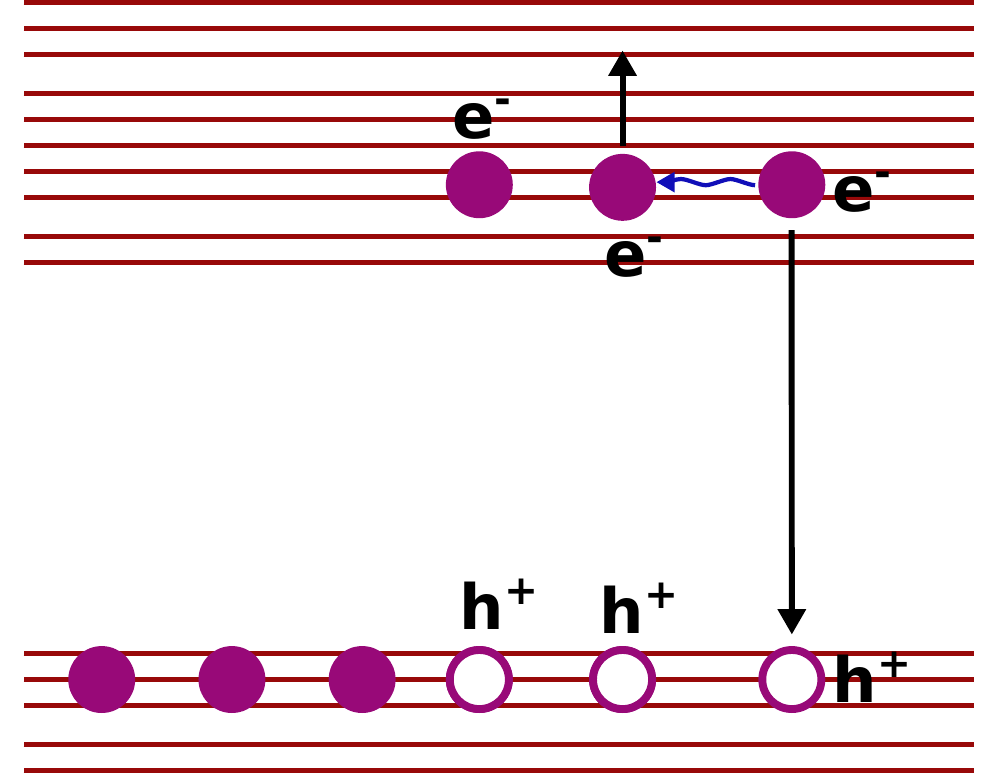}}\\
	\subfloat[]{\label{fig:AugerR3-F}\includegraphics[width=0.3\linewidth]
		{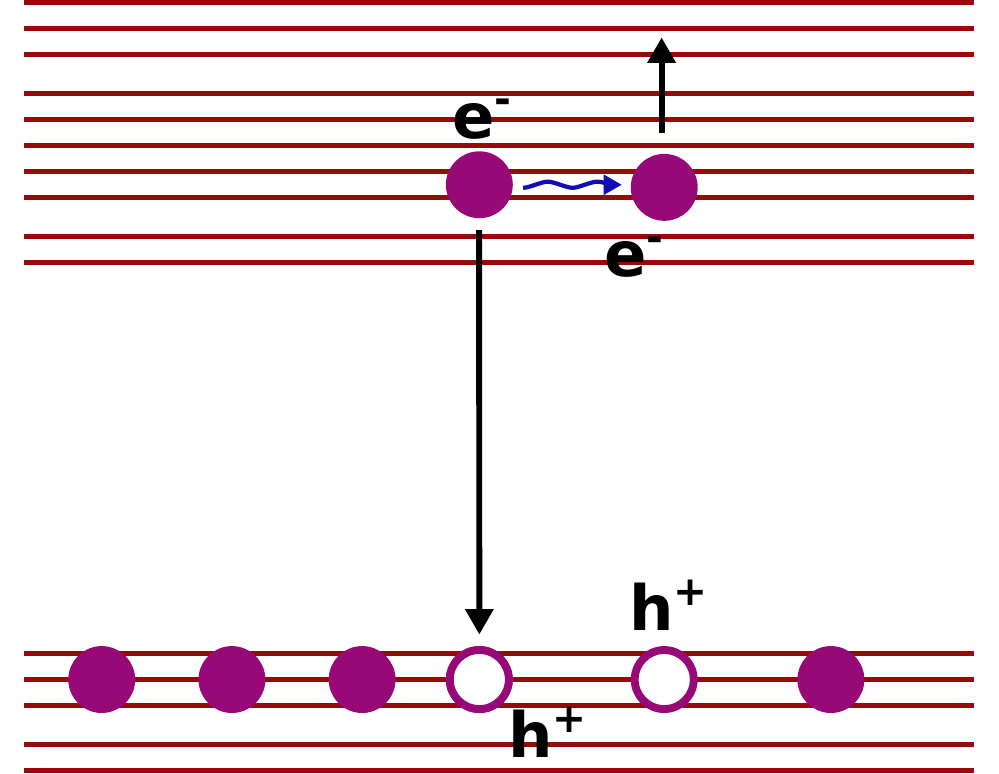}}\qquad
	\subfloat[]{\label{fig:AugerR4-F}\includegraphics[width=0.3\linewidth]
		{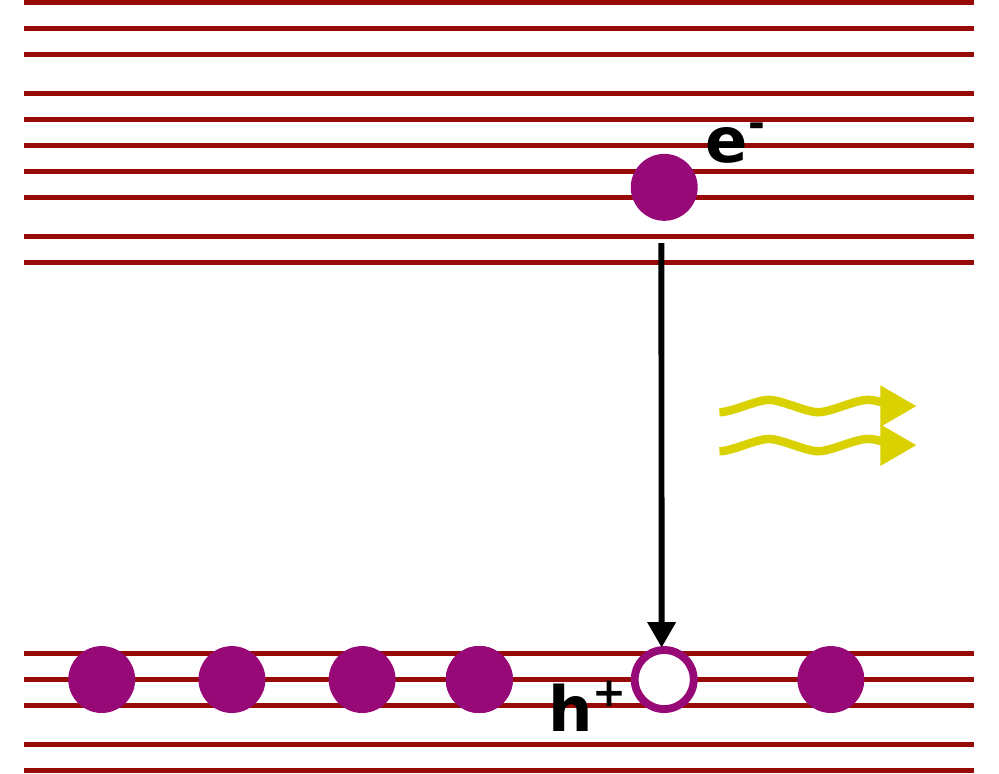}}
	\caption{Mecanisme de recombinaison Auger: \protect\subref{fig:AugerR1-F}~Le
	semi-conducteur est excité par un laser et plusieurs excitons sont créés.
		\protect\subref{fig:AugerR2-F} et \protect\subref{fig:AugerR3-F}~tant  que plusieurs excitons sont présents, un exciton peut
		relaxer en cédant son énergie aux autres ; enfin
		\protect\subref{fig:AugerR4-F}~le dernier exciton ne peut que relaxer
		radiativement en émettant un seul photon.
	}
	\label{fig:AugerR-F}
\end{figure}
\subsubsection{Saturation}
\begin{figure}[tb]
	\centering
	\includegraphics[width=0.5\linewidth]{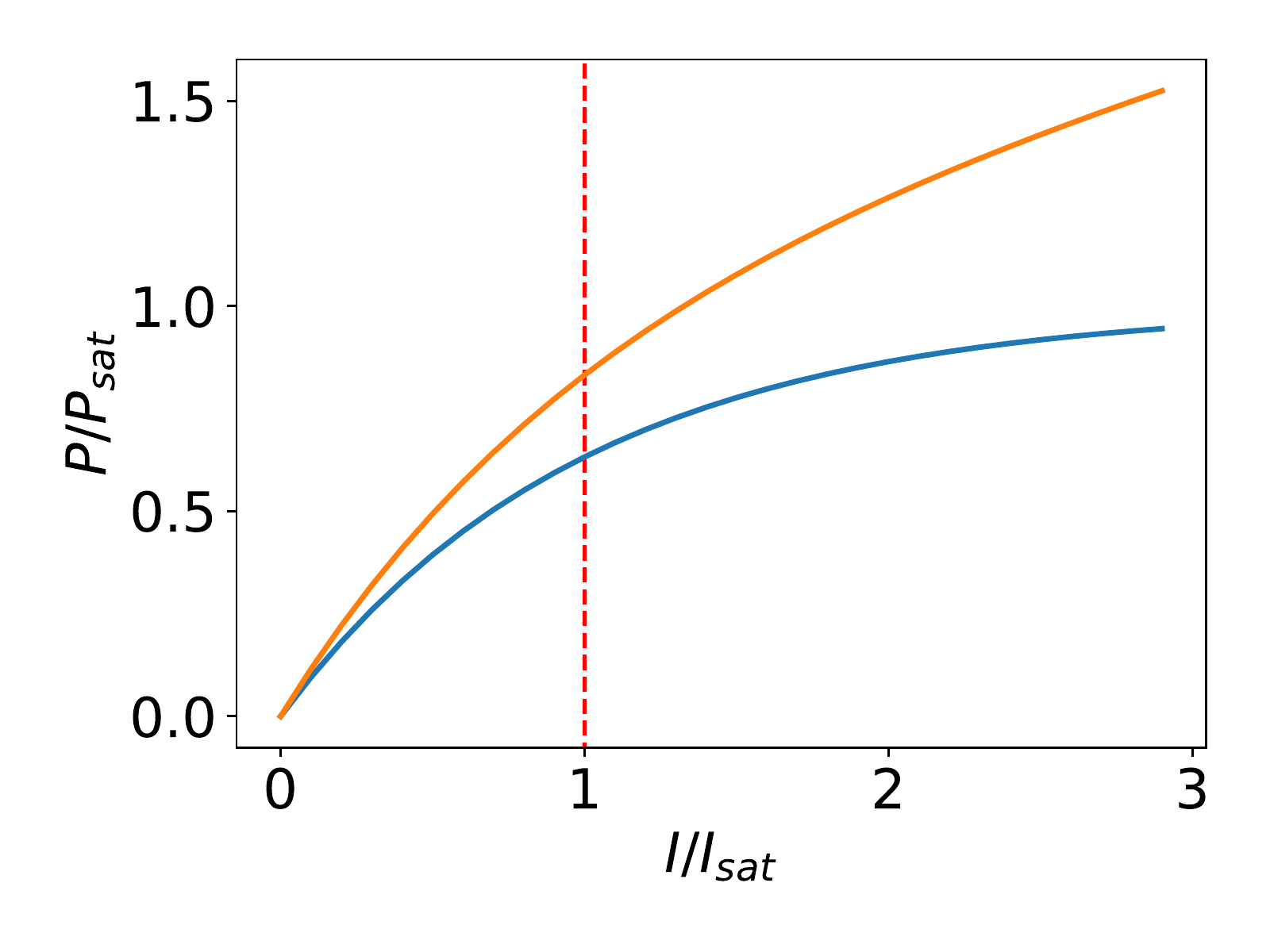}
	\caption{En bleu, la courbe de saturation décrite par
		l'équation~\protect\eqref{eq:sat-F} avec $B=0$, en orange la courbe
		décrite
		par la même équation quand $B=0.2 P_{sat}/I$.}
	\label{fig:ch1sat-F}
\end{figure}
En augmentant la puissance d'excitation, on observe un effet de saturation de
l'émission. Dans les boites quantiques, cet effet est observable grâce à la présence de la recombinaison Auger: en
effet, quand on augmente la puissance, plusieurs excitons sont créés simultanément mais seulement l'un d'entre eux relaxera de façon non radiative, en émettant un photon.
L'équation qui décrit ce comportement est la suivante:
\begin{equation}
	\label{eq:sat-F}
	P_{{PL}} \of{P}=P_{sat} \cdot \tonda{1-e^{-\frac{I}{I_{sat} }}}+ B \cdot I
\end{equation}
où $B$ est un paramètre qui rend compte du fait que l'effet
Auger n'est pas complètement efficace.
Des exemples de courbes de saturation sont montrés dans la
figure~\ref{fig:ch1sat-F}.

\subsubsection{Temps de vie de l'émission}
Les mesures des temps de vie ont un rôle important dans la caractérisation d'un
émetteur à photons uniques : en effet elles constituent une mesure directe du temps de
relaxation radiative. On reporte sur un histogramme les délais de collection
des photons après l'excitation et on obtient une décroissance exponentielle
reportée en figure~\ref{fig:lifeem5-F}.
\begin{figure}[tb]
	\centering
	\includegraphics[width=0.5\linewidth]{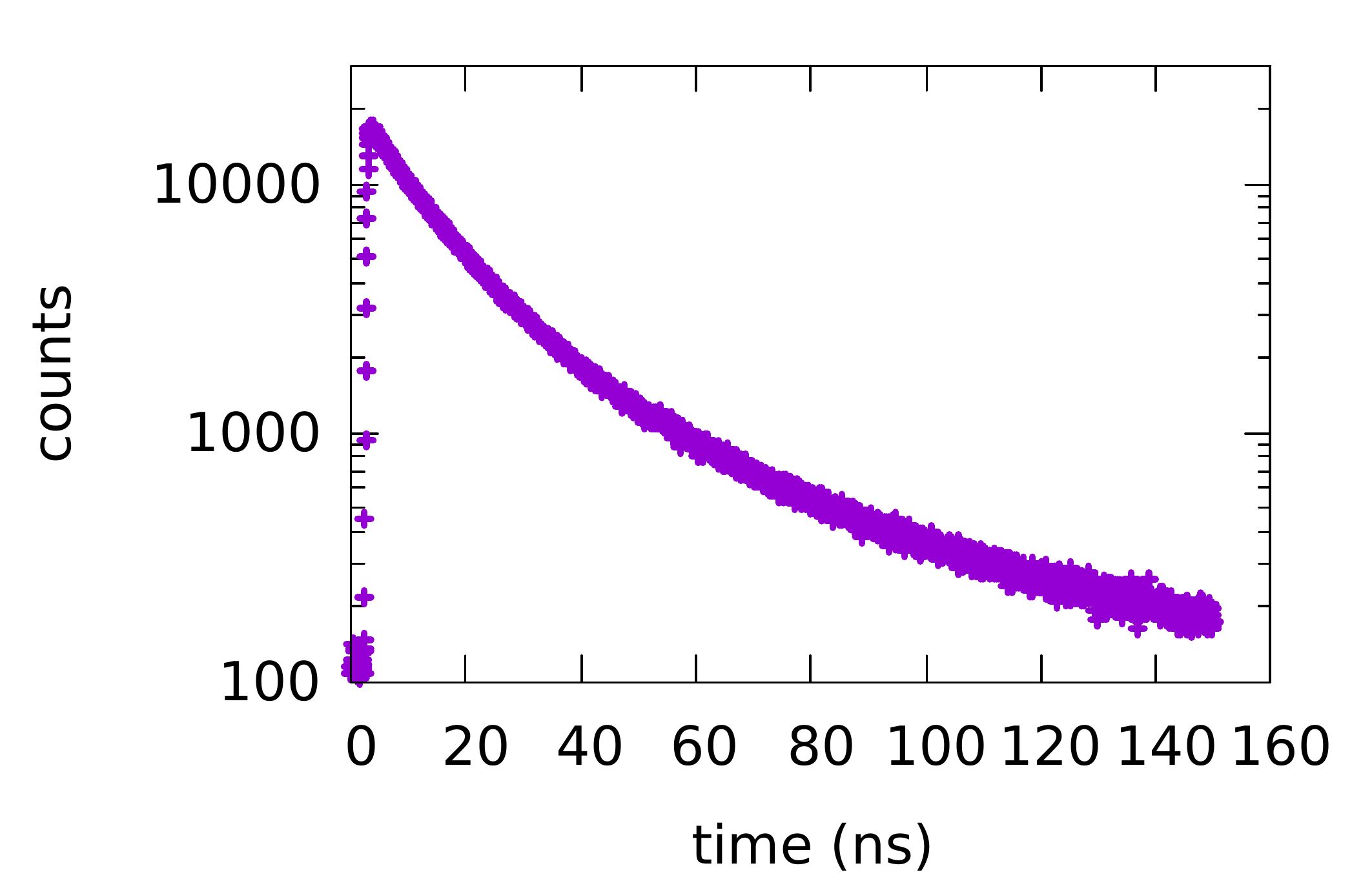}
	\caption{Histogramme du temps de vie; mesure effectuée sur l'émission d'un seul
	\ch{CdS}/\ch{CdSe} nanocristal dot-in-rod. Chaque point représente le nombre
	de photons qui sont arrivés sur le détecteur avec un certain retard.
	L'histogramme correspond à une fonction exponentielle décroissante
	(équation~\protect\eqref{eq:exp_dec_func_s-F}).}
	\label{fig:lifeem5-F}
\end{figure}
Dans le cas général où plusieurs états d'émission sont présents, on peut
décrire telle courbe à l'aide de l'équation suivante:
\begin{equation}
	\label{eq:exp_dec_func_s-F}
	I=\sum_{i} A_i \cdot e^{-t/\tau_i}
\end{equation}
\subsubsection{Clignotement}
Beaucoup d'émetteurs, à cause des effets de piégeage des charges, ont une
émission intermittente. Ce comportement peut constituer un sérieux problème pour les
applications et il est donc intéressant de l'étudier. Pour analyser le clignotement
il y
a plusieurs techniques différentes.

La plus simple prévoit d'utiliser un seuil sur la trace d'intensité ou trace de
clignotement: elle est créée en choisissant un intervalle temporel et en
comptant le nombre de photons qui arrivent dans cet intervalle. On peut ensuite définir un seuil et considérer que l'émetteur est dans un état
\textit{éteint} quand le nombre de photons collectés dans un certain intervalle
est inférieur au seuil et \textit{allumé} quand le nombre des photons collectés
dans un certain intervalle est supérieur au seuil. À ce stade on peut définir
la durée d'un état, \textit{allumé} ou \textit{éteint}, créer une distribution
cumulative et étudier ses caractéristiques.
Cette approche présente le problème d'être fortement influencée par le choix de
la durée de l'intervalle; en effet si le temps de clignotement est plus court que cette durée, ce type d'analyse n'est pas adéquat. De plus on ne peut pas choisir un intervalle trop
court, car on doit avoir suffisamment de comptages pour pouvoir sélectionner un
seuil.

Quand on ne peut pas séparer l'état \textit{éteint} de l'état \textit{allumé}
on utilise une approche différente. On peut en effet définir la FLID
(\foreignlanguage{english}{\textit{fluorescence life intensity distribution}}),
introduite pour la première fois par~\textcite{galland2011Two}. Ce type de
représentation consiste à calculer le temps moyen d'arrivée des photons pour
chaque intervalle de la trace d'intensité; ensuite la densité de
probabilité en fonction de l'intensité du signal et des temps d'arrivée de photons est reconstruite. Un exemple de cette représentation est montré en figure~\ref{fig:Galland-F}. Dans ce
cas, plutôt que de clignotement on parle de~\foreignlanguage{english}{\textit{flickering}}.
\begin{figure}[tbh]
	\centering
	\subfloat[]{\label{fig:GallandA-f}\includegraphics[height=0.3\linewidth]
		{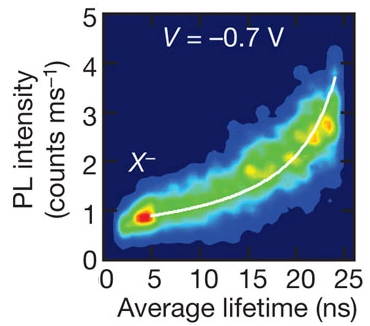}}\qquad\qquad
	\subfloat[]{\label{fig:GallandB-f}\includegraphics[height=0.3\linewidth]
		{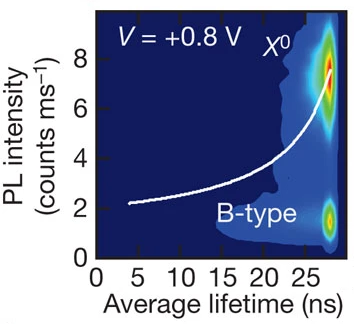}}
		\caption{FLID: \protect\subref{fig:GallandA-f}~en présence du
		clignotement de type~A, le temps de vie moyen dépend de l'intensité
		émise; au contraire, \protect\subref{fig:GallandB-f}~en présence du
		clignotement de type~B, le temps de vie moyen ne dépend pas de
		l'intensité émise.
		\ccredit{galland2011Two}}
	\label{fig:Galland-F}
\end{figure}
En général on peut distinguer deux types de clignotement, type~A et type~B.
Dans le cas d'un clignotement de type~A, le temps de vie dépend de l'intensité
d'émission. Dans le cas d'un clignotement de type~B, le temps de vie ne dépend
pas  de l'intensité de l'émission. D'un point de vue physique, les deux
différents types de clignotement correspondent à des mécanismes de relaxation différents.
\begin{itemize}
	\item Dans le cas du clignotement de type~A, la transition de l'état
	\textit{allumé} vers l'état \textit{éteint} a lieu quand un porteur
	de charge se déplace dans un état piégé. La transition inverse, de l'état
	\textit{éteint} vers l'état \textit{allumé}, a lieu quand il est libéré via
	un processus de relaxation. Ces types de clignotement sont représentés dans la
	figure~\ref{fig:mechanism-typea-F}.
	\item Dans le cas du clignotement de type~B, les fluctuations d'intensité
	sont dues à un piégeage rapide des électrons suivi d'une recombinaison non
	radiative. Ce mécanisme ne comporte pas une variation du temps de vie de
	l'émission.
\end{itemize}

\begin{figure}
	\centering
	\includegraphics[width=0.7\linewidth]{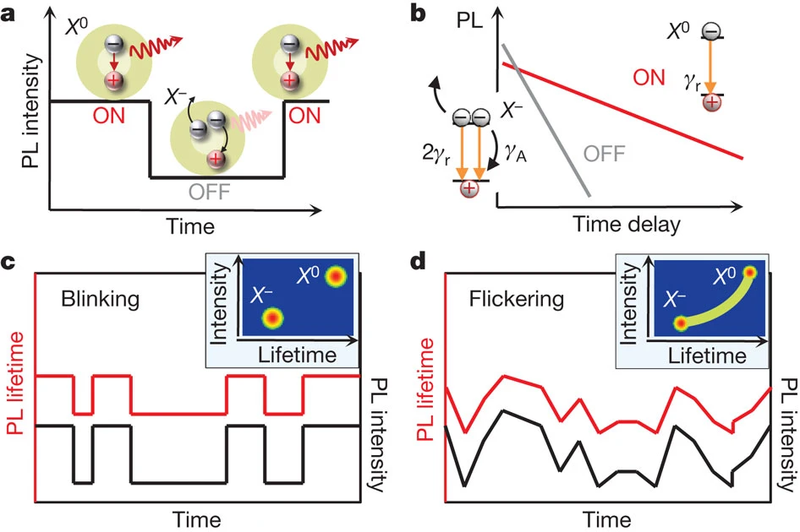}
	\caption{Mécanisme de clignotement de type~A:
		a)~l'état  \textit{allumé} correspond à l'état neutre, l'état
		\textit{éteint} correspond à l'état chargé négativement.
		b)~Des mécanismes différents correspondent à des temps de vie différents.
		c)~Le cas d'un clignotement ``propre'' correspond à deux points
		différents sur la FLID; le cas de \protect\foreignlanguage{english}{\it
		flickering} correspond à une courbe continue sur la FLID.
		\ccredit{galland2011Two}
	}
	\label{fig:mechanism-typea-F}
\end{figure}

\subsubsection{Polarisation}

La polarisation de la lumière émise par les émetteurs est une caractéristique importante: il est donc
utile de la mesurer. Il y a plusieurs façons différentes pour représenter la
polarisation de la lumière. Au niveau analytique, on utilise généralement les
paramètres de Stokes, qu'on peut indiquer $S_0$, $S_1$, $S_2$, $S_3$. Ils
représentent respectivement
\begin{itemize}
	\item[$S_0$] L'intensité,
	\item[$S_1$] le degrée de polarisation horizontal/vertical,
	\item[$S_2$] le degrée de polarisation à $\pm \SI{45}{\degree}$,
	\item[$S_3$] le degrée de polarisation circulaire.
\end{itemize}
Dans le cas d'une polarisation entièrement linéaire la relation suivante est valable:
\begin{equation}
	S_0^2=S_1^2+S_2^2+S_3^2
\end{equation}
Alternativement, on peut utiliser la sphère de Poincaré ou l'ellipse de
polarisation, qui sont deux méthodes pour représenter graphiquement la
polarisation.  Elles sont montrées dans la figure~\ref{fig:1graphpol-F}.
\begin{figure}
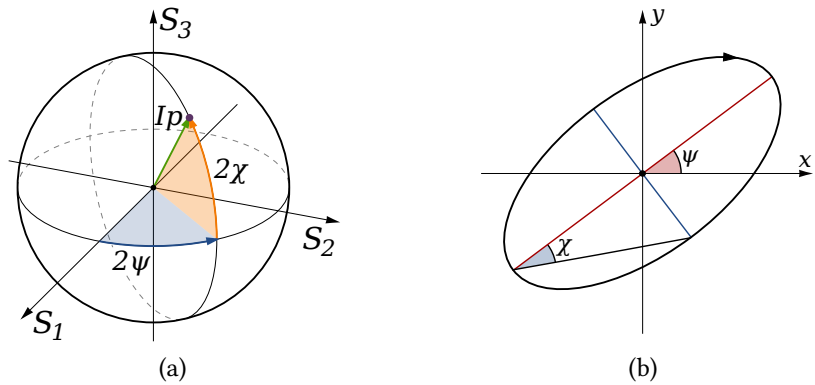

	\centering
	\subfloat[]{\label{fig:1polell-F}\includegraphics[height=0.3\linewidth]
		{Immagini/Poincare_sphere.pdf}}\qquad\qquad
	\subfloat[]{\label{fig:1polsphere-F}\includegraphics[height=0.3\linewidth]
		{Immagini/ellPol.pdf}}
	\caption{Représentations graphiques pour la polarisation.
		\protect\subref{fig:1polell}~Sphère de Poincaré:
		La polarisation est représentée comme un point sur la sphère, les
		cordonnées du point sont en relation avec le vecteur de Bloch selon les		équations~\protect\eqref{eq:sphere2stokes}.
		\protect\subref{fig:1polsphere}~Ellipse de polarisation: c'est une
		visualisation de la polarisation sur deux dimensions, elle est plus
		simple à reproduire sur un support plat. Elle contient la même quantité
		d'information que la sphère de Poincaré.
		\credit{Wikimedia Common}}
	\label{fig:1graphpol-F}
\end{figure}
\subsection{Optique guidée et integrée}
\begin{figure}
	\centering
	\subfloat[]{\label{fig:bancoOttico-F}\includegraphics[height=0.4\linewidth]
		{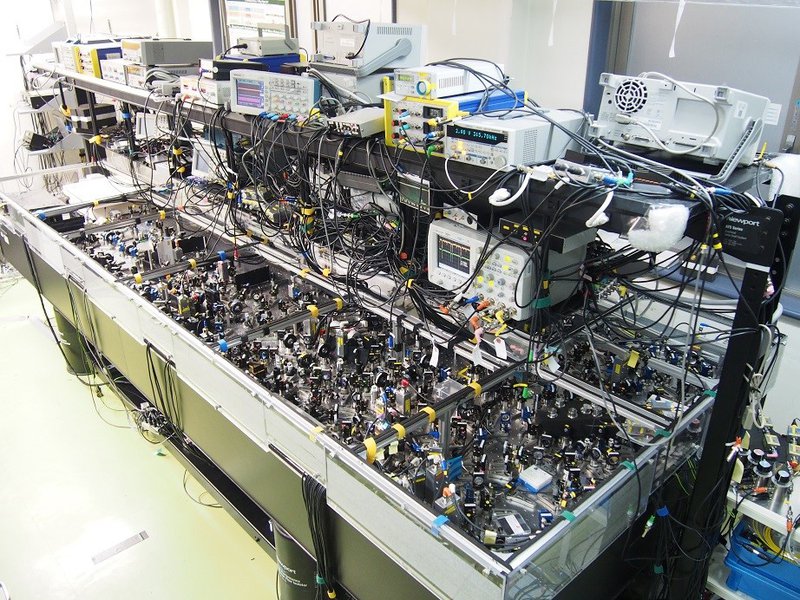}}\qquad
	\subfloat[]{\label{fig:chip-F}\includegraphics[height=0.4\linewidth]
		{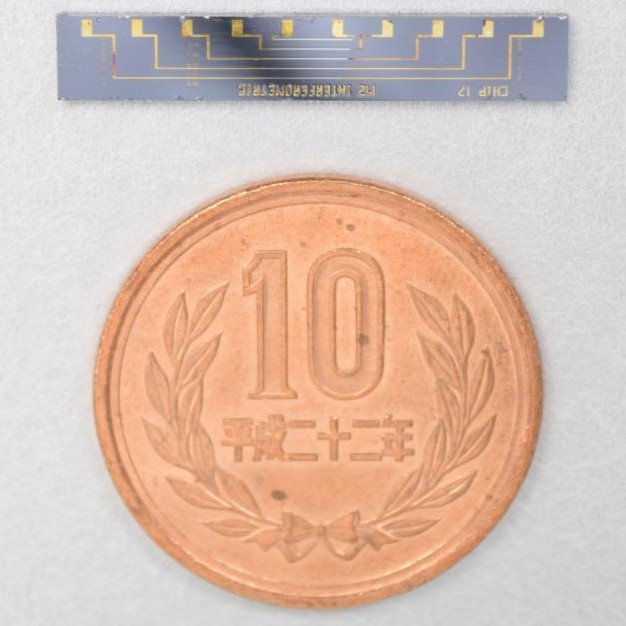}}
	\caption{\label{fig:bancoOtticovsChip-F}
		Différence entre espace libre~\protect\subref{fig:bancoOttico-F} et
		optique intégrée~\protect\subref{fig:chip-F}.
		L'optique en espace libre est flexible et bien
		connue, mais demande un alignement précis et de grandes surfaces; à
		l'opposé l'optique intégrée est stable et ne demande pas d'opération
		d'alignement.
		\credit{University of Bristol}}
\end{figure}
Quand on parle d'optique on peut avoir deux approches différentes. La première, bien
connue et utilisée depuis longtemps, est l'optique en espace libre: la lumière
est réfléchie par des miroirs et modifiée par des lentilles, des lames d'onde,
etc. L'optique en espace libre présente, par contre, l'inconvénient de
nécessiter d'une grande quantité d'espace, des réalignements réguliers et de ne pas
pouvoir être produite en série: des contraintes qui deviennent des
limites pour des applications.

Une deuxième approche possible est celle de l'optique intégrée qui présente
plusieurs avantages:
\begin{itemize}
	\item Elle est stable
	\item elle est compacte
	\item elle peut être reproduite industriellement en plusieurs exemplaires
	identiques
\end{itemize}
Ces avantages sont évidents si l'on regarde la
figure~\ref{fig:bancoOtticovsChip-F}.

Un exemple connu d'optique guidée est représenté par les fibres optiques, qui,
depuis leur découverte, ont trouvé beaucoup d'applications dans différents
domaines.
Pour les applications, il serait utile de coupler les sources de lumière à
des structures optiques guidées sans avoir besoin d'optique en espace libre,
comme le microscope. Dans le cadre de ma thèse, j'ai étudié le couplage
d'émetteurs de photons uniques avec des nanofibres optiques, couplage dû aux
effets du champ proche. J'ai également étudié le couplage des nanoémetteurs avec
des guides d'onde à échange ionique.

\section{Pérovskite}
Les cristaux de pérovskite sont des cristaux avec une structure similaire à
celle de l'Oxyde de Calcium et Titane, qui est le premier minéral de ce genre à
avoir été découvert. Leur structure est représentée dans la
figure~\ref{fig:PerovStruct-F}.
\begin{figure}
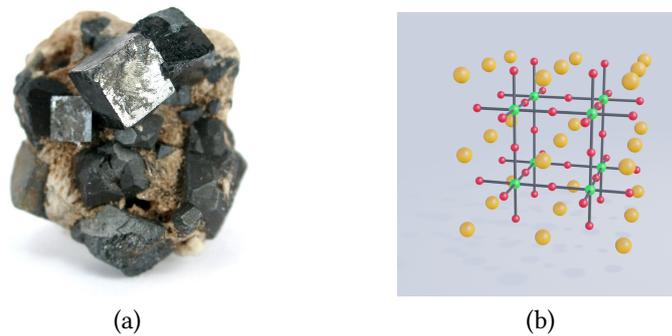

	\centering
	\subfloat[]{\label{fig:Naturalperovskite-F}\includegraphics[width=0.25\linewidth]
		{Immagini/Perovskite-155026}}\qquad\qquad
	\subfloat[]{\label{fig:PerovStruct-F}\includegraphics[width=0.25\linewidth]
		{Immagini/perovskite_struct3.png}}
	\caption{\protect\subref{fig:Naturalperovskite-F}~Cristaux de pérovskite,
		originaires de Magnet Cove, Arkansas, USA.
		\textit{(credits Wikimedia Commons)}
		\protect\subref{fig:PerovStruct-F}~Structure 3D d'une perovskite
		(\ch{ABX3}): en rouge les atomes \ch{X}, anions, en vert les atomes B
		et en jaune les atomes A. A et B sont des cations métalliques,
		avec les cations~B plus petits que les cations~A.}
\end{figure}
La structure des pérovskites est du type~\ch{ABX3}, ou A et B sont des cations
métalliques, avec les cations~B plus petits que les cations~A, et X un anion
métallique.

Les opticiens se sont intéressés aux pérovskites car leur structure
ordonnée leur confère un caractère de semi-conducteurs, et elles sont donc
intéressantes pour des applications dans le domaine du photo-voltaïque. Cependant, leur
caractère semi-conducteur les rend intéressants aussi pour la synthèse de
nanocristaux émetteurs de photons uniques.
\subsection{Nanocristaux de Pérovskite}
Les nanocristaux sont des boites quantiques intéressantes. Ils sont fabriqués
par voie chimique et ils émettent des photons uniques à température ambiante.
Les boites quantiques de pérovskite présentent les mêmes avantages que les
nanocristaux semi-conducteurs et y ajoutent la possibilité de contrôler la longueur d'onde d'émission en jouant sur la composition avec une flexibilité plus grande que dans les nanocristaux semi-conducteurs. Leur principale limitation vient de leur stabilité
optique très faible et c'est précisément sur l'amélioration de cet aspect que j'ai travaillé.

\subsection{Caractérisation expérimentale}
J'ai utilisé des nanocristaux fabriqués par Emmanuel Lhuillier à
l'``Institut de Nanosciences de Paris''. Leur synthèse demande seulement
quelques heures: cet avantage, que les pérovskites on en commun avec les
autres nanocristaux colloïdaux, est un atout considérable en comparaison d'autres émetteurs, tels que les défauts \ch{SiV}  dans les nanodiamants.
J'ai travaillé principalement avec deux échantillons différents:
\begin{description}
	\item[\bf échantillon~A] fabriqué avec la technique décrite
	par~\textcite{protesescu2015nanocrystals}
	\item[\bf échantillon~B] fabriqué avec une technique différente, utilisée
	pour la première fois dans le but de produire des émetteurs de photons uniques. Cette technique est décrite dans~\textcite{pierini2020Highly}
\end{description}

Pour étudier les nanocristaux, j'ai préparé les  échantillons en déposant la
solution contenant les émetteurs avec un appareil pour faire le
revêtement par centrifugation. Une goutte de solution contentant
les émetteurs est déposée sur une lamelle de microscope, mise en rotation par la
machine afin de distribuer la solution de façon homogène sur la lamelle, qui est ensuite regardée en utilisant un microscope confocal, comme décrit
dans la suite.
\subsubsection{Montage expérimental}
Le montage utilisé pour étudier les émetteurs est montré en
figure~\ref{fig:setup-f}
\begin{figure}
	\centering
	\includegraphics[width=\textwidth]{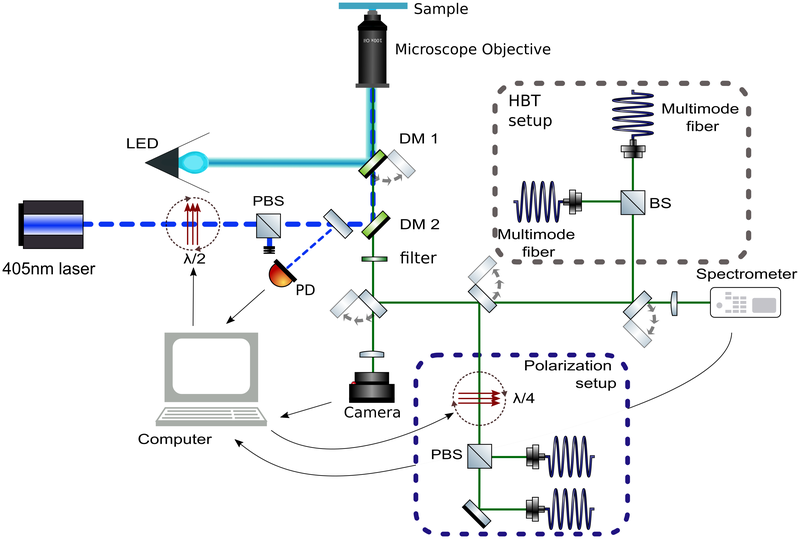}
	\caption{Montage expérimental pour l'analyse d'un nanocristal unique. Pour
	une vision en champ-large, la lumière de la LED est réfléchie par un filtre
	dichroïque et envoyée vers l'échantillon à travers l'objectif. Pour
	exciter un émetteur unique, le laser pulsé est réfléchi vers
	l'échantillon. La lumière collectée par la LED est filtrée pour enlever
	toute trace de la lumière d'excitation et peut ensuite être envoyée
	vers la camera, le spectromètre ou le montage pour mesurer le
	\gdt{} (montage HBT). DM: Miroir Dichroïque; BS:~miroir semi-réfléchissant;
	PBS: cube polariseur; Ph: photodiode; $\lambda/2$:  lame
	demi-onde tournante; $\lambda/4$:  lame quart d'onde tournante.
	}
	\label{fig:setup-f}
\end{figure}
L'échantillon est déposé sur le microscope. Afin de sélectionner un émetteur, la LED
est utilisée pour regarder un espace suffisamment large. Une image d'un
échantillon illuminé avec un spot de grande dimension est montrée en
figure~\ref{fig:nopsstartscale-f}
\begin{figure}
	\centering
	\includegraphics[width=0.5\linewidth]{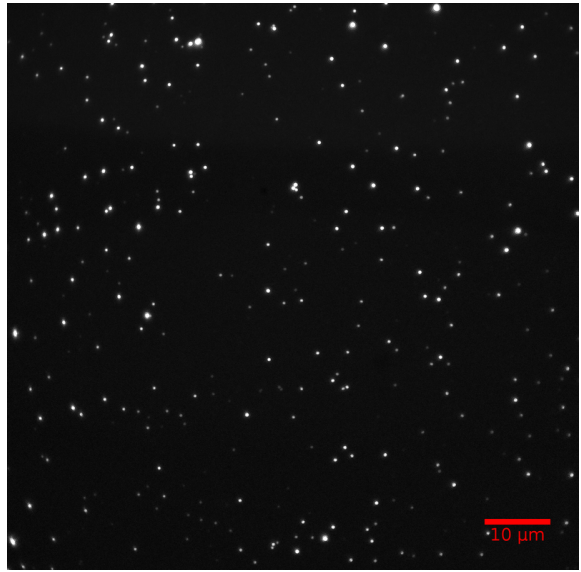}
	\caption{Exemple d'une image obtenue avec la camera quand l'échantillon est
	illuminé avec la LED. Les points blancs sont les émetteurs, le fond est
	noir car la lumière d'excitation est filtrée.}
	\label{fig:nopsstartscale-f}
\end{figure}

En utilisant des translations motorisées, l'échantillon est déplacé pour placer
l'émetteur à l'intérieur du faisceau laser. Ensuite, la LED
est éteinte et le laser est utilisé pour exciter l'émetteur et effectuer les mesures: spectre,
intensité de saturation, \gdt, clignotement, polarisation.

\subsection{Mesures}
\subsubsection{Spectres}
On peut mesurer les spectres avec le spectromètre. La largeur spectrale et la forme de
l'émission sont intéressantes: en effet pour de nombreuses applications telles que le couplage avec des cavités optiques ou encore la conversion de fréquence, on recherche de raies d'émission étroites. En plus, avec une statistique sur les spectres on peut avoir des  informations sur les différences entre émetteurs.
Par exemple, un spectre est représenté en
figure~\ref{fig:spettroemissione-f}.
\begin{figure}
	\centering
	\includegraphics[width=0.4\linewidth]
	{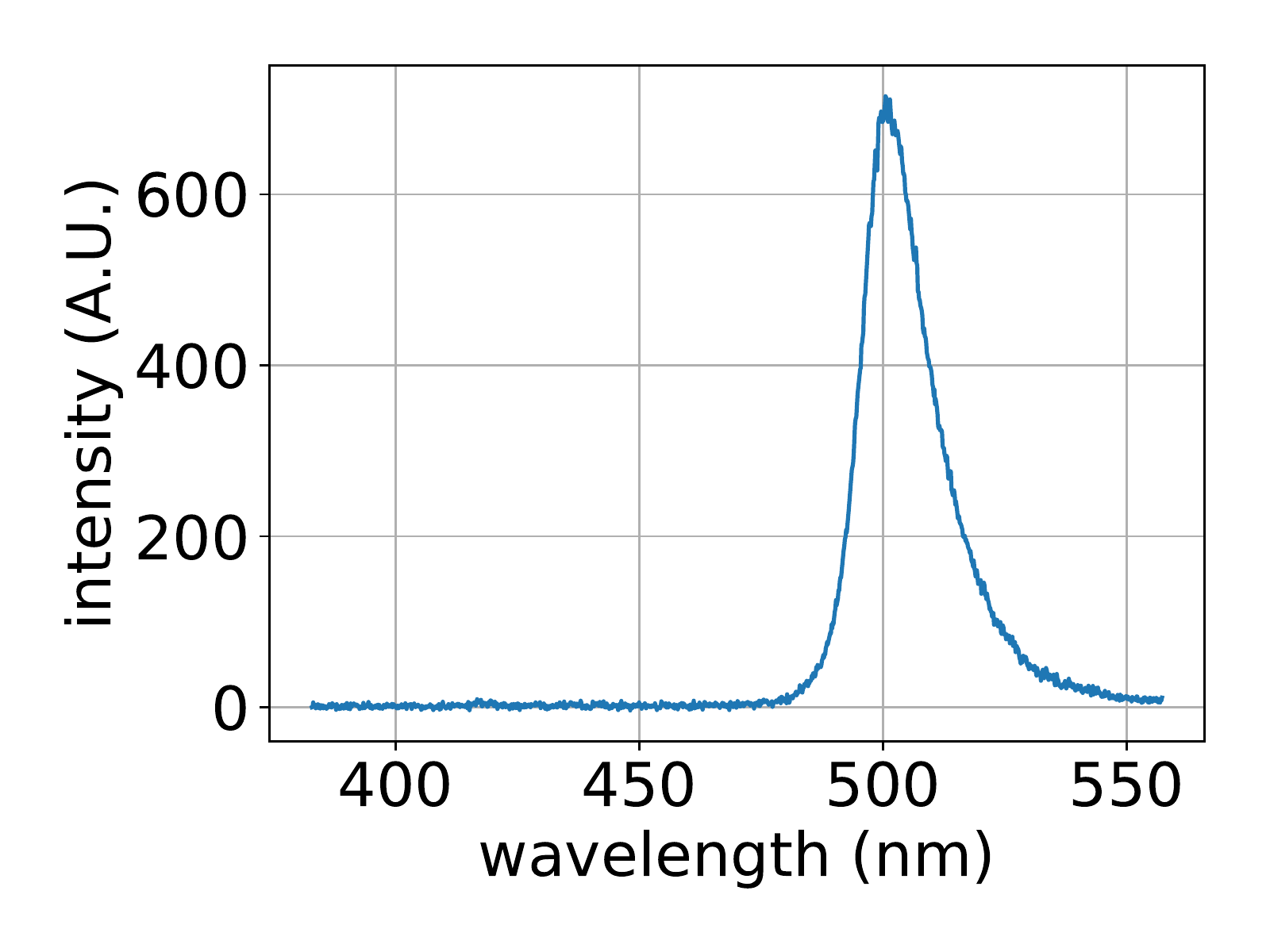}
	\caption{Spectre d'émission d'un nanocristal unique de pérovskite. Dans cet
	exemple on observe un pic d'émission à \SI{500}{\nano \meter}.}
	\label{fig:spettroemissione-f}
\end{figure}
\begin{figure}
	\centering
	\subfloat[]{\label{fig:wave_vs_FWHM_A-f}\includegraphics[scale=0.4]
		{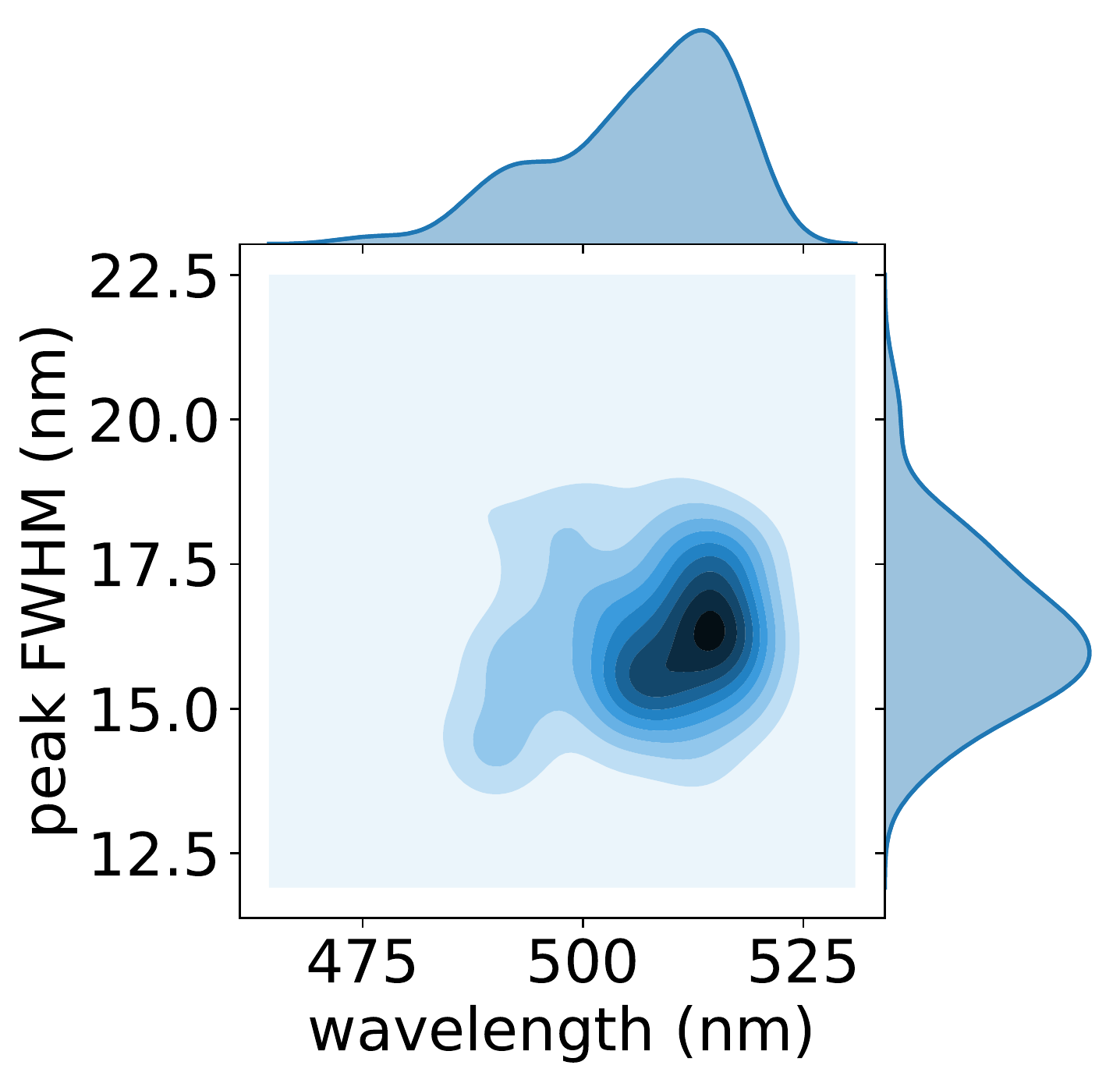}}\qquad
	\subfloat[]{\label{fig:wave_vs_FWHM_B-f}\includegraphics[scale=0.4]
		{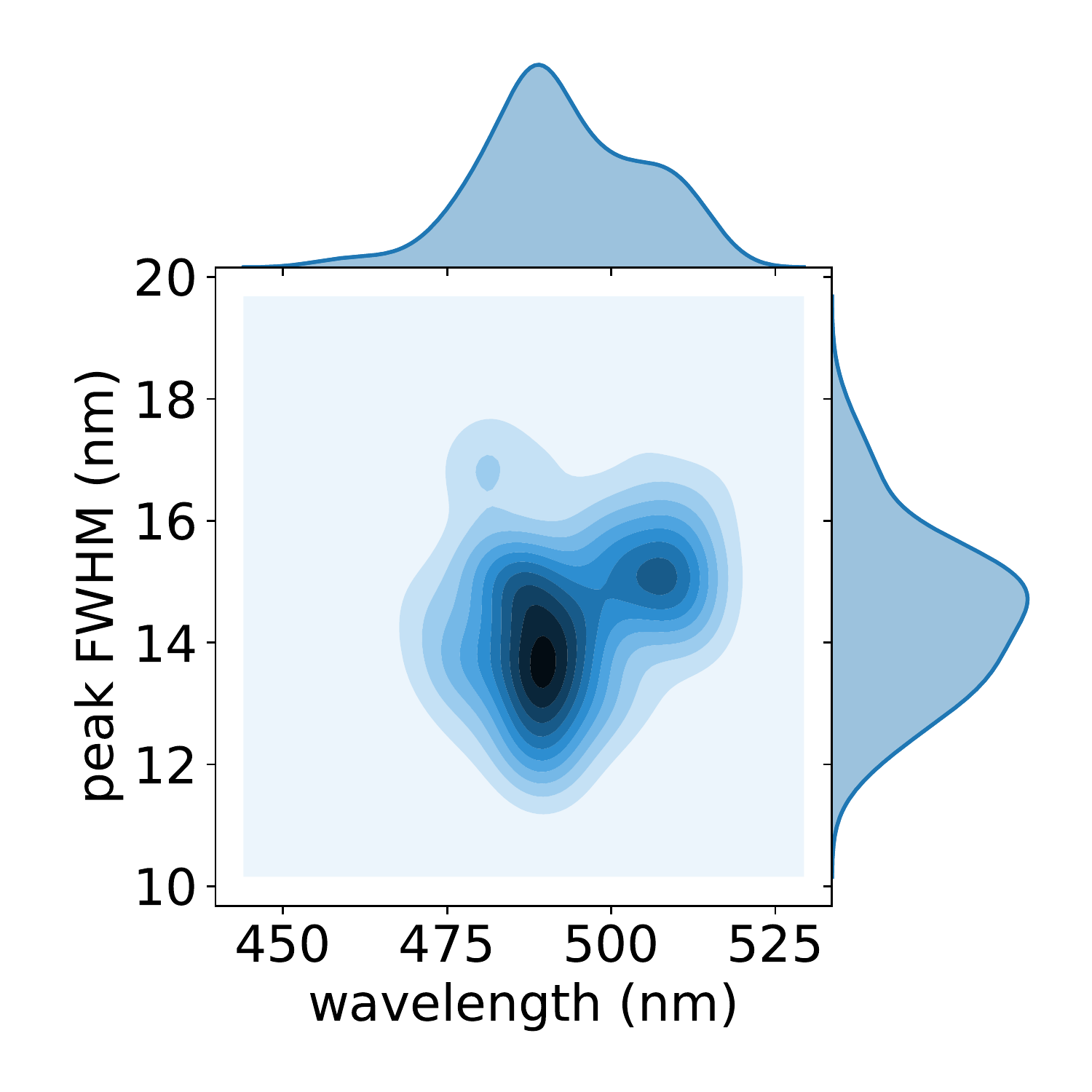}}
	\caption{\label{fig:wave_vs_FWHM-f} Distribution de la longueur d'onde
	d'émission centrale et de la largeur à mi hauteur pour un
	échantillon~A~\protect\subref{fig:wave_vs_FWHM_A-f} (calcul fait avec
	\SI{135}{}~émetteurs) et B~\protect\subref{fig:wave_vs_FWHM_B-f} (calcul
	fait avec \SI{135}{}~{émetteurs}).
	}
\end{figure}
Dans la figure~\ref{fig:wave_vs_FWHM-f} est reportée une analyse statistique
pour le spectre d'émission et pour la longueur d'onde d'émission centrale. On
peut s'attendre à un plus grand confinement dans l'échantillon~B car la longueur
d'onde d'émission est inférieure.

\subsection{Saturation}
\begin{figure}[p]
	\centering
	\includegraphics[width=0.5\linewidth]{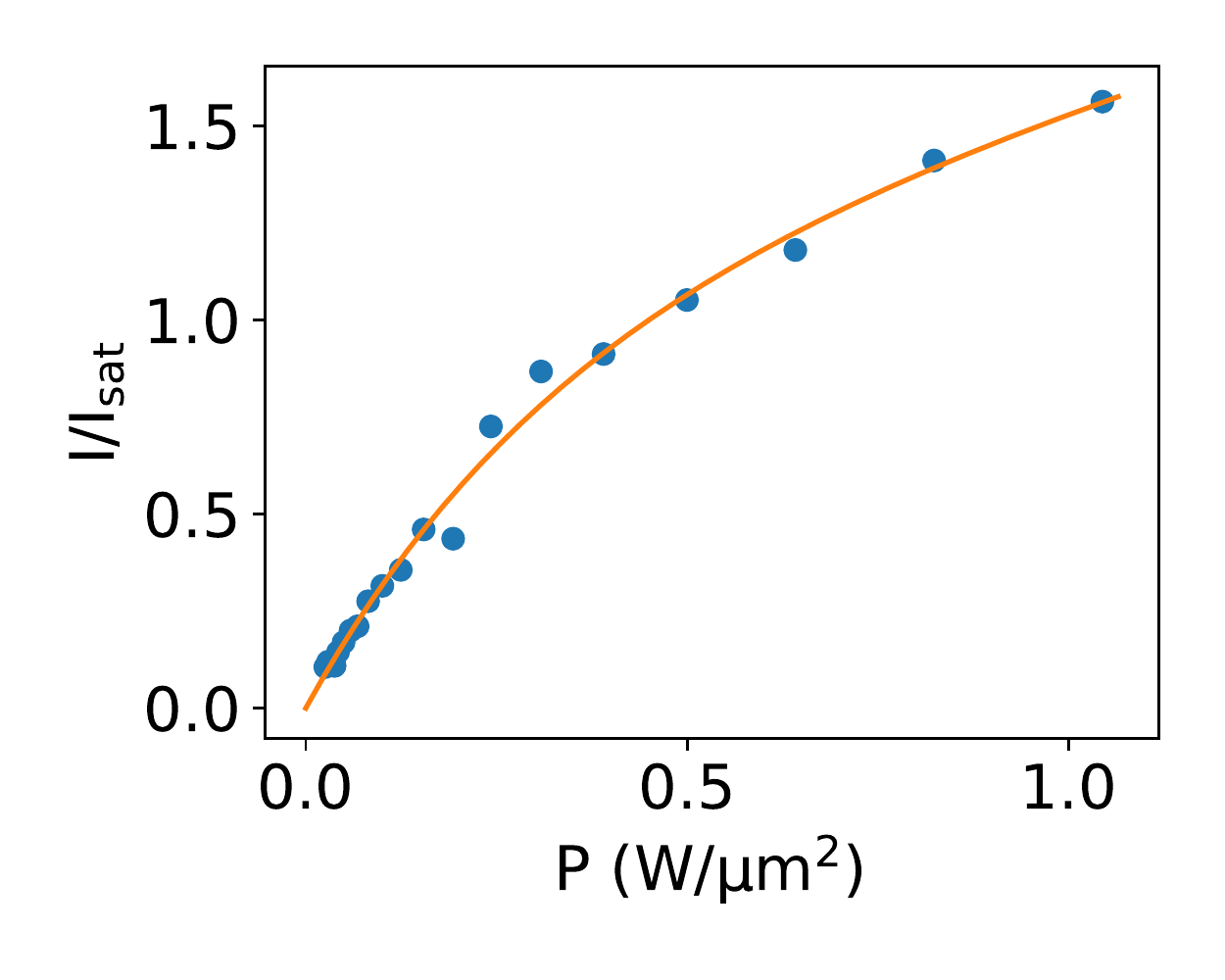}
	\caption{Exemple de mesure de saturation sur un nanocristal unique de
		pérovskite}
	\label{fig:satintensity-f}
\end{figure}
\begin{figure}[tbp]
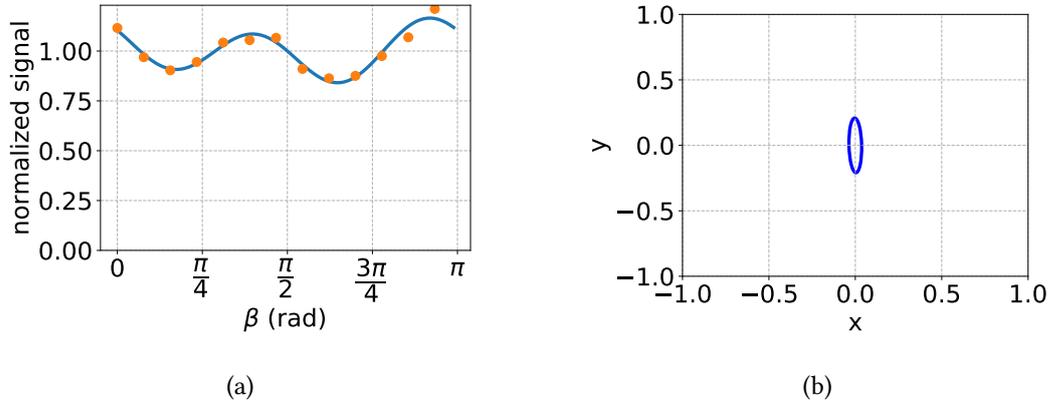

	\centering
	\subfloat[]{\label{fig:polplot_measure-f}\includegraphics[width=0.45\linewidth]
		{Immagini/pol_plot.pdf}}\qquad
	\subfloat[]{\label{fig:polplot_ell-f}\includegraphics[width=0.45\linewidth]
		{Immagini/pol_ellipse.pdf}}
	\caption{\protect\subref{fig:polplot_measure-f}~Exemple de mesure des
		paramètres de Stokes pour la lumière émise par un nanocristal de
		pérovskite. Les points rouges sont les données expérimentales, la courbe
		bleue
		est décrite par l'équation \protect\eqref{eq:berry-F}.
		\protect\subref{fig:polplot_ell-f}~Ellipse de polarisation. Le degrée de
		polarisation est~0.18 et le degrée de  polarisation linéaire est~0.17, cela montre une émission partiellement polarisée.}
	\label{fig:polplot-f}
\end{figure}
La saturation est importante car elle nous permet de comparer les émetteurs: en
effet, on a fait le choix d'exciter tous les émetteurs à l'intensité de
saturation. À cause de la présence du clignotement, pour pouvoir faire une mesure de
saturation correcte, il est important de prendre en compte d'éventuelles variations
de luminosité: pour cela on effectue plusieurs mesures à la même
puissance et on considère seulement celle qui détecte l'émission la plus intense.
On suppose en effet que ce soit celle où l'émetteur était toujours dans l'état
allumé.
Un exemple de mesure de saturation sur des pérovskites est représenté en
figure~\ref{fig:satintensity-f}. La courbe orange représente le fit avec la
fonction définie précédemment:
\begin{equation}
	P=A \cdot \quadra{ 1-e^{-\frac{I}{I_{sat}}}} +B \cdot I
\end{equation}

\subsection{Mesures de polarisation}
Pour mesurer la polarisation on tourne une lame quart d'onde et on polarise
ensuite
la lumière avec un cube polariseur. En mesurant l'intensité
transmise par le cube en fonction de la position de la lame
quart d'onde, il est possible de remonter aux paramètres de Stokes.
La relation qui connecte les paramètres de Stokes $S_0$, $S_1$, $S_2$, $S_3$,
l'intensité transmise $I_T$, l'angle de rotation du polariseur $\alpha$ (dans
notre cas le cube), l'angle de rotation de l'axe rapide de la
lame d'onde $\beta$ et le retard introduit par la lame d'onde $\delta$ est la
suivante:
\begin{equation}
	\label{eq:berry-F}
	\begin{split}
		I_{T}(\alpha, \beta, \delta)&=\frac{1}{2}\left[S_0+\left(\frac{S_1}{2}
		\cos 2
		\alpha+\frac{S_2}{2} \sin 2 \alpha\right)(1+\cos \delta)\right]
		+\\&+\frac{1}{2}[S_3 \sin \delta \cdot \sin (2 \alpha-2 \beta)]+\\&+
		\frac{1}{4}[(S_1 \cos
		2
		\alpha-S_2 \sin 2 \alpha) \cos 4 \beta
		+\\&+(S_1 \sin 2 \alpha+S_2 \cos 2 \alpha) \sin 4 \beta](1-\cos \delta).
	\end{split}
\end{equation}
Avec cette relation, en mesurant un nombre suffisant de points et en utilisant le
développement en série de Fourier, il est possible d'obtenir les valeurs des
coefficients $S_0$, $S_1$, $S_2$, $S_3$ de façon
analytique~\cite{berry1977Measurement}. La figure~\ref{fig:polplot-f} présente un
exemple de mesure des paramètres de Stokes sur un nanocristal unique de
pérovskites.

\subsubsection{Mesures de \gd{}}
D'un point de vue expérimental, la mesure de l'émission des photons uniques se fait grâce à une
mesure d'autocorrélation. Le signal est divisé en deux parties et chacune d'elles
est envoyée vers un module compteur des photons; on regarde ensuite la
corrélation entre le signal sur un détecteur et le signal sur l'autre. D'un
point de vue expérimental, ce type de mesures présente plusieurs difficultés:
\begin{enumerate}
	\item tout d'abord, il s'agit d'une mesure qui demande de travailler dans l'obscurité, car chaque photon parasite qui arrive sur le détecteur peut la perturber. Cependant, on peut
	essayer de réduire l'effet des \eng{dark counts} en corrigeant le signal après la mesure.
	\item une grande quantité de données est collectée pendant la mesure, et donc
	une grande quantité de données est à élaborer par la suite.
	\item Le système de détection présente la limite de ne pas pouvoir détecter deux
	photons en même temps. L'intérêt de la mesure est exactement de vérifier
	si, dans les deux parties il y a deux photons au même temps. On peut
	résoudre le problème en ajoutant une ligne de retard.
\end{enumerate}
Un exemple de mesure de \gd{} est reporté en figure~\ref{fig:gd-f}.
\begin{figure}[tbh]
	\centering
	\subfloat[]{\label{fig:g2raw-f}\includegraphics[width=0.45\linewidth]
		{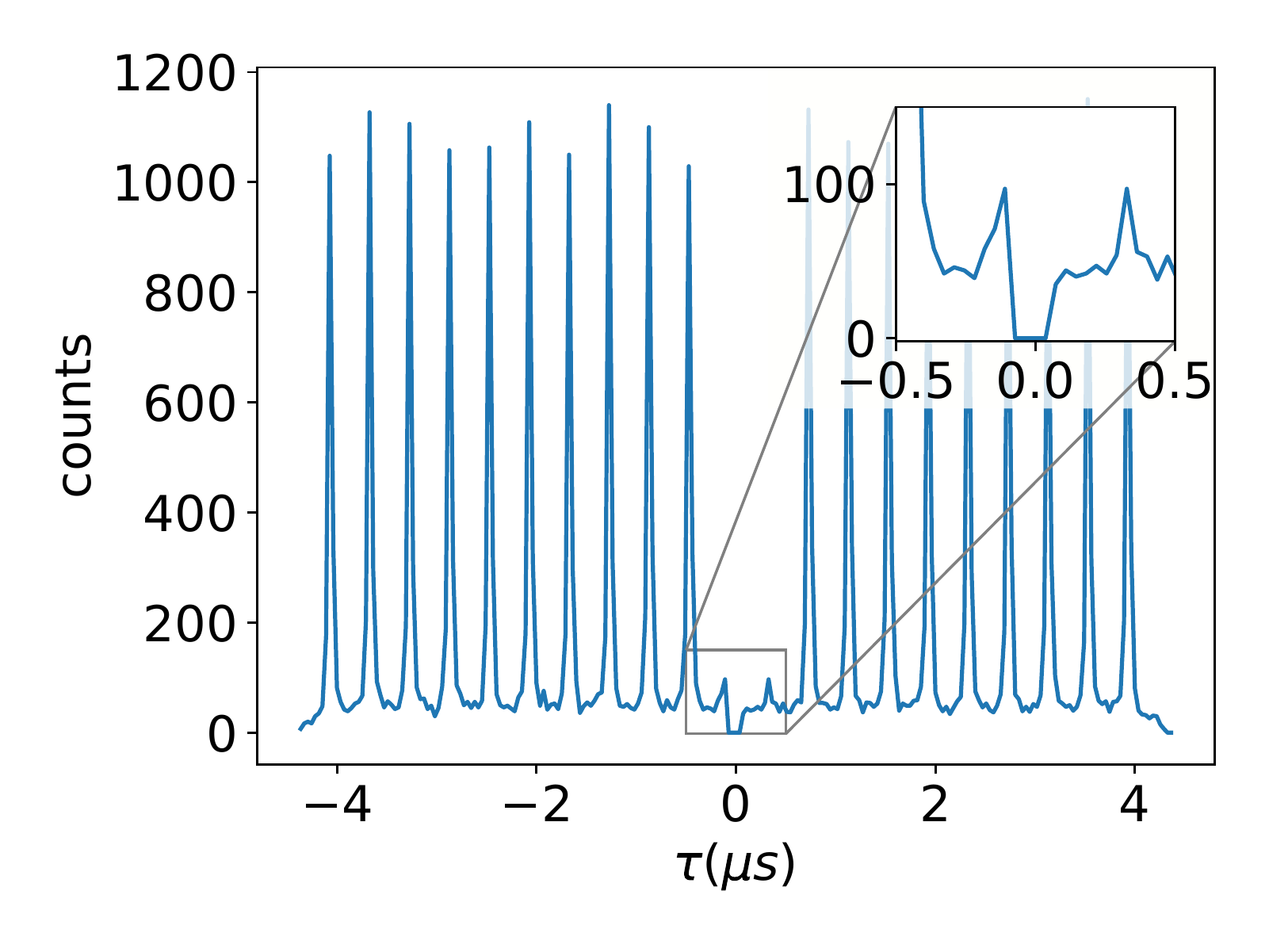}}\qquad
	\subfloat[]{\label{fig:g2clean-f}\includegraphics[width=0.45\linewidth]
		{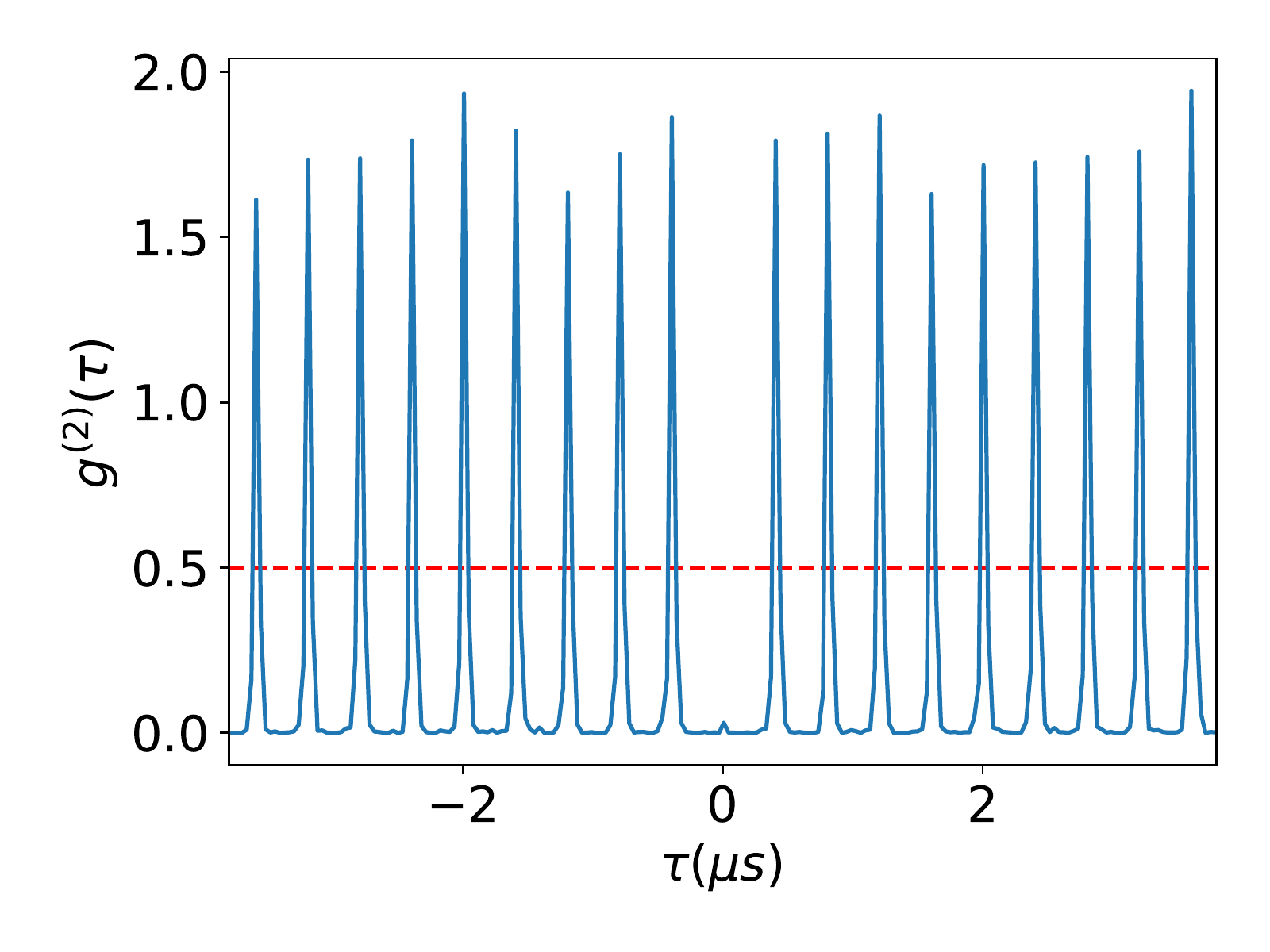}}
	\caption{%\label{fig:g2_example}
		Mesures de \protect\subref{fig:g2raw-f}~\gdt{} avant traitement,
		dans l'encart le temps mort de l'instrument est bien visible.
		\protect\subref{fig:g2clean-f}~\gdt{} après le traitement. L'émetteur
		montre un $ \textrm{g}^{(2)}\of{0}\approx0$, qui signifie une très
		bonne émission de photons uniques.
		\label{fig:gd-f}
	}
\end{figure}
En particulier, en figure \ref{fig:g2raw-f} est montré le signal de \gd{}
sans aucun traitement. On voit au centre le temps mort de
l'instrument, qui correspond à un espace sans coïncidences, et le petit pic
qui correspond au délai nul, $\tau=0$, translaté sur la droite grâce à la ligne
à retard. Sur la figure~\ref{fig:g2clean-f} on voit la mesure de \gdt{} après la
procédure de correction du signal, qui enlève l'effet de la ligne à retard et
le problème dû au temps mort de l'instrument.

\subsection{Stabilité}
Un des problèmes bien connus des pérovskites est leur photo-stabilité : en effet
leur structure tend à être endommagée par un effet combiné de la lumière et de
l'humidité, ce qui comporte une dégradation de leur émission jusqu'à la
disparition. Plusieurs approches ont été testées pour résoudre le problème:
j'en ai utilisé deux différentes. La première consiste à envelopper les
pérovskites dans une matrice de polystyrène pour éviter le contact
avec l'humidité ambiante, la deuxième à modifier la procédure de fabrication pour
obtenir des matériaux plus robustes.

\begin{figure}[tbh]
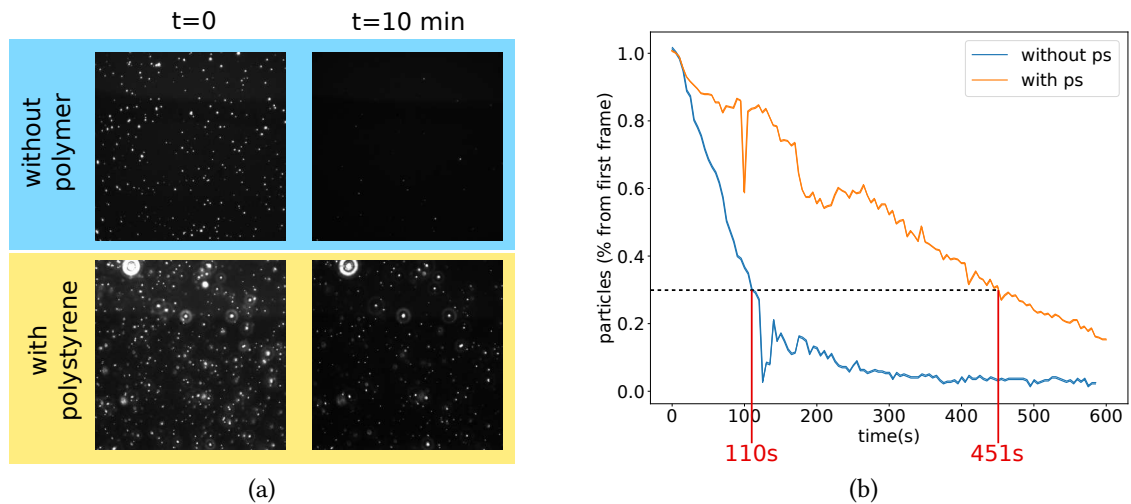

	\centering
	\subfloat[]{\label{fig:bleach_withpsornoti-f}\includegraphics[height=0.40\linewidth]
		{Immagini/withPSorNot_text.pdf}} \qquad
	\subfloat[]{\label{fig:bleach_withpsornott-f}\includegraphics[height=0.40\linewidth]
		{Immagini/PS_vs_noPS_time.pdf}}
	\caption{\protect\subref{fig:bleach_withpsornoti}~
    Images des échantillons au début de la mesure ($t=0$) et après dix minutes ($t=\SI{10}{min}$) sous un fort éclairage LED, sans et avec le polystyrène. \protect\subref{fig:bleach_withpsornott}~Analyse quantitative de la mesure : la stabilité est augmentée d'un facteur quatre.}
	\label{fig:bleach_withpsornot-f}
\end{figure}
Les résultats de la première approche sont montrés dans la
figure~\ref{fig:bleach_withpsornot-f}. On peut voir que les nanocristaux
enveloppés dans le polystyrène sont capables d'émettre de la lumière pendant un temps  quatre
fois plus long que les émetteurs nus. On voit donc qu’envelopper les nanocristaux dans un
polymère, et en particulier dans le polystyrène, peut être une méthode efficace pour en
améliorer la stabilité. Le test a été effectué avec l'échantillon~A.
Cette approche a l'inconvénient de nécessiter la présence d'un polymère dans la
solution, ce qui pourrait perturber le couplage avec le champ proche et rendre
difficile la déposition sur la nanofibre.

On a donc étudié l'effet des ligands présents dans la solution et le rôle de la
préparation de l'échantillon.
\begin{figure}[p]
	\centering
	\includegraphics[width=0.6\linewidth]{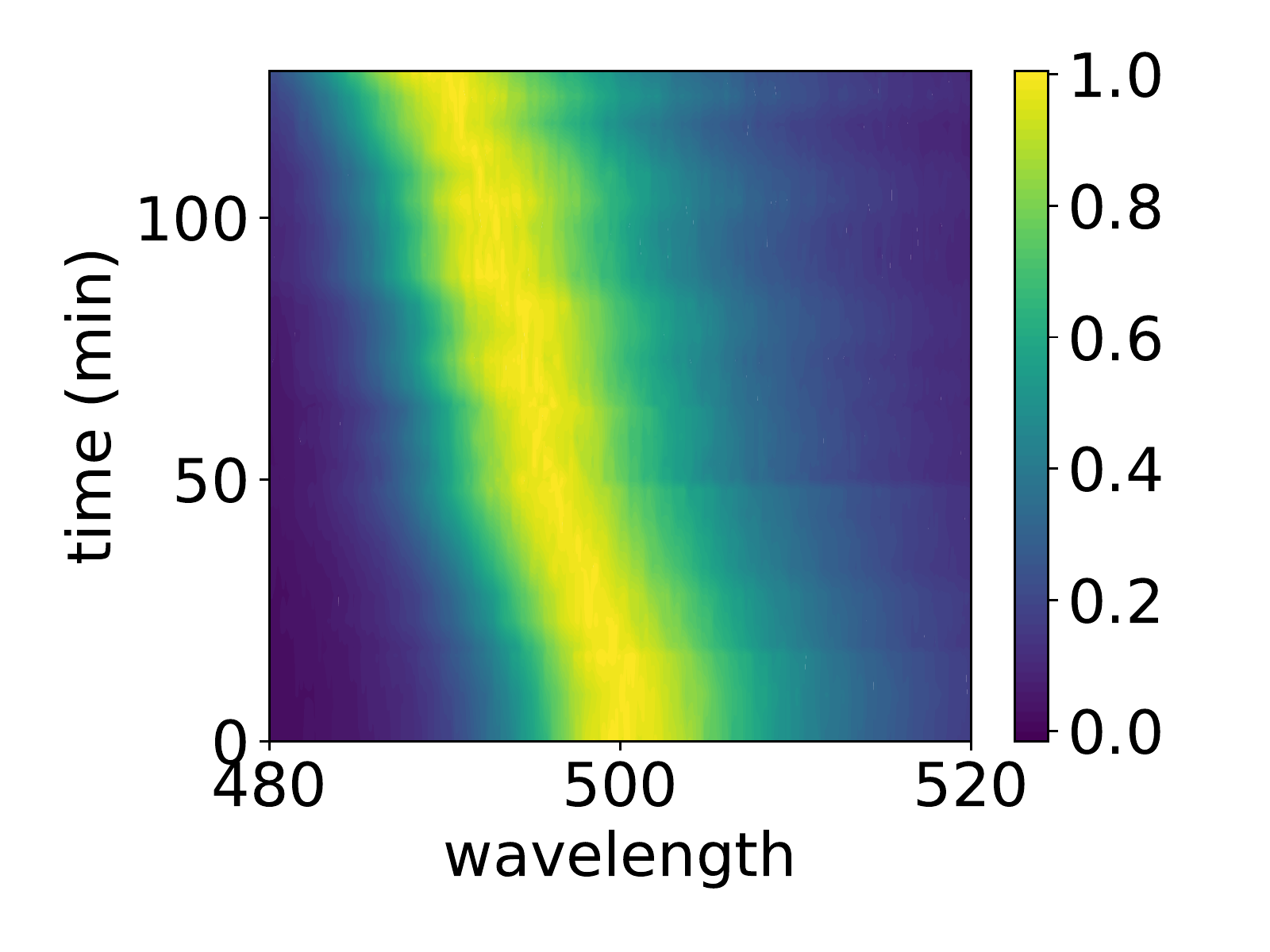}
	\caption{Évolution du spectre d'émission d'un émetteur de photons uniques de
	l'échantillon~B sur le temps d'illumination. L'émetteur était excité à
	l'intensité de saturation. Chaque spectre est normalisé.
	Une dérive d'approximativement  \SI{10}{\nano \meter} est visible après
	deux heures de mesure.
	}
	\label{fig:spectradegrationn-f}
\end{figure}
\begin{figure}[p]
	\centering
	\includegraphics[width=0.7\linewidth]{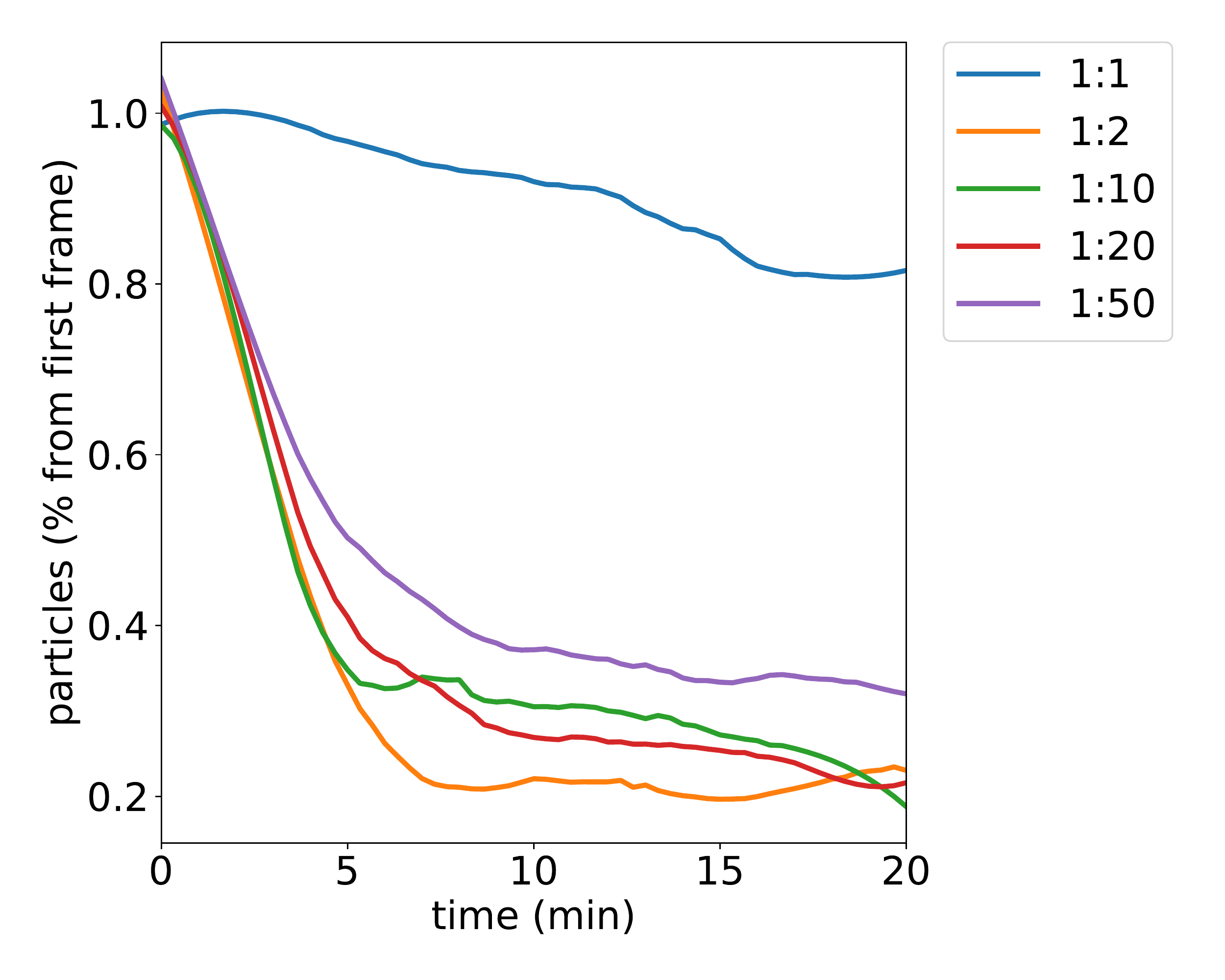}
	\caption{Mesure de la robustesse de l'échantillon de pérovskites avec des
	concentrations différentes. Les échantillons ont été obtenus en diluant la
	solution originale avec du toluène.}
	\label{fig:resistenza_diluizione-f}
\end{figure}
En figure~\ref{fig:spectradegrationn-f} est montrée la dégradation du spectre
d'émission après deux heures de mesure avec l'échantillon~B. La dégradation
est d'une dizaine de nanomètres, très inférieure
aux observations précédentes en littérature \textcite{raino2019Underestimated}. Dans ce travail une dérive équivalente avait été observée en seulement quelques dizaines de secondes.
Ce qui est intéressant est aussi l'effet de la dilution, la
figure~\ref{fig:resistenza_diluizione-f} montre qu'en diluant l'échantillon, on en dégrade rapidement les propriétés de stabilité.
La concentration joue donc un rôle crucial dans la stabilité.

\subsection{Distribution du \gdz{} en fonction de la longueur d'onde}
Avec plusieurs mesures de \gdz{} on peut vérifier s’il y a un effet de la
longueur d'onde sur l'émission.
\begin{figure}
	\centering
	\includegraphics[width=0.6\linewidth]{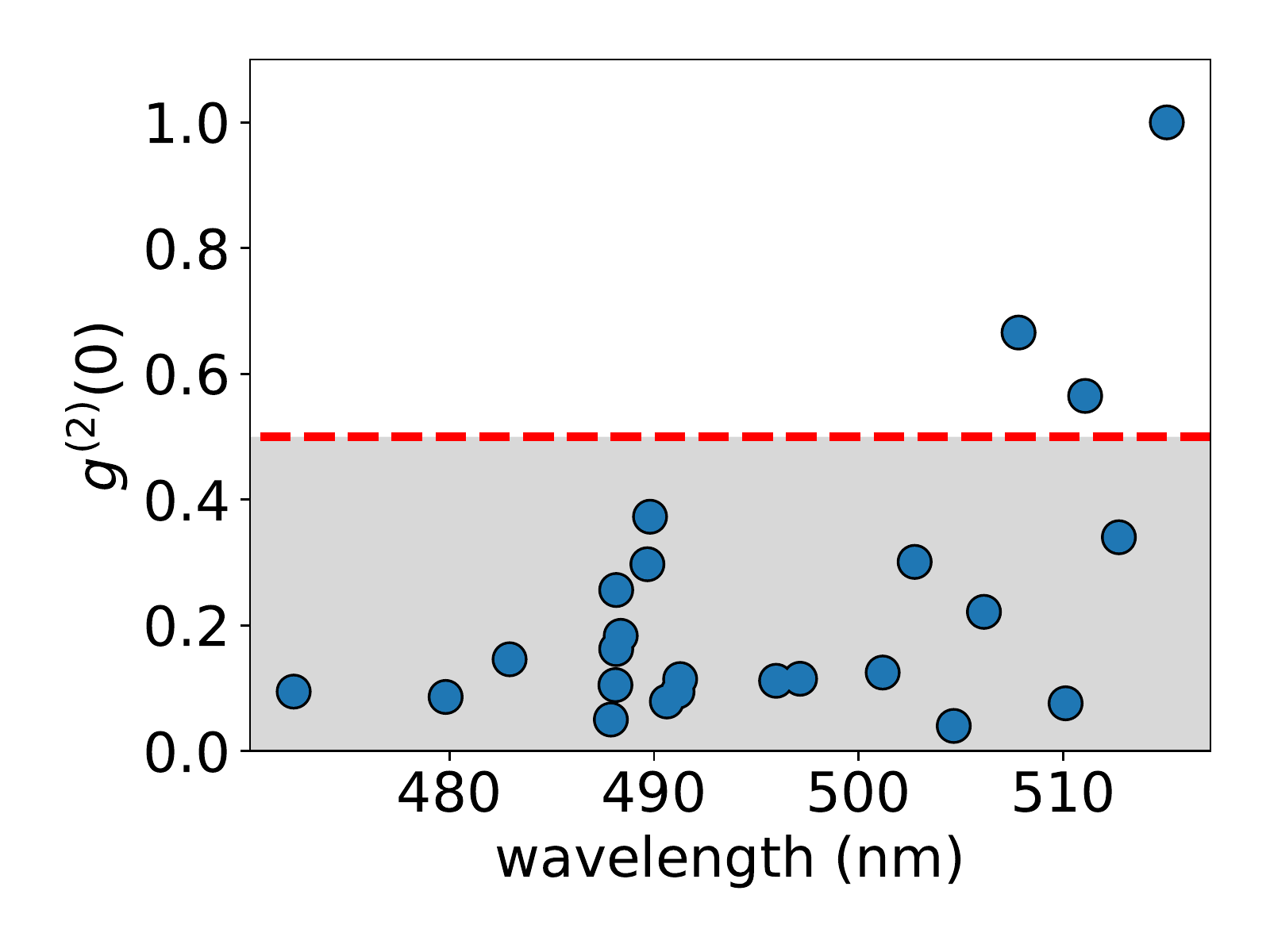}
	\caption{Distribution de \gdz{} en fonction de la longueur d'onde
	d'émission centrale pour 33 émetteurs différents excités à l'intensité de
	saturation}
	\label{fig:g2vslambdapsat-f}
\end{figure}
En particulier on peut voir dans la figure~\ref{fig:g2vslambdapsat-f} où le
\gdz{} pour trente-trois émetteurs différents est reporté, qu’en général on a
une détérioration du \gdz{} quand la longueur d'onde d'émission augmente. Cela est
dû au fait que les grandes longueurs d'onde d'émission correspondent aux émetteurs plus grands, pour lesquels l'effet du confinement quantique est inférieur, d'où la détérioration de l'émission de photons uniques.

\section{Nanofibres}
Les fibres optiques sont un moyen privilégié pour guider la lumière. Depuis
leur découverte, elles ont été améliorées et utilisées dans une grande quantité
d'applications différentes. Le guidage dans une fibre optique est dû
principalement au phénomène de la réflexion totale, ce qui comporte que, si la
surface est de bonne qualité (et on peut donc négliger la diffusion), les pertes dans la fibre sont minimales.

Les nanofibres optiques ont un diamètre inférieur à la
longueur d'onde de la lumière guidée, ce qui facilite le couplage de la
lumière émise par un émetteur directement dans la nanofibre. Généralement elles sont
obtenues en étirant une fibre optique standard de façon contrôlée: en
particulier dans notre expérience on utilise une flamme pure d'oxygène et
hydrogène pour réchauffer la fibre avant de l'étirer, mais l'utilisation d'un
laser est aussi possible. Le profil de zone de transition entre la fibre et la
nanofibre est important, pour réduire au minimum les pertes de lumière dans la propagation.

Le montage utilisé pour étirer les nanofibres est montré en
figure~\ref{fig:fiberpulling-f}.
\begin{figure}[tbp]
	\centering
	\includegraphics[width=0.7\linewidth]{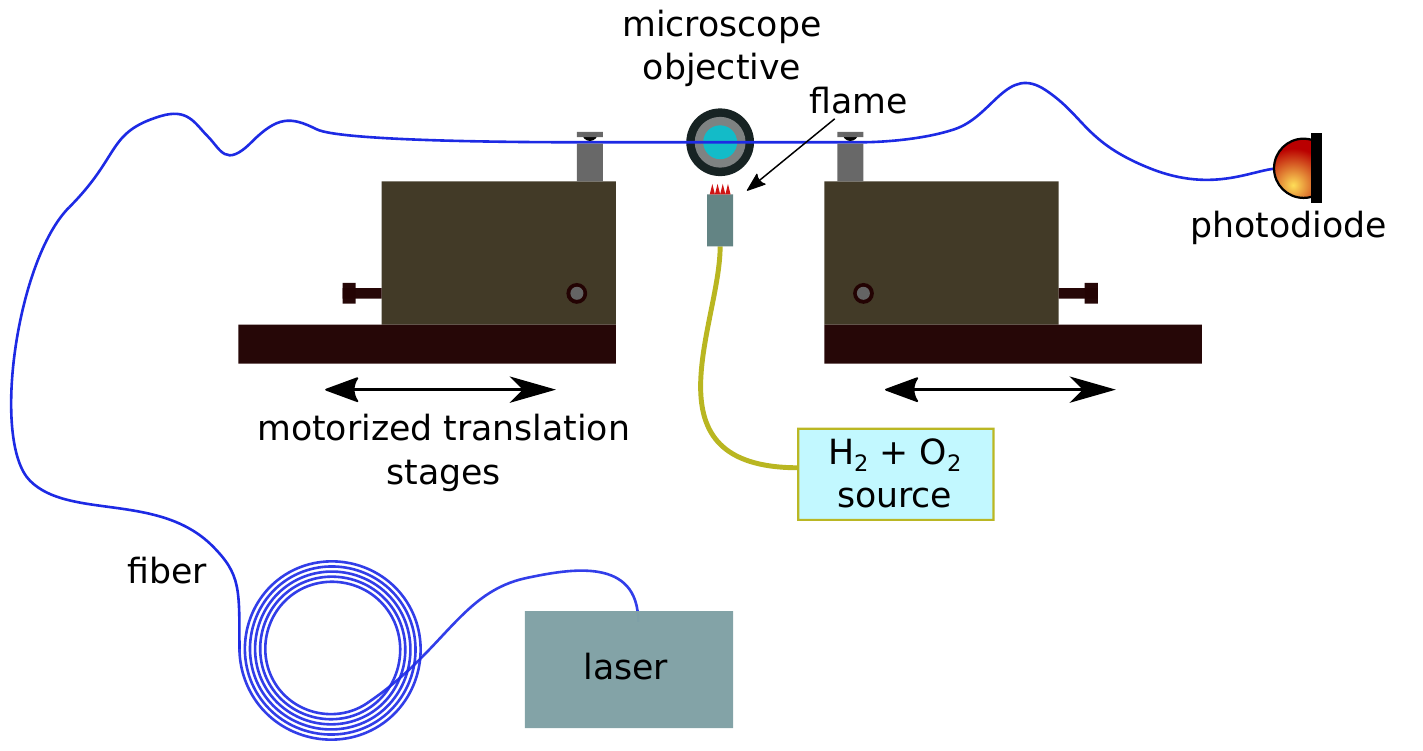}
	\caption{Montage expérimental pour étirer les nanofibres. Le laser est
	utilisé pour un surveillance constant de la transmission.}
	\label{fig:fiberpulling-f}
\end{figure}
\begin{figure}[tbp]
	\centering
	\includegraphics[width=0.8\linewidth]{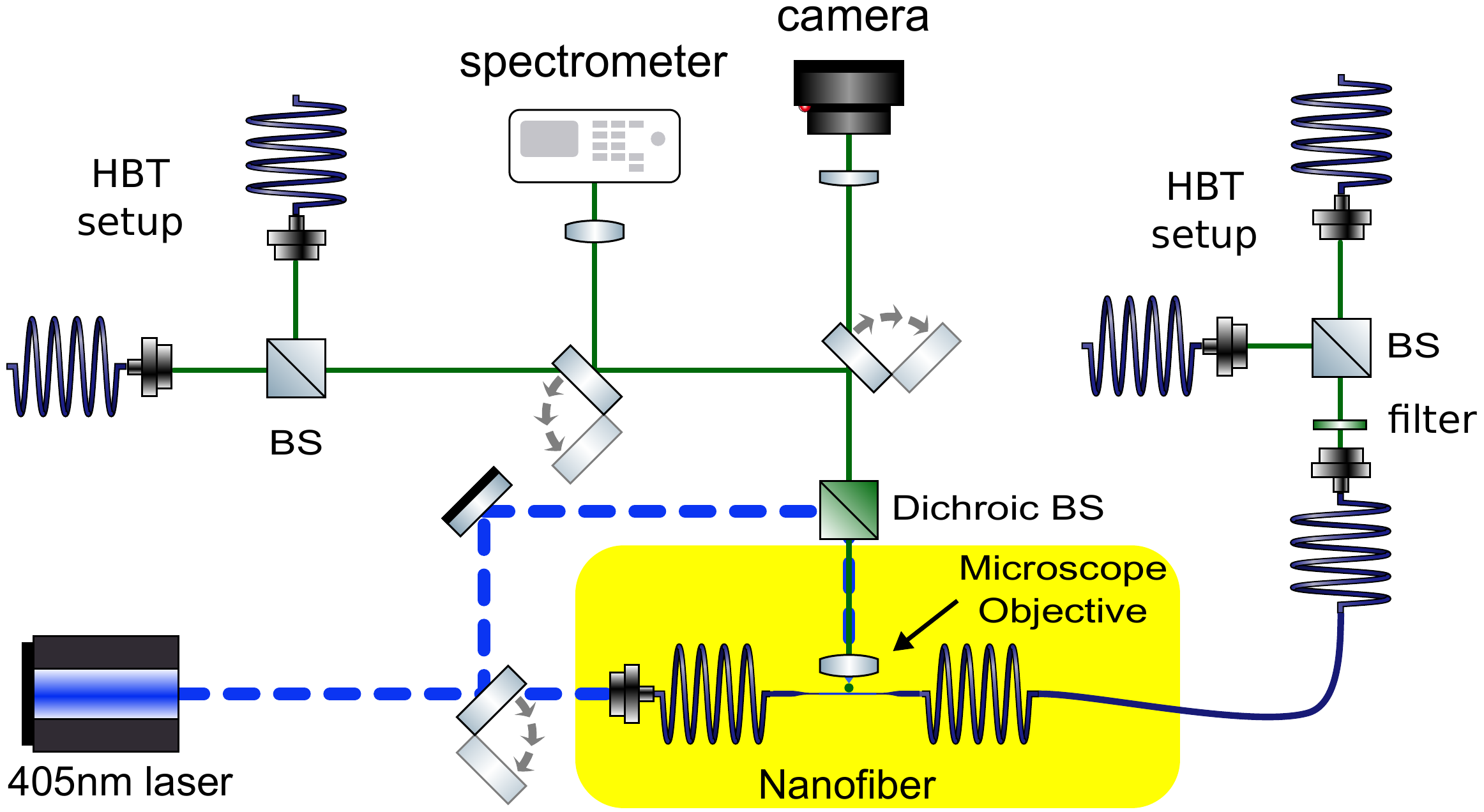}
	\caption{Montage pour l'étude d'un émetteur sur une nanofibre: le laser
		pulsé bleu peut exciter le nanoémetteur par la nanofibre ou par
		l'espace
		libre. L'image peut être prise par la caméra, le spectre et le \gd{}
		peuvent entre mesurés soit en espace libre soit en utilisant la
		lumière
		émise directement dans la fibre. BS: diviseur de faisceaux non
		polariseur,
		HBT: Hanbury Brown
		and Twiss.}
	\label{fig:setupfibra-f}
\end{figure}
Une flamme d'oxygène et hydrogène est utilisé pour réchauffer la nanofibre. Les
étages de translation, contrôlés par ordinateur, étirent et déplacent la
nanofibre pour obtenir le profil désiré.
Les mouvements nécessaires pour obtenir ce profil sont calculés par ordinateur
en utilisant la technique développé par~\textcite{birks1992Shape}. Dans ce
calcul on tient compte de la transmission souhaitée ainsi que des contraintes
expérimentales dont la plus importante est la longueur maximale de la région de
transition, limitée à quelques centimètres.

Une fois la fibre étirée, on la déplace sur le montage de l'expérience, pour la coupler avec
les émetteurs. Le montage expérimental utilisé est montré en
figure~\ref{fig:setupfibra-f}. Ce type de montage permet, entre autre, de
mesurer le \gd{} en utilisant des photons émis directement dans la fibre via le couplage de
champ proche.

La déposition d'un émetteur sur la nanofibre s'effectue en la touchant
délicatement avec une goutte de solution contenant les émetteurs. Un émetteur peut
alors rester collé à la fibre. On peut alors exciter directement le nanoémetteur via la nanofibre pour mesurer le \gd{} et
vérifier si l'on a effectivement un émetteur unique couplé avec la nanofibre.
Un exemple de \gd{} obtenu sur des photons émis par un nanocristal de
pérovskite directement couplé à la nanofibre est montré dans la
figure~\ref{fig:fig9pltem3norm-f}.
\begin{figure}
	\centering
	\includegraphics[width=0.5\linewidth]{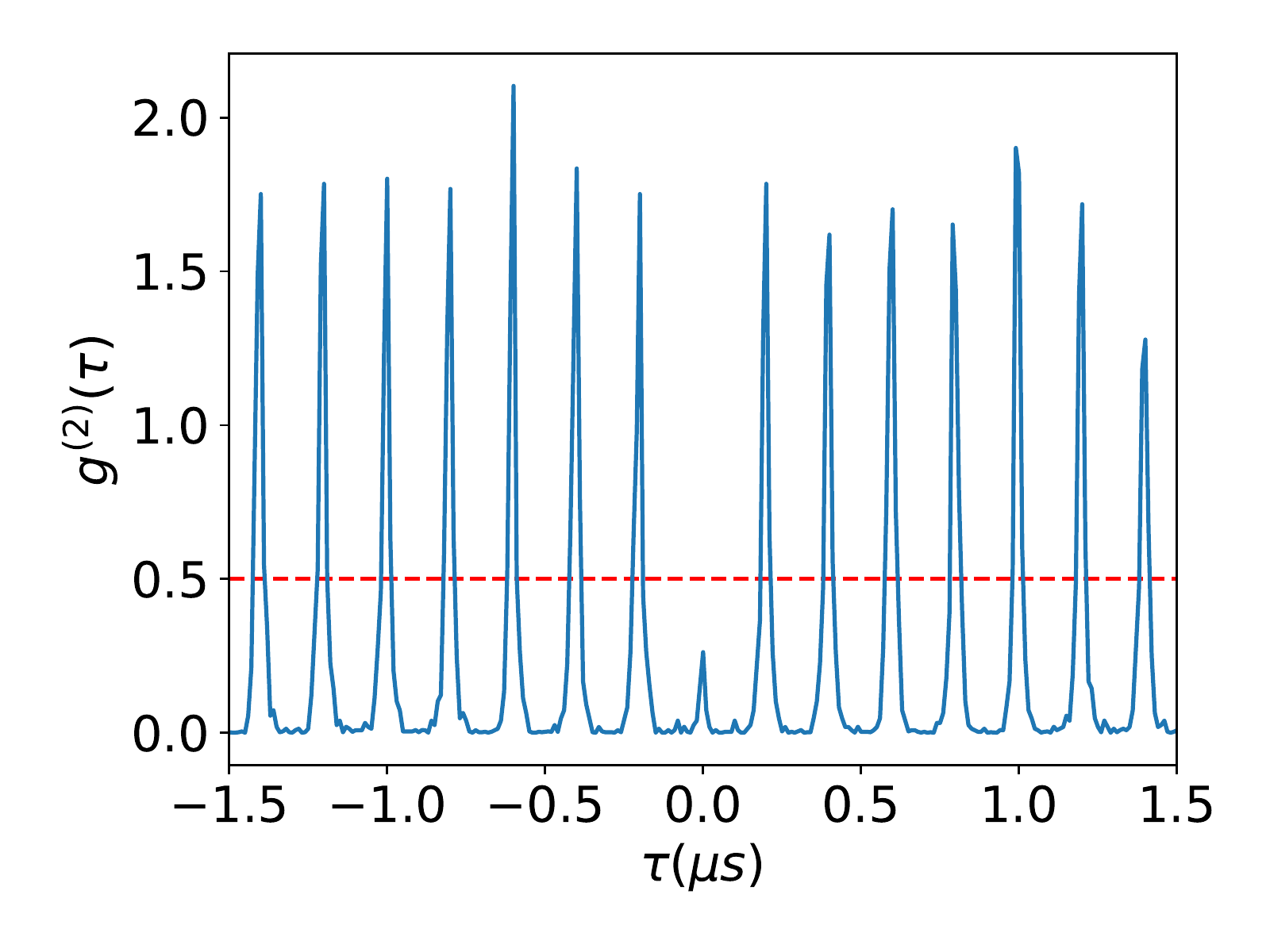}
	\caption{Fonction d'autocorrélation \gd{} d'un nanocristal unique de pérovskite déposé sur
	une nanofibre: les photons sont émis directement dans la nanofibre et
	mesurés à sa sortie. On obtient $ g^{(2)}\of{0}=0.24$.}
	\label{fig:fig9pltem3norm-f}
\end{figure}

\section{Perspectives}
Dans ma thèse je me suis concentré sur l'étude des nanocristaux de pérovskites
couplés à des nanofibres optiques. Ce système présente cependant des limites,
surtout dus, d'une part, à l'instabilité des pérovskites et, d'autre part, à la
fragilité des fibres optiques.

Concernant ce dernier aspect, j'ai eu l'occasion d'étudier des systèmes différents basés sur de guides d'onde à échange ionique qui
pourraient résoudre le problème.

\subsection{Emetteurs}
Pour ce qui concerne les émetteurs, les défauts réticulaires dans les
nanodiamants sont une option qui pourrait résoudre le problème de stabilité des
pérovskites.
La présence d'un défaut va créer dans le diamant un système de bandes
similaire à celui des semi-conducteurs, et il a été montré que ces défauts peuvent émettre des photons uniques. En particulier, les centres \ch{SiV} ont un
spectre d'émission fin et la plupart de leur émission (\SI{70}{\percent}) est dans
la \eng{zero phonon
line}, même à température ambiante. S'ils présentent l'avantage d'être robustes, leur fabrication est beaucoup plus
complexe de celle des nanocristaux de pérovskites. Dans l'échantillon que j'ai
pu analyser, je n'ai pas trouvé d'émetteurs uniques.

\subsection{Guides d'onde à échange ionique}
Une solution pour résoudre le problème de la fragilité des nanofibres peut être d'utiliser des structures
de guidage de la lumière inscrites dans le verre. J'ai eu l'occasion d'étudier
les guides d'onde à échange ionique. En particulier, en travaillant avec Giuseppe
Lio, nous avons étudié un système pour déposer des émetteurs sur le guide~\cite{lio2019Integration}.

En ajoutant une photorésine à la solution qui contient les
émetteurs et en déposant une couche de cette solution sur le verre qui contient
les guides, il est possible d'utiliser la polymérisation induite
 par le champ proche et il est ainsi possible de durcir la couche proche du guide.
 En lavant ensuite l'échantillon avec un solvant adapté, seuls les émetteurs présents sur le guide d'onde
 vont rester. En effet, on a vérifié que l'émission des émetteurs était
 présente à la fin du processus et nous avons caractérisé l'intensité de lumière et le temps nécessaire pour faire la
 couche la plus fine possible.
\end{otherlanguage}

\printbibliography

@article{glauber_quantum_1963,
	title = {The {Quantum} {Theory} of {Optical} {Coherence}},
	volume = {130},
	issn = {0031-899X},
	url = {https://link.aps.org/doi/10.1103/PhysRev.130.2529},
	doi = {10.1103/PhysRev.130.2529},
	language = {en},
	number = {6},
	urldate = {2019-11-08},
	journal = {Physical Review},
	author = {Glauber, Roy J.},
	month = jun,
	year = {1963},
	pages = {2529--2539},
}

@incollection{teich_i_1988,
	title = {I {Photon} {Bunching} and {Antibunching}},
	volume = {26},
	isbn = {978-0-444-87096-4},
	url = {https://linkinghub.elsevier.com/retrieve/pii/S0079663808701744},
	language = {en},
	urldate = {2019-11-25},
	booktitle = {Progress in {Optics}},
	publisher = {Elsevier},
	author = {Teich, Malvin C. and Saleh, Bahaa E.A.},
	year = {1988},
	doi = {10.1016/S0079-6638(08)70174-4},
	pages = {1--104},
}

@book{loudon2000quantum,
  title={The quantum theory of light},
  author={Loudon, Rodney},
  year={2000},
  publisher={OUP Oxford}
}

@article{katzPerovskitea,
	author = {Katz, Eugene  A.},
	title = {Perovskite: Name Puzzle and German-Russian Odyssey of Discovery},
	journal = {Helvetica Chimica Acta},
	pages = {},
	doi = {10.1002/hlca.202000061},
}

@article{jonker1950Ferromagnetica,
	title = {Ferromagnetic Compounds of Manganese with Perovskite Structure},
	author = {Jonker, G. H. and Van Santen, J. H.},
	year = {1950},
	month = mar,
	volume = {16},
	pages = {337--349},
	issn = {0031-8914},
	doi = {10.1016/0031-8914(50)90033-4},
	journal = {Physica},
	language = {en},
	number = {3}
}

@article{chung2012Allsolidstate,
	title = {All-Solid-State Dye-Sensitized Solar Cells with High Efficiency},
	author = {Chung, In and Lee, Byunghong and He, Jiaqing and Chang, Robert P.
	H. and Kanatzidis, Mercouri G.},
	year = {2012},
	month = may,
	volume = {485},
	pages = {486--489},
	publisher = {{Nature Publishing Group}},
	issn = {1476-4687},
	doi = {10.1038/nature11067},
	copyright = {2012 Nature Publishing Group, a division of Macmillan
	Publishers Limited. All Rights Reserved.},
	journal = {Nature},
	language = {en},
	number = {7399}
}

@article{deschler2019Perovskite,
	title = {Perovskite Semiconductors for next Generation Optoelectronic
	Applications},
	author = {Deschler, Felix and Neher, Dieter and {Schmidt-Mende}, Lukas},
	year = {2019},
	month = aug,
	volume = {7},
	pages = {080401},
	publisher = {{American Institute of Physics}},
	doi = {10.1063/1.5119744},
	journal = {APL Materials},
	number = {8}
}

@article{geng2016Localised,
	title = {Localised Excitation of a Single Photon Source by a
	Nanowave\-guide},
	author = {Geng, Wei and Manceau, Mathieu and Rahbany, Nancy and Sallet,
	Vincent and Vittorio, Massimo De and Carbone, Luigi and Glorieux, Quentin
	and Bramati, Alberto and Couteau, Christophe},
	year = {2016},
	month = jan,
	volume = {6},
	pages = {1--9},
	publisher = {{Nature Publishing Group}},
	issn = {2045-2322},
	doi = {10.1038/srep19721},
	copyright = {2016 The Author(s)},
	journal = {Scientific Reports},
	language = {en},
	number = {1}
}

@article{shafi2020Occurrence,
	title = {Occurrence Control of Charged Exciton for a Single {{CdSe}}
	Quantum Dot at Cryogenic Temperatures on an Optical Nanofiber},
	author = {Shafi, K. Muhammed and Iida, Kazunori and Tsutsumi, Emi and
	Miyanaga, Akiharu and Hakuta, Kohzo},
	year = {2020},
	month = mar,
	archivePrefix = {arXiv},
	eprint = {2003.10619},
	eprinttype = {arxiv},
	journal = {arXiv:2003.10619 [cond-mat, physics:quant-ph]},
	language = {en},
	primaryClass = {cond-mat, physics:quant-ph}
}

@article{park2015Room,
	title = {Room {{Temperature Single}}-{{Photon Emission}} from {{Individual
	Perovskite Quantum Dots}}},
	author = {Park, Young-Shin and Guo, Shaojun and Makarov, Nikolay S. and
	Klimov, Victor I.},
	year = {2015},
	month = oct,
	volume = {9},
	pages = {10386--10393},
	issn = {1936-0851, 1936-086X},
	doi = {10.1021/acsnano.5b04584},
	journal = {ACS Nano},
	language = {en},
	number = {10}
}

@article{raino2019Underestimated,
title = {Underestimated {{Effect}} of a {{Polymer Matrix}} on the {{Light
Emission}} of {{Single CsPbBr}} {\textsubscript{3}} {{Nanocrystals}}},
author = {Rain{\`o}, Gabriele and Landuyt, Annelies and Krieg, Franziska and
Bernasconi, Caterina and Ochsenbein, Stefan T. and Dirin, Dmitry N. and
Bodnarchuk, Maryna I. and Kovalenko, Maksym V.},
year = {2019},
month = jun,
volume = {19},
pages = {3648--3653},
issn = {1530-6984, 1530-6992},
doi = {10.1021/acs.nanolett.9b00689},
journal = {Nano Letters},
language = {en},
number = {6}
}

@article{protesescu2015nanocrystals,
	title={Nanocrystals of cesium lead halide perovskites (\ch{CsPbX3}, X= Cl,
	Br, and I): novel optoelectronic materials showing bright emission with
	wide color gamut},
	author={Protesescu, Loredana and Yakunin, Sergii and Bodnarchuk, Maryna I
	and Krieg, Franziska and Caputo, Riccarda and Hendon, Christopher H and
	Yang, Ruo Xi and Walsh, Aron and Kovalenko, Maksym V},
	journal={Nano letters},
	volume={15},
	number={6},
	pages={3692--3696},
	year={2015},
	publisher={ACS Publications}
}

@article{weidman2016Highly,
	title = {Highly {{Tunable Colloidal Perovskite Nanoplatelets}} through
	{{Variable Cation}}, {{Metal}}, and {{Halide Composition}}},
	author = {Weidman, Mark C. and Seitz, Michael and Stranks, Samuel D. and
	Tisdale, William A.},
	year = {2016},
	month = aug,
	volume = {10},
	pages = {7830--7839},
	publisher = {{American Chemical Society}},
	issn = {1936-0851},
	doi = {10.1021/acsnano.6b03496},
	journal = {ACS Nano},
	number = {8}
}

@misc{tarafdar2018Droplet,
	title = {Droplet {{Drying Patterns}} on {{Solid Substrates}}: {{From
	Hydrophilic}} to {{Superhydrophobic Contact}} to {{Levitating Drops}}},
	shorttitle = {Droplet {{Drying Patterns}} on {{Solid Substrates}}},
	author = {Tarafdar, Sujata and Tarasevich, Yuri Yu and Dutta Choudhury,
	Moutushi and Dutta, Tapati and Zang, Duyang},
	year = {2018},
	volume = {2018},
	pages = {e5214924},
	publisher = {{Hindawi}},
	issn = {1687-8108},
	doi = {10.1155/2018/5214924},
	journal = {Advances in Condensed Matter Physics},
	language = {en},
	type = {Review {{Article}}}
}

@article{berry1977Measurement,
	title = {Measurement of the Stokes Parameters of Light},
	author = {Berry, H. G. and Gabrielse, G. and Livingston, A. E.},
	year = {1977},
	month = dec,
	volume = {16},
	pages = {3200--3205},
	issn = {2155-3165},
	doi = {10.1364/AO.16.003200},
	journal = {Applied Optics},
	language = {EN},
	number = {12}
}

@article{shannon1949Communication,
	title = {Communication in the {{Presence}} of {{Noise}}},
	author = {Shannon, C.E.},
	year = {1949},
	month = jan,
	volume = {37},
	pages = {10--21},
	issn = {2162-6634},
	doi = {10.1109/JRPROC.1949.232969},
	journal = {Proceedings of the IRE},
	number = {1}
}

@article{andreev2017Selfcalibrating,
	title = {A Self-Calibrating Polarimeter to Measure {{Stokes}} Parameters},
	author = {Andreev, V. and Panda, C. D. and Hess, P. W. and Spaun, B. and
	Gabrielse, G.},
	year = {2017},
	month = feb,
	archivePrefix = {arXiv},
	eprint = {1703.00963},
	eprinttype = {arxiv},
	journal = {arXiv:1703.00963 [physics]},
	primaryClass = {physics}
}

@article{galland2011Two,
	title = {Two Types of Luminescence Blinking Revealed by
	Spectroelectrochemistry of Single Quantum Dots},
	author = {Galland, Christophe and Ghosh, Yagnaseni and Steinbr{\"u}ck,
	Andrea and Sykora, Milan and Hollingsworth, Jennifer A. and Klimov, Victor
	I. and Htoon, Han},
	year = {2011},
	month = nov,
	volume = {479},
	pages = {203--207},
	publisher = {{Nature Publishing Group}},
	issn = {1476-4687},
	doi = {10.1038/nature10569},
	copyright = {2011 Nature Publishing Group, a division of Macmillan
	Publishers Limited. All Rights Reserved.},
	journal = {Nature},
	language = {en},
	number = {7372}
}

@book{mach2013Principles,
	title={The principles of physical optics: an historical and philosophical
	treatment},
	author={Mach, Ernst},
	year={2013},
	publisher={Courier Corporation}
}

@article{colladon1842reflections,
	title={On the reflections of a ray of light inside a parabolic liquid
	stream},
	author={Colladon, Daniel},
	journal={Comptes Rendus},
	volume={15},
	pages={800--802},
	year={1842}
}

@book{saleh2007Fundamentals,
	title = {Fundamentals of Photonics},
	author = {Saleh, Bahaa E. A. and Teich, Malvin Carl},
	year = {2007},
	edition = {2nd ed},
	publisher = {{Wiley Interscience}},
	address = {{Hoboken, N.J}},
	isbn = {978-0-471-35832-9},
	lccn = {TA1520 .S24 2007},
	series = {Wiley Series in Pure and Applied Optics}
}

@article{tong2004Singlemode,
	title = {Single-Mode Guiding Properties of Sub\-wa\-ve\-length-Diameter
	Silica
	and Silicon Wire Waveguides},
	author = {Tong, Limin and Lou, Jingyi and Mazur, Eric},
	year = {2004},
	month = mar,
	volume = {12},
	pages = {1025--1035},
	publisher = {{Optical Society of America}},
	issn = {1094-4087},
	doi = {10.1364/OPEX.12.001025},
	copyright = {\&\#169; 2004 Optical Society of America},
	journal = {Optics Express},
	language = {EN},
	number = {6}
}

@article{birks1992Shape,
	title = {The Shape of Fiber Tapers},
	author = {Birks, T.A. and Li, Y.W.},
	year = {1992},
	month = apr,
	volume = {10},
	pages = {432--438},
	issn = {07338724},
	doi = {10.1109/50.134196},
	journal = {Journal of Lightwave Technology},
	number = {4}
}

@article{dewynne1989mathematical,
	title={On a mathematical model for fiber tapering},
	author={Dewynne, J and Ockendon, JR and Wilmott, P},
	journal={SIAM Journal on Applied Mathematics},
	volume={49},
	number={4},
	pages={983--990},
	year={1989},
	publisher={SIAM}
}

@article{nagai2014Ultralowloss,
	title = {Ultra-Low-Loss Tapered Optical Fibers with Minimal Lengths},
	author = {Nagai, Ryutaro and Aoki, Takao},
	year = {2014},
	month = nov,
	volume = {22},
	pages = {28427},
	issn = {1094-4087},
	doi = {10.1364/OE.22.028427},
	journal = {Optics Express},
	language = {en},
	number = {23}
}

@book{griffiths2018introduction,
	title={Introduction to quantum mechanics},
	author={Griffiths, David J and Schroeter, Darrell F},
	year={2018},
	publisher={Cambridge University Press}
}

@book{sakurai2011Modern,
	title = {Modern Quantum Mechanics},
	author = {Sakurai, J. J. and Napolitano, Jim},
	year = {2011},
	edition = {2nd ed},
	publisher = {{Addison-Wesley}},
	address = {{Boston}},
	isbn = {978-0-8053-8291-4},
	lccn = {QC174.12 .S25 2011}
}

@article{wootters1982Single,
	title = {A Single Quantum Cannot Be Cloned},
	author = {Wootters, W. K. and Zurek, W. H.},
	year = {1982},
	month = oct,
	volume = {299},
	pages = {802--803},
	publisher = {{Nature Publishing Group}},
	issn = {1476-4687},
	doi = {10.1038/299802a0},
	copyright = {1982 Nature Publishing Group},
	journal = {Nature},
	language = {en},
	number = {5886}
}

@article{arute2019Quantum,
	title = {Quantum Supremacy Using a Programmable Superconducting Processor},
	author = {Arute, Frank and Arya, Kunal and Babbush, Ryan and Bacon, Dave
	and Bardin, Joseph C. and Barends, Rami and Biswas, Rupak and Boixo, Sergio
	and Brandao, Fernando G. S. L. and Buell, David A. and Burkett, Brian and
	Chen, Yu and Chen, Zijun and Chiaro, Ben and Collins, Roberto and Courtney,
	William and Dunsworth, Andrew and Farhi, Edward and Foxen, Brooks and
	Fowler, Austin and Gidney, Craig and Giustina, Marissa and Graff, Rob and
	Guerin, Keith and Habegger, Steve and Harrigan, Matthew P. and Hartmann,
	Michael J. and Ho, Alan and Hoffmann, Markus and Huang, Trent and Humble,
	Travis S. and Isakov, Sergei V. and Jeffrey, Evan and Jiang, Zhang and
	Kafri, Dvir and Kechedzhi, Kostyantyn and Kelly, Julian and Klimov, Paul V.
	and Knysh, Sergey and Korotkov, Alexander and Kostritsa, Fedor and
	Landhuis, David and Lindmark, Mike and Lucero, Erik and Lyakh, Dmitry and
	Mandr{\`a}, Salvatore and McClean, Jarrod R. and McEwen, Matthew and
	Megrant, Anthony and Mi, Xiao and Michielsen, Kristel and Mohseni, Masoud
	and Mutus, Josh and Naaman, Ofer and Neeley, Matthew and Neill, Charles and
	Niu, Murphy Yuezhen and Ostby, Eric and Petukhov, Andre and Platt, John C.
	and Quintana, Chris and Rieffel, Eleanor G. and Roushan, Pedram and Rubin,
	Nicholas C. and Sank, Daniel and Satzinger, Kevin J. and Smelyanskiy, Vadim
	and Sung, Kevin J. and Trevithick, Matthew D. and Vainsencher, Amit and
	Villalonga, Benjamin and White, Theodore and Yao, Z. Jamie and Yeh, Ping
	and Zalcman, Adam and Neven, Hartmut and Martinis, John M.},
	year = {2019},
	month = oct,
	volume = {574},
	pages = {505--510},
	publisher = {{Nature Publishing Group}},
	issn = {1476-4687},
	doi = {10.1038/s41586-019-1666-5},
	copyright = {2019 The Author(s), under exclusive licence to Springer Nature
	Limited},
	journal = {Nature},
	language = {en},
	number = {7779}
}

@article{fitch2003Photonnumber,
	title = {Photon-Number Resolution Using Time-Multiplexed Single-Pho\-ton
	Detectors},
	author = {Fitch, M. J. and Jacobs, B. C. and Pittman, T. B. and Franson, J.
	D.},
	year = {2003},
	month = oct,
	volume = {68},
	pages = {043814},
	publisher = {{American Physical Society}},
	doi = {10.1103/PhysRevA.68.043814},
	journal = {Physical Review A},
	number = {4}
}

@article{jeffrey2004Periodic,
	title = {Towards a Periodic Deterministic Source of Arbitrary Single-Photon
	States},
	author = {Jeffrey, Evan and Peters, Nicholas A. and Kwiat, Paul G.},
	year = {2004},
	month = jul,
	volume = {6},
	pages = {100--100},
	publisher = {{IOP Publishing}},
	issn = {1367-2630},
	doi = {10.1088/1367-2630/6/1/100},
	journal = {New Journal of Physics},
	language = {en}
}

@article{migdall2002Tailoring,
	title = {Tailoring Single-Photon and Multiphoton Probabilities of a
	Single-Photon on-Demand Source},
	author = {Migdall, A. L. and Branning, D. and Castelletto, S.},
	year = {2002},
	month = nov,
	volume = {66},
	pages = {053805},
	publisher = {{American Physical Society}},
	doi = {10.1103/PhysRevA.66.053805},
	journal = {Physical Review A},
	number = {5}
}

@article{shapiro2007Ondemand,
	title = {On-Demand Single-Photon Generation Using a Modular Array of
	Parametric Downconverters with Electro-Optic Polarization Controls},
	author = {Shapiro, Jeffrey H. and Wong, Franco N.},
	year = {2007},
	month = sep,
	volume = {32},
	pages = {2698--2700},
	publisher = {{Optical Society of America}},
	issn = {1539-4794},
	doi = {10.1364/OL.32.002698},
	copyright = {\&\#169; 2007 Optical Society of America},
	journal = {Optics Letters},
	language = {EN},
	number = {18}
}

@article{eisaman2011Inviteda,
	ids = {eisaman2011Invited},
	title = {Invited {{Review Article}}: {{Single}}-Photon Sources and
	Detectors},
	shorttitle = {Invited {{Review Article}}},
	author = {Eisaman, M. D. and Fan, J. and Migdall, A. and Polyakov, S. V.},
	year = {2011},
	month = jul,
	volume = {82},
	pages = {071101},
	publisher = {{American Institute of Physics}},
	issn = {0034-6748, 1089-7623},
	doi = {10.1063/1.3610677},
	journal = {Review of Scientific Instruments},
	language = {en},
	number = {7}
}

@article{aoki2009Efficient,
	title = {Efficient {{Routing}} of {{Single Photons}} by {{One Atom}} and a
	{{Microtoroidal Cavity}}},
	author = {Aoki, Takao and Parkins, A. S. and Alton, D. J. and Regal, C. A.
	and Dayan, Barak and Ostby, E. and Vahala, K. J. and Kimble, H. J.},
	year = {2009},
	month = feb,
	volume = {102},
	pages = {083601},
	publisher = {{American Physical Society}},
	doi = {10.1103/PhysRevLett.102.083601},
	journal = {Physical Review Letters},
	number = {8}
}

@article{dayan2008Photon,
	title = {A {{Photon Turnstile Dynamically Regulated}} by {{One Atom}}},
	author = {Dayan, Barak and Parkins, A. S. and Aoki, Takao and Ostby, E. P.
	and Vahala, K. J. and Kimble, H. J.},
	year = {2008},
	month = feb,
	volume = {319},
	pages = {1062--1065},
	publisher = {{American Association for the Advancement of Science}},
	issn = {0036-8075, 1095-9203},
	doi = {10.1126/science.1152261},
	chapter = {Report},
	copyright = {American Association for the Advancement of Science},
	journal = {Science},
	language = {en},
	number = {5866},
	pmid = {18292335}
}

@article{hennrich2004Photon,
	title = {Photon Statistics of a Non-Stationary Periodically Driven
	Single-Photon Source},
	author = {Hennrich, M. and Legero, T. and Kuhn, A. and Rempe, G.},
	year = {2004},
	month = jul,
	volume = {6},
	pages = {86--86},
	publisher = {{IOP Publishing}},
	issn = {1367-2630},
	doi = {10.1088/1367-2630/6/1/086},
	journal = {New Journal of Physics},
	language = {en}
}

@article{hijlkema2007Singlephoton,
	title = {A Single-Photon Server with Just One Atom},
	author = {Hijlkema, Markus and Weber, Bernhard and Specht, Holger P. and
	Webster, Simon C. and Kuhn, Axel and Rempe, Gerhard},
	year = {2007},
	month = apr,
	volume = {3},
	pages = {253--255},
	publisher = {{Nature Publishing Group}},
	issn = {1745-2481},
	doi = {10.1038/nphys569},
	journal = {Nature Physics},
	language = {en},
	number = {4}
}

@article{kuhn2002Deterministic,
	title = {Deterministic {{Single}}-{{Photon Source}} for {{Distributed
	Quantum Networking}}},
	author = {Kuhn, Axel and Hennrich, Markus and Rempe, Gerhard},
	year = {2002},
	month = jul,
	volume = {89},
	pages = {067901},
	publisher = {{American Physical Society}},
	doi = {10.1103/PhysRevLett.89.067901},
	journal = {Physical Review Letters},
	number = {6}
}

@article{mckeever2004Deterministic,
	title = {Deterministic {{Generation}} of {{Single Photons}} from {{One Atom
	Trapped}} in a {{Cavity}}},
	author = {McKeever, J. and Boca, A. and Boozer, A. D. and Miller, R. and
	Buck, J. R. and Kuzmich, A. and Kimble, H. J.},
	year = {2004},
	month = mar,
	volume = {303},
	pages = {1992--1994},
	publisher = {{American Association for the Advancement of Science}},
	issn = {0036-8075, 1095-9203},
	doi = {10.1126/science.1095232},
	chapter = {Report},
	copyright = {American Association for the Advancement of Science},
	journal = {Science},
	language = {en},
	number = {5666},
	pmid = {14988512}
}

@article{wilk2007PolarizationControlled,
	title = {Polarization-{{Controlled Single Photons}}},
	author = {Wilk, T. and Webster, S. C. and Specht, H. P. and Rempe, G. and
	Kuhn, A.},
	year = {2007},
	month = feb,
	volume = {98},
	pages = {063601},
	publisher = {{American Physical Society}},
	doi = {10.1103/PhysRevLett.98.063601},
	journal = {Physical Review Letters},
	number = {6}
}

@incollection{purcell1995spontaneous,
	title={Spontaneous emission probabilities at radio frequencies},
	author={Purcell, Edward Mills},
	booktitle={Confined Electrons and Photons},
	pages={839--839},
	year={1995},
	publisher={Springer}
}

@article{vitanov2017Stimulated,
	title = {Stimulated {{Raman}} Adiabatic Passage in Physics, Chemistry, and
	Beyond},
	author = {Vitanov, Nikolay V. and Rangelov, Andon A. and Shore, Bruce W.
	and Bergmann, Klaas},
	year = {2017},
	month = mar,
	volume = {89},
	pages = {015006},
	publisher = {{American Physical Society}},
	doi = {10.1103/RevModPhys.89.015006},
	journal = {Reviews of Modern Physics},
	number = {1}
}

@article{home2009Complete,
	title = {Complete {{Methods Set}} for {{Scalable Ion Trap Quantum
	Information Processing}}},
	author = {Home, Jonathan P. and Hanneke, David and Jost, John D. and Amini,
	Jason M. and Leibfried, Dietrich and Wineland, David J.},
	year = {2009},
	month = sep,
	volume = {325},
	pages = {1227--1230},
	publisher = {{American Association for the Advancement of Science}},
	issn = {0036-8075, 1095-9203},
	doi = {10.1126/science.1177077},
	chapter = {Report},
	copyright = {Copyright \textcopyright{} 2009, American Association for the
	Advancement of Science},
	journal = {Science},
	language = {en},
	number = {5945},
	pmid = {19661380}
}

@article{kielpinski2002Architecture,
	title = {Architecture for a Large-Scale Ion-Trap Quantum Computer},
	author = {Kielpinski, D. and Monroe, C. and Wineland, D. J.},
	year = {2002},
	month = jun,
	volume = {417},
	pages = {709--711},
	publisher = {{Nature Publishing Group}},
	issn = {1476-4687},
	doi = {10.1038/nature00784},
	copyright = {2002 Macmillan Magazines Ltd.},
	journal = {Nature},
	language = {en},
	number = {6890}
}

@article{riebe2008Deterministic,
	title = {Deterministic Entanglement Swapping with an Ion-Trap Quantum
	Computer},
	author = {Riebe, M. and Monz, T. and Kim, K. and Villar, A. S. and
	Schindler, P. and Chwalla, M. and Hennrich, M. and Blatt, R.},
	year = {2008},
	month = nov,
	volume = {4},
	pages = {839--842},
	publisher = {{Nature Publishing Group}},
	issn = {1745-2481},
	doi = {10.1038/nphys1107},
	journal = {Nature Physics},
	language = {en},
	number = {11}
}

@article{hadden2018Integrated,
	title = {Integrated Waveguides and Deterministically Positioned Nitrogen
	Vacancy Centers in Diamond Created by Femtosecond Laser Writing},
	author = {Hadden, J. P. and Bharadwaj, V. and Sotillo, B. and Rampini, S.
	and Osellame, R. and Witmer, J. D. and Jayakumar, H. and Fernandez, T. T.
	and Chiappini, A. and Armellini, C. and Ferrari, M. and Ramponi, R. and
	Barclay, P. E. and Eaton, S. M.},
	year = {2018},
	month = aug,
	volume = {43},
	pages = {3586--3589},
	publisher = {{Optical Society of America}},
	issn = {1539-4794},
	doi = {10.1364/OL.43.003586},
	copyright = {\&\#169; 2018 Optical Society of America},
	journal = {Optics Letters},
	language = {EN},
	number = {15}
}

@article{lindnerStrongly,
	title = {Strongly Inhomogeneous Distribution of Spectral Properties of
	Silicon-Vacancy Color Centers in Nanodiamonds},
	author = {Lindner, Sarah and Bommer, Alexander and Muzha, Andreas and
	Gines, Laia and Mandal, Soumen and Williams, Oliver and Gali, Adam and
	Becher, Christoph},
	pages = {24},
	language = {en}
}

@article{neu2011Single,
	title = {Single Photon Emission from Silicon-Vacancy Colour Centres in
	Chemical Vapour Deposition Nano-Diamonds on Iridium},
	author = {Neu, Elke and Steinmetz, David and {Riedrich-M{\"o}ller}, Janine
	and Gsell, Stefan and Fischer, Martin and {Matthias Schreck} and Becher,
	Christoph},
	year = {2011},
	volume = {13},
	pages = {025012},
	issn = {1367-2630},
	doi = {10.1088/1367-2630/13/2/025012},
	journal = {New Journal of Physics},
	language = {en},
	number = {2}
}

@article{wang2006Single,
	title = {Single Photon Emission from {{SiV}} Centres in Diamond Produced by
	Ion Implantation},
	author = {Wang, Chunlang and Kurtsiefer, Christian and Weinfurter, Harald
	and Burchard, Bernd},
	year = {2006},
	volume = {39},
	pages = {37},
	issn = {0953-4075},
	doi = {10.1088/0953-4075/39/1/005},
	journal = {Journal of Physics B: Atomic, Molecular and Optical Physics},
	language = {en},
	number = {1}
}

@article{jelezko2006Single,
	title = {Single Defect Centres in Diamond: {{A}} Review},
	shorttitle = {Single Defect Centres in Diamond},
	author = {Jelezko, F. and Wrachtrup, J.},
	year = {2006},
	month = oct,
	volume = {203},
	pages = {3207--3225},
	issn = {1862-6319},
	doi = {10.1002/pssa.200671403},
	copyright = {Copyright \textcopyright{} 2006 WILEY-VCH Verlag GmbH \& Co.
	KGaA, Weinheim},
	journal = {physica status solidi (a)},
	language = {en},
	number = {13}
}

@article{meitner1922Ueber,
	title = {{\"Uber die Entstehung der {$\beta$}-Strahl-Spektren radioaktiver
	Substanzen}},
	author = {Meitner, Lise},
	year = {1922},
	month = dec,
	volume = {9},
	pages = {131--144},
	issn = {0044-3328},
	doi = {10.1007/BF01326962},
	journal = {Zeitschrift f\"ur Physik},
	language = {de},
	number = {1}
}

@article{klimov2000Quantization,
	title = {Quantization of {{Multiparticle Auger Rates}} in {{Semiconductor
	Quantum Dots}}},
	author = {Klimov, V. I. and Mikhailovsky, A. A. and McBranch, D. W. and
	Leatherdale, C. A. and Bawendi, M. G.},
	year = {2000},
	month = feb,
	volume = {287},
	pages = {1011--1013},
	publisher = {{American Association for the Advancement of Science}},
	issn = {0036-8075, 1095-9203},
	doi = {10.1126/science.287.5455.1011},
	chapter = {Report},
	journal = {Science},
	language = {en},
	number = {5455},
	pmid = {10669406}
}

@article{robel2009Universal,
	title = {Universal {{Size}}-{{Dependent Trend}} in {{Auger Recombination}}
	in {{Direct}}-{{Gap}} and {{Indirect}}-{{Gap Semiconductor Nanocrystals}}},
	author = {Robel, Istv{\'a}n and Gresback, Ryan and Kortshagen, Uwe and
	Schaller, Richard D. and Klimov, Victor I.},
	year = {2009},
	month = may,
	volume = {102},
	pages = {177404},
	publisher = {{American Physical Society}},
	doi = {10.1103/PhysRevLett.102.177404},
	journal = {Physical Review Letters},
	number = {17}
}

@phdthesis{vezzoli2013Experimental,
	title = {Experimental Study of Nanocrystals as Single Photon Sources},
	author = {Vezzoli, Stefano},
	year = {2013},
	month = jan,
	collaborator = {Bramati, Alberto and Cialdi, Simone},
	copyright = {Licence Etalab},
	school = {Paris 6},
	type = {These de Doctorat}
}

@article{manceau2018CdSe,
	title = {{{CdSe}}/{{CdS Dot}}-in-{{Rods Nanocrystals Fast Blinking
	Dynamics}}.},
	author = {Manceau, M. and Vezzoli, S. and Glorieux, Q. and Giacobino, E.
	and Carbone, L. and De Vittorio, M. and Hermier, J.-P. and Bramati, A.},
	year = {2018},
	volume = {19},
	pages = {3288--3295},
	issn = {1439-7641},
	doi = {10.1002/cphc.201800694},
	journal = {ChemPhysChem},
	language = {en},
	number = {23}
}

@article{euler1738progressionibus,
	title={De progressionibus transcendentibus seu quarum termini generales
	algebraice dari nequeunt},
	author={Euler, Leonhard},
	journal={Commentarii academiae scientiarum Petropolitanae},
	pages={36--57},
	year={1738}
}

@book{davies1994Properties,
	title = {Properties and Growth of Diamond},
	editor = {Davies, Gordon and {Institution of Electrical Engineers}},
	year = {1994},
	publisher = {{INSPEC, the Institution of Electrical Engineers}},
	address = {{London}},
	isbn = {978-0-85296-875-8},
	language = {eng},
	number = {9},
	series = {{{EMIS}} Datareviews Series}
}

@article{zeleneev2020Nanodiamonds,
	title = {Nanodiamonds with {{SiV}} Colour Centres for Quantum Technologies},
	author = {Zeleneev, A. I. and Bolshedvorskii, S. V. and Soshenko, V. V. and
	Rubinas, O. R. and Garanina, A. S. and Lyapin, S. G. and Agafonov, V. N.
	and Uzbekov, R. E. and Kudryavtsev, O. S. and Sorokin, V. N. and
	Smolyaninov, A. N. and Davydov, V. A. and Akimov, A. V.},
	year = {2020},
	month = mar,
	volume = {50},
	pages = {299},
	publisher = {{IOP Publishing}},
	issn = {1063-7818},
	doi = {10.1070/QEL17189},
	journal = {Quantum Electronics},
	language = {en},
	number = {3}
}

@article{tervonen2011Ionexchanged,
	title = {Ion-Exchanged Glass Wave\-guide Technology: A Review},
	shorttitle = {Ion-Exchanged Glass Waveguide Technology},
	author = {Tervonen, Ari and Honkanen, Seppo K. and West, Brian R.},
	year = {2011},
	month = jul,
	volume = {50},
	pages = {071107},
	issn = {0091-3286, 1560-2303},
	doi = {10.1117/1.3559213},
	journal = {Optical Engineering},
	number = {7}
}

@article{walker1983Integrated,
	title = {Integrated Optical Waveguiding Structures Made by Silver
	Ion-Exchange in Glass. 1: {{The}} Propagation Characteristics of Stripe
	Ion-Exchanged Waveguides; a Theoretical and Experimental Investigation},
	shorttitle = {Integrated Optical Waveguiding Structures Made by Silver
	Ion-Exchange in Glass. 1},
	author = {Walker, R. G. and Wilkinson, C. D. W. and Wilkinson, J. a. H.},
	year = {1983},
	month = jun,
	volume = {22},
	pages = {1923--1928},
	publisher = {{Optical Society of America}},
	issn = {2155-3165},
	doi = {10.1364/AO.22.001923},
	copyright = {\&\#169; 1983 Optical Society of America},
	journal = {Applied Optics},
	language = {EN},
	number = {12}
}

@article{weiss1995Determination,
	title = {Determination of Ion-Exchanged Channel Waveguide Profile
	Parameters by Mode-Index Measurements},
	author = {Weiss, Martin N. and Srivastava, Ramakant},
	year = {1995},
	month = jan,
	volume = {34},
	pages = {455--458},
	publisher = {{Optical Society of America}},
	issn = {2155-3165},
	doi = {10.1364/AO.34.000455},
	copyright = {\&\#169; 1995 Optical Society of America},
	journal = {Applied Optics},
	language = {EN},
	number = {3}
}

@article{beltranmadrigal2016Hybrid,
	title = {Hybrid Integrated Optical Waveguides in Glass for Enhanced Visible
	Photoluminescence of Nanoemitters},
	author = {Beltran Madrigal, Josslyn and {Tellez-Limon}, Ricardo and
	Gardillou, Florent and Barbier, Denis and Geng, Wei and Couteau, Christophe
	and {Salas-Montiel}, Rafael and Blaize, Sylvain},
	year = {2016},
	month = dec,
	volume = {55},
	pages = {10263},
	issn = {0003-6935, 1539-4522},
	doi = {10.1364/AO.55.010263},
	journal = {Applied Optics},
	language = {en},
	number = {36}
}

@article{lio2019Integration,
	title = {Integration of {{Nanoemitters}} onto {{Photonic Structures}} by
	{{Guided Evanescent}}-{{Wave Nano}}-{{Photopolymerization}}},
	author = {Lio, Giuseppe Emanuele and Madrigal, Josslyn Beltran and Xu,
	Xiaolun and Peng, Ying and Pierini, Stefano and Couteau, Christophe and
	Jradi, Safi and Bachelot, Renaud and Caputo, Roberto and Blaize, Sylvain},
	year = {2019},
	month = jun,
	volume = {123},
	pages = {14669--14676},
	publisher = {{American Chemical Society}},
	issn = {1932-7447},
	doi = {10.1021/acs.jpcc.9b03716},
	journal = {The Journal of Physical Chemistry C},
	number = {23}
}

@article{cumpston1999Twophoton,
	title = {Two-Photon Polymerization Initiators for Three-Di\-men\-sio\-nal
	Optical
	Data Storage and Microfabrication},
	author = {Cumpston, Brian H. and Ananthavel, Sundaravel P. and Barlow,
	Stephen and Dyer, Daniel L. and Ehrlich, Jeffrey E. and Erskine, Lael L.
	and Heikal, Ahmed A. and Kuebler, Stephen M. and Lee, I.-Y. Sandy and
	{McCord-Maughon}, Dianne and Qin, Jinqui and R{\"o}ckel, Harald and Rumi,
	Mariacristina and Wu, Xiang-Li and Marder, Seth R. and Perry, Joseph W.},
	year = {1999},
	month = mar,
	volume = {398},
	pages = {51--54},
	publisher = {{Nature Publishing Group}},
	issn = {1476-4687},
	doi = {10.1038/17989},
	copyright = {1999 Macmillan Magazines Ltd.},
	journal = {Nature},
	language = {en},
	number = {6722}
}

@article{maruo1997Threedimensional,
	title = {Three-Dimensional Microfabrication with Two-Photon-Absorbed
	Photopolymerization},
	author = {Maruo, Shoji and Nakamura, Osamu and Kawata, Satoshi},
	year = {1997},
	month = jan,
	volume = {22},
	pages = {132--134},
	publisher = {{Optical Society of America}},
	issn = {1539-4794},
	doi = {10.1364/OL.22.000132},
	copyright = {\&\#169; 1997 Optical Society of America},
	journal = {Optics Letters},
	language = {EN},
	number = {2}
}

@article{duocastella2017Improving,
	title = {Improving the {{Spatial Resolution}} in {{Direct Laser Writing
	Lithography}} by {{Using}} a {{Reversible Cationic Photoinitiator}}},
	author = {Duocastella, Mart{\'i} and Vicidomini, Giuseppe and
	Korobchevskaya, Kseniya and Pydzi{\'n}ska, Katarzyna and Zi{\'o}{\l}ek,
	Marcin and Diaspro, Alberto and {de Miguel}, Gustavo},
	year = {2017},
	month = aug,
	volume = {121},
	pages = {16970--16977},
	publisher = {{American Chemical Society}},
	issn = {1932-7447},
	doi = {10.1021/acs.jpcc.7b03591},
	journal = {The Journal of Physical Chemistry C},
	number = {31}
}

@article{ecoffet1998Photopolymerization,
	title = {Photopolymerization by {{Evanescent Waves}}: {{A New Method}} to
	{{Obtain Nanoparts}}},
	shorttitle = {Photopolymerization by {{Evanescent Waves}}},
	author = {Ecoffet, Carole and Espanet, Anne and Lougnot, Daniel J.},
	year = {1998},
	volume = {10},
	pages = {411--414},
	issn = {1521-4095},
	doi = {10.1002/(SICI)1521-4095(199803)10:5<411::AID-ADMA411>3.0.CO;2-P},
	copyright = {\textcopyright{} 1998 WILEY-VCH Verlag GmbH, Weinheim, Fed.
	Rep. of Germany},
	journal = {Advanced Materials},
	number = {5}
}

@article{pierini2020Hybrid,
% 	title = {Hybrid Device for Quantum Nanophotonics},
% 	author = {Pierini, S. and D'Amato, M. and Joos, M. and Glorieux, Q. and
% 	Giacobino, E. and Lhuillier, E. and Couteau, C. and Bramati, A.},
% 	year = {2020},
% 	month = jan,
% 	archivePrefix = {arXiv},
% 	eprint = {2001.10480},
% 	eprinttype = {arxiv},
% 	journal = {arXiv:2001.10480 [physics, physics:quant-ph]},
% 	keywords = {Physics - Optics,Quantum Physics},
% 	primaryClass = {physics, physics:quant-ph}
% }

@article{pierini2020Highly,
	title = {Highly {{Photostable Perovskite Nanocubes}}: {{Toward Integrated
	Single Photon Sources Based}} on {{Tapered Nanofibers}}},
	shorttitle = {Highly {{Photostable Perovskite Nanocubes}}},
	author = {Pierini, Stefano and D'Amato, Marianna and Goyal, Mayank and
	Glorieux, Quentin and Giacobino, Elisabeth and Lhuillier, Emmanuel and
	Couteau, Christophe and Bramati, Alberto},
	year = {2020},
	month = aug,
	volume = {7},
	pages = {2265--2272},
	publisher = {{American Chemical Society}},
	doi = {10.1021/acsphotonics.0c00820},
	journal = {ACS Photonics},
	number = {8}
}

@article{yan2019Ultrastablea,
	title = {Ultrastable {{CsPbBr}} {\textsubscript{3}} {{Perovskite Quantum
	Dot}} and {{Their Enhanced Amplified Spontaneous Emission}} by {{Surface
	Ligand Modification}}},
	author = {Yan, Dongdong and Shi, Tongchao and Zang, Zhigang and Zhou,
	Tingwei and Liu, Zhengzheng and Zhang, Zeyu and Du, Juan and Leng, Yuxin
	and Tang, Xiaosheng},
	year = {2019},
	month = apr,
	pages = {1901173},
	issn = {1613-6810, 1613-6829},
	doi = {10.1002/smll.201901173},
	journal = {Small},
	language = {en}
}

@article{robledo2011spin,
	title={Spin dynamics in the optical cycle of single nitrogen-vacancy
	centres in diamond},
	author={Robledo, Lucio and Bernien, Hannes and Van Der Sar, Toeno and
	Hanson, Ronald},
	journal={New Journal of Physics},
	volume={13},
	number={2},
	pages={025013},
	year={2011},
	publisher={IOP Publishing}
}

@article{becker2016Ultrafast,
	title = {Ultrafast All-Optical Coherent Control of Single Silicon Vacancy
	Colour Centres in Diamond},
	author = {Becker, Jonas Nils and G{\"o}rlitz, Johannes and Arend, Carsten
	and Markham, Matthew and Becher, Christoph},
	year = {2016},
	month = dec,
	volume = {7},
	pages = {13512},
	issn = {2041-1723},
	doi = {10.1038/ncomms13512},
	journal = {Nature Communications},
	language = {en},
	number = {1}
}

@article{trupke2011Enhancing,
	title = {Enhancing Photon Collection from Quantum Emitters in Diamond},
	author = {Trupke, Michael and Munro, William J. and Nemoto, Kae and
	Schmiedmayer, Jorg},
	year = {2011},
	month = mar,
	pages = {33},
	issn = {1349-8614, 1349-8606},
	doi = {10.2201/NiiPi.2011.8.4},
	journal = {Progress in Informatics},
	language = {en},
	number = {8}
}

@article{becker2018Brighta,
  title = {Bright Triplet Excitons in Caesium Lead Halide Perovskites},
  author = {Becker, Michael A. and Vaxenburg, Roman and Nedelcu, Georgian and Sercel, Peter C. and Shabaev, Andrew and Mehl, Michael J. and Michopoulos, John G. and Lambrakos, Samuel G. and Bernstein, Noam and Lyons, John L. and St{\"o}ferle, Thilo and Mahrt, Rainer F. and Kovalenko, Maksym V. and Norris, David J. and Rain{\`o}, Gabriele and Efros, Alexander L.},
  year = {2018},
  month = jan,
  volume = {553},
  pages = {189--193},
  publisher = {{Nature Publishing Group}},
  issn = {1476-4687},
  doi = {10.1038/nature25147},
  copyright = {2018 Macmillan Publishers Limited, part of Springer Nature. All rights reserved.},
  journal = {Nature},
  language = {en},
  number = {7687}
}

@article{sercel2019Exciton,
  ids = {sercel2019Excitona},
  title = {Exciton {{Fine Structure}} in {{Perovskite Nanocrystals}}},
  author = {Sercel, Peter C. and Lyons, John L. and Wickramaratne, Darshana and Vaxenburg, Roman and Bernstein, Noam and Efros, Alexander L.},
  year = {2019},
  month = jun,
  volume = {19},
  pages = {4068--4077},
  publisher = {{American Chemical Society}},
  issn = {1530-6984},
  doi = {10.1021/acs.nanolett.9b01467},
  journal = {Nano Letters},
  number = {6}
}

@article{lethiec2014Measurement,
  title = {Measurement of Three-Dimensional Dipole Orientation of a Single Fluorescent Nanoemitter by Emission Polarization Analysis},
  author = {Lethiec, Clotilde and Laverdant, Julien and Vallon, Henri and Javaux, Cl{\'e}mentine and Dubertret, Beno{\^i}t and Frigerio, Jean-Marc and Schwob, Catherine and Coolen, Laurent and Ma{\^i}tre, Agn{\`e}s},
  year = {2014},
  month = may,
  volume = {4},
  pages = {021037},
  doi = {10.1103/PhysRevX.4.021037},
  journal = {Physical Review X},
  number = {2}
}

@article{pisanello2010Room,
  title = {Room Temperature-Dipolelike Single Photon Source with a Colloidal Dot-in-Rod},
  author = {Pisanello, Ferruccio and Martiradonna, Luigi and Lem{\'e}nager, Godefroy and Spinicelli, Piernicola and Fiore, Angela and Manna, Liberato and Hermier, Jean-Pierre and Cingolani, Roberto and Giacobino, Elisabeth and De Vittorio, Massimo and Bramati, Alberto},
  year = {2010},
  month = jan,
  volume = {96},
  pages = {033101},
  publisher = {{American Institute of Physics}},
  issn = {0003-6951},
  doi = {10.1063/1.3291849},
  journal = {Applied Physics Letters},
  number = {3}
}

@article{pisanello2010Dots,
  title = {Dots in Rods as Polarized Single Photon Sources},
  author = {Pisanello, F. and Martiradonna, L. and Spinicelli, P. and Fiore, A. and Hermier, J. P. and Manna, L. and Cingolani, R. and Giacobino, E. and De Vittorio, M. and Bramati, A.},
  year = {2010},
  month = jan,
  volume = {47},
  pages = {165--169},
  issn = {0749-6036},
  doi = {10.1016/j.spmi.2009.06.009},
  journal = {Superlattices and Microstructures},
  language = {en},
  number = {1},
  series = {Proceedings of the 9th {{International Conference}} on {{Physics}} of {{Light}}-{{Matter Coupling}} in {{Nanostructures}}, {{PLMCN}} 2009 ({{Lecce}} - {{Italy}})}
}

@article{thylen2014Integrated,
  title = {Integrated Photonics in the 21st Century},
  author = {Thyl{\'e}n, Lars and Wosinski, Lech},
  year = {2014},
  month = apr,
  volume = {2},
  pages = {75--81},
  publisher = {{Optical Society of America}},
  issn = {2327-9125},
  doi = {10.1364/PRJ.2.000075},
  copyright = {\&\#169; 2014 Chinese Laser Press},
  journal = {Photonics Research},
  language = {EN},
  number = {2}
}

@article{block2014Bloch,
  title = {Bloch Oscillations in Plasmonic Waveguide Arrays},
  author = {Block, A. and Etrich, C. and Limboeck, T. and Bleckmann, F. and Soergel, E. and Rockstuhl, C. and Linden, S.},
  year = {2014},
  month = may,
  volume = {5},
  pages = {3843},
  publisher = {{Nature Publishing Group}},
  issn = {2041-1723},
  doi = {10.1038/ncomms4843},
  journal = {Nature Communications},
  language = {en},
  number = {1}
}

@article{garcia-ruperez2010Labelfree,
  title = {Label-Free Antibody Detection Using Band Edge Fringes in {{SOI}} Planar Photonic Crystal Waveguides in the Slow-Light Regime},
  author = {{Garc{\'i}a-Rup{\'e}rez}, Jaime and Toccafondo, Veronica and Ba{\~n}uls, Mar{\'i}a Jos{\'e} and Castell{\'o}, Javier Garc{\'i}a and Griol, Amadeu and {Peransi-Llopis}, Sergio and Maquieira, {\'A}ngel},
  year = {2010},
  month = nov,
  volume = {18},
  pages = {24276--24286},
  publisher = {Optical Society of America},
  issn = {1094-4087},
  doi = {10.1364/OE.18.024276},
  copyright = {\&\#169; 2010 OSA},
  journal = {Optics Express},
  language = {EN},
  number = {23}
}

@misc{Qt,
	title = {About {{Qt}} - {{Qt Wiki}}},
	howpublished = {https://wiki.qt.io/About\_Qt\#What\_is\_Qt.3F},
	note = "[Online; accessed 13-Ott-2020]"
}

@misc{Qwt,
	title = {Qwt User's Guide: Qwt - Qt Widgets for Technical
	Applications},
	howpublished = {https://\-qwt.source\-forge.\-io\-/index.html},
	note = "[Online; accessed 13-Ott-2020]"
}

@article{akkerman2016Solution,
  title = {Solution {{Synthesis Approach}} to {{Colloidal Cesium Lead Halide Perovskite Nanoplatelets}} with {{Monolayer}}-{{Level Thickness Control}}},
  author = {Akkerman, Quinten A. and Motti, Silvia Genaro and Srimath Kandada, Ajay Ram and Mosconi, Edoardo and D'Innocenzo, Valerio and Bertoni, Giovanni and Marras, Sergio and Kamino, Brett A. and Miranda, Laura and De Angelis, Filippo and Petrozza, Annamaria and Prato, Mirko and Manna, Liberato},
  year = {2016},
  month = jan,
  volume = {138},
  pages = {1010--1016},
  publisher = {{American Chemical Society}},
  issn = {0002-7863},
  doi = {10.1021/jacs.5b12124},
  journal = {Journal of the American Chemical Society},
  number = {3}
}

@article{bekenstein2015Highly,
  title = {Highly {{Luminescent Colloidal Nanoplates}} of {{Perovskite Cesium Lead Halide}} and {{Their Oriented Assemblies}}},
  author = {Bekenstein, Yehonadav and Koscher, Brent A. and Eaton, Samuel W. and Yang, Peidong and Alivisatos, A. Paul},
  year = {2015},
  month = dec,
  volume = {137},
  pages = {16008--16011},
  publisher = {{American Chemical Society}},
  issn = {0002-7863},
  doi = {10.1021/jacs.5b11199},
  journal = {Journal of the American Chemical Society},
  number = {51}
}

@article{marseglia2018Bright,
  title = {Bright Nanowire Single Photon Source Based on {{SiV}} Centers in Diamond},
  author = {Marseglia, L. and Saha, K. and Ajoy, A. and Schr{\"o}der, T. and Englund, D. and Jelezko, F. and Walsworth, R. and Pacheco, J. L. and Perry, D. L. and Bielejec, E. S. and Cappellaro, P.},
  year = {2018},
  month = jan,
  volume = {26},
  pages = {80--89},
  issn = {1094-4087},
  doi = {10.1364/OE.26.000080},
  copyright = {\&\#169; 2018 Optical Society of America},
  journal = {Optics Express},
  language = {EN},
  number = {1}
}

@article{neu2012Photophysics,
  title = {Photophysics of Single Silicon Vacancy Centers in Diamond: Implications for Single Photon Emission},
  shorttitle = {Photophysics of Single Silicon Vacancy Centers in Diamond},
  author = {Neu, Elke and Agio, Mario and Becher, Christoph},
  year = {2012},
  month = aug,
  volume = {20},
  pages = {19956--19971},
  publisher = {{Optical Society of America}},
  issn = {1094-4087},
  doi = {10.1364/OE.20.019956},
  copyright = {\&\#169; 2012 OSA},
  journal = {Optics Express},
  language = {EN},
  number = {18}
}

@article{haussler2017Photoluminescence,
  title = {Photoluminescence Excitation Spectroscopy of {{SiV}} {\textsuperscript{-}} and {{GeV}} {\textsuperscript{-}} Color Center in Diamond},
  author = {H{\"a}u{\ss}ler, Stefan and Thiering, Gerg{\H o} and Dietrich, Andreas and Waasem, Niklas and Teraji, Tokuyuki and Isoya, Junichi and Iwasaki, Takayuki and Hatano, Mutsuko and Jelezko, Fedor and Gali, Adam and Kubanek, Alexander},
  year = {2017},
  month = jun,
  volume = {19},
  pages = {063036},
  issn = {1367-2630},
  doi = {10.1088/1367-2630/aa73e5},
  journal = {New Journal of Physics},
  language = {en},
  number = {6}
}

@article{boudou2013Fluorescent,
  title = {Fluorescent Nanodiamonds Derived from {{HPHT}} with a Size of Less than 10nm},
  author = {Boudou, Jean-Paul and Tisler, Julia and Reuter, Rolf and Thorel, Alain and Curmi, Patrick A. and Jelezko, Fedor and Wrachtrup, Joerg},
  year = {2013},
  month = aug,
  volume = {37},
  pages = {80--86},
  issn = {0925-9635},
  doi = {10.1016/j.diamond.2013.05.006},
  journal = {Diamond and Related Materials},
  language = {en}
}

@article{schirhagl2014NitrogenVacancy,
  title = {Nitrogen-{{Vacancy Centers}} in {{Diamond}}: {{Nanoscale Sensors}} for {{Physics}} and {{Biology}}},
  shorttitle = {Nitrogen-{{Vacancy Centers}} in {{Diamond}}},
  author = {Schirhagl, Romana and Chang, Kevin and Loretz, Michael and Degen, Christian L.},
  year = {2014},
  volume = {65},
  pages = {83--105},
  doi = {10.1146/annurev-physchem-040513-103659},
  journal = {Annual Review of Physical Chemistry},
  number = {1},
  pmid = {24274702}
}

@article{holzgrafe2019Error,
  title = {Error Corrected Spin-State Readout in a Nanodiamond},
  author = {Holzgrafe, Jeffrey and Beitner, Jan and Kara, Dhiren and Knowles, Helena S. and Atat{\"u}re, Mete},
  year = {2019},
  month = feb,
  volume = {5},
  pages = {1--6},
  publisher = {{Nature Publishing Group}},
  issn = {2056-6387},
  doi = {10.1038/s41534-019-0126-2},
  copyright = {2019 The Author(s)},
  journal = {npj Quantum Information},
  language = {en},
  number = {1}
}

@article{knowles2014Observing,
  title = {Observing Bulk Diamond Spin Coherence in High-Purity Nanodiamonds},
  author = {Knowles, Helena S. and Kara, Dhiren M. and Atat{\"u}re, Mete},
  year = {2014},
  month = jan,
  volume = {13},
  pages = {21--25},
  publisher = {{Nature Publishing Group}},
  issn = {1476-4660},
  doi = {10.1038/nmat3805},
  journal = {Nature Materials},
  language = {en},
  number = {1}
}

@article{schell2011Scanning,
  title = {A Scanning Probe-Based Pick-and-Place Procedure for Assembly of Integrated Quantum Optical Hybrid Devices},
  author = {Schell, Andreas W. and Kewes, G{\"u}nter and Schr{\"o}der, Tim and Wolters, Janik and Aichele, Thomas and Benson, Oliver},
  year = {2011},
  month = jul,
  volume = {82},
  pages = {073709},
  publisher = {{American Institute of Physics}},
  issn = {0034-6748},
  doi = {10.1063/1.3615629},
  journal = {Review of Scientific Instruments},
  number = {7}
}

@article{verberk2002Simple,
  title = {Simple Model for the Power-Law Blinking of Single Semiconductor Nanocrystals},
  author = {Verberk, Rogier and {van Oijen}, Antoine M. and Orrit, Michel},
  year = {2002},
  month = dec,
  volume = {66},
  pages = {233202},
  publisher = {{American Physical Society}},
  doi = {10.1103/PhysRevB.66.233202},
  journal = {Physical Review B},
  number = {23}
}

@article{verberk2003Photon,
  ids = {verberk2003Photona},
  title = {Photon Statistics in the Fluorescence of Single Molecules and Nanocrystals: {{Correlation}} Functions versus Distributions of on- and off-Times},
  shorttitle = {Photon Statistics in the Fluorescence of Single Molecules and Nanocrystals},
  author = {Verberk, Rogier and Orrit, Michel},
  year = {2003},
  month = jul,
  volume = {119},
  pages = {2214--2222},
  publisher = {{American Institute of Physics}},
  issn = {0021-9606},
  doi = {10.1063/1.1582848},
  journal = {The Journal of Chemical Physics},
  number = {4}
}

@article{kaiser2009Polarization,
  title = {Polarization Properties of Single Photons Emitted by Nitrogen-Vacancy Defect in Diamond at Low Temperature},
  author = {Kaiser, F. and Jacques, V. and Batalov, A. and Siyushev, P. and Jelezko, F. and Wrachtrup, J.},
  year = {2009},
  month = jun,
  archivePrefix = {arXiv},
  eprint = {0906.3426},
  eprinttype = {arxiv},
  journal = {arXiv:0906.3426 [quant-ph]},
  primaryClass = {quant-ph}
}

@article{dutta2016Coupling,
  title = {Coupling Light in Photonic Crystal Waveguides: {{A}} Review},
  shorttitle = {Coupling Light in Photonic Crystal Waveguides},
  author = {Dutta, Hemant Sankar and Goyal, Amit Kumar and Srivastava, Varun and Pal, Suchandan},
  year = {2016},
  month = jul,
  volume = {20},
  pages = {41--58},
  issn = {1569-4410},
  doi = {10.1016/j.photonics.2016.04.001},
  journal = {Photonics and Nanostructures - Fundamentals and Applications},
  language = {en}
}

@article{engelen2006Effect,
  title = {The Effect of Higher-Order Dispersion on Slow Light Propagation in Photonic Crystal Waveguides},
  author = {Engelen, R. J. P. and Sugimoto, Y. and Watanabe, Y. and Korterik, J. P. and Ikeda, N. and van Hulst, N. F. and Asakawa, K. and Kuipers, L.},
  year = {2006},
  month = feb,
  volume = {14},
  pages = {1658--1672},
  publisher = {{Optical Society of America}},
  issn = {1094-4087},
  doi = {10.1364/OE.14.001658},
  copyright = {\&\#169; 2006 Optical Society of America},
  journal = {Optics Express},
  language = {EN},
  number = {4}
}

@article{fang2015Nanoplasmonic,
  title = {Nanoplasmonic Waveguides: Towards Applications in Integrated Nanophotonic Circuits},
  shorttitle = {Nanoplasmonic Waveguides},
  author = {Fang, Yurui and Sun, Mengtao},
  year = {2015},
  month = jun,
  volume = {4},
  pages = {e294-e294},
  publisher = {{Nature Publishing Group}},
  issn = {2047-7538},
  doi = {10.1038/lsa.2015.67},
  copyright = {2015 The Author(s)},
  journal = {Light: Science \& Applications},
  language = {en},
  number = {6}
}

@article{gloge1971Weakly,
  title = {Weakly {{Guiding Fibers}}},
  author = {Gloge, D.},
  year = {1971},
  month = oct,
  volume = {10},
  pages = {2252--2258},
  publisher = {{Optical Society of America}},
  issn = {2155-3165},
  doi = {10.1364/AO.10.002252},
  copyright = {\&\#169; 1971 Optical Society of America},
  journal = {Applied Optics},
  language = {EN},
  number = {10}
}

@book{okamoto2006Fundamentals,
  title = {Fundamentals of Optical Waveguides},
  author = {Okamoto, Katsunari},
  year = {2006},
  edition = {2nd ed},
  publisher = {{Elsevier}},
  address = {{Amsterdam ; Boston}},
  isbn = {978-0-12-525096-2},
  language = {en},
  lccn = {TA1800 .O37 2006}
}

@article{petersen2014Chirala,
  title = {Chiral Nanophotonic Waveguide Interface Based on Spin-Orbit Interaction of Light},
  author = {Petersen, Jan and Volz, J{\"u}rgen and Rauschenbeutel, Arno},
  year = {2014},
  month = oct,
  volume = {346},
  pages = {67--71},
  issn = {0036-8075, 1095-9203},
  doi = {10.1126/science.1257671},
  journal = {Science},
  language = {en},
  number = {6205}
}

@article{mitsch2014Quantuma,
  title = {Quantum State-Controlled Directional Spontaneous Emission of Photons into a Nanophotonic Waveguide},
  author = {Mitsch, R. and Sayrin, C. and Albrecht, B. and Schneeweiss, P. and Rauschenbeutel, A.},
  year = {2014},
  month = dec,
  volume = {5},
  pages = {5713},
  publisher = {{Nature Publishing Group}},
  issn = {2041-1723},
  doi = {10.1038/ncomms6713},
  copyright = {2014 The Author(s)},
  journal = {Nature Communications},
  language = {en},
  number = {1}
}

@article{love1991Tapered,
  title = {Tapered Single-Mode Fibres and Devices. {{Part}} 1: {{Adiabaticity}} Criteria},
  shorttitle = {Tapered Single-Mode Fibres and Devices. {{Part}} 1},
  author = {Love, J.D. and Henry, W.M. and Stewart, W.J. and Black, R.J. and Lacroix, S. and Gonthier, F.},
  year = {1991},
  volume = {138},
  pages = {343},
  issn = {02673932},
  doi = {10.1049/ip-j.1991.0060},
  journal = {IEE Proceedings J Optoelectronics},
  language = {en},
  number = {5}
}

@article{fujiwara2015Ultrathin,
  title = {Ultrathin Fiber-Taper Coupling with Nitrogen Vacancy Centers in Nanodiamonds at Cryogenic Temperatures},
  author = {Fujiwara, Masazumi and Zhao, Hong-Quan and Noda, Tetsuya and Ikeda, Kazuhiro and Sumiya, Hitoshi and Takeuchi, Shigeki},
  year = {2015},
  month = dec,
  volume = {40},
  pages = {5702--5705},
  issn = {1539-4794},
  doi = {10.1364/OL.40.005702},
  copyright = {\&\#169; 2015 Optical Society of America},
  journal = {Optics Letters},
  language = {EN},
  number = {24}
}

@article{liebermeister2014Tapered,
  title = {Tapered Fiber Coupling of Single Photons Emitted by a Deterministically Positioned Single Nitrogen Vacancy Center},
  author = {Liebermeister, Lars and Petersen, Fabian and v. M{\"u}nchow, Asmus and Burchardt, Daniel and Hermelbracht, Juliane and Tashima, Toshiyuki and Schell, Andreas W. and Benson, Oliver and Meinhardt, Thomas and Krueger, Anke and Stiebeiner, Ariane and Rauschenbeutel, Arno and Weinfurter, Harald and Weber, Markus},
  year = {2014},
  month = jan,
  volume = {104},
  pages = {031101},
  issn = {0003-6951, 1077-3118},
  doi = {10.1063/1.4862207},
  journal = {Applied Physics Letters},
  language = {en},
  number = {3}
}

@article{schroder2012Nanodiamondtapered,
  title = {A Nanodiamond-Tapered Fiber System with High Single-Mode Coupling Efficiency},
  author = {Schr{\"o}der, Tim and Fujiwara, Masazumi and Noda, Tetsuya and Zhao, Hong-Quan and Benson, Oliver and Takeuchi, Shigeki},
  year = {2012},
  month = may,
  volume = {20},
  pages = {10490--10497},
  issn = {1094-4087},
  doi = {10.1364/OE.20.010490},
  copyright = {\&\#169; 2012 OSA},
  journal = {Optics Express},
  language = {EN},
  number = {10}
}

@article{nahra2020Single,
  title = {Single Germanium Vacancy Centres in Nanodiamonds with Bulk-like Spectral Stability},
  author = {Nahra, Mackrine and Alshamaa, Daniel and Deturche, Regis and Davydov, Valery and Kulikova, Ludmila and Agafonov, Viatcheslav and Couteau, Christophe},
  year = {2020},
  month = oct,
  archivePrefix = {arXiv},
  eprint = {2010.15475},
  eprinttype = {arxiv},
  journal = {arXiv:2010.15475 [quant-ph]},
  language = {en},
  primaryClass = {quant-ph}
}

@article{ge2020Hybrid,
  title = {Hybrid Plasmonic Nano-Emitters with Controlled Single Quantum Emitter Positioning on the Local Excitation Field},
  author = {Ge, Dandan and Marguet, Sylvie and Issa, Ali and Jradi, Safi and Nguyen, Tien Hoa and Nahra, Mackrine and B{\'e}al, J{\'e}remie and Deturche, R{\'e}gis and Chen, Hongshi and Blaize, Sylvain and Plain, J{\'e}r{\^o}me and Fiorini, C{\'e}line and Douillard, Ludovic and Soppera, Olivier and Dinh, Xuan Quyen and Dang, Cuong and Yang, Xuyong and Xu, Tao and Wei, Bin and Sun, Xiao Wei and Couteau, Christophe and Bachelot, Renaud},
  year = {2020},
  month = jul,
  volume = {11},
  pages = {3414},
  publisher = {{Nature Publishing Group}},
  issn = {2041-1723},
  doi = {10.1038/s41467-020-17248-8},
  copyright = {2020 The Author(s)},
  journal = {Nature Communications},
  language = {en},
  number = {1}
}

@article{nayak2008Single,
  title = {Single Atoms on an Optical Nanofibre},
  author = {Nayak, K. P. and Hakuta, K.},
  year = {2008},
  month = may,
  volume = {10},
  pages = {053003},
  publisher = {{IOP Publishing}},
  issn = {1367-2630},
  doi = {10.1088/1367-2630/10/5/053003},
  journal = {New Journal of Physics},
  language = {en},
  number = {5}
}

@article{nayak2009Antibunching,
  title = {Antibunching and Bunching of Photons in Resonance Fluorescence from a Few Atoms into Guided Modes of an Optical Nanofiber},
  author = {Nayak, K. P. and Le Kien, Fam and Morinaga, M. and Hakuta, K.},
  year = {2009},
  month = feb,
  volume = {79},
  pages = {021801},
  issn = {1050-2947, 1094-1622},
  doi = {10.1103/PhysRevA.79.021801},
  journal = {Physical Review A},
  language = {en},
  number = {2}
}

@article{park2015perovskite,
	title={Perovskite solar cells: an emerging photovoltaic technology},
	author={Park, Nam-Gyu},
	journal={Materials today},
	volume={18},
	number={2},
	pages={65--72},
	year={2015},
	publisher={Elsevier}
}

@article{rainoSingleCesiumLead2016,
	title = {Single {{Cesium Lead Halide Perovskite Nanocrystals}} at {{Low
	Temperature}}: {{Fast Single}}-{{Photon Emission}}, {{Reduced Blinking}},
	and {{Exciton Fine Structure}}},
	shorttitle = {Single {{Cesium Lead Halide Perovskite Nanocrystals}} at
	{{Low Temperature}}},
	author = {Rain{\`o}, Gabriele and Nedelcu, Georgian and Protesescu,
	Loredana and Bodnarchuk, Maryna I. and Kovalenko, Maksym V. and Mahrt,
	Rainer F. and St{\"o}ferle, Thilo},
	year = {2016},
	month = feb,
	volume = {10},
	pages = {2485--2490},
	issn = {1936-0851, 1936-086X},
	doi = {10.1021/acsnano.5b07328},
	journal = {ACS Nano},
	language = {en},
	number = {2}
}

@article{huo2020Optical,
	title = {Optical {{Spectroscopy}} of {{Single Colloidal \ch{CsPbBr3}
	Perovskite Nanoplatelets}}},
	author = {Huo, Caixia and Fong, Chee Fai and Amara, Mohamed-Raouf and
	Huang, Yuqing and Chen, Bo and Zhang, Hua and Guo, Lingjun and Li, Hejun
	and Huang, Wei and Diederichs, Carole and Xiong, Qihua},
	year = {2020},
	month = mar,
	publisher = {{American Chemical Society}},
	issn = {1530-6984},
	doi = {10.1021/acs.nanolett.0c00611},
	journal = {Nano Letters}
}

@article{chenInfluencePMMAAllInorganic2019,
	title = {Influence of {PMMA} on {All}-{Inorganic} {Halide} {Perovskite}
	\ch{CsPbBr3} {Quantum} {Dots} {Combined} with {Polymer} {Matrix}},
	volume = {12},
	copyright = {http://creativecommons.org/licenses/by/3.0/},
	url = {https://www.mdpi.com/1996-1944/12/6/985},
	doi = {10.3390/ma12060985},
	language = {en},
	number = {6},
	urldate = {2019-09-23},
	journal = {Materials},
	author = {Chen, Lung-Chien and Tien, Ching-Ho and Tseng, Zong-Liang and
	Dong, Yu-Shen and Yang, Shengyi},
	month = jan,
	year = {2019},
	pages = {985}
}

@article{panAirStableSurfacePassivatedPerovskite2015,
	title = {Air-{{Stable Surface}}-{{Passivated Perovskite Quantum Dots}} for
	{{Ultra}}-{{Robust}}, {{Single}}- and {{Two}}-{{Photon}}-{{Induced
	Amplified Spontaneous Emission}}},
	author = {Pan, Jun and Sarmah, Smritakshi P. and Murali, Banavoth and
	Dursun, Ibrahim and Peng, Wei and Parida, Manas R. and Liu, Jiakai and
	Sinatra, Lutfan and Alyami, Noktan and Zhao, Chao and Alarousu, Erkki and
	Ng, Tien Khee and Ooi, Boon S. and Bakr, Osman M. and Mohammed, Omar F.},
	year = {2015},
	month = dec,
	volume = {6},
	pages = {5027--5033},
	issn = {1948-7185},
	doi = {10.1021/acs.jpclett.5b02460},
	journal = {The Journal of Physical Chemistry Letters},
	language = {en},
	number = {24}
}

@article{le2005spontaneous,
	title={Spontaneous emission of a cesium atom near a nanofiber: Efficient
	coupling of light to guided modes},
	author={Le Kien, Fam and Gupta, S Dutta and Balykin, VI and Hakuta, K},
	journal={Physical Review A},
	volume={72},
	number={3},
	pages={032509},
	year={2005},
	publisher={APS}
}

@article{klimov2004spontaneous,
	title={Spontaneous emission rate of an excited atom placed near a
	nanofiber},
	author={Klimov, VV and Ducloy, Martial},
	journal={Physical Review A},
	volume={69},
	number={1},
	pages={013812},
	year={2004},
	publisher={APS}
}

@article{fujiwara2011highly,
	title={Highly efficient coupling of photons from nanoemitters into
	single-mode optical fibers},
	author={Fujiwara, Masazumi and Toubaru, Kiyota and Noda, Tetsuya and Zhao,
	Hong-Quan and Takeuchi, Shigeki},
	journal={Nano letters},
	volume={11},
	number={10},
	pages={4362--4365},
	year={2011},
	publisher={ACS Publications}
}

@article{joos2018polarization,
	title={Polarization control of linear dipole radiation using an optical
	nanofiber},
	author={Joos, Maxime and Ding, Chengjie and Loo, Vivien and Blanquer,
	Guillaume and Giacobino, Elisabeth and Bramati, Alberto and Krachmalnicoff,
	Valentina and Glorieux, Quentin},
	journal={Physical Review Applied},
	volume={9},
	number={6},
	pages={064035},
	year={2018},
	publisher={American Physical Society}
}

@article{vorobyovCouplingSingleNV2016,
	title = {Coupling of Single {{NV}} Center to Adiabatically Tapered
	Opticalsingle Mode Fiber},
	author = {Vorobyov, Vadim V. and Soshenko, Vladimir V. and Bolshedvorskii,
	Stepan V. and Javadzade, Javid and Lebedev, Nikolay and Smolyaninov, Andrey
	N. and Sorokin, Vadim N. and Akimov, Alexey V.},
	year = {2016},
	month = dec,
	volume = {70},
	pages = {269},
	issn = {1434-6079},
	doi = {10.1140/epjd/e2016-70099-3},
	journal = {The European Physical Journal D},
	language = {en},
	number = {12}
}

@article{yalla2012Fluorescence,
  title = {Fluorescence Photon Measurements from Single Quantum Dots on an Optical Nanofiber},
  author = {Yalla, Ramachandrarao and Nayak, K. P. and Hakuta, K.},
  year = {2012},
  month = jan,
  volume = {20},
  pages = {2932--2941},
  publisher = {{Optical Society of America}},
  issn = {1094-4087},
  doi = {10.1364/OE.20.002932},
  copyright = {\&\#169; 2012 OSA},
  journal = {Optics Express},
  language = {EN},
  number = {3}
}
\end{document}